\documentclass[12pt]{article}
\usepackage{epsfig}
\usepackage{amsmath}
\usepackage{hhline}
\usepackage{amssymb}
\usepackage{times}
\usepackage{cite}
\usepackage{rotating}
\usepackage{multirow}
\usepackage{dcolumn}
\newlength{\dinwidth}
\newlength{\dinmargin}
\begin{document}
\newcommand{\ssss}{Schildknecht, Schuler and Surrow}
\newcommand{\mrtt}{Martin, Ryskin and Teubner}
\newcommand{\rcc}{Royen and Cudell}
\newcommand{\ikk}{Ivanov and Kirschner}
\newcommand{\mphi}{\mbox{$m_{\phi}$}}        %  for use in B_W
\newcommand{\Gphi}{\mbox{$\Gamma_{\phi}$}}   %  for use in B_W
\newcommand{\Gm}{\mbox{$\Gamma(m)$}}   %  for use in B_W
\newcommand{\GM}{\mbox{$\Gamma(M)$}}   %  for use in B_W
\newcommand{\cosths}{\mbox{$\cos \theta$}}
\newcommand{\cosdelta}{\mbox{$\cos \delta$}}
\def\mbig#1{\mbox{\rule[-2. mm]{0 mm}{6 mm}#1}}

%%%%%%%%%%%%%%%%%%%%%%%%%%%%%
\newcommand{\invfb}{\ensuremath{\mathrm{fb^{-1}}}}
\newcommand{\invpb}{\ensuremath{\mathrm{pb^{-1}}}}
\newcommand{\invnb}{\ensuremath{\mathrm{nb^{-1}}}}
\newcommand{\abt} {\mbox{$|t|$}}
\newcommand{\rhoprim}{\mbox{$\rho^\prime$}}
\newcommand{\tprim}{\mbox{$t^\prime$}}
\newcommand{\rhop}{\mbox{$\rho^\prime$}}
\newcommand{\zet}{\mbox{$\zeta$}}
\newcommand{\rfivecomb}{\mbox{$r^5_{00} + 2 r^5_{11}$}}
\newcommand{\ronecomb}{\mbox{$r^1_{00} + 2 r^1_{11}$}}
\newcommand{\gstarVM} {\mbox{$\gamma^* +  p \rightarrow V + Y$}}
\newcommand{\gstarVMel} {\mbox{$\gamma^* \ \!p \rightarrow V\ \!p$}}
\newcommand{\gsrel} {\mbox{$\gamma^* \ \!p \rightarrow \rho \ \!p$}}
\newcommand{\gsrpd} {\mbox{$\gamma^* \ \!p \rightarrow \rho \ \!Y$}}
\newcommand{\gspel} {\mbox{$\gamma^* \ \!p \rightarrow \phi \ \!p$}}
\newcommand{\gsppd} {\mbox{$\gamma^* \ \!p \rightarrow \phi \ \!Y$}}
\newcommand{\gstarp} {\mbox{$\gamma^*\ \!p$}}
\newcommand{\mv} {\mbox{$M_V$}}
\newcommand{\mvsq} {\mbox{$M_V^2$}}
\newcommand{\msq} {\mbox{$M_V^2$}}
\newcommand{\qsqplmsq} {\mbox{($Q^2 \!+ \!M_V^2$})}
\newcommand{\qqsqplmsq} {\mbox{$Q^2 \!+ \!M_V^2$}}
\newcommand{\alprim}{\mbox{$\alpha^\prime$}}
\newcommand{\alphaz}{\mbox{$\alpha(0)$}}
\newcommand{\alpomz}{\mbox{$\alpha_{\PO}(0)$}}
\newcommand{\hence}{\mbox{$=>$}}
\newcommand{\vm}{\mbox{$V\!M$}}
\newcommand{\sur}{\mbox{\ \! / \ \!}}
\newcommand{\tzz} {\mbox{$T_{00}$}}
\newcommand{\tuu} {\mbox{$T_{11}$}}
\newcommand{\tzu} {\mbox{$T_{01}$}}
\newcommand{\tuz} {\mbox{$T_{10}$}}
\newcommand{\tmuu} {\mbox{$T_{-11}$}}
\newcommand{\tumu} {\mbox{$T_{1-1}$}}
\newcommand{\ralpha} {\mbox{$\tuu / \tzz$}}
\newcommand{\rbeta} {\mbox{$\tzu / \tzz$}}
\newcommand{\rdelta} {\mbox{$\tuz / \tzz$}}
\newcommand{\reta} {\mbox{$\tmuu / \tzz$}}
\newcommand{\abstzz} {\mbox{$|T_{00}|$}}
\newcommand{\abstuu} {\mbox{$|T_{11}|$}}
\newcommand{\abstzu} {\mbox{$|T_{01}|$}}
\newcommand{\abstuz} {\mbox{$|T_{10}|$}}
\newcommand{\abstmuu} {\mbox{$|T_{-11}|$}}
\newcommand{\rralpha} {\mbox{\abstuu / \abstzz}}
\newcommand{\rrbeta} {\mbox{\abstzu / \abstzz}}
\newcommand{\rrdelta} {\mbox{\tuz / \tzz}}
\newcommand{\rreta} {\mbox{\abstmuu / \abstzz}}
\newcommand{\averm} {\mbox{$\av {M}$}}
\newcommand{\rapproch} {\mbox{$R_{SCHC+T_{01}}$}}
\newcommand{\chisq} {\mbox{$\chi^2 / {\rm d.o.f.}$}}

\newcommand{\scaleqsqplmsq} {\mbox{$\qsqplmsq /4$}}

\newcolumntype{d}{D{.}{.}{-1}}

%%%%%%%%%%%%%%%%%%%%%%%%%%%%%

% \include{abb}
%%%%%%%%%%%%%%%%%%%%%%%%%%%%%%%%%%%%%%%%%%%%%%%%%%%%%%%%%%%%%%%%%%%%%%
%                                                                    %
%                                                                    %
%           DEFINITIONS OF NEW COMMANDS IN LATEX                     %
%                                                                    %
%                                                                    %
%%%%%%%%%%%%%%%%%%%%%%%%%%%%%%%%%%%%%%%%%%%%%%%%%%%%%%%%%%%%%%%%%%%%%%
%
%=====================================================================
% kinematics
%
\newcommand{\s}{\mbox{$s$}}
\newcommand{\ttra}{\mbox{$t$}}
\newcommand{\modt}{\mbox{$|t|$}}
\newcommand{\eminpz}{\mbox{$E-p_z$}}
\newcommand{\eminpzs}{\mbox{$\Sigma(E-p_z)$}}
\newcommand{\rap}{\ensuremath{\eta^*} }
\newcommand{\W}{\mbox{$W$}}
\newcommand{\w}{\mbox{$W$}}
\newcommand{\Q}{\mbox{$Q$}}
\newcommand{\q}{\mbox{$Q$}}
\newcommand{\xB}{\mbox{$x$}}  % Bjorken x
\newcommand{\xF}{\mbox{$x_F$}}  % Feynman x
\newcommand{\xg}{\mbox{$x_g$}}  % x_g
\newcommand{\xbj}{x}
\newcommand{\xpom}{x_{\PO}}
\newcommand{\y}{\mbox{$y~$}}
\newcommand{\Qsq}{\mbox{$Q^2$}}
\newcommand{\qsq}{\mbox{$Q^2$}}
\newcommand{\kjet}{\mbox{$k_{T\rm{jet}}$}}
\newcommand{\xjet}{\mbox{$x_{\rm{jet}}$}}
\newcommand{\Ejet}{\mbox{$E_{\rm{jet}}$}}
\newcommand{\thjet}{\mbox{$\theta_{\rm{jet}}$}}
\newcommand{\pjet}{\mbox{$p_{T\rm{jet}}$}}
\newcommand{\et}{\mbox{$E_T~$}}
\newcommand{\kt}{\mbox{$k_T~$}}
\newcommand{\ptrans}{\mbox{$p_T~$}}
\newcommand{\pth}{\mbox{$p_T^h~$}}
\newcommand{\pte}{\mbox{$p_T^e~$}}
\newcommand{\ptsq}{\mbox{$p_T^{\star 2}~$}}
\newcommand{\as}{\mbox{$\alpha_s~$}}
\newcommand{\ycut}{\mbox{$y_{\rm cut}~$}}
\newcommand{\gx}{\mbox{$g(x_g,Q^2)$~}}
\newcommand{\xpart}{\mbox{$x_{\rm part~}$}}
\newcommand{\mrsdm}{\mbox{${\rm MRSD}^-~$}}
\newcommand{\mrsdmp}{\mbox{${\rm MRSD}^{-'}~$}}
\newcommand{\mrsdn}{\mbox{${\rm MRSD}^0~$}}
\newcommand{\lambdams}{\mbox{$\Lambda_{\rm \bar{MS}}~$}}
%
%=====================================================================
% section efficace abbreviations
%
\newcommand{\gp}{\ensuremath{\gamma}p }
\newcommand{\gammasp}{\ensuremath{\gamma}*p }
\newcommand{\gammap}{\ensuremath{\gamma}p }
\newcommand{\gsp}{\ensuremath{\gamma^*}p }
\newcommand{\dsiget}{\ensuremath{{\rm d}\sigma_{ep}/{\rm d}E_t^*} }
\newcommand{\dsigrap}{\ensuremath{{\rm d}\sigma_{ep}/{\rm d}\eta^*} }
% ep 
\newcommand{\epem}{\mbox{$e^+e^-$}}
\newcommand{\ep}{\mbox{$ep~$}}
\newcommand{\epl}{\mbox{$e^{+}$}}
\newcommand{\emi}{\mbox{$e^{-}$}}
\newcommand{\epm}{\mbox{$e^{\pm}$}}
\newcommand{\se}{section efficace}
\newcommand{\ses}{sections efficaces}
%
%=====================================================================
% elastic VM abbreviations 
%
% VM
\newcommand{\phib}{\mbox{$\varphi$}}
\newcommand{\rh}{\mbox{$\rho$}}
\newcommand{\rhz}{\mbox{$\rh^0$}}
\newcommand{\ph}{\mbox{$\phi$}}
\newcommand{\om}{\mbox{$\omega$}}
\newcommand{\jpsi}{\mbox{$J/\psi$}}
\newcommand{\pipi}{\mbox{$\pi^+\pi^-$}}
\newcommand{\pip}{\mbox{$\pi^+$}}
\newcommand{\pim}{\mbox{$\pi^-$}}
\newcommand{\kk}{\mbox{K^+K^-$}}
% b parameter
\newcommand{\bsl}{\mbox{$b$}}
\newcommand{\alp}{\mbox{$\alpha^\prime$}}
\newcommand{\alpom}{\mbox{$\alpha_{\PO}$}}
\newcommand{\alpomp}{\mbox{$\alpha_{\PO}^\prime$}}
% polarisation
\newcommand{\rzzzz}{\mbox{$r_{00}^{04}$}}
\newcommand{\rzqzz}{\mbox{$r_{00}^{04}$}}
\newcommand{\rzquz}{\mbox{$r_{10}^{04}$}}
\newcommand{\rzqumu}{\mbox{$r_{1-1}^{04}$}}
\newcommand{\ruuu}{\mbox{$r_{11}^{1}$}}
\newcommand{\ruzz}{\mbox{$r_{00}^{1}$}}
\newcommand{\ruuz}{\mbox{$r_{10}^{1}$}}
\newcommand{\ruumu}{\mbox{$r_{1-1}^{1}$}}
\newcommand{\rduz}{\mbox{$r_{10}^{2}$}}
\newcommand{\rdumu}{\mbox{$r_{1-1}^{2}$}}
\newcommand{\rcuu}{\mbox{$r_{11}^{5}$}}
\newcommand{\rczz}{\mbox{$r_{00}^{5}$}}
\newcommand{\rcuz}{\mbox{$r_{10}^{5}$}}
\newcommand{\rcumu}{\mbox{$r_{1-1}^{5}$}}
\newcommand{\rsuz}{\mbox{$r_{10}^{6}$}}
\newcommand{\rsumu}{\mbox{$r_{1-1}^{6}$}}
\newcommand{\rzqik}{\mbox{$r_{ik}^{04}$}}
\newcommand{\rhzik}{\mbox{$\rh_{ik}^{0}$}}
\newcommand{\rhqik}{\mbox{$\rh_{ik}^{4}$}}
\newcommand{\rhaik}{\mbox{$\rh_{ik}^{\alpha}$}}
\newcommand{\rhzzz}{\mbox{$\rh_{00}^{0}$}}
\newcommand{\rhqzz}{\mbox{$\rh_{00}^{4}$}}
\newcommand{\raik}{\mbox{$r_{ik}^{\alpha}$}}
\newcommand{\razz}{\mbox{$r_{00}^{\alpha}$}}
\newcommand{\rauz}{\mbox{$r_{10}^{\alpha}$}}
\newcommand{\raumu}{\mbox{$r_{1-1}^{\alpha}$}}

\newcommand{\R}{\mbox{$R$}}
\newcommand{\rzero}{\mbox{$r_{00}^{04}$}}
\newcommand{\rone}{\mbox{$r_{1-1}^{1}$}}
\newcommand{\costh}{\mbox{$\cos\theta$}}
\newcommand{\cosp}{\mbox{$\cos\psi$}}
\newcommand{\costop}{\mbox{$\cos(2\psi)$}}
\newcommand{\cosd}{\mbox{$\cos\delta$}}
\newcommand{\cossqp}{\mbox{$\cos^2\psi$}}
\newcommand{\cossqt}{\mbox{$\cos^2\theta^*$}}
\newcommand{\sint}{\mbox{$\sin\theta^*$}}
\newcommand{\sintot}{\mbox{$\sin(2\theta^*)$}}
\newcommand{\sinsqt}{\mbox{$\sin^2\theta^*$}}
\newcommand{\costhst}{\mbox{$\cos\theta^*$}}
\newcommand{\vep}{\mbox{$V p$}}
% mass
\newcommand{\mpipi}{\mbox{$m_{\pi^+\pi^-}$}}
\newcommand{\mkk}{\mbox{$m_{KK}$}}
\newcommand{\mkaka}{\mbox{$m_{K^+K^-}$}}
\newcommand{\mpp}{\mbox{$m_{\pi\pi}$}}       %  for use in B_W
\newcommand{\mppsq}{\mbox{$m_{\pi\pi}^2$}}   %  for use in B_W
\newcommand{\mpi}{\mbox{$m_{\pi}$}}          %  for use in B_W
\newcommand{\mrho}{\mbox{$m_{\rho}$}}        %  for use in B_W
\newcommand{\mrhosq}{\mbox{$m_{\rho}^2$}}    %  for use in B_W
% width
\newcommand{\Gmpp}{\mbox{$\Gamma (\mpp)$}}   %  for use in B_W
\newcommand{\Gmppsq}{\mbox{$\Gamma^2(\mpp)$}}%  for use in B_W
\newcommand{\Grho}{\mbox{$\Gamma_{\rho}$}}   %  for use in B_W
\newcommand{\grho}{\mbox{$\Gamma_{\rho}$}}   %  for use in B_W
\newcommand{\Grhosq}{\mbox{$\Gamma_{\rho}^2$}}   %  for use in B_W
%
%=====================================================================
% units
%
\newcommand{\cm}{\mbox{\rm cm}}
\newcommand{\GeV}{\mbox{\rm GeV}}
\newcommand{\gev}{\mbox{\rm GeV}}
\newcommand{\GeVx}{\rm GeV}
\newcommand{\gevx}{\rm GeV}
\newcommand{\GeVc}{\rm GeV}
\newcommand{\gevc}{\rm GeV}
\newcommand{\MeVc}{\rm MeV}
\newcommand{\mevc}{\rm MeV}
\newcommand{\MeV}{\mbox{\rm MeV}}
\newcommand{\mev}{\mbox{\rm MeV}}
\newcommand{\MeVx}{\mbox{\rm MeV}}
\newcommand{\mevx}{\mbox{\rm MeV}}
\newcommand{\GeVsq}{\mbox{${\rm GeV}^2$}}
\newcommand{\gevsq}{\mbox{${\rm GeV}^2$}}
\newcommand{\gevsqc}{\mbox{${\rm GeV^2}$}}
\newcommand{\gevcsq}{\mbox{${\rm GeV}$}}
\newcommand{\mevcsq}{\mbox{${\rm MeV}$}}
\newcommand{\GeVsqm}{\mbox{${\rm GeV}^{-2}$}}
\newcommand{\gevsqm}{\mbox{${\rm GeV}^{-2}$}}
\newcommand{\nb}{\mbox{${\rm nb}$}}
\newcommand{\nbinv}{\mbox{${\rm nb^{-1}}$}}
\newcommand{\pbinv}{\mbox{${\rm pb^{-1}}$}}
\newcommand{\mm}{\mbox{$\cdot 10^{-2}$}}
\newcommand{\mmm}{\mbox{$\cdot 10^{-3}$}}
\newcommand{\mmmm}{\mbox{$\cdot 10^{-4}$}}
\newcommand{\degr}{\mbox{$^{\circ}$}}
%
%=====================================================================
% F2
%
\newcommand{\F}{$ F_{2}(x,Q^2)\,$}  
\newcommand{\Fc}{$ F_{2}\,$}    
\newcommand{\XP}{x_{{I\!\!P}/{p}}}       
\newcommand{\TOSS}{x_{{i}/{\PO}}}        
\newcommand{\un}[1]{\mbox{\rm #1}} 
\newcommand{\LO}{Leading Order}
\newcommand{\NLO}{Next to Leading Order}
\newcommand{\ft}{$ F_{2}\,$}
%
%=====================================================================
%  latex command abbreviations 
%
\newcommand{\mc}{\multicolumn}
\newcommand{\bce}{\begin{center}}
\newcommand{\ece}{\end{center}}
\newcommand{\beq}{\begin{equation}}
\newcommand{\eeq}{\end{equation}}
\newcommand{\bea}{\begin{eqnarray}}
\newcommand{\eea}{\end{eqnarray}}
%
%=====================================================================
% inequation symbols
%
%less than or approx. symbol
\def\lsim{\mathrel{\rlap{\lower4pt\hbox{\hskip1pt$\sim$}}
    \raise1pt\hbox{$<$}}}         
%greater than or approx. symbol
\def\gsim{\mathrel{\rlap{\lower4pt\hbox{\hskip1pt$\sim$}}
    \raise1pt\hbox{$>$}}}         
%
%=====================================================================
% diffrative abbreviations : F2d3
%
\newcommand{\pom}{{I\!\!P}}
\newcommand{\PO}{I\!\!P}
\newcommand{\slowpi}{\pi_{\mathit{slow}}}
\newcommand{\fiidiii}{F_2^{D(3)}}
\newcommand{\fiidiiiarg}{\fiidiii\,(\beta,\,Q^2,\,x)}
\newcommand{\n}{1.19\pm 0.06 (stat.) \pm0.07 (syst.)}
\newcommand{\nz}{1.30\pm 0.08 (stat.)^{+0.08}_{-0.14} (syst.)}
\newcommand{\fiidiiiful}{F_2^{D(4)}\,(\beta,\,Q^2,\,x,\,t)}
\newcommand{\fiipom}{\tilde F_2^D}
\newcommand{\ALPHA}{1.10\pm0.03 (stat.) \pm0.04 (syst.)}
\newcommand{\ALPHAZ}{1.15\pm0.04 (stat.)^{+0.04}_{-0.07} (syst.)}
\newcommand{\fiipomarg}{\fiipom\,(\beta,\,Q^2)}
\newcommand{\pomflux}{f_{\pom / p}}
\newcommand{\nxpom}{1.19\pm 0.06 (stat.) \pm0.07 (syst.)}
\newcommand {\gapprox}
   {\raisebox{-0.7ex}{$\stackrel {\textstyle>}{\sim}$}}
\newcommand {\lapprox}
   {\raisebox{-0.7ex}{$\stackrel {\textstyle<}{\sim}$}}
\newcommand{\pomfluxarg}{f_{\pom / p}\,(x_\pom)}
\newcommand{\dsf}{\mbox{$F_2^{D(3)}$}}
\newcommand{\dsfva}{\mbox{$F_2^{D(3)}(\beta,Q^2,x_{I\!\!P})$}}
\newcommand{\dsfvb}{\mbox{$F_2^{D(3)}(\beta,Q^2,x)$}}
\newcommand{\dsfpom}{$F_2^{I\!\!P}$}
\newcommand{\gap}{\stackrel{>}{\sim}}
\newcommand{\lap}{\stackrel{<}{\sim}}
\newcommand{\fem}{$F_2^{em}$}
\newcommand{\tsnmp}{$\tilde{\sigma}_{NC}(e^{\mp})$}
\newcommand{\tsnm}{$\tilde{\sigma}_{NC}(e^-)$}
\newcommand{\tsnp}{$\tilde{\sigma}_{NC}(e^+)$}
\newcommand{\st}{$\star$}
\newcommand{\sst}{$\star \star$}
\newcommand{\ssst}{$\star \star \star$}
\newcommand{\sssst}{$\star \star \star \star$}
\newcommand{\tw}{\theta_W}
\newcommand{\sw}{\sin{\theta_W}}
\newcommand{\cw}{\cos{\theta_W}}
\newcommand{\sww}{\sin^2{\theta_W}}
\newcommand{\cww}{\cos^2{\theta_W}}
\newcommand{\trm}{m_{\perp}}
\newcommand{\trp}{p_{\perp}}
\newcommand{\trmm}{m_{\perp}^2}
\newcommand{\trpp}{p_{\perp}^2}
%
%=====================================================================
% alpha s
%
\newcommand{\sqrts}{$\sqrt{s}$}
\newcommand{\Oa}{$O(\alpha_s)$}
\newcommand{\Oaa}{$O(\alpha_s^2)$}
\newcommand{\PT}{p_{\perp}}
\newcommand{\sh}{\hat{s}}
\newcommand{\uh}{\hat{u}}
\newcommand{\ttbs}{\char'134}
\newcommand{\xpomlo}{3\times10^{-4}}
\newcommand{\xpomup}{0.05}
\newcommand{\llq}{$\alpha_s \ln{(\qsq / \Lambda_{QCD}^2)}$}
\newcommand{\llqx}{$\alpha_s \ln{(\qsq / \Lambda_{QCD}^2)} \ln{(1/x)}$}
\newcommand{\llx}{$\alpha_s \ln{(1/x)}$}
%
%=====================================================================
% name groups
%
\newcommand{\Brodsky}{Brodsky {\it et al.}}
\newcommand{\FKS}{Frankfurt, Koepf and Strikman}
\newcommand{\Kop}{Kopeliovich {\it et al.}}
\newcommand{\Ginzburg}{Ginzburg {\it et al.}}
\newcommand{\Ryskin}{\mbox{Ryskin}}
\newcommand{\Kaidalov}{Kaidalov {\it et al.}}
%
%=====================================================================
% journals
%
\def\ar#1#2#3   {{\em Ann. Rev. Nucl. Part. Sci.} {\bf#1} (#2) #3}
\def\epj#1#2#3  {{\em Eur. Phys. J.} {\bf#1} (#2) #3}
\def\err#1#2#3  {{\it Erratum} {\bf#1} (#2) #3}
\def\ib#1#2#3   {{\it ibid.} {\bf#1} (#2) #3}
\def\ijmp#1#2#3 {{\em Int. J. Mod. Phys.} {\bf#1} (#2) #3}
\def\jetp#1#2#3 {{\em JETP Lett.} {\bf#1} (#2) #3}
\def\mpl#1#2#3  {{\em Mod. Phys. Lett.} {\bf#1} (#2) #3}
\def\nim#1#2#3  {{\em Nucl. Instr. Meth.} {\bf#1} (#2) #3}
\def\nc#1#2#3   {{\em Nuovo Cim.} {\bf#1} (#2) #3}
\def\np#1#2#3   {{\em Nucl. Phys.} {\bf#1} (#2) #3}
\def\pl#1#2#3   {{\em Phys. Lett.} {\bf#1} (#2) #3}
\def\prep#1#2#3 {{\em Phys. Rep.} {\bf#1} (#2) #3}
\def\prev#1#2#3 {{\em Phys. Rev.} {\bf#1} (#2) #3}
\def\prl#1#2#3  {{\em Phys. Rev. Lett.} {\bf#1} (#2) #3}
\def\ptp#1#2#3  {{\em Prog. Th. Phys.} {\bf#1} (#2) #3}
\def\rmp#1#2#3  {{\em Rev. Mod. Phys.} {\bf#1} (#2) #3}
\def\rpp#1#2#3  {{\em Rep. Prog. Phys.} {\bf#1} (#2) #3}
\def\sjnp#1#2#3 {{\em Sov. J. Nucl. Phys.} {\bf#1} (#2) #3}
\def\spj#1#2#3  {{\em Sov. Phys. JEPT} {\bf#1} (#2) #3}
\def\zp#1#2#3   {{\em Zeit. Phys.} {\bf#1} (#2) #3}
%
%=====================================================================
% non classified
%
\newcommand{\clearemptydoublepage}{\newpage{\pagestyle{empty}\cleardoublepage}}
\newcommand{\scaption}[1]{\caption{\protect{\footnotesize  #1}}}
\newcommand{\proc}[2]{\mbox{$ #1 \rightarrow #2 $}}
\newcommand{\average}[1]{\mbox{$ \langle #1 \rangle $}}
\newcommand{\av}[1]{\mbox{$ \langle #1 \rangle $}}

\begin{titlepage}

%\noindent

\noindent
DESY 09-093   \hfill   ISSN 0418-9833 \\
June 2009
\vspace{2cm}
 
\begin{center}
\begin{Large}
 
%%%%%%%%%%%%%%%%%%%%%%%%%%%%%%%%%%%%%%%%%%%%%%%%
\boldmath
{\bf Diffractive Electroproduction of \rh\ and \ph\  Mesons} \\
{\bf  at HERA}
\unboldmath
%%%%%%%%%%%%%%%%%%%%%%%%%%%%%%%%%%%%%%%%%%%%%%%%
 
\vspace {0.5 cm}
 
H1 Collaboration
 
\end{Large}
\end{center}

\vspace{0.5 cm}

%%%%%%%%%%%%%%%%%%%%%%%%%%%%%%%%%%%%%%%%%%%%%%%%
%%%%%%%%%%%%%%%%%%%%%%%%%%%%%%%%%%%%%%%%%%%%%%%%
%%%%%%%%%%%%%%%%%%%%%%%%%%%%%%%%%%%%%%%%%%%%%%%%
%%%%%%%%%%%%%%%%%%%%%%%%%%%%%%%%%%%%%%%%%%%%%%%%

\begin{abstract}
%%%%%%%%%%%%%%%%%%%%%%%%%%%%%%%%%%%%%%%%%%%%%%%%
\noindent

Diffractive electroproduction of \rh\ and \ph\ mesons 
is measured at HERA with the H1 detector 
in the elastic and proton dissociative channels.
The data correspond to an integrated luminosity of $51~\invpb$.
About $10500$ \rh\ and $2000$ \ph\ events are analysed
in the kinematic range of  squared photon virtuality $2.5 \leq \qsq  \leq 60~\gevsq$, photon-proton 
centre of mass energy $35 \leq W \leq 180~\gev$ and squared four-momentum transfer to the 
proton $|t| \leq 3~\gevsq$.
The total, longitudinal and transverse cross sections are measured as a function of \qsq, $W$
and \modt.
The measurements show a transition to a dominantly ``hard" behaviour, typical of high gluon 
densities and small $q \bar q$ dipoles, for \qsq\ larger than $10$ to $20~\gevsq$.
They support flavour independence of the diffractive exchange, expressed in terms of the scaling 
variable \scaleqsqplmsq, and proton vertex factorisation.
The spin density matrix elements are measured as a function of kinematic variables.
The ratio of the longitudinal to transverse cross sections, the ratio 
of the helicity amplitudes and their relative phases are extracted. 
Several of these measurements have not been performed before and bring new information on 
the dynamics of diffraction in a QCD framework.
The measurements are discussed in the context of models using generalised parton distributions 
or universal dipole cross sections.

\end{abstract}
%%%%%%%%%%%%%%%%%%%%%%%%%%%%%%%%%%%%%%%%%%%%%%%%

\vspace{1.cm}

\begin{center}
Accepted by {\it JHEP} \\
\end{center}
 
\end{titlepage}
 
\newpage

%%%%%%%%%%%%%%%%%%%%%%%%%%%%%%%%%%%%%%%%%%%%%%%%
%%%%%%%%%%%%%%%%%%%%%%%%%%%%%%%%%%%%%%%%%%%%%%%%
%%%%%%%%%%%%%%%%%%%%%%%%%%%%%%%%%%%%%%%%%%%%%%%%
%%%%%%%%%%%%%%%%%%%%%%%%%%%%%%%%%%%%%%%%%%%%%%%%

\begin{flushleft}
  % \input{h1auts}
%-- H1AUTS Author list by names 
%-- Status: Tue Jun  2 10:22:10 CEST 2009  Number of authors = 247 

F.D.~Aaron$^{5,49}$,           %BUCH-PD        11/06           Aaron               
M.~Aldaya~Martin$^{11}$,       %DESY-PD        10/08           Aldayamartin        
C.~Alexa$^{5}$,                %BUCH-PD        06/06           Alexa               
V.~Andreev$^{25}$,             %LPI -PD        8/88            Andreev             
B.~Antunovic$^{11}$,           %DESY-LEFT      12/08           Antunovic           
A.~Asmone$^{33}$,              %ROME-ST        07/2            Asmone              
S.~Backovic$^{30}$,            %PODG-PD        03/2            Backovic            
A.~Baghdasaryan$^{38}$,        %YERE-PD        09/03           Baghdasaryana       
E.~Barrelet$^{29}$,            %PARI-PD        11/99           Barrelet            
W.~Bartel$^{11}$
% ...

%-- H1 Institutes 
\bigskip{\it
% ...
$ ^{5}$ National Institute for Physics and Nuclear Engineering (NIPNE) ,
          Bucharest, Romania \\
% ...
$ ^{11}$ DESY, Hamburg, Germany \\
% ...
$ ^{25}$ Lebedev Physical Institute, Moscow, Russia$^{ e}$ \\
% ...
$ ^{29}$ LPNHE, Universit\'{e}s Paris VI and VII, CNRS/IN2P3,
          Paris, France \\
 $ ^{30}$ Faculty of Science, University of Montenegro,
          Podgorica, Montenegro$^{ e}$ \\
% ...
$ ^{33}$ Dipartimento di Fisica Universit\`a di Roma Tre
          and INFN Roma~3, Roma, Italy \\
% ...
$ ^{38}$ Yerevan Physics Institute, Yerevan, Armenia \\
% ...
$ ^{49}$ Also at Faculty of Physics, University of Bucharest,
          Bucharest, Romania \\
% ...
$ ^e$ Supported by the Deutsche Forschungsgemeinschaft \\
% ...
}
\end{flushleft}
\newpage

%%%%%%%%%%%%%%%%%%%%%%%%%%%%%%%%%%%%%%%%%%%%%%%%
%%%%%%%%%%%%%%%%%%%%%%%%%%%%%%%%%%%%%%%%%%%%%%%%
%%%%%%%%%%%%%%%%%%%%%%%%%%%%%%%%%%%%%%%%%%%%%%%%
%%%%%%%%%%%%%%%%%%%%%%%%%%%%%%%%%%%%%%%%%%%%%%%%

%\newpage

%%%%%%%%%%%%%%%%%%%%%%%%%%%%%%%%%%%%%%%%%%%%%%%%
%%%%%%%%%%%%%%%%%%%%%%%%%%%%%%%%%%%%%%%%%%%%%%%%

\section{Introduction}  
                                                                                       \label{sect:introduction}

%%%%%%%%%%%%%%%%%%%%%%%%%%%%%%%%%%%%%%%%%%%%%%%%

Diffractive scattering is characterised, in high energy hadron interactions, by final states 
consisting of two systems well separated in rapidity, which carry the quantum numbers 
of the initial state hadrons.
The process is related through unitarity to inelastic scattering and governs the high 
energy behaviour of total cross sections.
It is described in Regge theory~\cite{regge} by the exchange of the vacuum singularity,
called the ``pomeron", and may be interpreted as the differential absorption of the various 
virtual components of the interacting systems~\cite{good-walker}.
It is a challenge for Quantum Chromodynamics (QCD) to explain diffraction in terms 
of quark and gluon interactions.

Most diffractive phenomena -- which include elastic scattering -- are governed by large 
distance, ``soft" processes, which in general are not accessible to perturbative QCD (pQCD) 
calculations.
However, for short distance processes, the presence of a ``hard" scale offers the possibility
to use perturbative techniques to calculate diffractive amplitudes.
Alternatively, at high energy the interaction properties of colour fields are 
invoked in models which characterize the incident particles as a superposition of 
colour dipoles with various size to calculate diffractive and total cross sections.

%-----------------------------------------------------------------------------
\begin{figure}[htbp]
\begin{center}
\setlength{\unitlength}{1.0cm}
\begin{picture}(6.0,5)   
\put(0.0,0.0){\epsfig{file=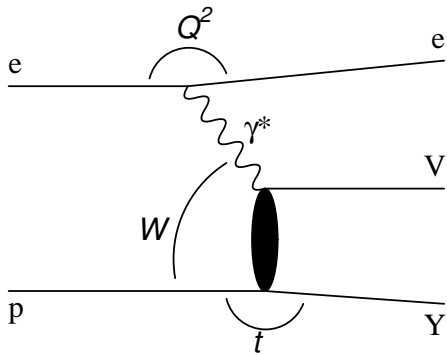  ,width=6.0cm}}
\end{picture}
\caption{Diffractive vector meson electroproduction.}
    \label{fig:VM}
\end{center}
\end{figure}
%-----------------------------------------------------------------------------

An important testing ground for calculations in diffraction is provided by the 
study of exclusive vector meson (VM) production $e + p \to e + V + Y$.
This process is illustrated in Fig.~\ref{fig:VM}: the intermediate photon of 
four-momentum $q$ converts into a diffractively scattered VM (\rh, \om, \ph, \jpsi, ...) 
of mass $M_V$, while the incoming proton is scattered into a system $Y$ of mass $M_Y$,
which can be a proton (``elastic" scattering) or a diffractively excited system (``proton
dissociation").
In VM production, a hard scale can be provided by the photon 
virtuality $Q$, with $Q^2 = - q^2$, 
the four-momentum transfer $\sqrt{\modt}$ from the proton, 
or by the quark mass (for heavy VM production). 
The reaction energy is defined by the photon-proton 
centre of mass energy $W$, with  $W^2 \simeq \qsq / x$, where $x$ is the Bjorken scaling 
variable.
The high energy electron-proton collider HERA offers access to all these scales, over a
wide range of values.

The present publication is devoted to the study of the diffractive electroproduction of \rh\ 
and \ph\ mesons, both for elastic and proton dissociative scattering. The data were 
taken at HERA with the H1 detector in the period from 1996 to 2000.
A common analysis of the four channels is performed.
Measurements of the production cross sections and of the 
spin density matrix elements, which give access to the helicity amplitudes, are presented 
as a function of 
the kinematic variables \qsq\ (including the \qsq\ dependence of the polarised cross sections), 
$W$, $t$ and, for \rh\ mesons, the dipion mass.

The measurement of kinematic dependences and the comparison between different 
VMs provide tests of a large spectrum of predictions.
The data cover the interesting transition from the low \qsq\ domain, dominated by soft diffraction, 
%to the domain of \scaleqsqplmsq\ above $3$ to $5~\gevsq$, where hard diffraction is
to the higher \qsq\ domain where hard diffraction is
expected to be dominant.
This offers the opportunity to test the relevance of soft physics 
features present in the photon and VM wave function, and to study the development of 
features predicted by pQCD calculations.
Quantitative tests of pQCD and colour dipole calculations are provided by the comparison 
with the data of various model predictions.
Two important aspects of diffraction are tested: the flavour independence of the diffractive
process and the factorisation of the process into a hard scattering contribution
at the photon vertex and soft diffractive scattering at the proton vertex (``Regge factorisation").
In addition, valuable information is provided by precise measurements of empirical 
parameters, in particular the \qsq\ and $t$ dependences of the cross sections and the ratio 
of the proton dissociative to elastic cross sections, as well as the contributions of 
various backgrounds.

The present studies confirm with increased precision previous H1 measurements 
on \rh~\cite{h1-rho-jpsi-94,h1-pdis-phi-94,h1-rho-95-96,h1-rho-large-t-97}
and \ph~\cite{h1-pdis-phi-94,h1-phi-95-96} electroproduction,
mainly in the elastic channel but 
also in proton dissociation~\cite{h1-pdis-phi-94,h1-rho-large-t-97}.
The samples analysed here include data taken in 1996 and 1997, and the present results
supersede those presented in~\cite{h1-rho-95-96,h1-phi-95-96, h1-rho-large-t-97}.
Thanks to the larger statistics, the scope of the investigation is significantly extended.

This analysis complements other H1 measurements of exclusive diffractive processes:
production 
of real photons, in photoproduction  ($\qsq\ \simeq 0$) at large \modt~\cite{h1-gamma-larget} 
and in electroproduction at small \modt\ 
(deeply virtual Compton scattering --~DVCS)~\cite{h1-dvcs},
production of \rh\ mesons in photoproduction at low~\cite{h1-rho-photoprod-94} and large 
\modt~\cite{h1-rho-photoprod-large-t},
of \jpsi\ mesons in photo- and electroproduction at low~\cite{h1-jpsi-hera1} and 
large~\modt~\cite{h1-jpsi-photoprod-large-t},
of $\psi(2s)$~\cite{h1-psi2s} and $\Upsilon$~\cite{h1-upsilon} mesons in photoproduction.

The ZEUS collaboration at HERA has performed measurements of
DVCS~\cite{z-dvcs}, \rh~\cite{z-rho-photoprod,z-rho-older,z-rho,z-high-t}, 
$\omega$~\cite{z-omega-photoprod,z-omega}, 
\ph~\cite{z-phi-photoprod,z-phi,z-high-t}, 
\jpsi~\cite{z-jpsi-photoprod,z-jpsi-elprod,z-high-t,z-high-t-jpsi} and $\Upsilon$~\cite{z-upsilon} production.
Results at lower energy have been published, in particular for \rh\ electroproduction, by
the DESY-Glasgow~\cite{joos}, CHIO~\cite{chio}, NMC~\cite{NMC}, 
E665~\cite{E665} and HERMES~\cite{hermes} collaborations.
The experimental and theoretical status of diffractive VM production before the high energy
fixed target and HERA experiments is presented in detail in the review~\cite{bauer}.

The paper is organised as follows.
The theoretical context and the models which will be compared to the data are
presented in section~\ref{sect:th_context}.
The H1 detector and the event selection criteria are summarised in section~\ref{sect:selection}, 
where the kinematic and angular variables are defined.
The various signal samples are defined in 
section~\ref{sect:extraction}, which also contains a detailed discussion of the backgrounds, 
a description of the Monte Carlo (MC) simulations used for the analyses and a discussion of the 
systematic errors affecting the measurements.
In section~\ref{sect:cross_sections}, the measurements of the VM line shapes and of 
the elastic and proton dissociative cross sections 
are presented, and 
VM universality and proton vertex factorisation are discussed.
Section~\ref{sect:polarisation} is devoted to the 
polarisation characteristics of the reactions and their kinematic dependence.
A summary of the results and conclusions are given in section~\ref{sect:conclusions}.

%%%%%%%%%%%%%%%%%%%%%%%%%%%%%%%%%%%%%%%%%%%%%%%%
%%%%%%%%%%%%%%%%%%%%%%%%%%%%%%%%%%%%%%%%%%%%%%%%
%%%%%%%%%%%%%%%%%%%%%%%%%%%%%%%%%%%%%%%%%%%%%%%%
%%%%%%%%%%%%%%%%%%%%%%%%%%%%%%%%%%%%%%%%%%%%%%%%

%\newpage

%%%%%%%%%%%%%%%%%%%%%%%%%%%%%%%%%%%%%%%%%%%%%%%%
%%%%%%%%%%%%%%%%%%%%%%%%%%%%%%%%%%%%%%%%%%%%%%%%
\section{Theoretical Context}
                                                                                          \label{sect:th_context}
%%%%%%%%%%%%%%%%%%%%%%%%%%%%%%%%%%%%%%%%%%%%%%%%

Since the first observation of high \qsq\ inclusive 
diffraction~\cite{inclusive_diffr} and of VM production at HERA, a large number of 
theoretical studies has been published on diffractive VM production
(see 
e.g.~\cite{cfs,DLrho,KNNZ,NNZ,NNPZ,ryskin,brodsky,FKS,DGKP,mrt,mrt2,gotsman,ik,royen,
sss,MFGS,caldwell,DGS,FS-m2g,fs-phi,fss,ISSK,ivanov-NLO,kmw,df,kroll,soyez,diehl-NLL}).
%e.g.~\cite{cfs,KNNZ,NNZ,NNPZ,ryskin,brodsky,FKS,DGKP,mrt,mrt2,ik,royen,
%sss,caldwell,DGS,FS-m2g,fss,ISSK,ivanov-NLO,kmw,kroll,soyez,diehl-NLL}).
Reviews of theoretical predictions confronted by the data have been published 
recently~\cite{fsw,ins}.

%%%%%%%%%%%%%%%%%%%%%%%%%%%%%%%%%%%%%%%%%%%%%%%%
\subsection {Cross section calculations}

\begin{figure}[htbp]
\begin{center}
\setlength{\unitlength}{1.0cm}
\begin{picture} (14.2,4.)   
\put(-.5,0.0){\epsfig{file=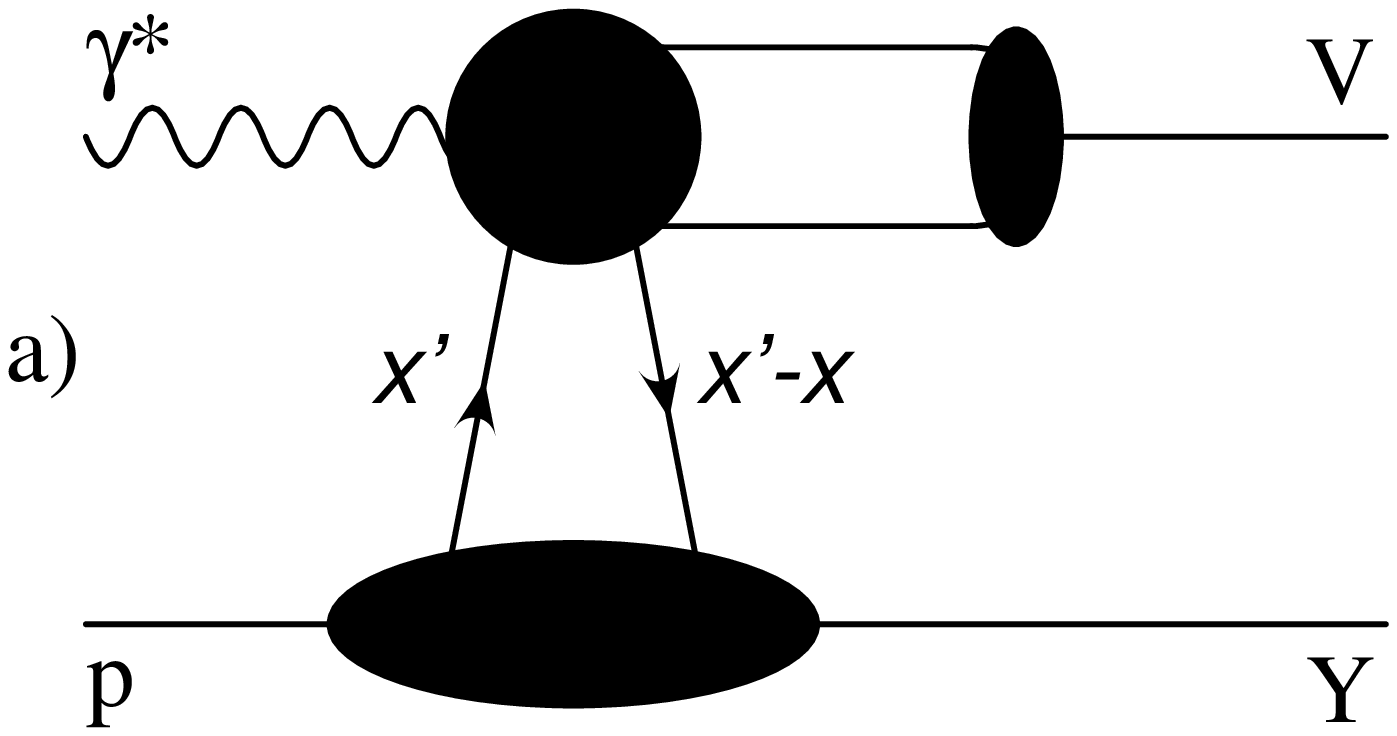,height=3.5cm}}
\put(6.5,0.0){\epsfig{file=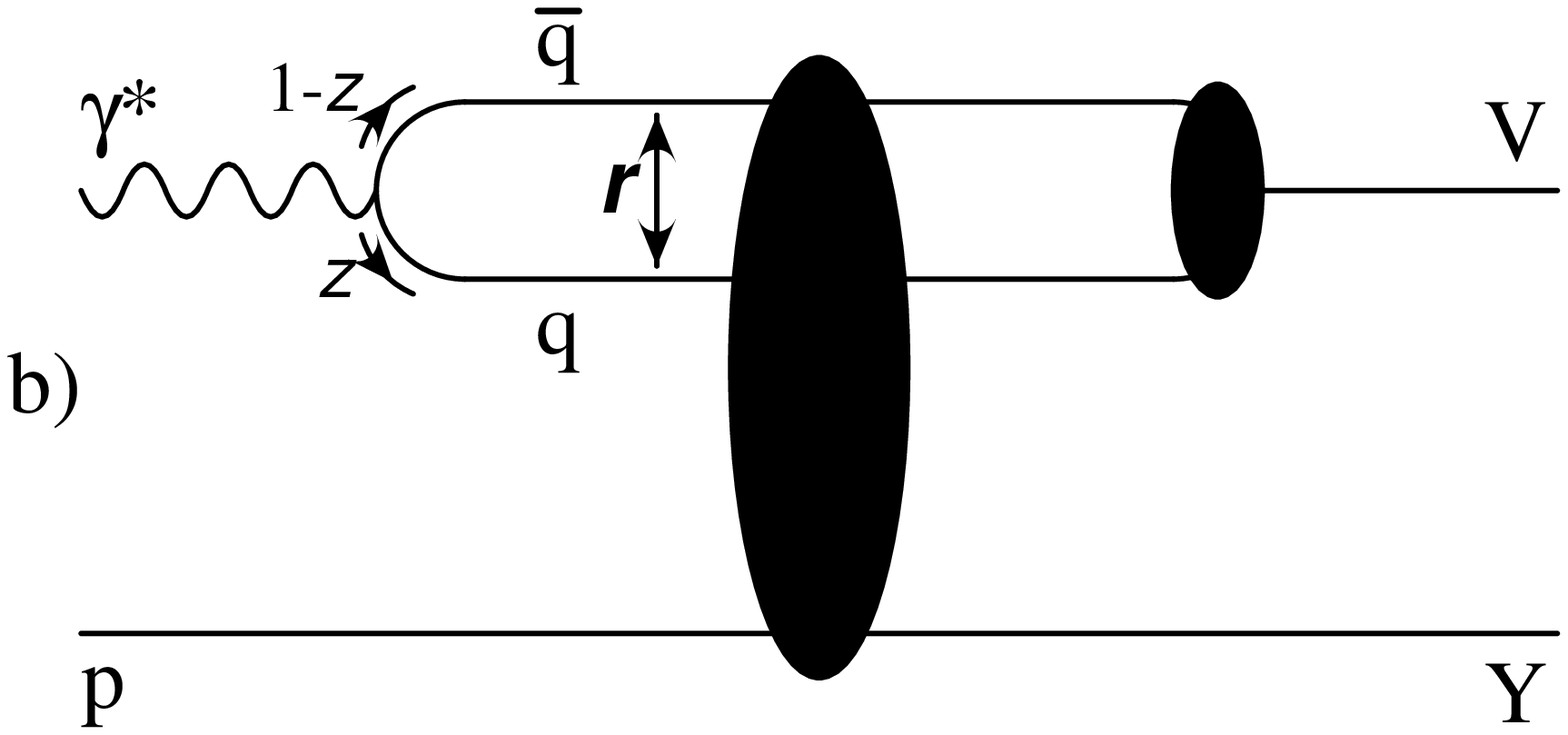,height=3.86cm}}
\end{picture}
\caption{Representative diagrams for diffractive VM electroproduction: 
a) the collinear factorisation, GPD approach;
b) the high energy, low $x$ colour dipole approach.}
    \label{fig:diag}
\end{center}
\end{figure}

Calculations are performed following two main approaches, sketched in Fig.~\ref{fig:diag}.
The approach based on collinear factorisation, illustrated in Fig.~\ref{fig:diag}(a),
describes VM production using the parton content
of the proton, in the presence of a hard scale.
The colour dipole picture of Fig.~\ref{fig:diag}(b) provides a complementary
way to describe high energy scattering.

%-----------------------------------------------------------------------------
%-----------------------------------------------------------------------------

%%%%%%%%%%%%%%%
\paragraph {Collinear factorisation}
%%%%%%%%%%%%%%%

In a pQCD framework, a collinear factorisation theorem~\cite{cfs} has been proven for 
the production of longitudinally polarised VMs in the kinematic domain 
with $W^2 \gg M_V^2$, $\qsq \gg \Lambda^2_{QCD}$ and $\modt\ \lapprox\ \Lambda^2_{QCD}$,
for leading powers of $Q$ and for all values of $x$.
The longitudinal amplitude, sketched in Fig.~\ref{fig:diag}(a), is given by
\begin{equation}
T_L^{\gamma ^{\star}p \rightarrow Vp}(x;t)
= \Sigma_{i,j} \int _0^1 {\rm d}z \int {\rm d}x^\prime \ f_{i/p}(x^\prime,x^\prime-x,t;\mu) \cdot
   H_{i,j}(Q^2x^\prime/x,Q^2,z;\mu) \cdot \Psi^V_j(z;\mu),
                                                                           \label{eq:GPD-amplitude}
\end{equation}
where $f_{i/p}(x^\prime,x^\prime-x,t;\mu)$ is the generalised 
parton distribution function (GPD) for parton $i$ in the proton and $\mu$ is the 
factorisation  and renormalisation scale, of the order of $Q$.
The GPDs (see e.g.~\cite{rev-gpd}), which are an extension of ordinary parton distribution 
functions (PDF), include correlations between partons with longitudinal 
momenta $x$ and $x^\prime$ and transverse momenta $t$;
they describe the off-diagonal kinematics ($x^\prime \neq x$) implied by the different 
squared four-momenta of the incoming photon and outgoing VM.
The $H_{i,j}$ matrix elements describe the hard scattering from the parton $i$ in the 
proton to the parton $j$ in the meson with wave function $\Psi^V_j(z;\mu)$, where
$z$ is the fraction of the photon longitudinal momentum carried by one of the
quarks.
The scale evolution is calculated, in the HERA kinematic domain, 
 using the DGLAP equations, and 
higher order corrections have been calculated~\cite{diehl-NLL,ISSK,ivanov-NLO}.
Collinear factorisation holds for heavy VMs~\cite{cfs}, and its validity is extended to 
transverse amplitudes at sufficiently high \qsq\ (see e.g.~\cite{cfs,ins,mrt,ik}).

%%%%%%%%%%%%%%%
\paragraph {Dipole approach}
%%%%%%%%%%%%%%%

At high energy (small $x$) and small \modt, VM production is conveniently 
studied in the proton rest frame, for all values of \qsq.
It is described as three factorising contributions, characterised by
widely different time scales~\cite{Mueller,NZ}, as illustrated in Fig.~\ref{fig:diag}(b):
the fluctuation of the virtual photon into a $q \bar q$ colour dipole, with a coupling depending
only on the quark charge, 
the dipole-proton scattering (either elastic or proton dissociative scattering), and 
the $q \bar q$ recombination into the final state VM.
The amplitude is
\begin{equation}
T^{\gamma ^{\star}p \rightarrow Vp}(x;t)
= \int _0^1 {\rm d}z \int {\rm d}^2{\bold r}\ \Psi^\gamma(z,{\bold r}) \cdot \sigma^{q \bar q - p}(x,{\bold r};t)
                   \cdot  \Psi^V(z,{\bold r}),
                                                                           \label{eq:dipole-amplitude}
\end{equation}
where 
${\bold r}$ is the transverse distance between the quark and the antiquark, and
$\Psi^\gamma(z,{\bold r})$ and $\Psi^V(z,{\bold r})$ are the photon
and the VM wave functions, respectively.
The diffractive dipole-proton cross section $\sigma^{q \bar q - p}(x,{\bold r};t)$
is expected to be flavour independent and to depend only on the dipole transverse size
(the impact parameter between the dipole and the proton is integrated over).
Photons with large virtuality and fluctuations into heavy quarks are dominated by 
dipoles with small transverse size.
In this case, the two quarks tend at large distance to screen each other's colour 
(``colour transparency"~\cite{CT}), which explains the small cross section.
In several models~\cite{KNNZ,FKS}, the convolution of the VM wave function with the dipole is 
expected to play a significant role in VM production, by selecting specific dipoles.
It can be noted that the Generalised Vector Meson Dominance model~\cite{bauer,sss} 
is related to the dipole approach.

Dipole-proton scattering is modeled at lowest order (LO) in pQCD
through the exchange of a gluon pair in a colour singlet state~\cite{low-nussinov}, and
in the leading logarithm approximation (LL~$1/x$) as the exchange of a 
BFKL-type gluon ladder.
In a ($z, {\bold k_t}$) representation, where ${\bold k_t}$ represents the quark (or antiquark)
momentum component transverse to the photon direction (i.e. the Fourier transform of the
dipole transverse size), $k_t$-unintegrated gluon 
distribution functions are used. 
The contributions of gluons with small $k_t$ are of a non-perturbative 
nature, whereas at large $k_t$ they can be obtained from the \qsq\ logarithmic 
derivative of the usual, integrated, gluon distribution, $G(x,Q^2)$.
In the LO and LL~$1/x$ approximations both gluons emitted from the proton carry the same 
fraction $x$ of the proton 
longitudinal momentum and the cross section is proportional to the square of the gluon 
density~\cite{ryskin,brodsky}.
Calculations beyond the LL~$1/x$ approximation take into account the difference
between the longitudinal momentum fractions carried by the two gluons 
(``skewing" effects)~\cite{mrt2,ins,shuvaev}.

At low $x$, VM production can be calculated~\cite{soyez,fss,kmw,caldwell}, in 
the absence of a hard scale, using universal dipole-proton cross sections 
obtained from deep inelastic scattering (DIS) measurements~\cite{munier}.
This approach automatically incorporates soft, non-perturbative contributions.
Such models often involve parton saturation effects, expected from the recombination of 
high density gluons~\cite{GBW,CGC} as inferred from the observation of geometric 
scaling~\cite{CGC,soyez}.
DGLAP evolution can also be included, for instance in the model~\cite{kmw}.

%%%%%%%%%%%%%%%%%%%%%%%%%%%%%%%%%%%%%%%%%%%%%%%%
\subsection {Kinematic dependences and 
                     \boldmath {$ \sigma_L / \sigma_T$} }
                                                          \label{sect:th_diff_cross_sections}

The photon-proton cross section can be decomposed into a longitudinal and
transverse part, $\sigma_L$ and $\sigma_T$, respectively.
At LO and for $t = 0$, the dependences $\sigma_L \propto 1/Q^6$ and 
$\sigma_T \propto 1/Q^8$ are predicted~\cite{brodsky}, and the ratio 
$R \equiv \sigma_L / \sigma_T$ is predicted to be $R = \qsq / M_V^2$.
Modifications to these dependences are expected (see e.g.~\cite{FKS}),
due to the \qsq\ dependence of  the gluon density, the quark transverse movement (Fermi motion)
and quark virtuality~\cite{royen}, and the \qsq\ dependence of the strong coupling constant 
$\alpha_s$. 

In the dipole approach, the square of the scale $\mu$ of the interaction is of the order of
\begin{equation}
\mu^2 \simeq  z (1-z) Q^2 + k_t^2 +m_q^2 \simeq z (1-z) (Q^2 + M_V^2),
                                                                              \label{eq:mu}
\end{equation}
$m_q$ being the current quark mass.
It is related to the inverse of the relevant ``scanning radius"~\cite{KNNZ,ins,FKS} 
in the dipole-proton interaction.

For longitudinally polarised photons or for heavy quark production, the 
$q \bar q$ wave function $\Psi^\gamma(z,{\bold r})$ is concentrated around 
$z \simeq 1-z \simeq 1/2$.
This suggests that a universal hard scale, $\mu$, following from the transverse size of the 
dominant dipoles, can be of the order of $\mu^2 \simeq \scaleqsqplmsq$.
For transverse photons fluctuating into light quarks, in contrast, the wave function is non-zero at 
the end-points $z \simeq 0$ or $1$.
These contributions correspond to small $k_t$ values of the quarks forming the dipole, 
and hence to a large transverse distance between them.
The scale $\mu$ is therefore damped to smaller values than for longitudinal photons with
the same virtuality, soft contributions may be significant and formal divergences 
appear in pQCD calculations for $z \to 0, 1$~\cite{FKS,DGKP,ins}.
For moderate $\qsq$ values, the $z$ distribution of light quark
pairs from longitudinal photons can present a non-negligible smearing 
around the value $z = 1/2$, which results in a contamination of soft, ``finite size" 
effects~\cite{fsw}.
It is estimated that the fully perturbative QCD regime is reached for light VM production by
longitudinal photons for \qsq\ above $20$ to $30~\gevsq$~\cite{fsw,ins}.

The $W$ dependence of VM production is governed by the $x^{-\lambda}$ 
evolution of the gluon distribution, with
$\lambda$ increasing from $\approx 0.16$ for $\qsq = 2~\gevsq$ to $\approx 0.26$ for 
$\qsq = 20~\gevsq$, as measured in the total DIS cross sections at 
HERA~\cite{hera-lambda}.
For heavy VMs and for longitudinally polarised light VMs at sufficiently high \qsq, a 
strong (``hard") $W$ dependence of the cross section is thus expected, fixed for all
VMs by the scale \scaleqsqplmsq.
In contrast, the $W$ dependence of the transverse cross section is expected to be 
milder than for longitudinal photons, since the $\lambda$ parameter is taken
at a smaller value of the effective scale.
This may result in a $W$ dependence of the cross section ratio
$R =  \sigma_L / \sigma_T$.
In the framework of Regge theory, the existence of two pomerons~\cite{DL-two-pomeron} 
is postulated to describe both the soft and hard behaviours of the cross section~\cite{DGS}. 

At low $\modt$ ($\modt\ \lapprox\ 0.5-0.6~\gevsq$ for elastic scattering), the $t$ 
dependence of VM production is well described by an exponentially falling distribution 
with slope $b$, ${\rm d}\sigma / {\rm d}t\  \propto e^{-b~\! |t|}$
(predictions for the \modt\ dependence are also given e.g. in~\cite{royen,FS-m2g,soyez}).
In an optical model approach, the slope $b$ is given by the sum of the transverse sizes of 
the scattered system $Y$, of the $q \bar q$ dipole and of the exchanged system, with
possibly in addition a VM form factor.
Neglecting the latter, the $t$ slopes for heavy VMs and for light VM
production by longitudinally polarised photons are expected to take
universal values, depending only on \scaleqsqplmsq, whereas
the production of light VMs by transverse photons, which is dominated by dipoles with larger 
transverse size, is expected to exhibit steeper $t$ distributions~\cite{FKS,fss}.
This may result in a $t$ dependence of $\sigma_L / \sigma_T$.

%%%%%%%%%%%%%%%%%%%%%%%%%%%%%%%%%%%%%%%%%%%%%%%%
\subsection {Helicity amplitudes}

The helicity amplitudes $T_{\lambda_V \lambda_{\gamma}}$, where
$\lambda_V$ and $ \lambda_{\gamma}$ are the VM and photon helicities, respectively,
have been calculated in perturbative QCD for the electroproduction of light VMs with 
$\modt \ll \qsq$~\cite{ins,ik,soyez, royen}.
In this domain, the dominant amplitude is the $s$-channel helicity conserving 
(SCHC) \tzz\ amplitude, which describes the transition 
from a longitudinal photon to a longitudinal VM.
Other amplitudes are damped by powers of $Q$.
Those leading to the production of a transverse VM, of which the SCHC \tuu\ 
amplitude is largest, contain an additional factor $\propto 1/Q$.
SCHC violation implies for single helicity flip amplitudes an additional factor 
$\propto \sqrt{\modt} / Q$, to be squared for the double flip \tmuu\ 
amplitude.
This leads, in the kinematic range studied here, to the following hierarchy of amplitude 
intensities (assuming natural parity exchange):
$  |T_{00}| > |T_{11}| > |T_{01}| > |T_{10}|, |T_{-11}|$.

%%%%%%%%%%%%%%%%%%%%%%%%%%%%%%%%%%%%%%%%%%%%%%%%
%%%%%%%%%%%%%%%%%%%%%%%%%%%%%%%%%%%%%%%%%%%%%%%%
\subsection {Comparison of models with the data}
%%%%%%%%%%%%%%%%%%%%%%%%%%%%%%%%%%%%%%%%%%%%%%%%

Predictions for VM production are available from a large number of models.
Quantitative calculations generally imply the choice of PDF or GPD parameterisations or, 
in colour dipole models, of dipole-proton cross section parameterisations. 
Model calculations also generally imply the choice of VM wave function parameterisations, often 
taken as following a Gaussian 
shape, with several variants~\cite{NNZ,DGKP,NNPZ,fss,munier,ins}.
In view of the large number of models, no attempt is made in this paper to provide exhaustive
comparisons to the data.
Instead, a few models and parameterisations, representative of recent 
approaches, are compared to various choices of observables.
Examples of the uncertainties on the predictions, due to the choice of parton distribution 
functions and wave function parameterisations, are given for two of the models.

\begin {itemize} 

\item The GPD model of Goloskokov and Kroll (GK~\cite{kroll}) provides predictions 
within the handbag factorisation scheme for the longitudinal and transverse amplitudes 
in the SCHC approximation.
Soft physics is described by a GPD parameterisation of the proton structure, 
constructed from standard PDFs with adequate skewing features and $t$ dependences.
The end-point singularities are
removed with the aid of a specific model for the VM wave function.
Error bands are provided with the model predictions.

\item The model of Martin, Ryskin and Teubner 
(MRT~\cite{mrt}) for \rh\ meson production is based on parton-hadron duality.
Open $q \bar{q}$ production is calculated in an appropriate spin-angular
state and in a specific invariant mass interval, which is then assumed to
saturate \rh\ production, thus neglecting any VM wave function effects.
The \qsq\ dependence of the gluon density, described by the anomalous dimension 
$\gamma$ with $G(x,Q^2) \propto (Q^2/Q_0^2)^\gamma$, is used to
calculate the longitudinal and transverse cross sections.
Skewing effects are parameterised~\cite{mrt2,shuvaev} without explicit use of GPDs.
Predictions using two alternative PDFs are compared with the present data:
CTEQ6.5M~\cite{cteq} and MRST-2004-NLO~\cite{mrst}.

\item The model presented in the review of Ivanov, Nikolaev and 
Savin (INS~\cite{ins}) is framed in the $k_t$-factorisation dipole approach.
The helicity amplitudes are calculated perturbatively and then extended into
the soft region by constructing parameterisations of the off-forward
unintegrated gluon density. 
The \qsq\ and $W$ dependences of the cross sections and the full set of spin density 
matrix elements are predicted.
Two wave function models, ``compact" and ``large", are used for \rh\ mesons, corresponding to 
two extreme cases for describing the $\rho \to e^+ e^-$ decay width.

\item The $k_t$-factorisation calculations of Ivanov and Kirschner (IK~\cite{ik}) 
provide predictions for the full set of helicity amplitudes, including helicity
flip transitions.
Similar to the MRT approach, the relevance of pQCD for transverse amplitude 
calculations is justified by the scale behaviour $\propto (Q^2/Q_0^2)^\gamma$ of the 
gluon distribution, which avoids divergences for $z \to 0,1$.

\item The dipole approach of Kowalski, Motyka and Watt (KMW~\cite{kmw})
uses an impact parameter dependent description of the dipole cross section 
in the non-forward direction~\cite{bgp}, within the saturation models of 
Golec-Biernat and W\"{u}sthoff (GW~\cite{GBW}) and of Iancu {\it et~al}. 
(Colour Glass Condensate -- CGC~\cite{CGC}).
The \qsq\ and $W$ dependences of the SCHC longitudinal and transverse amplitudes
are predicted using the DGLAP evolution equations for $\modt\ \lapprox\ 0.5~\gevsq$.

\item The dipole approach of Marquet, Peschanski and Soyez (MPS~\cite{soyez})
proposes an extension of the saturation model~\cite{CGC}, geometric scaling 
being extended to non-forward amplitudes with a linear $t$ dependence of the
saturation scale.
The exponential $t$ dependence at the proton vertex is parameterised with a 
universal slope obtained from previous VM measurements.

\end {itemize}

%%%%%%%%%%%%%%%%%%%%%%%%%%%%%%%%%%%%%%%%%%%%
%%%%%%%%%%%%%%%%%%%%%%%%%%%%%%%%%%%%%%%%%%%%
%%%%%%%%%%%%%%%%%%%%%%%%%%%%%%%%%%%%%%%%%%%%
%%%%%%%%%%%%%%%%%%%%%%%%%%%%%%%%%%%%%%%%%%%%

%\newpage

%%%%%%%%%%%%%%%%%%%%%%%%%%%%%%%%%%%%%%%%%%%%
\section{Experimental Conditions and Variable Definitions}  
                                                                                       \label{sect:selection}
%%%%%%%%%%%%%%%%%%%%%%%%%%%%%%%%%%%%%%%%%%%%

The diffractive production and decay of \rh\ and \ph\ mesons is identified 
using the following reactions:
\begin{eqnarray}
  e + p & \rightarrow & e + V + Y,                       \nonumber  \\
  \rho   & \rightarrow & \pi^+ + \pi^-  \ \ \ \ \ \    {({\cal BR} \simeq 100\% )},   \nonumber \\
   \phi  & \rightarrow & K^+ + K^-     \ \ \ \    {({\cal BR} = 49.2 \pm 0.6\% )}.
                                                                                             \label{eq:process}
\end{eqnarray}

The events are selected by requiring the detection  
of the scattered electron and of a pair of oppositely charged particles, and by requiring 
the absence of additional activity
in the detector, except in the region close 
to the outgoing proton beam, where proton dissociation can 
contribute.

The kinematic domain of the measurements is:
\begin{eqnarray}
2.5 \ \leq \ \qsq  & \leq &    60~\gevsq ,     \nonumber \\
35 \ \leq \  W     & \leq &    180~\mbox{${\rm GeV}$} ,       \nonumber \\
             |t|     & \leq &3~\gevsq ,  \nonumber \\
            M_Y  &  <   & 5~\gevcsq.                        \label{eq:kin_range}
\end{eqnarray}

The large values of $W^2$ compared to $Q^2$, $M_Y^2$, $M_V^2$ and \modt\ ensure that the 
process is diffractive, i.e. due to pomeron exchange. 
The variable $\xpom = (\qsq + \msq + \modt) / (W^2 + \qsq - M_Y^2)$, which corresponds
to the proton energy loss, is always smaller than $10^{-2}$.

%%%%%%%%%%%%%%%%%%%%%%%%%%%%%%%%%%%%%%%%%%%%
\subsection{Data sets}  
                                                                                           \label{sect:data-sets}
%%%%%%%%%%%%%%%%%%%%%%%%%%%%%%%%%%%%%%%%%%%%

The data studied here were taken with $27.5~\gev$ energy electrons or 
positrons colliding with $820$ or $920~\gev$ protons
(in the rest of this paper the term ``electron'' is used generically to refer 
to both electrons and positrons).
The data sets are summarised in Table~\ref{table:data_sets}, where $\sqrt{s} $ 
is the $ep$ centre of mass energy and the lepton beam type is specified.
The integrated luminosity of $51~\invpb$ corresponds to running periods with 
all relevant parts of the detector fully operational. The periods with high 
prescaling of the triggers relevant for the present analyses are discarded.
The published results with $1 \leq \qsq \leq 2.5~\gevsq$~\cite{h1-rho-95-96,h1-phi-95-96} 
are also presented in Table~\ref{table:data_sets} (``H1 SV").
They were obtained in 1995 in a special run of $125~\invnb$, with the $ep$ interaction point shifted 
by $70~{\rm cm}$ in the outgoing $p$ beam direction.
This data set is not re-analysed in the present publication.

%-----------------------------------------------------------------------------
\begin{table}[htbp]
\begin{center}
\begin{tabular}{|c|c|c|c|c|} 
\multicolumn{5}{c} { }  \\
\hline
   Data taking &   lepton    & proton energy   & $\sqrt{s} $  &    luminosity   \\
    year          &  beam   &     (\mbox{${\rm GeV}$})          &     (\mbox{${\rm GeV}$})    &  (\invpb)             \\
\hline 
\hline 
  1995 (SV) &      $e^+$     &      $820$           &   $300$  &   $0.125$          \\
\hline 
  1996  &    $e^+$              &       $820$          &    $300$  &      $4.0 $                   \\      
  1997  &    $e^+$              &       $820$          &    $300$  &      $9.8  $                  \\      
  1999  &    $e^-$              &       $920$          &    $320$  &      $4.8  $                 \\      
  1999  &    $e^+$              &       $920$          &    $320$  &      $4.6  $                 \\      
  2000  &    $e^+$              &       $920$          &    $320$  &      $28.1$                   \\      
 \hline
\end{tabular} 
\caption{Characteristics of the data taken in 1995 with a shifted vertex (SV) and of the
data sets used in the present paper (1996-2000).}
  \label{table:data_sets}
  \end{center}
  \end{table}
%-----------------------------------------------------------------------------
%

%%%%%%%%%%%%%%%%%%%%%%%%%%%%%%%%%%%%%%%%%%%%
\subsection{The H1 detector and triggers}  
                                                                                           \label{sect:detector}
%%%%%%%%%%%%%%%%%%%%%%%%%%%%%%%%%%%%%%%%%%%%

A detailed description of the H1 detector can be found in \cite{nim}.  
Only the components  essential to the present analysis are described here. 
The origin of the H1 coordinate system is the nominal $ep$ interaction point, with
the positive $z$--axis (forward direction) along the direction of the proton beam.
The polar angles $\theta$ and the particle transverse momenta are defined with respect 
to this axis, and the pseudorapidity is $\eta= -\log{\tan (\theta/2)}$.

A system of two large coaxial cylindrical drift chambers (CJC) of $2~{\rm m}$ length 
and~$0.85~{\rm m}$ 
external radius, with wires parallel to the beam direction, is located in a $1.16~{\rm T}$ uniform 
magnetic field.
This provides a measurement of the transverse momentum of charged particles with 
resolution $\Delta p_{t} / p_{t} \simeq 0.006 \ p_{t} \oplus 0.015$ ($p_t$ measured 
in~\gevc), for particles emitted from the nominal interaction point with polar angle 
$ 20 \leq \theta \leq 160^{\rm \circ}$. 
Drift chambers with wires perpendicular to the beam direction, located inside the inner CJC 
and between the two CJC chambers, provide measurements of $z$ coordinates.
Track measurements are improved by the use of the central silicon tracker~\cite{CST}
(from 1997 onward). 
The interaction vertex is reconstructed from the tracks.

The liquid argon (LAr) calorimeter, located inside the magnet and surrounding the 
central tracker, covers the angular range $4 \leq \theta \leq 154^{\rm \circ}$.
The backward electromagnetic calorimeter Spacal 
($ 153  \leq \theta \leq 177.5 ^{\rm \circ}$) is used to identify scattered 
electrons.
In front of the Spacal, the backward drift chamber (BDC) provides a precise 
electron direction measurement.

The ``forward detectors" are sensitive to energy flow close to the outgoing 
proton beam direction.
They consist of the proton remnant tagger (PRT), a set of scintillators placed $24~{\rm m}$ 
downstream of the interaction point and covering the angles 
$0.06 \leq \theta \leq 0.17^{\rm \circ}$, and the forward muon detector (FMD), 
a system of drift chambers covering the angular region $3 \leq \theta \leq 17^{\rm \circ}$.
The PRT and the three layers of the FMD situated closer to the main calorimeter detect 
secondary particles produced in interactions with the beam collimators 
or the beam pipe walls of elastically scattered protons at large \modt\ and of decay 
products of diffractively excited systems $Y$ with $M_Y\ \gapprox\ 1.6~\gevcsq$.

For the data collected in 1996 and 1997, events with $\qsq \geq 2.5~\gevsq$ were selected 
by inclusive triggers requesting an electromagnetic energy deposit in the Spacal.
For the years 1999 and 2000, diffractive VM events with $\qsq \geq 5~\gevsq$ were 
registered using several inclusive triggers;
in addition, a special trigger was dedicated to elastic \ph\ production
with $\qsq > 2~\gevsq$. 

To reduce the data recording rate to an acceptable level, data selected by certain triggers
have been dowscaled. 
In the following, the accepted events are weighted accordingly.

%%%%%%%%%%%%%%%%%%%%%%%%%%%%%%%%%%%%%%%%%%%%
\subsection{Event selection}  
                                                                                           \label{sect:ev-sel}
%%%%%%%%%%%%%%%%%%%%%%%%%%%%%%%%%%%%%%%%%%%%
For the present analyses, 
the scattered electron candidate is identified as an electromagnetic cluster with energy larger than 
$17~\gev$ reconstructed in the Spacal calorimeter. 
This energy threshold reduces to a negligible level the background of photoproduction events 
with a wrongly identified electron candidate in the Spacal.
The electron direction is calculated from the position of the measured interaction vertex
and from the BDC signals, when their 
transverse distance to the cluster barycentre is less than $3~{\rm cm}$; if no such BDC signal
is registered, the cluster centre is used.

The VM candidate selection requires the reconstruction in the central tracking detector 
of the trajectories of two, and only two, oppositely charged particles.
They must originate from a common vertex lying within $30~\rm{cm}$ in $z$ of the nominal 
$ep$  interaction point, and must have transverse momenta larger than $0.15~\gevc$ and 
polar angles within the interval $20 \leq \theta \leq 160^{\rm \circ}$.
This ensures a difference in pseudorapidity 
of at least two units between the most forward track and the most forward cell
of the LAr calorimeter.
The VM momentum is calculated as the vector sum of the two charged particle momenta.
 
The existence of a gap in rapidity between the VM and the forward system $Y$ is further 
ensured by two veto conditions:
that there is in the central tracker no additional track, except if it is
associated to the electron candidate, and that there is in the
LAr calorimeter no cluster with energy above noise level, $E > 400~\mev$, unless it is
associated to the VM candidate.
These requirements reduce to negligible level the contamination from 
non-diffractive DIS interactions, which are characterised by the 
absence of a significant gap in rapidity in the fragmentation process. 
They imply that the mass of the diffractively excited proton system is restricted to 
$M_Y\ \lapprox\ 5~\gevcsq$.
They also contribute to the 
suppression of backgrounds due to the diffractive production of systems subsequently decaying 
into a pair of charged particles and additional neutral particles.
Energy deposits unrelated to the VM event and noise in the calorimeter are monitored from 
randomly triggered readouts of the detector.
The energy threshold of $400~\mev$ leads to an average loss of $13^{+3}_{-5} \%$ 
of the diffractive VM events.

A cut  is applied to the difference between the sum of energies 
and the sum of longitudinal momenta of the scattered electron and VM candidate,
$\Sigma (E-p_z) > 50~\gev$. 
For events where all particles except the forward going system $Y$ are detected,  this quantity 
is close to twice the incident electron beam energy, $55~\gev$.
The cut reduces the QED radiation and background contributions in which additional 
particles remain undetected.

%%%%%%%%%%%%%%%%%%%%%%%%%%%%%%%%%%%%%%%%%%%%
\subsection{Kinematic and angular variables}  
                                                                                           \label{sect:variables}
%%%%%%%%%%%%%%%%%%%%%%%%%%%%%%%%%%%%%%%%%%%%

To optimise measurements in the selected domain, 
the kinematic variables are reconstructed from the measured quantities following the 
algorithms detailed in~\cite{h1-rho-95-96}.
In addition to the nominal beam energies, they make use of well measured quantities 
in the H1 detector: the electron and VM directions and the VM momentum.

The variable \qsq\ is reconstructed from the polar angles of the electron and of the VM 
(``double angle" method~\cite{double-angle}).
The modulus of the variable $t$ is to very good precision equal to the square of the 
transverse momentum of the scattered system $Y$, which is calculated as the vector sum 
$\vec{p}_{t, miss} = - (\vec{p}_{t,V} + \vec{p}_{t, e})$ of the transverse momenta of
the VM candidate and of the scattered electron\footnote{More precisely, the quantity 
$|\vec{p}_{t, miss}|^2$ is a measure of $t^\prime =  |t| - |t|_{min}$, where $|t|_{min}$ is the 
minimum value 
of $|t|$ kinematically required for the VM and the system $Y$ to be produced
on shell through longitudinal momentum transfer. At HERA energies and for the relevant 
values of $M_V$ and $M_Y$, $|t|_{min}$ is negligibly small compared to \modt.
In the following the notations $|t|$ is used for $ t^\prime$.}.
The electron transverse momentum, $\vec{p}_{t, e}$, is determined using the electron 
energy obtained from the ``double angle" method.
The variable $W$ is reconstructed from the VM energy and longitudinal 
momentum~\cite{jb}.
The electron energy measured in the Spacal is used only for the calculation of the 
variable $\Sigma (E-p_z)$.

%-----------------------------------------------------------------------------
\begin{figure}[htbp]
\begin{center}
\setlength{\unitlength}{1.0cm}
\begin{picture}(10.8,7.0)   
\put(0.0,0.0){\epsfig{file=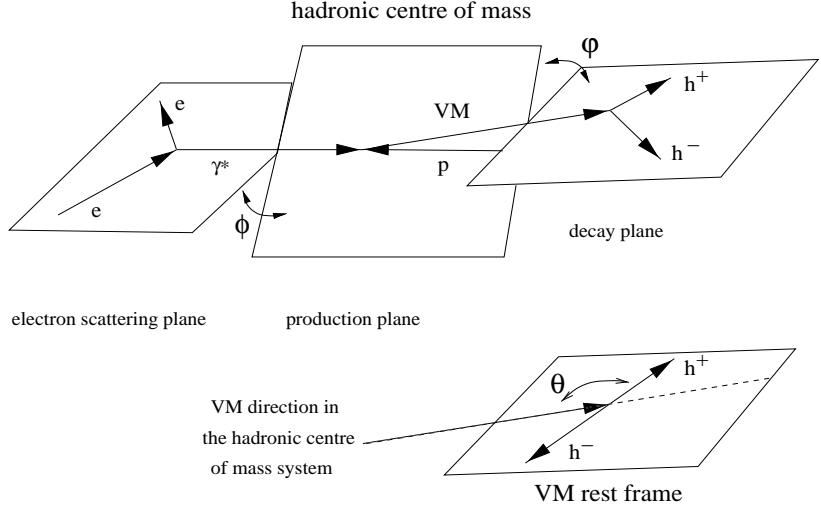  ,height=7.0cm,width=10.8cm}}
\end{picture}
\caption{Definition of the angles characterising diffractive VM production 
and decay in the helicity system.  }
    \label{fig:dec_ang}
    \end{center}
\end{figure}
%-----------------------------------------------------------------------------

Three angles characterise VM electroproduction and two-body decay 
(Fig.~\ref{fig:dec_ang}). 
In the helicity frame used for the present measurements, they are chosen as follows.
The azimuthal angle 
$\phi$ is defined in the hadronic centre of mass system as the angle between 
the electron scattering plane and the VM production plane, which is formed by the 
directions of the virtual photon and the VM.
The two other angles, which describe VM decay, are chosen in the VM rest frame
as the polar angle $\theta$ and the azimuthal angle $\phib$ of the positively charged decay 
particle, $h^+$, the quantization axis being opposite to the direction of the outgoing system $Y$.

%%%%%%%%%%%%%%%%%%%%%%%%%%%%%%%%%%%%%%%%%%%%
%%%%%%%%%%%%%%%%%%%%%%%%%%%%%%%%%%%%%%%%%%%%
%%%%%%%%%%%%%%%%%%%%%%%%%%%%%%%%%%%%%%%%%%%%
%%%%%%%%%%%%%%%%%%%%%%%%%%%%%%%%%%%%%%%%%%%%

%\newpage

%%%%%%%%%%%%%%%%%%%%%%%%%%%%%%%%%%%%%%%%%%%%
\section{Data Analysis}  
                                                                                       \label{sect:extraction}
%%%%%%%%%%%%%%%%%%%%%%%%%%%%%%%%%%%%%%%%%%%%

This section first defines the analysis samples.
The backgrounds are then discussed, the Monte Carlo simulations used to extract the 
signals are introduced, and the predictions are compared to the distributions of the 
hadronic invariant mass and of other observables.
Finally, systematic uncertainties are discussed.

%%%%%%%%%%%%%%%%%%%%%%%%%%%%%%%%%%%%%%%%%%%%
\subsection{Analysis samples}  
                                                                                       \label{sect:samples}
%%%%%%%%%%%%%%%%%%%%%%%%%%%%%%%%%%%%%%%%%%%%

Four event samples, which correspond approximately to the four processes studied 
in this paper, are selected following the conditions 
summarised in Tables~\ref{table:signal_VM} and~\ref{table:signal_diffr}.
These conditions are chosen to minimize background contributions.

%-----------------------------------------------------------------------------
\begin{table}[htbp]
\begin{center}
\begin{tabular}{|c|c|} 
\multicolumn{2}{c} { }  \\
\hline
   Vector meson             &     mass range                              \\
\hline
   \rh\ sample                 &   $0.6 \leq \mpp \leq 1.1~\gevcsq$         \\
   \ph\ sample                &    $  1.00 \leq \mkk \leq 1.04~\gevcsq$  \\
\hline
\end{tabular} 
\caption{Sample definition for the two VM selection.} 
  \label{table:signal_VM}
  \end{center}
  \end{table}
%-----------------------------------------------------------------------------

The VM identification relies on the invariant mass of the two particles with trajectories 
reconstructed in the central tracker; no decay particle identification is performed.
For the \rh\ sample, the mass \mpp\ calculated under the pion mass hypothesis
is required to lie in the range $0.6 \leq \mpp \leq 1.1~\gevcsq$. 
For the \ph\ sample, the range $1.00 \leq \mkk \leq 1.04~\gevcsq$ is selected, the
invariant mass \mkk\ being calculated under the kaon hypothesis.

%-----------------------------------------------------------------------------
\begin{table}[htb]
\begin{center}
\renewcommand{\arraystretch}{1.15}
\begin{tabular}{|c|c|c|} 
\multicolumn{3}{c} { }  \\
\hline
   Diffractive process   & forward detector selection               &           $t$ range          \\
\hline
   notag sample          & no signal above noise                     &   $|t| \leq 0.5~\gevsq$  \\
   tag sample             & signal detected above noise            &   $|t| \leq 3.0 ~\gevsq$   \\
\hline
\end{tabular} 
\caption{Sample definition for the two diffractive processes.} 
  \label{table:signal_diffr}
  \end{center}
  \end{table}
%-----------------------------------------------------------------------------

The events in the \rh\ and \ph\ samples are further classified in two 
categories, ``notag" and ``tag", according to the absence or the presence of 
activity above noise levels in the forward detectors, respectively.
Elastic production is studied in the notag sample with $|t| \leq 0.5~\gevsq$  
whereas the tag sample with $|t| \leq 3~\gevsq$ is used for proton dissociative
studies.

%-----------------------------------------------------------------------------
\begin{table}[htb]
\begin{center}
\begin{tabular}{|c|c| c c c c c|c c c c c|} 
\multicolumn{8}{c} { }  \\
\hline
   Year         & VM        & \multicolumn{5}{c| } {\qsq\ range (\gevsq)}  & \multicolumn{5}{c| } {$W$ range (\mbox{${\rm GeV}$}) }  \\
\hline 
\hline 
1995 - SV    &\rh, \ph     & $\phantom{0}1.0$ & $\leq$ &\qsq & $\leq$ &$\phantom{0}2.5$       &  $40$ & $\leq$ &$W$& $\leq$& $140$    \\
\hline 
  1996-1997 &  \rh, \ph   &   $\phantom{0}2.5$ & $\leq $&\qsq & $<$ & $\phantom{0}4.9$      &   $35$ &$\leq$& $W$&$\leq$&$ 100$     \\
            &		   &   $\phantom{0}4.9$ & $\leq $&\qsq & $<$ &$\phantom{0}9.8$	  &   $40$ &$\leq$& $W $&$\leq$&$120$	  \\
            &		   &   $\phantom{0}9.8$ & $\leq $&\qsq & $<$ &$15.5$	 &   $50$ &$\leq$& $W $&$\leq$&$140$	 \\
            &		   &  $15.5$ & $\leq $&\qsq & $<$ &$27.3$     &   $50$ &$\leq$& $W $&$\leq$&$150$     \\
            &		   &  $27.3$ & $\leq $&\qsq  & $\leq$ &$60.0$   &   $60$ &$\leq$& $W $&$\leq$&$150$     \\
  1999-2000 &\ph\ notag &  $\phantom{0}2.5$ & $\leq $&\qsq & $<$ &$\phantom{0}4.9$       &    $35$ &$\leq$& $W $&$\leq$&$100$     \\
            &  \rh, \ph  &  $\phantom{0}4.9$ & $\leq $&\qsq & $<$ &$\phantom{0}9.8$	    &	 $40$ &$\leq$& $W $&$\leq$&$120$     \\
            &		   &  $\phantom{0}9.8$ & $\leq $&\qsq & $<$ &$15.5$      &   $50$ &$\leq$& $W $&$\leq$&$140$     \\
            &		   &  $15.5$ & $\leq$ &\qsq & $<$ &$27.3$     &   $50$ &$\leq$& $W $&$\leq$&$160$     \\
            &		   &  $27.3$ & $\leq$ &\qsq  & $\leq$ &$60.0$   &   $60$ &$\leq$& $W $&$\leq$&$180$     \\
\hline
\end{tabular} 
\caption{Kinematic range of the measurements.} 
  \label{table:signal_kin_range}
  \end{center}
  \end{table}
%-----------------------------------------------------------------------------

The kinematic domain of the measurements is summarised in 
Table~\ref{table:signal_kin_range}.
It is determined by the detector geometry, the beam energies and the triggers, with
the requirement of a reasonably uniform acceptance.
The accepted \qsq\ range depends on the data taking period; for the notag \ph\ sample
in 1999-2000 it extends to smaller values than for the tag \ph\ sample and for the \rh\ 
samples, due to the special elastic \ph\ trigger. 
For $W$, the regions with good acceptance are determined by the track requirement; 
the accepted $W$ values increase with \qsq\ and with $\sqrt{s}$.

The acceptance increases with \qsq, mostly because of the non-uniform geometric 
acceptance of the electron trigger for $\qsq\ \lapprox\ 20~\gevsq$.
Monte Carlo studies show that the total acceptance increases from $15\%$ ($18\%$) for 
\rh\ (\ph) elastic production at $\qsq = 2.5~\gevsq$ to about $50\%$ at $\qsq = 8$ ($6$)~\gevsq\ 
and to more than $60\%$ for $\qsq = 12$ ($10$)~\gevsq, and that they are essentially 
independent of $W$ in the measurement domain.

%-----------------------------------------------------------------------------
\begin{table}[tb]
\begin{center}
\begin{tabular}{|c|c|c|c|c|} 
\multicolumn{5}{c} { }  \\
\hline
\multirow{2}{*}{Numbers of events} 
                          & \multicolumn{2}{c|} {\rh\ sample} &  \multicolumn{2}{c|} {\ph\ sample} \\
\cline{2-5}
                           & raw               &weighted             &  raw               &weighted               \\
\hline
\hline 
notag sample        &    $7793$  & $11775$             &   $1574$  & $1976$              \\
tag sample          &    $2760$  & $\phantom{0}3824$   &   $\phantom{0}416$  &  $\phantom{0}495$          \\
\hline 
\end{tabular} 
\caption{Events in the different data samples: raw numbers and numbers weighted to
account for the downscaling applied to certain triggers.}
  \label{table:nb-of-events}
  \end{center}
  \end{table}
%-----------------------------------------------------------------------------

The raw numbers of events selected in the four samples defined by 
Tables~\ref{table:signal_VM}-\ref{table:signal_kin_range} are given in 
Table~\ref{table:nb-of-events}, together with the numbers weighted to account for the 
downscaling applied to certain triggers.

%%%%%%%%%%%%%%%%%%%%%%%%%%%%%%%%%%%%%%%%%%%%
\subsection{Backgrounds}  
                                                                                        \label{sect:bg}
%%%%%%%%%%%%%%%%%%%%%%%%%%%%%%%%%%%%%%%%%%%%

Several background processes, which affect differently the four data samples 
and depend on the 
kinematic domain, are discussed in this section.
Their contributions are summarised in Table~\ref{table:percent-of-bg}.
The non-resonant $\pi \pi$ contribution to the \rh\ signal, which contributes essentially 
through interference, is discussed separately in section~\ref{sect:mass-rho}.
The $e^+e^-$ and $\mu^+ \mu^-$ backgrounds were found, using 
the GRAPE simulation~\cite{grape}, to be completely negligible.

%-----------------------------------------------------------------------------
\begin{table}[tb]
\begin{center}
\begin{tabular}{|c|c c c|c c c|c c c|c c c|} 
\multicolumn{13}{c} { }  \\
\hline
                                   & \multicolumn{3}{c|} {\rh\ notag}                  & \multicolumn{3}{c|} {\rh\ tag}
                                   & \multicolumn{3}{c|} {\ph\ notag}                 & \multicolumn{3}{c|} {\ph\ tag}              \\
                                   &\multicolumn{3}{c|} {$|t| \leq 0.5~\gevsq$}    & \multicolumn{3}{c|} {$|t| \leq 3~\gevsq$}
                                   &\multicolumn{3}{c|} {$|t| \leq 0.5~\gevsq$}    & \multicolumn{3}{c|} {$|t| \leq 3~\gevsq$}  \\
\hline
p. diss. events              & $10.7$ &$\pm$& $0.3 \%$&             &$-$     &               & $9.7$ &$\pm$ & $0.7 \%$ &            &$-$     &              \\
el. events                 &           &$-$     &              & $13.1$  &$\pm$ &$ 0.5 \%$ &          &$-$     &                & $11.8$ &$\pm$& $1.5 \%$ \\
$\pi^+ \pi^-$                  &           &$-$     &               &            &$-$     &               & $6.3$  &$\pm$& $0.5 \%$  & $\phantom{0}4.7$   &$\pm$& $0.9 \%$  \\
$\phi \rightarrow~3~\pi$& $\phantom{0}0.3$ &$\pm$& $0.1 \%$ &  $\phantom{0}0.4$ &$\pm$& $0.1 \%$  &          & $-$    &              &           &$-$      &              \\
$\omega$                    & $\phantom{0}0.6$ &$\pm$& $0.1 \%$ &  $\phantom{0}0.7$ &$\pm$& $0.1 \%$  & $1.7$ &$\pm$& $0.3 \%$ & $\phantom{0}2.8$ &$\pm$  & $0.7 \%$ \\
\rhop                           & $\phantom{0}4.0$ &$\pm$& $0.2 \%$ &  $\phantom{0}7.7$ &$\pm$& $0.4 \%$  & $3.6$ &$\pm$& $0.4 \%$ & $\phantom{0}9.2$ &$\pm$  & $1.3 \%$ \\
\hline
\end{tabular} 
\caption{Background contributions to the four data samples defined in
Tables~\ref{table:signal_VM}-\ref{table:signal_kin_range}.
The quoted errors are the statistical errors from the MC samples.} 
  \label{table:percent-of-bg}
  \end{center}
  \end{table}
%-----------------------------------------------------------------------------

%%%%%%%%%%%%%%%%%%%%%%%%%%%%%%%%%%%%%%%%%%%%
\subsubsection{Cross-contaminations between the elastic and proton dissociative processes}
                                                                                                 \label{sect:el-pd_bg}
%%%%%%%%%%%%%%%%%%%%%%%%%%%%%%%%%%%%%%%%%%%%

The notag and tag samples correspond roughly to the elastic and proton dissociative 
processes, respectively. 
However, cross-contaminations occur, due to the limited acceptance and efficiency of the 
forward detectors and to the presence of noise.
The response of these detectors is modeled using independent measurements, by comparing signals in 
the various PRT and FMD planes.

The cross-contaminations are determined for each VM species without {\it a priori} 
assumptions on the relative production rates of elastic and inelastic events.
In a first step, the contaminations are calculated from the numbers of tag and notag events 
and from the probabilities for elastic and proton dissociative events to deposit a signal in the 
forward detectors as obtained from the MC simulations. 
The crossed backgrounds are then determined in an iterative procedure from the 
simulations, after final tunings to the data.

%%%%%%%%%%%%%%%
\paragraph {Proton dissociative backgrounds in the notag samples}
%%%%%%%%%%%%%%%

Proton dissociative events produce a background to the elastic signals in the notag 
samples when the mass of the excited baryonic system is too low to give a signal
in the forward detectors ($M_Y \ \lapprox \ 1.6~\gevcsq$) or
because of inefficiencies of these detectors.
The background fraction increases strongly with $\modt$, because the proton dissociative 
cross sections have a shallower \modt\ distribution than the elastic cross 
sections.
In the notag samples with $\modt \leq 0.5~\gevsq$, the proton dissociative background 
amounts to $10.5\%$.

%%%%%%%%%%%%%%%
\paragraph {Elastic backgrounds in the tag samples}
%%%%%%%%%%%%%%%

Conversely, elastic background in the proton dissociative samples of tag events is
due to unrelated signal or noise in the forward detectors. 
For $\modt \leq 3~\gevsq$, it amounts to $12.5\%$, with larger contributions 
for small \modt\ values where the elastic to proton 
dissociative cross section ratio is larger.
In addition, when \modt\ is large enough for the scattered proton to hit the beam pipe walls 
or adjacent material ($\modt \gapprox~0.75~\gevsq$), elastic events may give signal in 
the forward detectors.

%%%%%%%%%%%%%%%%%%%%%%%%%%%%%%%%%%%%%%%%%%%%
\boldmath
\subsubsection{Cross-contaminations between \rh\ and \ph\ samples}
                                                                                                 \label{sect:rho-phi_bg}
\unboldmath
%%%%%%%%%%%%%%%%%%%%%%%%%%%%%%%%%%%%%%%%%%%%

For \rh\ production, the contribution from the $\phi \rightarrow K^+ K^-$ channel is 
removed
by the requirement $\mpp \geq 0.6~\gevcsq$, which also suppresses the contribution of the 
$ \phi \rightarrow K^0_S K^0_L$ channel (${\cal BR} =~34\%$) with the $K^0_S$ meson decaying 
into a pion pair close to the emission vertex and the $K^0_L$ being undetected in the 
calorimeter.

The largest background in the selected \ph\ samples is due to the low mass tail of 
$\pi^+ \pi^-$ pair production extending under the \ph\ peak.
It amounts to $6\%$ and depends on \qsq.
The shape of the $\pi^+ \pi^-$ distribution corresponding to small values of \mkk\ is discussed
in section~\ref{sect:MC}.

%%%%%%%%%%%%%%%%%%%%%%%%%%%%%%%%%%%%%%%%%%%%
\boldmath
\subsubsection{$\phi \rightarrow~3~\pi$ and $\omega$ backgrounds}
                                                                                                 \label{sect:phi-omega_bg}
\unboldmath
%%%%%%%%%%%%%%%%%%%%%%%%%%%%%%%%%%%%%%%%%%%%

A small \ph\ contamination in the \rh\ samples is due to the channel
       $ \phi \rightarrow \pi^+ \pi^- \pi^0 $ (${\cal BR} = 15\%$) 
when each photon from the $\pi^0$ decay remains undetected because it is emitted outside 
the LAr calorimeter acceptance, because the energy deposit in the LAr calorimeter 
does not pass the $400~\mev$ threshold, or because it is associated with one of the 
charged pions.
This background contributes to the \mpp\ distribution mostly below the selected 
mass range; 
it amounts to~$0.3\%$ of the selected \rh\ 
notag sample with $\modt \leq 0.5~\gevsq$ and~$0.4\%$ of the tag sample with 
$\modt \leq 3~\gevsq$.
The background rate increases with \modt\ because the non-detection of the $\pi^0$ decay
photons 
leads in general to an overestimate of the $p_t$ imbalance of the event, $\vec{p}_{t, miss}$,
which mimics a large \modt\ value.
The $\phi \rightarrow~3~\pi$ contribution below the $\phi \rightarrow~KK$ signal is negligible.

Similarly, the diffractive production of $\omega$ mesons decaying in the mode 
$\omega \rightarrow \pi^+ \pi^- \pi^0$ (${\cal BR} = 89\%$) gives background contributions to the \rh\ 
and \ph\ samples when the $\pi^0$ decay photons escape detection.
In addition, the $ \omega \rightarrow \pi^+ \pi^-$  (${\cal BR} = 1.7\%$) channel gives an irreducible
background to the \rh\ signal.
The background due to \om\ production contributes $0.6\%$ to the elastic and $0.7\%$ 
to the proton dissociative \rh\ samples, and $1.7$ and $2.8\%$ for the \ph\ samples, respectively. 
The non-detection of photons leads to large reconstructed \modt\ values for these contributions.
Note that for the cross sections quoted below, as for results in previous HERA papers, 
the $\omega - \rho$ interference is neglected: its contribution is small and cancels when 
integrated over the mass range.

%%%%%%%%%%%%%%%%%%%%%%%%%%%%%%%%%%%%%%%%%%%%
\boldmath
\subsubsection{\rhoprim\ background}
                                                                                                 \label{sect:rhop_bg}
\unboldmath
%%%%%%%%%%%%%%%%%%%%%%%%%%%%%%%%%%%%%%%%%%%%

The largest background to the \rh\ signal and the second largest background to the \ph\ 
signal is due to diffractive \rhop\ production\footnote{The detailed mass structure~\cite{pdg} 
of the states described in the past as the \rhoprim(1600) meson is not relevant for the present 
study. The name \rhoprim\ is used for all VM states with mass in the range 
$1.3 - 1.7~\gevcsq$.}.
The \rhop\ mesons decay mostly into a \rh\ meson and a pion pair, leading to final 
states with four charged pions ($\rho^\prime \rightarrow \rho^0 \pi^+ \pi^-$) or with two 
charged and two neutral pions 
($\rho^\prime \rightarrow \rho^{\pm} \pi^{\mp} \pi^0,\ \rho^{\pm} \rightarrow \pi^{\pm} \pi^0$).
The $\pi^+ \pi^- \pi^0 \pi^0$ events can mimic large \modt\ \rh\ or \ph\ production when the 
photons from the $\pi^0$ decays escape detection, which induces a $p_t$ imbalance in the 
event and a distortion of the $t$ distribution, similarly to the $\phi \rightarrow 3~\pi$ and 
$\omega \rightarrow 3~\pi$ backgrounds.
At high \modt, this background affects mostly the notag samples.
It is indeed distributed between the notag and tag samples following the elastic to 
proton dissociative production cross section ratio, whereas genuine high \modt\ \rh\ and 
\ph\ mesons are essentially produced with proton dissociation and thus contribute 
mainly to the tag samples.

No cross section measurement of diffractive \rhop\ production has been published in the 
relevant $Q^2$ range. 
The \rhoprim\ contribution to the \rh\ signal is thus determined from the data themselves, using a 
method presented in the H1 analysis of high \modt\ \rh\ 
electroproduction~\cite{h1-rho-large-t-97}.
The distribution of the variable $\zeta$, which is the cosine of the angle between the 
transverse components of the \rh\ candidate momentum, $\vec{p}_{t,\rho}$, and of the event 
missing momentum, $\vec{p}_{t, miss}$, is sensitive to the relative amounts of \rh\ signal 
and \rhop\ background.
The \rhop\ contribution gives a peak at $\zeta = +1$ and a negligible contribution at  
$\zeta = -1$, since the \rh\ and the missing $\pi^0$'s are all emitted roughly in the 
direction of the \rhop.
In contrast, the \rh\ signal gives peaks at $\zeta = +1$ and $\zeta = -1$.
However, for genuine \rh\ production, $\zeta$ is also correlated to the angle 
$\phi$ between the \rh\ production plane and the electron scattering plane,
which is distributed according to the a priori unknown
value of the combinations of spin density matrix elements \rfivecomb\ and \ronecomb\
(Eq.~(\ref{eq:angle_phi}) of the Appendix).

An iterative procedure is used to determine simultaneously the amounts of \rhop\ 
background in the notag and tag samples, the matrix element combinations \rfivecomb\ and 
\ronecomb\  (assumed to be identical for elastic and proton dissociative scattering),
and the \modt\ distributions of \rh\ elastic and proton dissociative production. 
It is found to converge after a few steps.
The  results are also used to calculate the \rhoprim\ background to the \ph\ signal.

The \rhop\ background is estimated to contribute $4\%$ to the 
notag samples with $\modt \leq 0.5~\gevsq$, and $8\%$ to the tag samples with
$\modt \leq 3~\gevsq$.

%%%%%%%%%%%%%%%%%%%%%%%%%%%%%%%%%%%%%%%%%%%%
\subsection{Monte Carlo simulations}  
                                                  \label{sect:MC}
%%%%%%%%%%%%%%%%%%%%%%%%%%%%%%%%%%%%%%%%%%%%

Monte Carlo simulations based on the DIFFVM program are used to describe \rh, \om, \ph\
and \rhop\ VM production and decay, detector response (acceptances, efficiencies and 
variable reconstruction) and radiative effects.

The DIFFVM program~\cite{diffvm} is based on Regge theory and Vector Meson 
Dominance~\cite{vdm}.
The $M_Y$ diffractive mass distribution for proton dissociative events contains an explicit 
simulation of baryonic resonance production for $M_Y < 1.9~\gevcsq$ and a dependence 
${\rm d} \sigma / {\rm d} M_Y^2 \propto 1/M_Y^{2.16}$ for larger masses~\cite{goulianos}, 
with quark and diquark fragmentation simulated using the JETSET 
programme~\cite{jetset}.

The \rh\ and \ph\ MC samples are reweighted according 
to the measurements of the \qsq, $W$ and \modt\ differential cross sections and of 
the angular VM production and decay distributions:
the angle $\theta$ is distributed according to the measurements of the \rzqzz\ matrix element
(Eq.~(\ref{eq:cosths})),
the angle \ph\ to those of the \rfivecomb\ and \ronecomb\ combinations
(Eq.~(\ref{eq:angle_phi})), and
the angle \phib\ to those of the \cosdelta\ parameter, which in the SCHC 
approximation fixes the $\psi = \phi - \phib$ distribution (Eq.~(\ref{eq:Wcosdelta})).

For the $\omega$ and \rhop\ backgrounds, the cross section dependences on the
kinematic variables \qsq, $W$ and 
\modt\ are taken to be the same as for \rh\ mesons at the same \scaleqsqplmsq\ value.
For the two-body $\omega$ decay, the angular distributions are taken as for \rh\ mesons.
For three-body $\omega$ and $\phi$ decays, the angular distributions are chosen to follow 
$\phi$ and \cosths\ distributions described by the same values of the matrix elements 
as for two-body decays.
For \rhop\ decays\footnote{In the dominant 
$\rhop \to \rho \pi \pi$ decay mode, the two pions do not form a \rh\ resonance and can 
be assumed to be in a spin~0 state.
The angular decay distribution thus includes the two possible polarisation states of the \rh\ 
meson, with the squared amplitude $|M_1(00)|^2$ ($|M_1(10)|^2$) corresponding to the 
probability that it is longitudinally (transversely) polarised, 
giving in the SCHC approximation, with the notations of the Appendix: \\
$ W(\theta, \psi) = \frac {3}{4 \pi} \ \frac {1} {1 + \varepsilon R}  
    \left\{ |M_1(00)|^2 \ \  [\ \frac {1} {2} \sin^2 \theta + \varepsilon R \cos^2 \theta 
    - \frac {K} {2} \sin 2 \theta \cos \psi \cos \delta + 
                              \frac {\varepsilon}{2} \sin^2 \theta  \cos 2 \psi \ ]  \right.$   \\
$   \left.       + \ |M_1(10)|^2 \ \ [ \ \frac {1} {2} (1 + \cos^2 \theta + \varepsilon R \sin^2 \theta
    + \frac {K} {2} \sin 2 \theta \cos \psi \cos \delta - 
                               \frac {\varepsilon}{2} \sin^2 \theta \cos 2 \psi \ ]  \right \} $, 
where $K = \sqrt {2 \varepsilon R  (1 + \varepsilon)}$. }, the parameters 
$M_1(00)$ and $M_1(10)$ describe the angular distributions~\cite{caro}.
The values $|M_1(00)|^2 = 0.5, |M_1(10)|^2 = 0.5$ are chosen for the present simulations. 

The ratio of proton dissociative to elastic cross sections is taken from the present \rh\
analysis and assumed to be the same for all VMs.
All kinematic and angular distributions are taken to be identical for elastic and proton 
dissociative scattering, as supported by the present data, except for the \modt\ 
dependence of the cross sections.

The \ph\ to \rh\ cross section ratio is set to that measured in this analysis.
The $\omega$ to \rh\ ratio is taken from ZEUS 
measurements~\cite{z-omega-photoprod,z-omega}.
For \rhop\ production, a \rhop\ to \rh\ ratio of $1.12$ is 
used\footnote{This number does not constitute a \rhop\ cross section 
measurement, but it is used as an empirical parameterisation for describing the \rhop\ 
background contribution under the \rh\ peak, for the \rhop\ mass and width chosen in the 
simulation; as a consequence, varying the latter values has negligible influence on the 
background subtraction.}, as a result of the procedure described in the 
previous section.

For \rh, \ph\ and $\omega$ mesons, the particle mass, width and decay branching ratios are 
taken from the PDG compilation~\cite{pdg}.
The mass and width of the \rhop\ resonance
are taken as $1450~\mevcsq$ and $300~\mevcsq$, respectively.
For \rh\ and \ph\ meson decays into two pseudoscalar mesons, the mass distributions 
are described by a relativistic Breit-Wigner function $BW(m)$ with momentum 
dependent width, as described in section~\ref{sect:mass_shapes}.
In addition, the \rh\ mass shape is skewed according to the parameterisation of Ross and
Stodolsky~\cite{rs},
\begin{equation}
\frac {{\rm d}N(\mpp)} {{\rm d}\mpp} \propto BW_{\rho}(m_{\pi \pi}) \
      \left( \frac {m_{\rho}} {\mpp} \right)^n,               
                                                                                             \label{eq:rs}
\end{equation}
with the \qsq\ dependent value of $n$ measured in this analysis.

The $\pi \pi$ background in the \ph\ mass region is taken from the skewed Breit-Wigner 
distribution for \rh\ mesons, modified for $\mpp < 0.6~\gevcsq$ according to the empirical 
form 
\begin{equation}
\frac {{\rm d}N(\mpp)} {{\rm d}\mpp} \propto BW_{\rho}(m_{\pi \pi}) \cdot 
                     \left( \frac {m_{\rho}} {0.6} \right)^n \cdot  \left[ 1 + \kappa  \sqrt{0.6 - \mpp} \ \right] ,
                                                                                            \label{eq:low-mpp}              
\end{equation}
with masses expressed in~\gev and the parameter $\kappa$ being taken to be $1.5$.
This parameterisation describes the low mass \mpp\ distribution well, as shown in
Figs.~\ref{fig:mass_distrib-mpp-NT}, \ref{fig:mass_distrib-mpp-T} and~\ref{fig:rh_mass_sod},
where the cut  $\mkk > 1.04~\gevcsq$ is applied to suppress genuine \ph\ production.

Radiative effects are calculated using the HERACLES program~\cite{heracles}.
Corrections for these effects in the selected kinematic range with 
$\Sigma (E-p_z) > 50~\gev$ are of the order of $1\%$.

All generated events are processed through the full GEANT~\cite{Brun:1987ma} based simulation 
of the H1 apparatus and are reconstructed using the same program chain as for the data.
Of particular relevance to the present analysis is the description of the forward detector 
response; the activity in these detectors, not related to VM production, is obtained 
from data taken independently of physics triggers, and is superimposed on generated 
events in the MC simulations.

%%%%%%%%%%%%%%%%%%%%%%%%%%%%%%%%%%%%%%%%%%%%
\subsection{Mass distributions}  
                                                  \label{sect:mass_control_plots}
%%%%%%%%%%%%%%%%%%%%%%%%%%%%%%%%%%%%%%%%%%%%

\begin{figure}[htbp]
\begin{center}
\setlength{\unitlength}{1.0cm}
\begin{picture}(12.0,12.0)   
\put(0.0,0.0){\epsfig{file=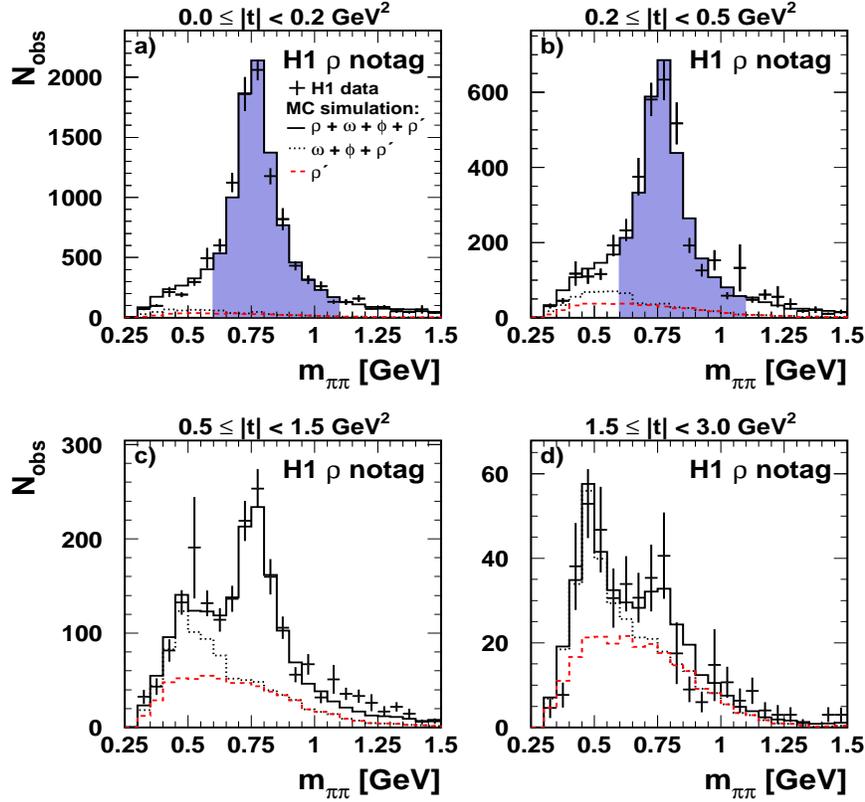  ,height=12.0cm,width=12.0cm}}
\end{picture}
\caption{Distributions of the invariant mass \mpp\ (with the cut $\mkk > 1.04~\gevcsq$ to
reject the $ \phi \rightarrow KK$ signal) in four domains in \modt, for the notag 
sample.
The dashed histograms show the MC predictions for the \rhoprim\ background, 
the dotted histograms the sum of the \rhoprim, \om\ and \ph\ backgrounds,
and the full histograms the \rh\ signal (including interference with $\pi \pi$ non-resonant 
production) and the sum of all backgrounds.
The mass and \modt\ domain where the cross section measurements are performed is 
shaded.}
\label{fig:mass_distrib-mpp-NT}
\end{center}
\end{figure}

\begin{figure}[htbp]
\begin{center}
\setlength{\unitlength}{1.0cm}
\begin{picture}(12.0,12.0)   
\put(0.0,0.0){\epsfig{file=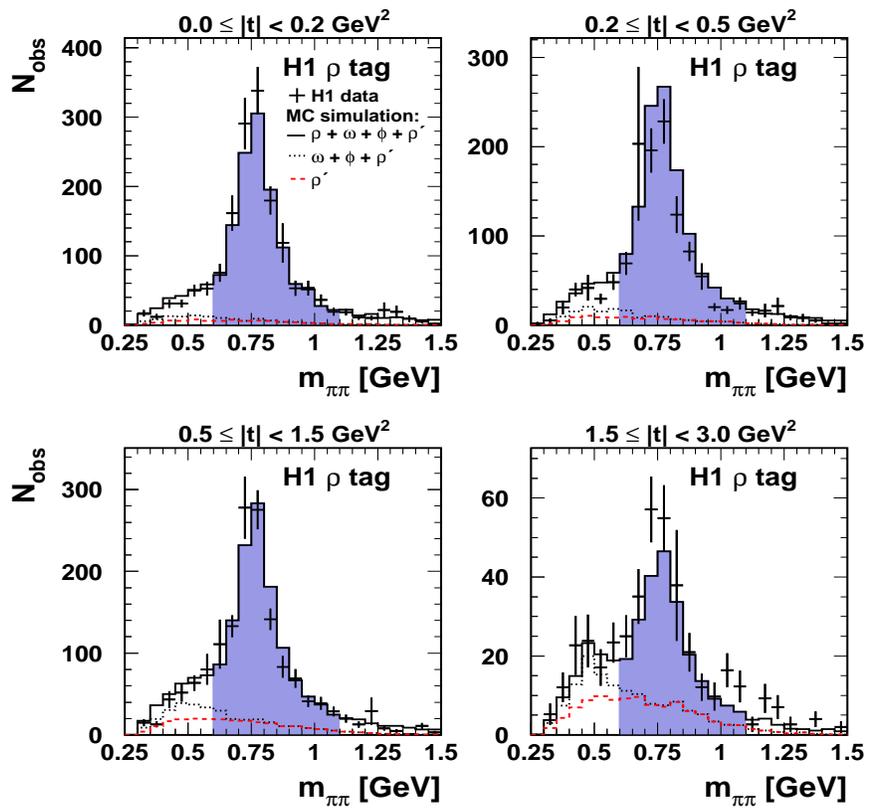  ,height=12.0cm,width=12.0cm}}
\end{picture}
\caption{Same as in Fig.~\ref{fig:mass_distrib-mpp-NT}, for the tag sample.}
\label{fig:mass_distrib-mpp-T}
\end{center}
\end{figure}

\begin{figure}[htbp]
\begin{center}
\setlength{\unitlength}{1.0cm}
\begin{picture}(12.0,12.0)   
\put(0.0,0.0){\epsfig{file=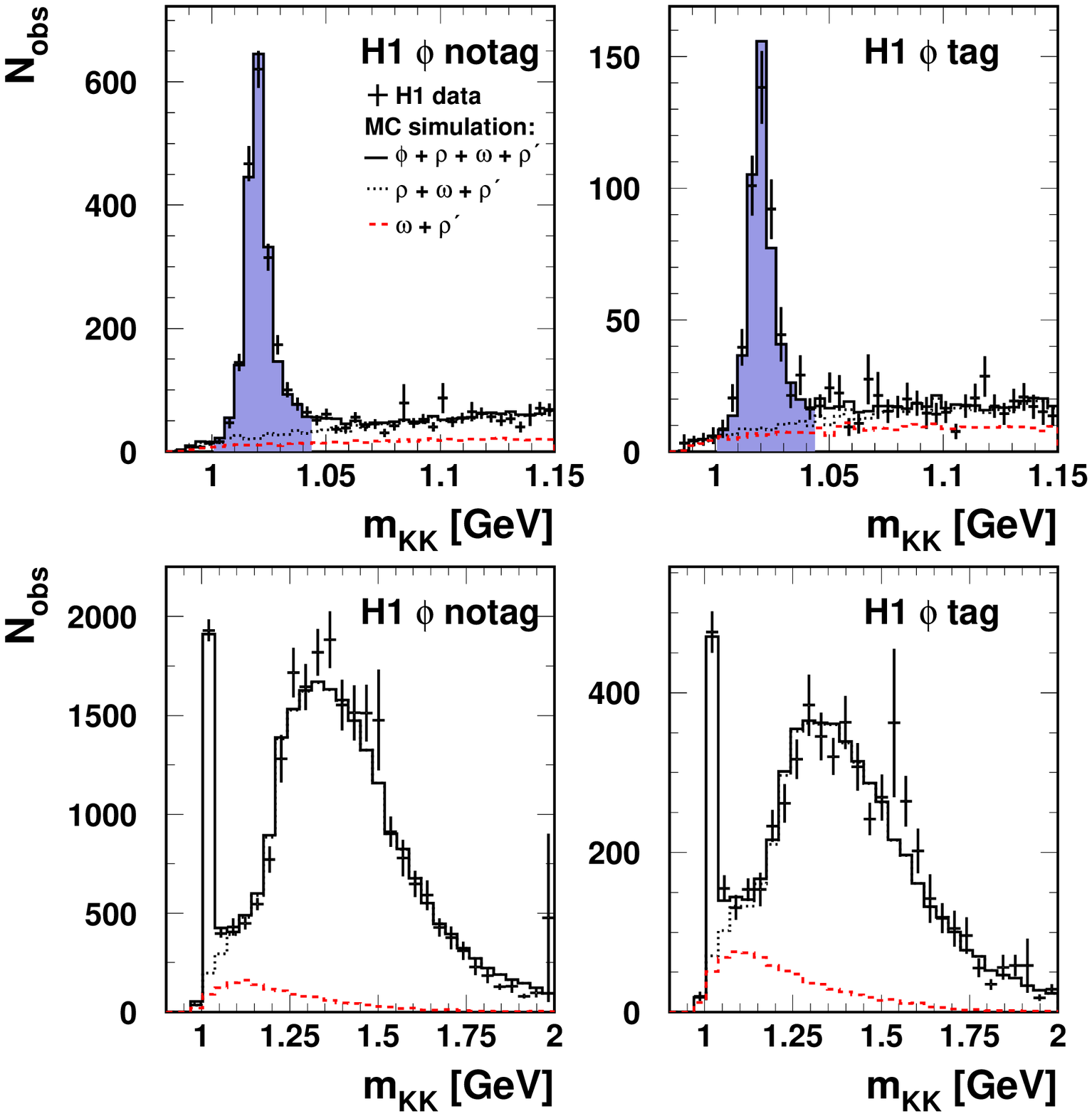  ,height=12.0cm,width=12.0cm}}
\end{picture}
\caption{Distributions of the invariant mass \mkk:
(upper plots) in the \ph\ mass region, for the notag and tag samples separately; 
(lower plots) over an extended mass range, showing  the \ph\ signal and the reflection of 
\rh\ production and the backgrounds.
The dashed histograms show the sum of the \rhoprim, \om\ and 
$\phi \rightarrow~3~\pi$ backgrounds, 
the dotted histograms show in addition the \rh\ and non-resonant $\pi \pi$ backgrounds, 
and the full histograms the $\phi \rightarrow~KK$ signal and the sum of all backgrounds.
In (a) and (b), the mass domain where the cross section measurements are performed is shaded.}
\label{fig:mass_distrib_mkk}
\end{center}
\end{figure}

The \mpp\ and \mkk\ mass distributions are shown in
Figs.~\ref{fig:mass_distrib-mpp-NT} to~\ref{fig:mass_distrib_mkk}, separately for the
notag and tag samples.
The results of the Monte Carlo simulations, comprising signal and backgrounds,
are also shown.
They are reweighted and normalised to the data as described in the previous section.

%-----------------------------------------------------------------------------
%-----------------------------------------------------------------------------

%-----------------------------------------------------------------------------
%-----------------------------------------------------------------------------

%-----------------------------------------------------------------------------
%-----------------------------------------------------------------------------

The mass spectra are presented from threshold to masses well above the actual 
measurement ranges defined in Table~\ref{table:signal_VM}.
The \mpp\ spectra in Figs.~\ref{fig:mass_distrib-mpp-NT} and~\ref{fig:mass_distrib-mpp-T}
are presented in four bins in \modt, with the cut $\mkk > 1.04~\gevcsq$.
The \mkk\ spectra in Fig.~\ref{fig:mass_distrib_mkk} exhibit the reflection of \rh\ 
production and of backgrounds. 

The \mpp\ mass distributions are well described from the
threshold at $2 m_{\pi}$ up to $1.5~\gevcsq$.
The backgrounds are small in the mass ranges selected for the physics analyses, 
shown as the shaded regions in the figures, but their contributions can be 
distinctly identified outside these domains.
In the \mpp\ distributions, they are particularly visible at low mass and,
as expected, they contribute mostly at large \modt, especially in the notag sample with 
$\modt > 0.5~\gevsq$ of Fig.~\ref{fig:mass_distrib-mpp-NT}.
A decrease of the background with increasing \qsq\ for the same ranges in \modt\ is also
observed (not shown here), which is explained by the larger transverse momentum of the 
virtual photon, resulting in larger $p_t$ values of the decay photons which thus pass the
detection threshold and lead to the rejection of the events.

In view of the small \rhop\ background in the final selected samples, an analysis 
of only the mass spectrum, performed in the restricted mass range 
$0.6 \leq \mpp \leq 1.1~\gevcsq$, 
is not sufficient to constrain the \rhop\ contribution.
Controlling this background is crucial for the measurements of the \modt\ slope 
and of the \rzqzz\ matrix element.
In the present analysis, the amount of \rhop\ background is obtained from the 
distribution of the variable \zet\ (defined in section~\ref{sect:rhop_bg}). 
The value determined within the mass range $0.6 \leq \mpp \leq 1.1~\gevcsq$ 
also gives a good description of the mass range $2 m_{\pi} < \mpp < 0.6~\gevcsq$, 
below the actual measurement.
This demonstrates the reliability of the background estimate.

The \mkk\ mass distribution shown in Fig.~\ref{fig:mass_distrib_mkk} is also very 
well described.
The $\pi \pi$ background under the \ph\ peak, which contains a \rhop\ contribution obtained 
from the \rh\ analysis, is small\footnote{For the notag sample with $\modt \leq 0.5~\gevsq$, 
the background under the \ph\ peak amounts to $20.5\%$ for $\qsq = 2.5~\gevsq$
($10\%$ from the $\pi^+ \pi^-$ low mass tail, $3\%$ from $\omega$ and $7.5\%$ from \rhop\ 
production), and to $5.5\%$ for $\qsq = 13~\gevsq$ ($2.5\%$, $0.5\%$ and $2.5\%$, respectively).
An empirical description of the background by ZEUS, using a simple power law shape, 
is in agreement with these detailed findings: it amounts to $18\%$ for $\qsq = 2.5~\gevsq$ 
and $5\%$ for $\qsq = 13~\gevsq$~\cite{z-phi}.}.

%%%%%%%%%%%%%%%%:q%%%%%%%%%%%%%%%%%%%%%%%%%%%%
\subsection{Kinematic and angular distributions}  
                                                  \label{sect:control_plots}
%%%%%%%%%%%%%%%%%%%%%%%%%%%%%%%%%%%%%%%%%%%%

\begin{figure}[htbp]
\begin{center}
\setlength{\unitlength}{1.0cm}
\begin{picture}(18.0,18.0)   
\put(-1.5,0.){\epsfig{file=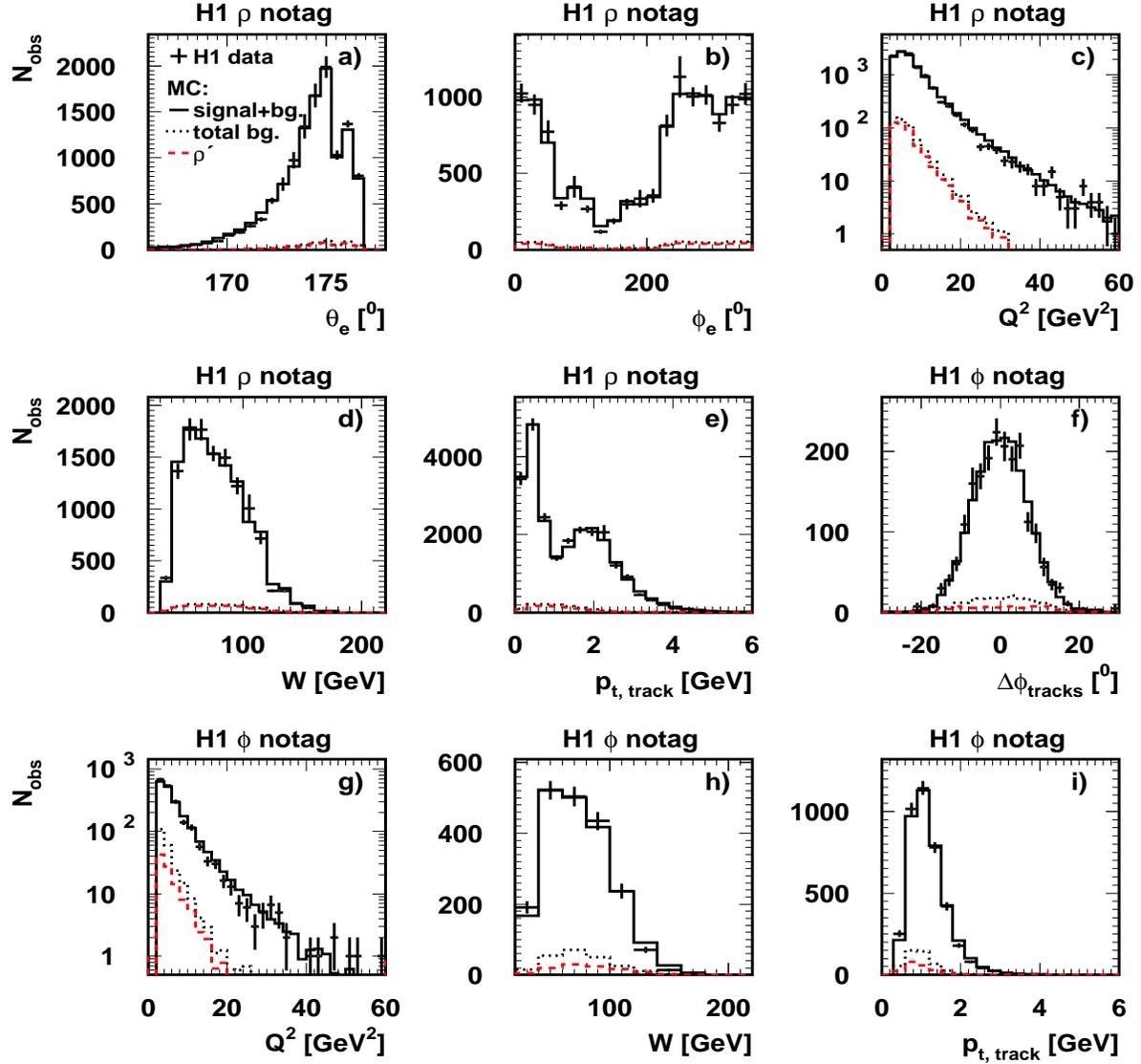  ,height=18.0cm,width=18.0cm}}
\end{picture}
\vspace{-2 cm}
\caption{Distributions 
of the polar angle $\theta_e$~(a) and azimuthal angle $\phi_e$~(b)
of the scattered 
electron, of the \qsq~(c) and $W$~(d) variables, and of the transverse momenta 
of the decay mesons~(e), 
for the \rh\ notag sample;
distributions of the difference between the azimuthal angles $\phi$ of the decay 
kaons~(f) and, in
(g)-(i), of the same observables as in (c)-(e), for the \ph\ notag sample.
In panels (a)-(e), the dashed histograms present the MC predictions for the 
distributions of the \rhoprim\ background, the dotted histograms 
in addition for the \om\ and \ph\ backgrounds, and the
full histograms for the \rh\ signal and the sum of all backgrounds;
in panels (f)-(i), the dashed histograms describe the \rhoprim\ and \om\ backgrounds, 
the dotted histograms in addition the $\pi \pi$ background, and the
full histograms the \ph\ signal and the sum of all backgrounds.}
\label{fig:control-plotsa}
\end{center}
\end{figure}

\begin{figure}[htbp]
\begin{center}
\setlength{\unitlength}{1.0cm}
\begin{picture}(18.0,18.0)   
\put(-1.5,0.0){\epsfig{file=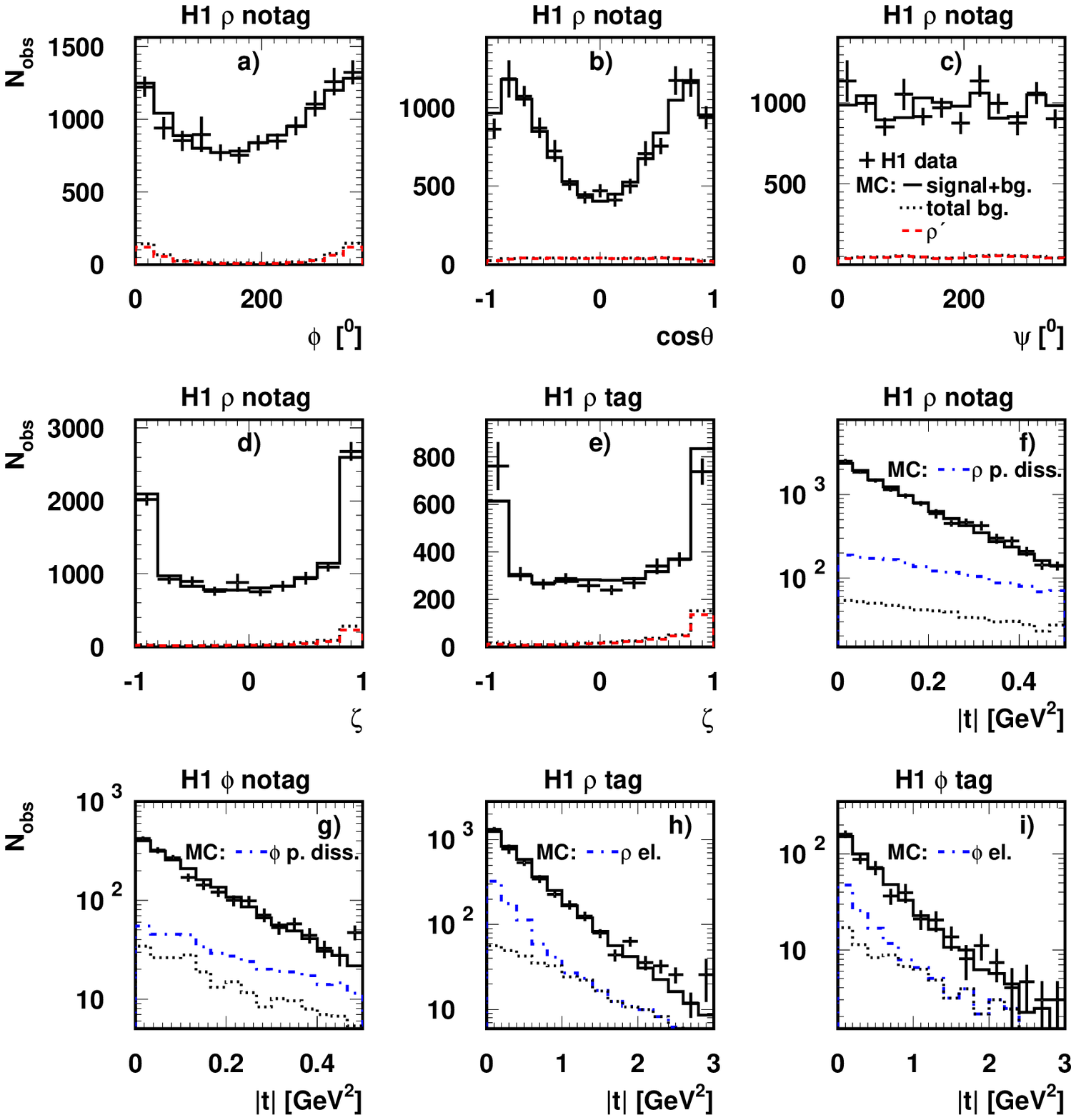  ,height=18.0cm,width=18.0cm}}
\end{picture}
\vspace{-2 cm}
\caption{Distributions 
of the VM production and decay angles $\phi$~(a), \cosths~(b)
and $\psi = \phi -\phib$~(c) for the \rh\ notag sample; 
of the \zet\ variable for the \rh\ notag~(d) and tag~(e) samples;
of the \modt\ variable for the \rh~(f) and \ph~(g) notag samples
and for the tag samples~(h)-(i).
In panels (a)-(e), the dashed histograms present the MC predictions for the 
distributions of the \rhoprim\ background, the dotted histograms show
in addition the \om\ and \ph\ backgrounds, and the
full histograms the \rh\ signal and the sum of all backgrounds;
in panels (f)-(i), the dotted histograms show the sum of the various VM backgrounds 
(\rhoprim, \om, \ph\ or $\rho + \pi \pi$), the dash-dotted histograms show in addition the
diffractive background (proton dissociation in panels (f)-(g) and elastic production 
in panels (h)-(i)), and the full histograms the signal and the sum of all backgrounds.}
\label{fig:control-plotsb}
\end{center}
\end{figure}

Figures~\ref{fig:control-plotsa} and~\ref{fig:control-plotsb} present 
several kinematic and angular variable distributions for the samples selected as
defined in Tables~\ref{table:signal_VM}-\ref{table:signal_kin_range}.
They demonstrate that the simulations, taking into account the detector acceptance
and response and the background contributions, correctly describe the data.

%-----------------------------------------------------------------------------
%-----------------------------------------------------------------------------

%-----------------------------------------------------------------------------
%-----------------------------------------------------------------------------

Figure~\ref{fig:control-plotsa} shows kinematic variable distributions of
the \rh\ and \ph\ notag samples.
The structure observed in the electron polar angle distribution (a) results from the 
different kinematic range selections for the different years.
The dip in the distribution~(b) of the laboratory azimuthal angle~$\phi_e$ of the electron 
is due to an asymmetric electron trigger acceptance.
The $p_t$ distributions of the decay mesons~(e), (i) reflect the VM mass and the  
decay angular distributions.
The good description of the difference between the azimuthal angles of the decay kaons 
in the \ph\ sample~(f) indicates that the reconstruction of pairs of tracks with small
differences in azimuthal angles is under control.
A description of similar quality is obtained for the tag samples.

Figures~\ref{fig:control-plotsb}(a)-(c) present distributions related to the spin density
matrix elements.
The \zet\ distributions (d)-(e) are sensitive to the values of the matrix element 
combinations \rfivecomb\ and \ronecomb\ and to the amount of 
\rhop\ background especially at high \modt\ as discussed in 
section~\ref{sect:rhop_bg}. 
The \modt\ distributions~(f)-(i) are sensitive to the amount of diffractive backgrounds
(proton dissociation for the notag sample, elastic scattering for the tag sample) and to the
values assumed for the exponential $t$ slopes.

%%%%%%%%%%%%%%%%%%%%%%%%%%%%%%%%%%%%%%%%%%%%
\subsection{Systematic errors}  
                                                  \label{sect:syst}
%%%%%%%%%%%%%%%%%%%%%%%%%%%%%%%%%%%%%%%%%%%%

Uncertainties on the detector response and background contributions are listed in 
Table~\ref{table:syst}.
They are estimated by variations in the MC simulations
within the indicated limits, which are in most cases determined from the data.
Global normalisation errors are given separately.

%
%-----------------------------------------------------------------------------
\begin{table}[htbp]
\begin{center}
\begin{tabular}{|c|c|c|c|} 
\hline
  {\bf Uncertainty source}          &   {\bf functional  dependence}    & {\bf VM}   & {\bf variation }        \\
\hline
\hline
   \multicolumn{4}{|c| } {Detector effects}  \\
\hline
  electron polar angle $ \theta_e$  &                                                & all VM       & $ \pm 1 $ mrad          \\
  Spacal energy scale                 &                                                  & all VM      & $ \pm 1\%$                \\
  noise threshold in LAr               &                                                 & all  VM      &  $ \pm 100~\mev$ \\ 
\hline
   \multicolumn{4}{|c| } {Cross section dependences}  \\
\hline
 ${\rm d} \sigma / {\rm d} Q^2$ & $ (Q^2 + M^2)^{-n}$               & all  VM & $n \pm 0.15$                 \\
 ${\rm d} \sigma / {\rm d} W$   &  $ W^{\delta}$                       & all  VM & $  \delta \pm 25\% $      \\
 ${\rm d} \sigma / {\rm d} t  $   &  $ e^{-b |t|}$, $b$ in~\gevsqm  &  \rh      &  el. :         $ b \pm 0.5~\gevsqm$ \\
                                             &                                            &            &  p. diss. : $ b \pm 0.3~\gevsqm$  \\  
                                             &                     &  \ph, \rhop,  $\omega$  &  el. :        $ b \pm 1.0~\gevsqm$  \\
                                             &                                            &            &  p. diss. : $ b \pm 0.7~\gevsqm$   \\ 
\hline
   \multicolumn{4}{|c| } {Backgrounds }  \\
\hline 
 proton dissoc. $/$ elastic          &                                                 & all  VM             &  $ \pm 0.10$  
                                                                                                           $( \approx \pm 20\%) $       \\
\rh\ shape skewing                & $ (m_{\rho}/\mpp)^n $           &  \rh                & $n \pm 0.15$      \\
VM cross sections                &                                                 & $\omega  / \rho $ & $ \pm 0.02$
                                                                                                              $( \approx \pm 20\%) $      \\
                                           &                                                  & $\phi  / \rho$        &  $ \pm 0.03$                  
                                                                                                              $( \approx \pm 15\%) $      \\
                                           &                                                 &  $\rho^\prime  / \rho$  &   $ \pm 0.40$     
                                                                                                              $( \approx \pm 35\%) $     \\
 \rhop\ decay                         & $M_1(00)$ and $M_1(10)$         & \rhop        &        see text            \\
\hline
   \multicolumn{4}{|c| } {\rh\ and \ph\ angular decay distributions}  \\
\hline
 \rzqzz
                                           &    $ f(Q^2) $                               & \rh, \ph      & $\pm 15\% $            \\
 \rfivecomb, \ronecomb\   
                                           &    $ f(\modt) $                             & \rh, \ph     &$\pm 30\% $          \\ 
 \cosdelta\                       
                                           &                                                  &  \rh, \ph     &$\pm 0.05 $           \\
\hline
\hline
   \multicolumn{4}{|c| } {Global normalisation}  \\
\hline
  luminosity                             &                                                  & all VM     &     $\pm 1.5\% $       \\
  trigger efficiency                  &                                                   & all  VM    &    $ \pm 1.0\% $       \\
  track rec. eff.  (per track) &                                                & all VM     &     $ \pm 2\%  $      \\   
width of rel. B.-W.                 &         see text                              &   \rh          &    $\pm 2\% $       \\ 
$\phi \rightarrow K K$ $\cal{BR}$   &          see~\cite{pdg}                   &   \ph         &    $\pm 1.2\%$          \\
$\pi \pi$ under \ph\ peak ($\kappa$ param.)  &                    &   \ph                   &     $ \pm 100\%$   \\
${\rm d}\sigma / {\rm d}M_Y^2$ &    $ 1/M_Y^{2n} $                    & all p. diss. & $n \pm 0.15$  \\
\hline
\end{tabular} 
\caption{Variations in MC simulations for the estimation of systematic 
uncertainties. Numbers between parentheses indicate the relative variations.} 
  \label{table:syst}
  \end{center}
  \end{table}
%-----------------------------------------------------------------------------

The error on the electron polar angle $\theta_e$, which affects the \qsq\ measurements and the 
acceptance calculations, is due to the uncertainty on the absolute positioning
of the BDC with respect to the CJC chambers, the uncertainty on the electron beam direction
in the interaction region and the error on the $z$ position of the interaction vertex. 

The uncertainty on the energy scale of the Spacal calorimeter affects the cross section 
measurements through the electron energy threshold of $17~\gev$ and the $\Sigma (E-p_z)$ cut.

The uncertainty on losses due to the rejection of events affected by noise in the LAr 
calorimeter or containing energy deposits unrelated to the diffractive event is estimated by 
varying the energy threshold, both in the data and in the simulation (where data taken from
random triggers are directly superimposed to the simulated events).

The uncertainties on the simulated cross section dependences on \qsq, $W$ and \modt\ affect 
the bin-to-bin migrations and the extrapolations from the average value of the kinematic variables 
in a bin to the position where they are presented (``bin centre corrections").

An absolute error of $\pm 0.10$ is used for the ratio of the proton dissociative (with 
$M_Y < 5~\gevcsq$) to elastic cross sections, which corresponds to about $20\%$ relative 
error.
It is estimated by varying by $\pm 0.15$ the parameter $n$ in the simulated dissociative 
mass distribution ${\rm d} \sigma / {\rm d} M_Y^2 \propto 1/M_Y^{2n}$, by varying the 
slope parameters of the exponential \modt\ distributions of elastic and proton dissociative 
events within the experimental limits, and by calculating the cross section ratio using only 
the PRT or only the FMD.
The latter covers uncertainties in the inefficiencies of these detectors.

For the \rh\ cross section measurements, the error due to the extraction of the non-resonant 
$\pi \pi$ background is estimated through the variation of the \qsq\ dependent skewing 
parameter $n$ of the Ross-Stodolsky parameterisation of Eq.~(\ref{eq:rs}).

The errors on the various cross section ratios are taken from the present analysis 
for the \ph\ to \rh\ and \rhop\ to \rh\ ratios, and 
from the ZEUS measurements of the $\omega / \rho$ ratio~\cite{z-omega}.

The errors due to the uncertainty on the \rhop\ decay angular distribution are estimated by
considering the two extreme cases $|M_1(00)|^2 = 1, |M_1(10)|^2 = 0$ and 
$|M_1(00)|^2 = 0, |M_1(10)|^2 = 1$ of the pair of variables defined in~\cite{caro}.  

The uncertainty on the angular distributions are described by varying the values of
the matrix element \rzqzz\ (for the angle $\theta$), 
of the combinations \rfivecomb\ and \ronecomb\ (for the angle \ph) 
and of the \cosdelta\ parameter (for the angle $\psi = \ph -\phib $).

The uncertainty on the choice of the momentum dependent width of \rh\ mesons results 
in normalisation uncertainties of~$2\%$ (see section~\ref{sect:mass_shapes}).

For \ph\ production, the uncertainty on the $\pi \pi$ background under the signal is estimated 
by varying the parameter $\kappa$ globally from~$0$ to~$3$ (Eq.~(\ref{eq:low-mpp}) in 
section~\ref{sect:MC}), leading to a normalisation error of $\pm 3\%$ on the cross 
section measurements.

For the proton dissociative cross sections, the error on the correction for the smearing 
through the experimental cut $M_Y < 5~\gevcsq$ is estimated by varying the parameter 
$n$ of the $M_Y$ distribution (${\rm d}\sigma / {\rm d}M_Y^2 \propto 1/M_Y^{2n}$, with 
$n \pm 0.15$), which leads to an additional normalisation error of $\pm 2.4\%$ on the 
proton dissociative cross section measurement.

The uncertainties on the luminosity measurement, on the triggers and on the track 
reconstruction efficiency are assumed to affect globally the normalisation only.

Systematic errors due to limited MC statistics are negligible compared to the
statistical precision of the measurements (the generated
samples correspond to at least ten times the data integrated luminosity). 

All systematic errors on the measurements presented in the rest of this paper are calculated 
from separate quadratic sums of positive and negative effects of the variations listed in
Table~\ref{table:syst}.
In all figures, measurements are shown with statistical errors (inner error bars) and statistical 
and systematic errors added in quadrature (full error bars). 
In tables, the errors are given separately: first the statistical, second the systematic errors.
Overall normalisation errors are not included in the error bars but are quoted in the relevant 
captions.

%%%%%%%%%%%%%%%%%%%%%%%%%%%%%%%%%%%%%%%%%%%%
%%%%%%%%%%%%%%%%%%%%%%%%%%%%%%%%%%%%%%%%%%%%
%%%%%%%%%%%%%%%%%%%%%%%%%%%%%%%%%%%%%%%%%%%%
%%%%%%%%%%%%%%%%%%%%%%%%%%%%%%%%%%%%%%%%%%%%

%\newpage

%%%%%%%%%%%%%%%%%%%%%%%%%%%%%%%%%%%%%%%%%%%%
\section{Cross Section Results}  
                                                  \label{sect:cross_sections}
%%%%%%%%%%%%%%%%%%%%%%%%%%%%%%%%%%%%%%%%%%%%

In this section, measurements of the \rh\ and \ph\ line shapes are presented first.
The elastic and proton dissociative cross sections are then measured as a function of \qsq\ 
(total and polarised cross sections), $W$ and $t$ (total cross sections); 
results for different VMs are compared.
Finally, elastic and proton dissociative scatterings are compared, including tests of
proton vertex factorisation.
Model predictions are compared to the data.

%%%%%%%%%%%%%%%%%%%%%%%%%%%%%%%%%%%%%%%%%%%%
\subsection{Measurement of cross sections}  
                                                  \label{sect:cross_sections_meas}
%%%%%%%%%%%%%%%%%%%%%%%%%%%%%%%%%%%%%%%%%%%%

The cross sections for \rh\ and \ph\ production presented in this paper are extracted from 
the numbers of events in the mass ranges $0.6 \leq \mpp \leq 1.1~\gevcsq$ and 
$1.00 \leq \mkk \leq 1.04~\gevcsq$, respectively. 
They are corrected for all backgrounds, including for \rh\ mesons the non-resonant 
dipion diffractive production (see section~\ref{sect:mass-rho}).
They include all corrections for detector acceptance and response.
When quoted at a fixed value of a kinematic variable, the cross sections are evolved
from the average value in the bin using dependences measured in this analysis.

The cross sections are quoted for the full resonance mass range from the two particle 
threshold up to the nominal mass plus five times the resonance width:
\begin{eqnarray}
   2 \ m_{\pi} \leq \mpp \leq \mrho + 5~\grho\ \simeq 1501~\mevcsq, \nonumber   \\ 
   2 \ m_{K} \leq \mkk \leq m_\phi + 5~\Gamma_{\phi}~\simeq 1041~\mevcsq.
                                                                                           \label{eq:window}
\end{eqnarray}
For \ph\ mesons, the cross sections take into account the branching ratio to the 
$K^+K^-$ channel.

Elastic and proton dissociative cross sections are given at the Born level (i.e. they are
corrected for QED radiation effects) in terms 
of $\gamma^{\star} p$ cross sections
(except for the mass shapes, which are given in terms of $ep$ cross sections). 
The $\gamma^{\star} p$ cross sections are extracted from the $ep$ cross sections in the 
Weizs\"acker-Williams equivalent photon approximation~\cite{WW} using 
the definition
\begin{equation}
  \sigma (\gamma^* + p \rightarrow V + Y) = \frac {1} {\Gamma}  \cdot
   \frac { {\rm d}^2 \sigma (e + p \rightarrow e + V + Y)}   
          { {\rm d} y \  {\rm d} \qsq}                                        \label{eq:gammastarp}
\end{equation}
where the flux $\Gamma$ of virtual photons~\cite{hand} and the inelasticity $y$ are given by
\begin{equation}
   \Gamma = \frac {\alpha_{em}} {\pi } \ \frac {1 - y + y^2 / 2 } { y \ \qsq} ,
   \ \ \ \ \ \ \ \     y = \frac {p \cdot q} {p \cdot k},                 
                                                                                       \label{eq:gamma}
\end{equation}
$\alpha_{em}$ being the fine structure constant and $p$ and  $k$ the four-momenta of 
the incident proton and electron, respectively.

%%%%%%%%%%%%%%%%%%%%%%%%%%%%%%%%%%%%%%%%%%%%
%%%%%%%%%%%%%%%%%%%%%%%%%%%%%%%%%%%%%%%%%%%%
\subsection{Vector meson line shapes}  
                                                                                     \label{sect:mass_shapes}
%%%%%%%%%%%%%%%%%%%%%%%%%%%%%%%%%%%%%%%%%%%%

The distribution of the invariant mass $m$ of the VM decay particles is analysed assuming 
the relativistic Breit-Wigner distribution $BW(m)$ with momentum dependent width 
$\Gamma(m)$~\cite{jackson}:
\begin{eqnarray}
    BW(m) &=& \frac {m\ \mv\ \Gamma(m)}
               {(\mvsq - m^2)^2 + \mvsq \ \Gamma(m)^2},               
                                                                                         \label{eq:b_w} \\
\Gamma(m) &=& \Gamma_{V} \ \left( \frac {q^*} {q_0^*} \right)^3 \
        \frac {\mv} {m},
                                                                                         \label{eq:GJ}
\end{eqnarray}
where \mv\ and $\Gamma_{V}$ are the nominal VM resonance mass and width, $q^*$ 
is the momentum of the decay particles in the rest frame of the pair with 
mass $m$, and 
$q^*_0$ is the value taken by $q^*$ when $m = \mv$.

For \rh\ mesons, the mass extrapolation from the measurement domain 
$0.6 \leq \mpp \leq 1.1~\gevcsq$ to the full range given by Eq.~(\ref{eq:window}), including the 
correction for skewing effects, implies a correction factor 
of $1.15$ with a systematic error of $2\%$ due to the theoretical uncertainty on the choice 
of the momentum dependent width~\cite{jackson}.
For \ph\ production, a very small extrapolation outside the measurement domain is 
required, with negligible related error.

%%%%%%%%%%%%%%%%%%%%%%%%%%%%%%%%%%%%%
\boldmath
\subsubsection{\rh\ mesons}
                                                                                \label{sect:mass-rho}
\unboldmath
%%%%%%%%%%%%%%%%%%%%%%%%%%%%%%%%%%%%%

\begin{figure}[htbp]
\begin{center}
\setlength{\unitlength}{1.0cm}
\begin{picture}(12.0,12.0)   
\put(0.0,0.0){\epsfig{file=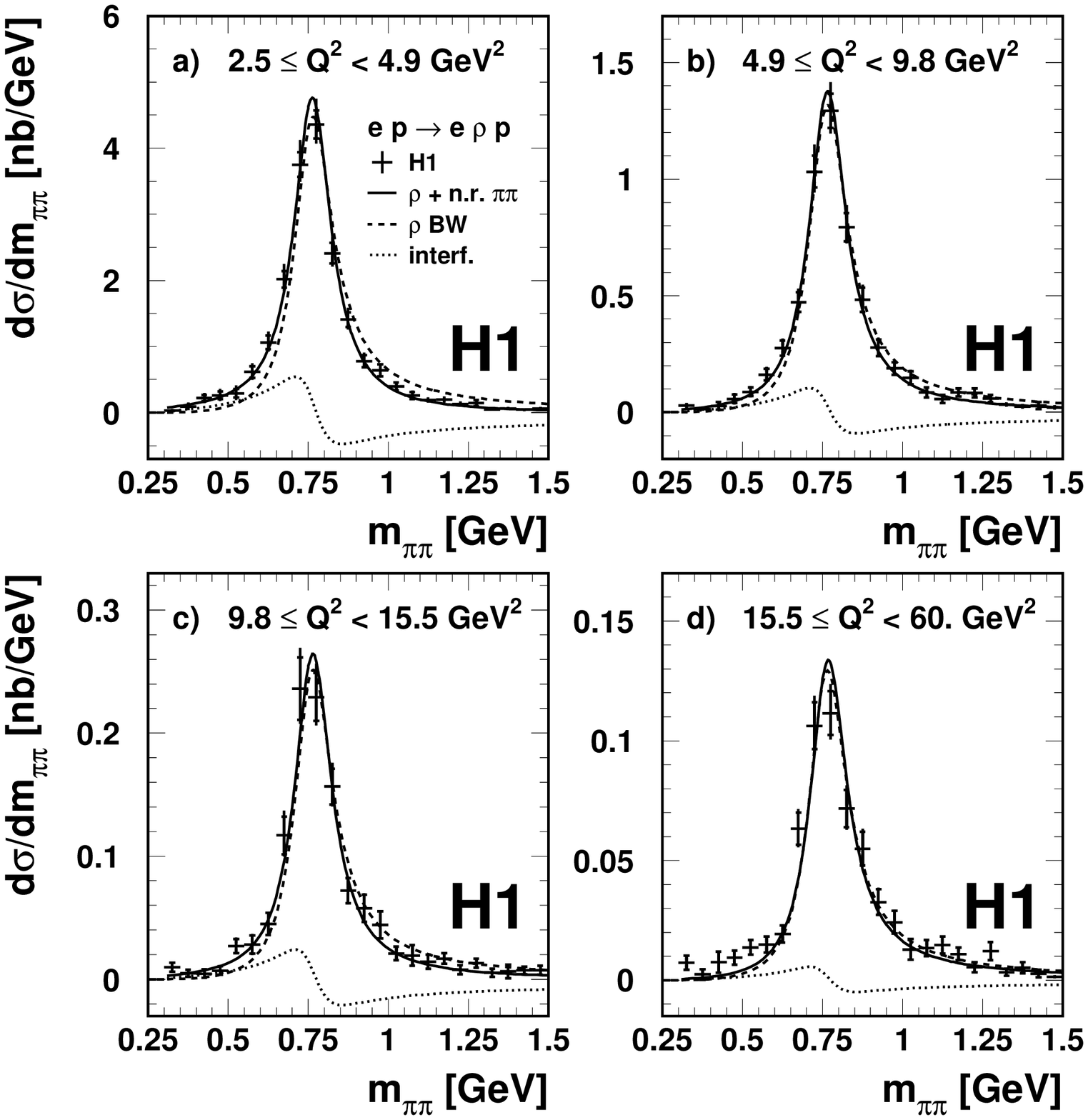  ,height=12.0cm,width=12.0cm}}
\end{picture}
\caption{Distributions of the \mpp\ mass for elastic \rh\ production
with $\modt < 0.5~\gevsq$, expressed as $ep$ cross sections, after 
experimental corrections and background subtraction,
for four ranges in \qsq\ and in the $W$ domains defined
in Table~\ref{table:signal_kin_range}.
The solid curves show the results of fits to the data in the mass range 
$0.6 \leq \mpp \leq 1.1~\gevcsq$ of the 
relativistic Breit-Wigner function with momentum dependent width defined in 
Eqs.~(\protect\ref{eq:b_w}-\protect\ref{eq:GJ}), with skewing of the mass distribution 
following the S\"{o}ding parameterisation given by Eq.~(\protect\ref{eq:sod}); 
the dashed curves correspond to a non-skewed relativistic Breit-Wigner function and the dotted 
curves to the interference between resonant and non-resonant amplitudes.}
\label{fig:rh_mass_sod}
\end{center}
\end{figure}

Distributions of the \mpp\ mass in the range $2\ m_\pi \leq \mpp \leq 1.5~\gevcsq$, 
with the cut $\mkk > 1.04~\gevcsq$ applied to suppress the \ph\ signal at low mass, 
are shown in Fig.~\ref{fig:rh_mass_sod} for elastically produced events in four
ranges in  \qsq, after subtraction of the proton dissociative, \ph, \om\ and \rhop\ 
backgrounds and corrections for detector and QED radiation effects.
The mass resolution in the \rh\ mass range, determined with the MC simulation, is 
about $10~\mevcsq$.

%-----------------------------------------------------------------------------
%-----------------------------------------------------------------------------

\paragraph {Skewing}

The mass distributions are skewed towards small masses, especially at low \qsq.
According to S\"{o}ding's analysis~\cite{soding}, this is due to the interference of
the \rh\ meson with background from $p$-wave Drell-type non-resonant $\pi \pi$ pair production, 
with positive interference for $\mpp < \mrho$ and negative interference for $\mpp > \mrho$.

Following one of the forms of skewing proposed in~\cite{z-rho-photoprod}, the \rh\ mass 
shape is described as
\begin{equation}
\frac {{\rm d}N(\mpp)} {{\rm d}\mpp} \propto 
      \left |  \frac { \sqrt{\mpp \ \mrho \ \Gmpp} } 
                             {\mrhosq - \mppsq + i\ \mrho \Gmpp } + \frac {f_I} {2} \right |^2,                    
                                                                                          \label{eq:sod}
\end{equation}
where resonant and non-resonant $\pi \pi$ production are supposed to be in 
phase.
The  interference is proportional to $f_I$, which is taken to be independent of the
\mpp\ mass; the very small purely non-resonant contribution is given by $f_I^2 / 4$.
Figure~\ref{fig:rh_mass_sod} shows that the \rh\ mass shape is well described 
by Eqs.~(\ref{eq:b_w}-\ref{eq:sod}) over the full range $2 m_{\pi} \leq \mpp \leq 1.5~\gevcsq$, 
with the skewing parameters fitted in the range $0.6 \leq \mpp \leq 1.1~\gevcsq$.
No indication is found for significant additional backgrounds, also outside the mass
domain used for the measurements. 
The \rh\ skewing effect is also often conveniently parameterised in the form proposed 
by Ross and Stodolsky~\cite{rs}, given by Eq.~(\ref{eq:rs}).

For a fit over the whole \qsq\ range with the parameterisation of Ross and Stodolsky, 
the values of the resonance mass and width are 
$769 \pm 4~{\rm (stat.)}~\mevcsq$
and $162 \pm 8~{\rm (stat.)}~\mevcsq$,
respectively.
The  S\"{o}ding parameterisation gives similar values, with larger errors.
This is in agreement with the world average values as obtained in photoproduction~\cite{pdg}:
$m_{\rho} = 768.5 \pm 1.1~\mevcsq$ and $\Gamma_{\rho} = 150.7 \pm 2.9~\mevcsq$.
Within errors, no difference is observed between the elastic and proton dissociative 
samples.

%-----------------------------------------------------------------------------
\begin{figure}[htbp]
\begin{center}
\setlength{\unitlength}{1.0cm}
\begin{picture}(10.0,5.0)   
\put(0.0,0.0){\epsfig{file=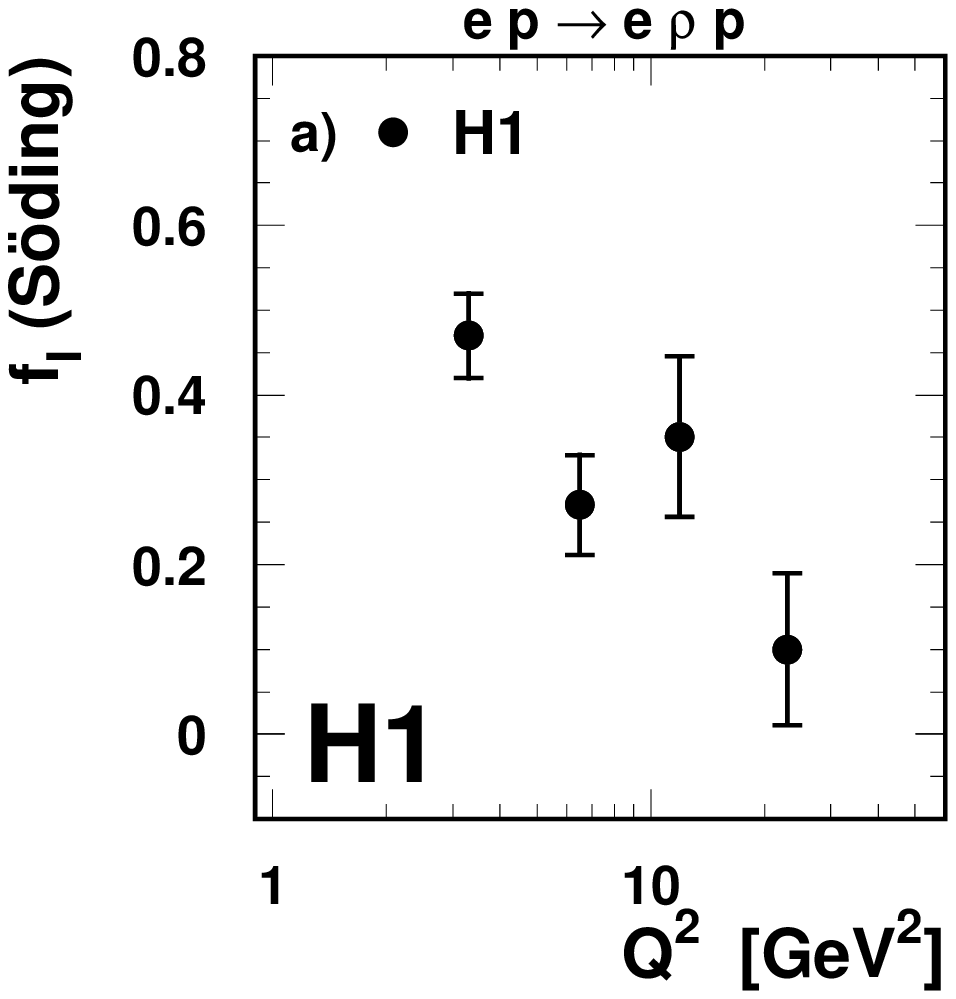,height=5.0cm,width=5.0cm}}
\put(6.0,0.0){\epsfig{file=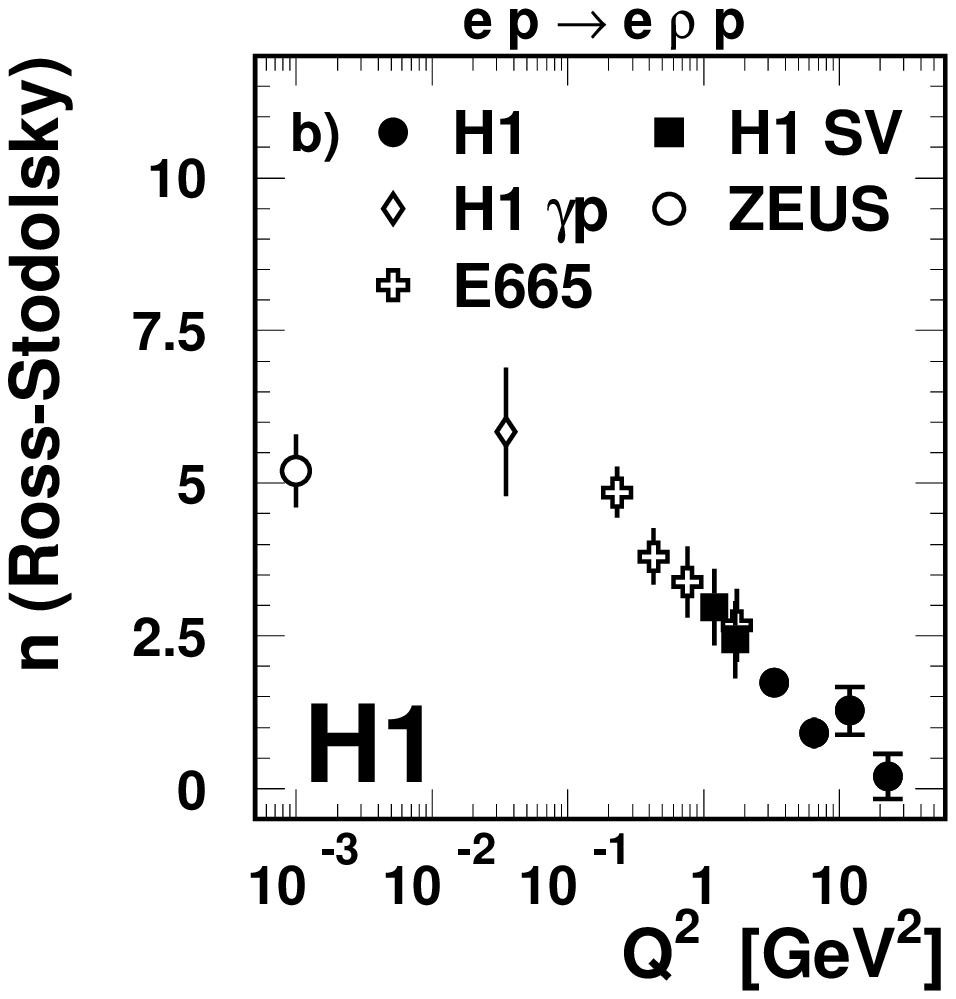,height=5.0cm,width=5.0cm}}
\end{picture}
\caption{\qsq\ dependence 
(a) of the S\"{o}ding skewing parameter $f_I$ defined in Eq.~(\ref{eq:sod});
(b) of the Ross-Stodolsky parameter $n$ defined in Eq.~(\ref{eq:rs}),
for \rh\ elastic production.
Measurements from H1~\protect\cite{h1-rho-photoprod-94} and 
ZEUS~\protect\cite{z-rho-photoprod} in photoproduction and
E665~\protect\cite{E665} 
in electroproduction are also shown.
The present measurements are given in Table~\ref{table:skew}.
}
\label{fig:skew}
\end{center}
\end{figure}
%-----------------------------------------------------------------------------

Figure~\ref{fig:skew} presents the \qsq\ dependence of the fitted values of the skewing 
parameters for elastic \rh\ production\footnote{The 
values of the parameters $f_I$ and $n$ slightly depend on the fit mass range.
At low mass, this is related to the shape uncertainties reflected by the uncertainty in the
parameterisation of Eq.~(\ref{eq:low-mpp}). 
For higher masses, the mass limit dependence may be due to additional interference 
of \rh\ mesons with heavier (\rhop) resonances~\cite{ryskin-shab-phi-rhop}.}, the mass and width of the resonance being fixed 
to the PDG values~\cite{pdg}.
The skewing effects decrease with increasing \qsq, showing that the non-resonant amplitude 
decreases faster with \qsq\ than the resonant amplitude, as expected on theoretical
grounds~\cite{ryskin-shab-skewing}.
No significant dependence of the skewing parameters is observed as a function of $W$ 
or \modt.

%%%%%%%%%%%%%%%%%%%%%%%%%%%%%%%%%%%%%
\boldmath
\subsubsection{\ph\ mesons}
                                                                                \label{sect:mass-phi}
\unboldmath
%%%%%%%%%%%%%%%%%%%%%%%%%%%%%%%%%%%%%

%-----------------------------------------------------------------------------
\begin{figure}[p]
\begin{center}
\setlength{\unitlength}{1.0cm}
\begin{picture}(6.0,6.0)   
\put(0.0,0.0){\epsfig{file=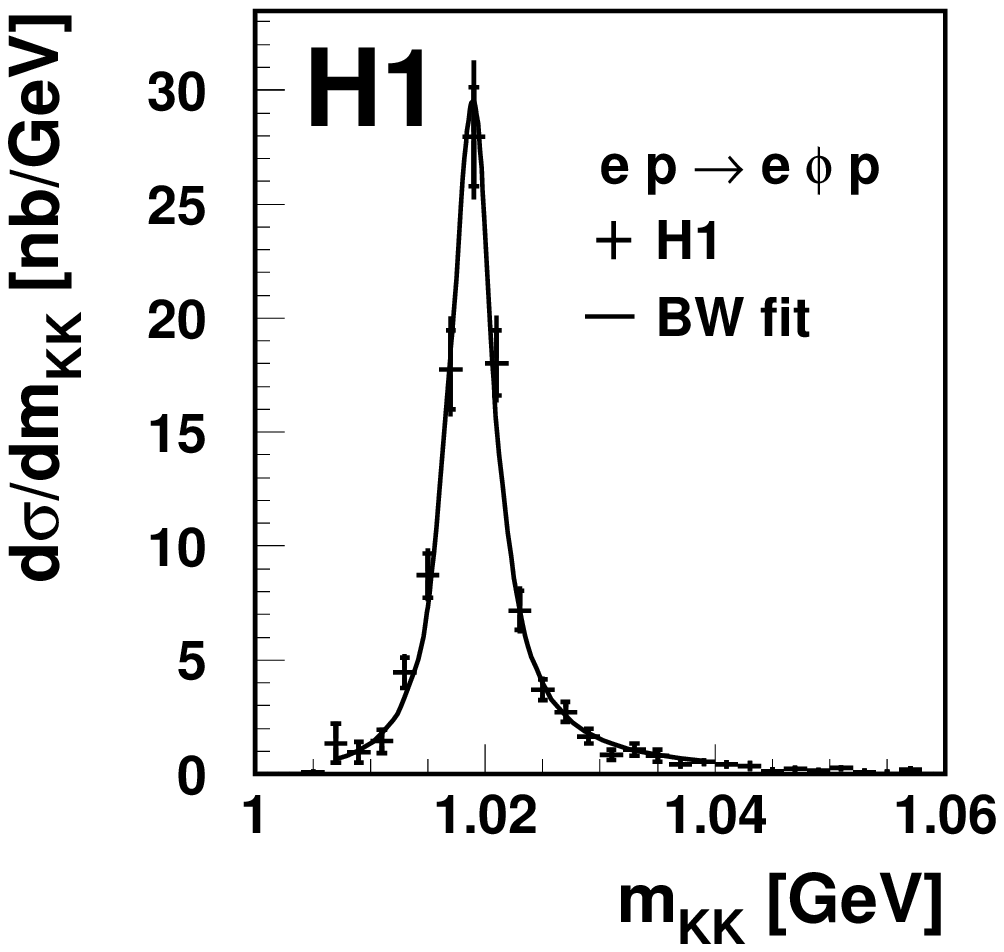  ,height=6.0cm,width=6.0cm}}
\end{picture}
\caption{Distribution of the \mkk\ mass for elastic \ph\ production
with $\modt < 0.5~\gevsq$, expressed as $ep$ cross section, after experimental 
corrections and background subtraction,
for the \qsq\ and $W$ domains defined in Table~\ref{table:signal_kin_range}.
The solid curve shows the result of a fit to the data in the mass range 
$1.00 \leq \mkk \leq 1.04~\gevcsq$ of a relativistic Breit-Wigner function with 
momentum dependent width defined in Eqs.~(\protect\ref{eq:b_w}-\protect\ref{eq:GJ}),
convoluted with the experimental resolution.}
\label{fig:ph_mass}
\end{center}
\end{figure}
%-----------------------------------------------------------------------------

The mass distribution for elastically produced kaon pairs is shown in Fig.~\ref{fig:ph_mass}, 
after background subtraction and corrections for detector and QED radiation effects.
It is described by the convolution of the Breit-Wigner function defined by 
Eqs.~(\ref{eq:b_w}-\ref{eq:GJ}) with a Gaussian function of width 
$\sigma = 2~\mevcsq$ describing the mass resolution, as evaluated using the 
MC simulation.
The mass and width of the resonance, fitted over the interval 
$1.006 \leq \mkk \leq 1.040~\gevcsq$, are 
$1018.9 \pm 0.2~{\rm (stat.)}~\mevcsq$ and 
$3.1 \pm 0.2~{\rm (stat.)}~\mevcsq$, respectively, 
reasonably close to the world average values of 
$1019.46 \pm 0.02~\mevcsq$ and $4.26 \pm 0.04~\mevcsq$~\cite{pdg}.
Conversely, when the \ph\ mass and width are fixed to the nominal values the fitted 
resolution, which is assumed to be Gaussian, is $1.0 \pm 0.1~\mevcsq$. This
value, which is slightly smaller 
than that obtained from simulations, is interpreted as to come from small systematic effects.
As expected~\cite{ryskin-shab-phi-rhop}, no indication is found for skewing effects 
due to interference with non-resonant $K^+ K^-$ production.

%%%%%%%%%%%%%%%%%%%%%%%%%%%%%%%%%%%%%%%%%%%%
%%%%%%%%%%%%%%%%%%%%%%%%%%%%%%%%%%%%%%%%%%%%
\boldmath
\subsection{\qsq\ dependence of the total cross sections}  
                                                                                \label{sect:sigma_tot_f_qsq}
\unboldmath
%%%%%%%%%%%%%%%%%%%%%%%%%%%%%%%%%%%%%%%%%%%%

%%%%%%%%%%%%%%%%%%%%%%%%%%%%%%%%%%%%%
\subsubsection{Cross section measurements}
                                                                                \label{sect:qsq-meas}
%%%%%%%%%%%%%%%%%%%%%%%%%%%%%%%%%%%%%

%-----------------------------------------------------------------------------
\begin{figure}[p]
\begin{center}
\setlength{\unitlength}{1.0cm}
\begin{picture}(16.0,8.0)   
\put(0.0,0.0){\epsfig{file=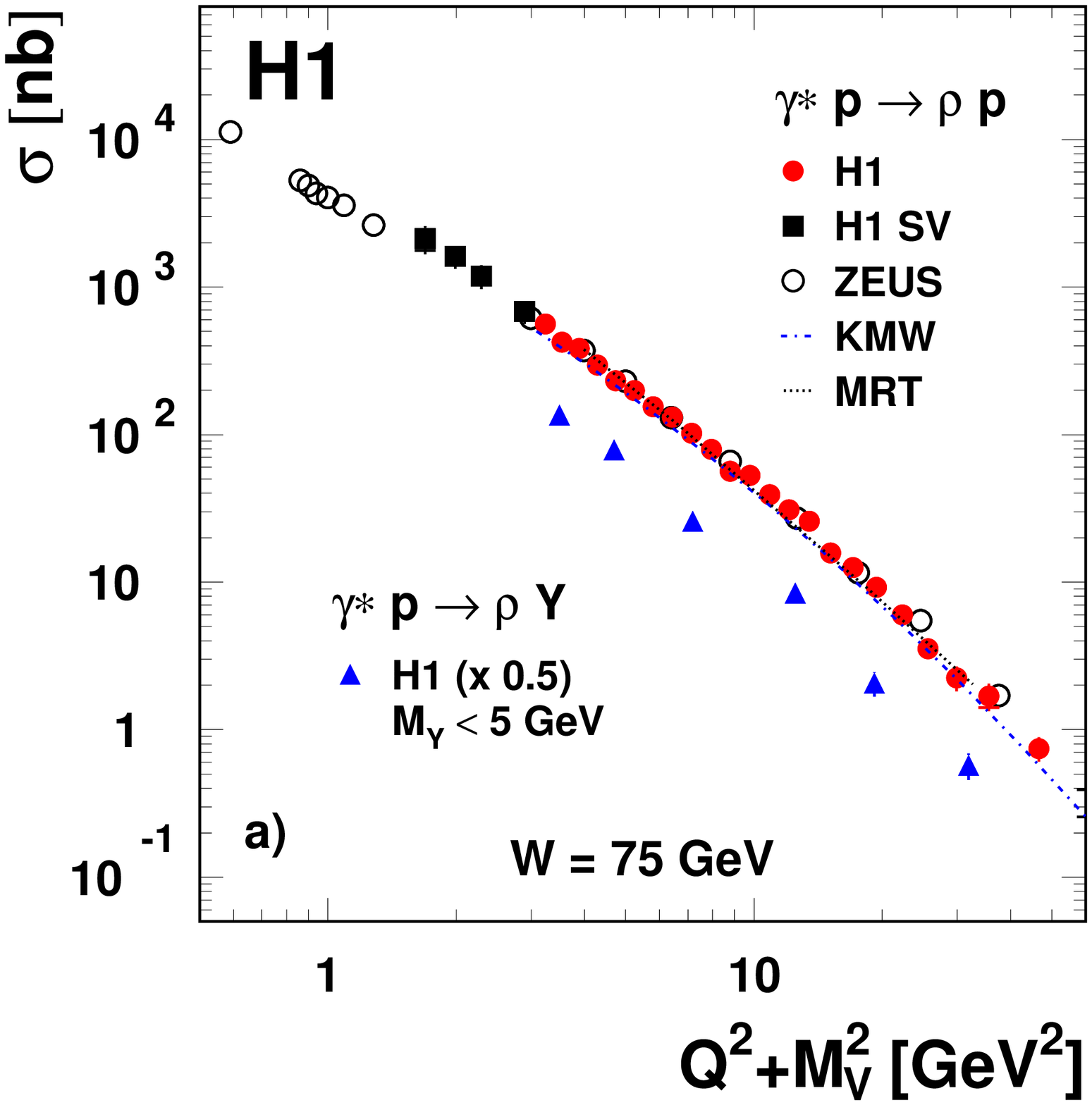,height=8.0cm,width=8.0cm}}
\put(8.0,0.0){\epsfig{file=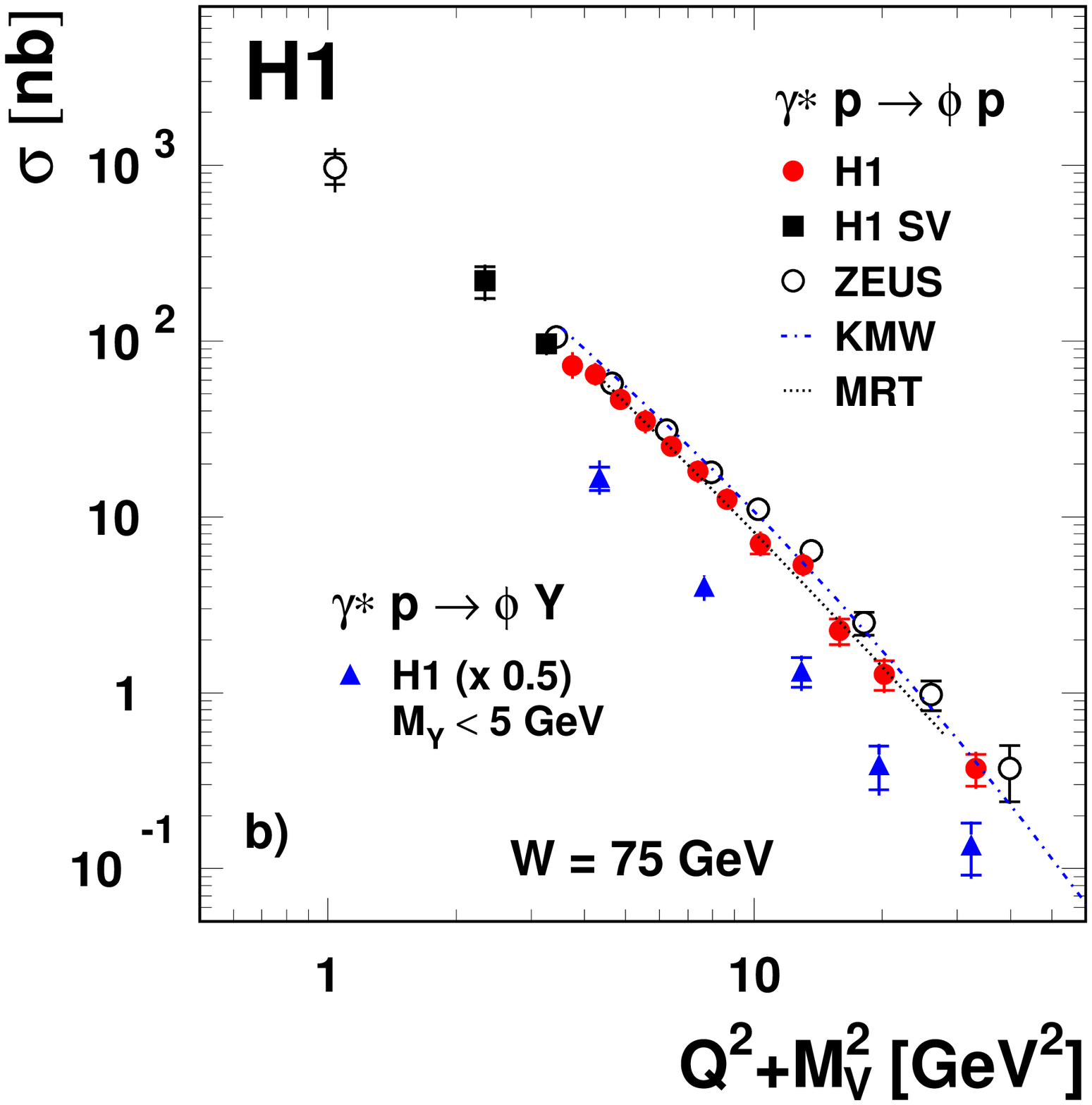,height=8.0cm,width=8.0cm}}
\end{picture}
\caption{\qsqplmsq\ dependence of the $\gstarp$ cross sections for $W = 75~\gev$: 
(a) \rh\ meson production; (b) \ph\ production. 
The upper points are for the elastic processes, the lower points for proton dissociative 
diffraction, divided by a factor~$2$ to improve the readability of the figures.
Overall normalisation errors of~$3.9\%$ ($4.6\%$) for elastic (proton dissociative) \rh\ 
production and~$4.7\%$ ($5.3\%$) for \ph\ production are not included in the error bars.
ZEUS measurements~\protect\cite{z-rho-photoprod,z-rho-older,z-rho,z-phi-photoprod,z-phi}
are also presented;
when needed, they were translated to $W = 75~\gev$ using the measured $W$ 
dependence.
The superimposed curves are from the KMW model~\protect\cite{kmw} with 
GW saturation~\protect\cite{GBW} (dash-dotted lines)
and from the MRT model~\protect\cite{mrt} with 
CTEQ6.5M PDFs~\protect\cite{cteq} (dotted lines).
The present measurements are given in 
Tables~\ref{table:rhoel_xsq2}-\ref{table:phipd_xsq2}.
}
\label{fig:sigma_f_qsqplmsq}
\end{center}
\end{figure}
%-----------------------------------------------------------------------------

The measurements of the $\gamma^*p$ cross sections for \rh\ and \ph\ meson elastic and 
proton dissociative production are presented in 
Fig.~\ref{fig:sigma_f_qsqplmsq} as a function of the scaling variable \qsqplmsq.
They are quoted for $W = 75~\gev$ using the $W$ dependences
parameterised as a function of \qsq\ following the measurements of
section~\ref{sect:W-dep_hardening}.
Using the fits of the \qsq\ dependence presented below, it is verified that the 
normalisations of the 1995 (SV) cross section measurement~\cite{h1-rho-95-96} 
and of the present measurement are in good agreement for \rh\ mesons (the 
ratio is $1.01 \pm 0.10$). For the \ph\ data, the 1995 SV measurement is 
slightly lower than extrapolated from the present result 
(the ratio is $0.84 \pm 0.11$).
This difference is attributed to the different treatments of the backgrounds.
ZEUS measurements of \rh\ and \ph\ electroproduction 
are also shown in Fig.~\ref{fig:sigma_f_qsqplmsq}.
Whereas the \rh\ measurements agree well, \ph\ measurements of ZEUS are a factor
$1.20$ above the present data.
When an improved estimation of the proton dissociative background, investigated for the latest 
ZEUS \rh\ production study~\cite{z-rho}, is used to subtract this background in 
their \ph\ analysis, the cross section ratio of the two experiments is reduced to $1.06$, 
which is within experimental errors~\cite{alessia}.

%-----------------------------------------------------------------------------
\renewcommand{\arraystretch}{1.15}
\begin{table}[htb]
\begin{center}
\begin{tabular}{|c|c|c|} 
\multicolumn{3}{c} { }  \\
\hline
                           \multicolumn{3}{|c| } { {\bf (a) {\boldmath $n$}   constant} }  \\
\hline
                              & \rh\ el.                                      & \rh\ p. diss.                   \\
\hline
$n$                        & $2.37 \pm 0.02~^{+0.06}_{-0.06}$      &$2.45 \pm 0.06~^{+0.10}_{-0.09}$  \\   
\chisq\                    &  $40.4 / 25$                                      &    $13.7 / 4$                            \\
\hline
                              &    \ph\ el.                                  & \ph\ p. diss.                    \\
\hline
$n$                         & $2.40 \pm 0.07~^{+0.07}_{-0.07}$     & $2.40 \pm 0.31~^{+0.14}_{-0.10}$   \\
\chisq\                    &     $11.3 / 13$                                   &   $0.67 / 3$                             \\
\hline
\hline
                                                  \multicolumn{3}{|c| } { {\bf (b)  
                                                                             {\boldmath $n = c_1 + c_2\ \qsqplmsq$} } } \\
\hline
                               & \rh\ el.                                    & \rh\ p. diss.                             \\
\hline
$c_1$                      & $2.09 \pm 0.07~^{+0.06}_{-0.07}$    &          $2.18 \pm 0.23~^{+0.13}_{-0.12}$  \\  
$c_2$   ($10^{-2}~\gevsqm$) & $0.73 \pm 0.18~^{+0.09}_{-0.08}$ &  $0.72 \pm 0.60~^{+0.12}_{-0.08}$    \\
\chisq\                      & $17.1 / 24$                                     &    $8.0 / 3 $                                   \\
\hline
                                &    \ph\ el.                                & \ph\ p. diss.                     \\
\hline
$c_1$                        &     $2.15 \pm 0.14~^{+0.10}_{-0.11}$ &   $2.45 \pm 0.52~^{+0.29}_{-0.20}$     \\
$c_2$   ($10^{-2}~\gevsqm$)  & $0.74 \pm 0.40~^{+0.23}_{-0.19}$ &  $0.11 \pm 1.04~^{+0.27}_{-0.39}$     \\
\chisq\                       &         $4.2 / 12$                               &  $ 0.65 / 2 $                                 \\
\hline
\end{tabular} 
\caption{\qsqplmsq\ dependence of the cross sections for \rh\ and \ph\ elastic
and proton dissociative production, parameterised in the form 
$1/\qsqplmsq^n$, with (a) $n$ constant and (b) $n$ parameterised as $n = c_1 + c_2\ 
\qsqplmsq$.
The 1995 (SV) measurements are normalised to those of 1996-2000.
} 
  \label{table:qsq_n_fits}
  \end{center}
  \end{table}
\renewcommand{\arraystretch}{1.0}
%-----------------------------------------------------------------------------

The total cross sections roughly follow power laws of the type $1/\qsqplmsq^n$ with 
values of $n$, fitted over the domain $1  \leq \qsq  \leq 60~\gevsq$,
given in Table~\ref{table:qsq_n_fits}(a).
These values are compatible for elastic and proton dissociative scattering.
They are also similar for \rh\ and \ph\ mesons, which supports the relevance of the scaling 
variable \qsqplmsq. 

The generally poor values of \chisq\ for fits with constant values of $n$ confirms the 
observation of~\cite{h1-rho-95-96}: 
compared to a simple power law, the cross section dependence is damped for small values 
of \qsqplmsq\ and steepens for larger values.
An empirical parameterisation $n = c_1 + c_2\ \qsqplmsq$ provides a significant 
improvement of the fit and a good description 
of the data (Table~\ref{table:qsq_n_fits}(b)).
It is interesting to note that the fitted values of the parameter $c_1$ are close to the
value $2$ expected in the Vector Meson Dominance model~\cite{vdm} for the exponent 
$n$ when $\qsq\ \to 0$.

%%%%%%%%%%%%%%%%%%%%%%%%%%%%%%%%%%%%%
\boldmath
\subsubsection{Comparison with models}
                                                                                \label{sect:qsq-dep_models}
\unboldmath
%%%%%%%%%%%%%%%%%%%%%%%%%%%%%%%%%%%%%

Predictions of the KMW dipole model~\cite{kmw} with GW saturation~\cite{GBW} 
are compared to the data in Fig.~\ref{fig:sigma_f_qsqplmsq}.
The shape of the \rh\ elastic cross section measurement is well described. 
The normalisation of the prediction is low by $10\%$,
while the overall normalisation error in the present measurement is of $4\%$.
Predictions using CGC saturation~\cite{CGC} (not shown) are nearly indistinguishable,
except for the highest bins in \qsq\ where, however, the limited precision of the data 
does not allow to discriminate.
The MRT model~\cite{mrt} does not provide normalisation predictions,
because of the uncertainty on the quark pair invariant mass window corresponding to the 
meson recombination. 
For this reason, the predictions for different PDF parameterisations are normalised 
to the data at $\qsq = 6~\gevsq$. 
Both the CTEQ6.5M~\cite{cteq} and the MRST 2004 NLO PDFs~\cite{mrst} (not shown) 
lead to predictions which are compatible with the \qsq\ dependence of the data. 
It should however be noted that the normalisation factors required to fit the data are 
about $1.1$ for CTEQ6.5M but larger than $2$ for the MRST04 NLO PDF (see 
also~\cite{h1-jpsi-hera1}).
This surprisingly large factor suggests that the gluon contribution in the 
MRST04 NLO PDFs is underestimated.

For elastic \ph\ production, the KMW predictions describe the shape of the distribution 
well, but are higher than the data by~$25\%$.
The MRT model gives a good description of the \qsq\ dependence of the cross section,
with normalisation factors similar to those for \rh\ mesons.

%%%%%%%%%%%%%%%%%%%%%%%%%%%%%%%%%%%%%
\subsubsection{Vector meson cross section ratios}
                                                                                \label{sect:VM-ratio}
%%%%%%%%%%%%%%%%%%%%%%%%%%%%%%%%%%%%%

%-----------------------------------------------------------------------------
\begin{figure}[htbp]
\begin{center}
\setlength{\unitlength}{1.0cm}
\begin{picture}(12.0,6.0)   
\put(0.0,0.0){\epsfig{file=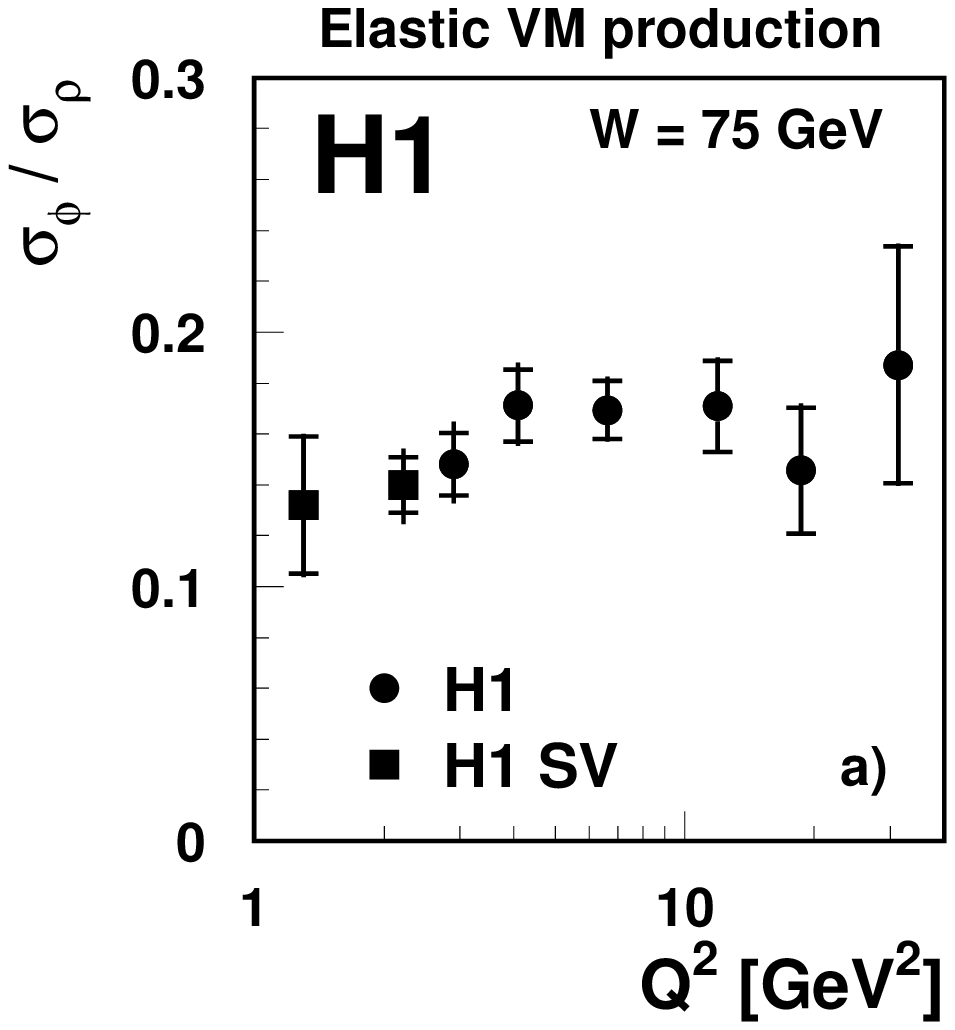,height=6.0cm,width=6.0cm}}
\put(6.0,0.0){\epsfig{file=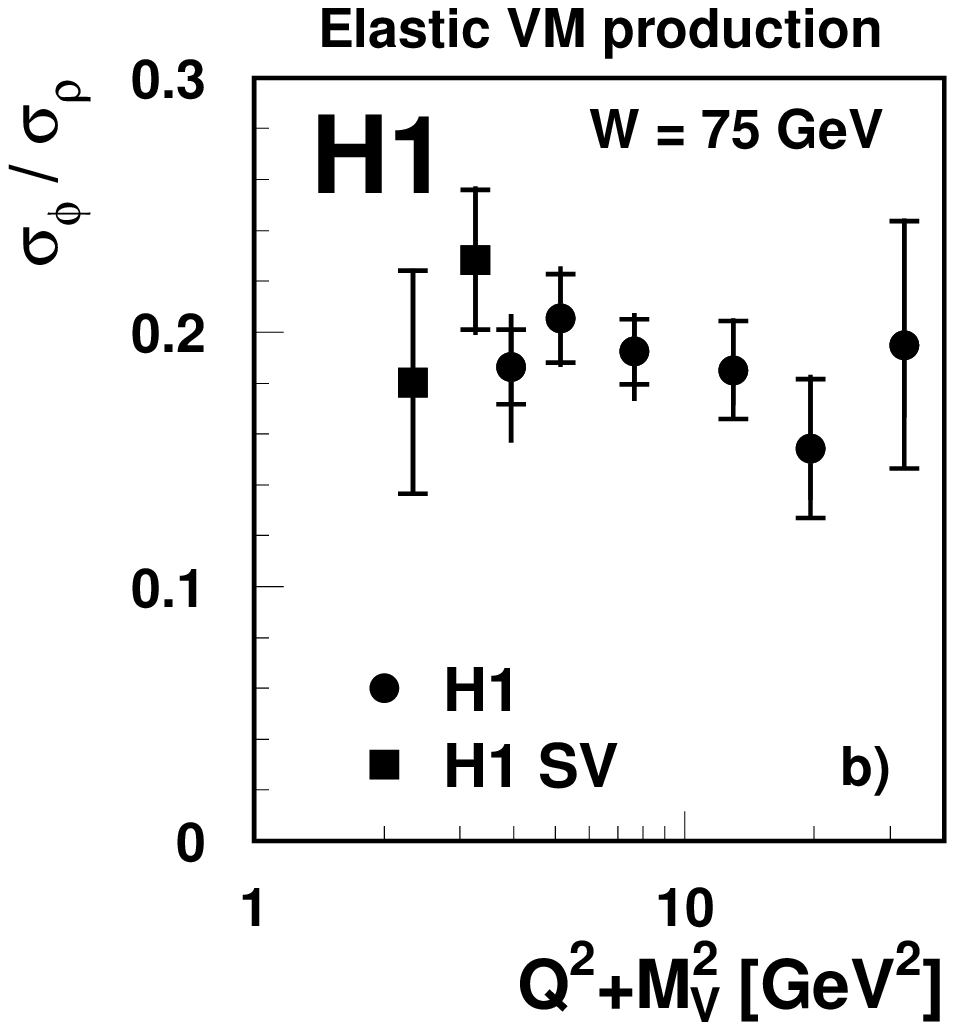,height=6.0cm,width=6.0cm}}
\end{picture}
\caption{Ratio of the \ph\ to \rh\ elastic production cross sections for $W = 75~\gev$: 
(a) as a function of \qsq; (b) as a function of \qsqplmsq.
The overall normalisation errors on the ratios, which are not included in the error bars,
are~$4.0\%$.
The present measurements are given in Table~\ref{table:p2r}.
}
\label{fig:phi_on_rh_f_qsq}
\end{center}
\end{figure}
%-----------------------------------------------------------------------------

Figures~\ref{fig:phi_on_rh_f_qsq}(a) and (b) present as a function of \qsq\ and \qsqplmsq, 
respectively, the ratio of the \ph\ to \rh\ elastic cross sections, for which several uncertainties 
cancel, in particular those related to the subtraction of the proton dissociative background.
The ratios are different because the same value of \qsq\ corresponds to different 
values of \qsqplmsq\ for \rh\ and \ph\ mesons, in view of the mass difference.
A slight increase of the ratio with \qsq\ is observed for $\qsq\ \lapprox\ 4~\gevsq$,
whereas the ratio is consistent with being constant when plotted as a function of \qsqplmsq.
Similar behaviours (not shown) are obtained for proton dissociative production.

The cross section ratios, computed for the same domains in
\qsqplmsq\ for rho and phi mesons and for $W = 75~\gev$, are
\begin{eqnarray}
\frac {\sigma (\phi)} {\sigma (\rho)} {\rm (el.)} &=& 0.191 \pm 0.007~{\rm (stat.)} 
                       ~^{+0.008}_{-0.006}~{\rm (syst.)} \pm 0.008~{\rm (norm.)}     \nonumber        \\
& &  \ \ \ \ \ \ \ \ \ \ \ \ \ \ \ \ \ \ \ \ \ \ \ \ \ \ \ \ \ \ \ \ \ \ \ \ \ \ \ \ \ \ \ \ \ 
                                                        (Q^2 \!+ \!M_V^2 \geq 2~\gevsq),         \nonumber        \\
\frac {\sigma (\phi)} {\sigma (\rho)} {\rm (p.~diss.)} &=& 0.178 \pm 0.015~{\rm (stat.)} 
                      ~^{+0.007}_{-0.010}~{\rm (syst.)} \pm 0.008~{\rm (norm.)}       \nonumber      \\
& &  \ \ \ \ \ \ \ \ \ \ \ \ \ \ \ \ \ \ \ \ \ \ \ \ \ \ \ \ \ \ \ \ \ \ \ \ \ \ \ \ \ \ \ \ \ 
                                                        (Q^2 \!+ \!M_V^2 \geq 3.5~\gevsq),
                                                                                \label{eq:phi_to_rh_ratio}
\end{eqnarray}
where the ratio of elastic cross sections includes the 1995 SV measurements 
($1 \leq \qsq \leq 2.5~\gevsq$).
The measurements are close to the value expected from quark charge counting 
$\phi / \rho=2:9$, but they tend to be slightly lower.

Qualitatively, the behaviour of the ratio is consistent with the dipole model.
At small \qsq, the influence of the meson mass on the transverse size of the $q \bar q$ pair is 
larger, which implies that colour screening is expected to be larger for \ph\ mesons than for \rh\ 
mesons.
In contrast, for $\qsq \gg \msq$, the transverse size of the dipole is given essentially
by \qsq\ and symmetry is expected to be restored.

%-----------------------------------------------------------------------------
\begin{figure}[bp]
\begin{center}
\setlength{\unitlength}{1.0cm}
\begin{picture}(16.0,8.0)   
\put(0.0,0.0){\epsfig{file=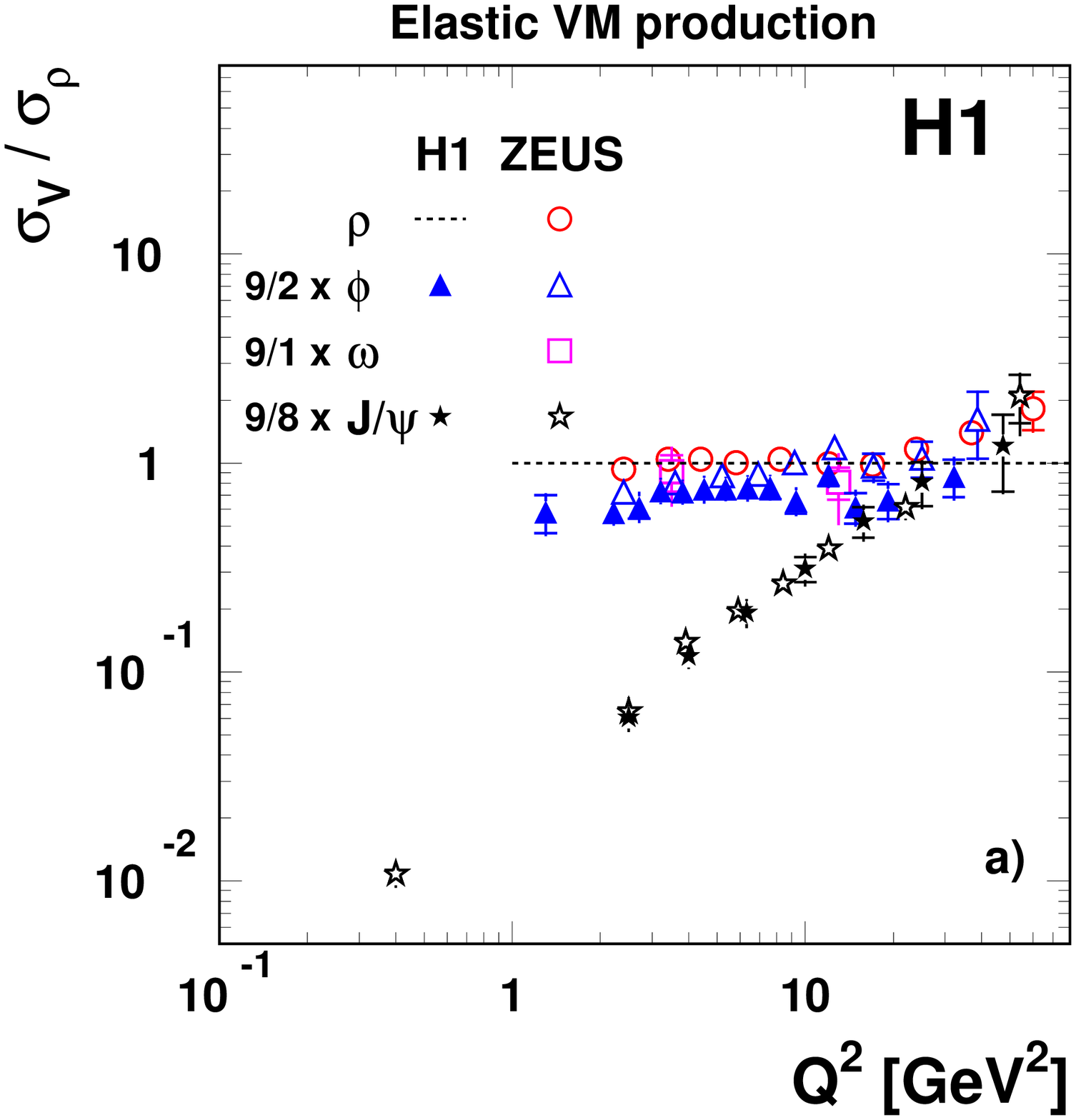,height=8.0cm,width=8.0cm}}
\put(8.0,0.0){\epsfig{file=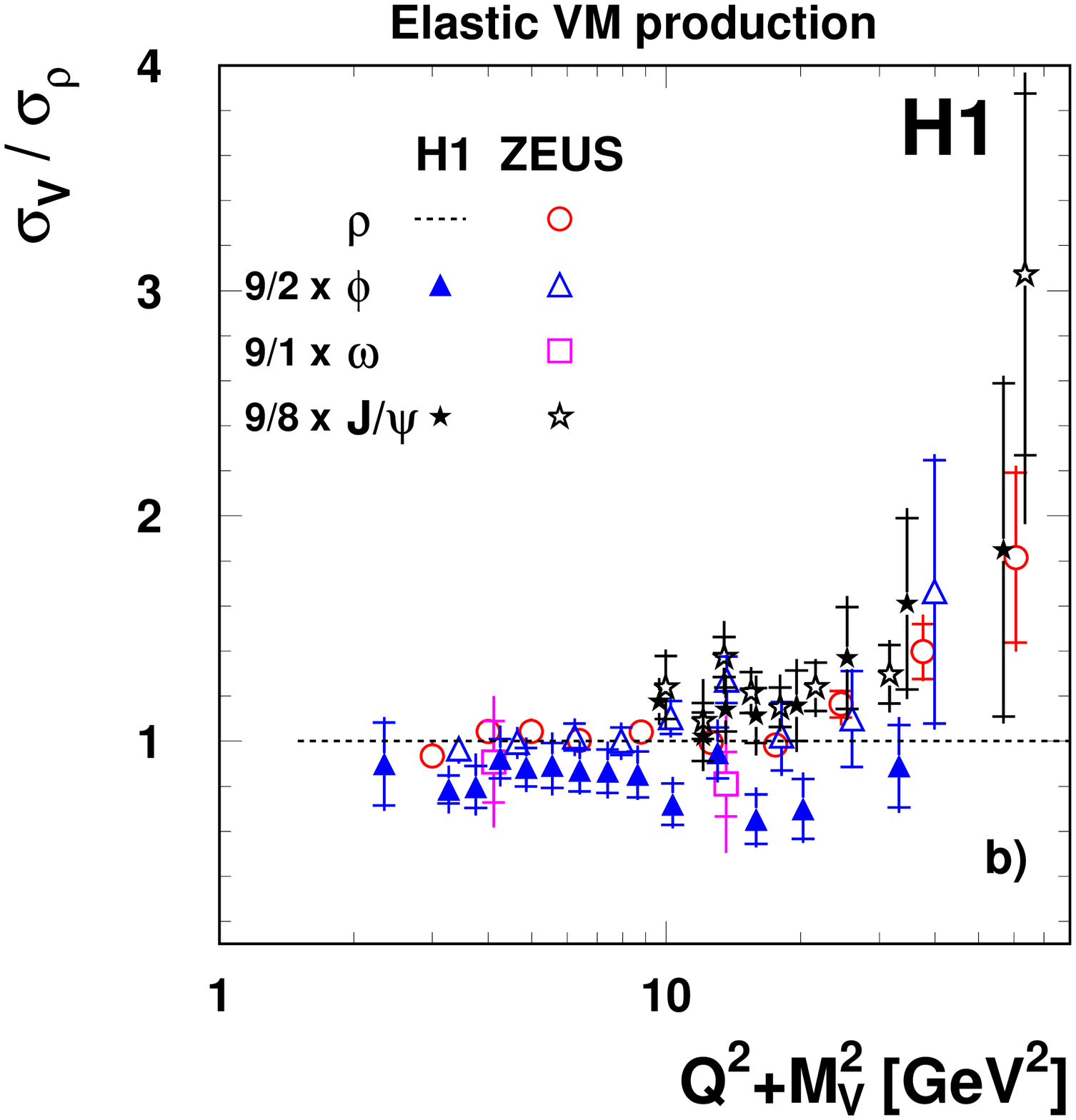,height=8.0cm,width=8.0cm}}
\end{picture}
\caption{Ratios of \om, \ph\ and \jpsi\ to \rh\ elastic production cross 
sections, scaled 
according to the quark charge contents, $\rh : \omega : \phi : \jpsi  = 9 : 1 : 2 : 8$,
plotted as a function of (a) \qsq; (b) \qsqplmsq.
The \rh\ cross section has been parameterised as described in 
Table~\ref{table:qsq_n_fits}(b).
The ratios are determined for the H1 \ph\ (this analysis) and \jpsi~\protect\cite{h1-jpsi-hera1}
measurements, and from the ZEUS 
\rh~\protect\cite{z-rho},
$\omega$~\protect\cite{z-omega},
\ph~\protect\cite{z-phi} and 
\jpsi~\protect\cite{z-jpsi-photoprod,z-jpsi-elprod} studies. }
\label{fig:SU5_sigma_f_qsqplmsq}
\end{center}
\end{figure}
%-----------------------------------------------------------------------------

The dipole size effect also explains the strong increase with \qsq\ of the \jpsi\ to \rh\ ratio,
scaled according to the quark charge content $\jpsi : \rh = 8 : 9$, 
as presented in Fig.~\ref{fig:SU5_sigma_f_qsqplmsq}(a), and the fact that 
the ratio is nearly constant and close to unity when studied as a function of 
\qsqplmsq, as shown in Fig.~\ref{fig:SU5_sigma_f_qsqplmsq}(b) (note the different
vertical scales).

Although striking, the agreement with SU(4) universality is however only qualitative, with the 
scaled \ph\ to \rh\ cross section ratios slightly below~$1$ and the scaled \jpsi\ to \rh\ ratios 
slightly above~$1$.
Scaling factors obtained from the VM decay widths into electrons~\cite{FKS,DGKP,ins} are
expected to encompass wave function and soft effects;
the use of the factors given in~\cite{ins} modifies the scaled \ph\ to \rh\ ratio very little and brings 
the scaled \jpsi\ to \rh\ ratio slightly below~$1$.

%%%%%%%%%%%%%%%%%%%%%%%%%%%%%%%%%%%%%%%%%%%%
%%%%%%%%%%%%%%%%%%%%%%%%%%%%%%%%%%%%%%%%%%%%
\boldmath
\subsection{\qsq\ dependence of the polarised cross sections}
                                                                                \label{sect:sigma_pol_f_qsq}
\unboldmath
%%%%%%%%%%%%%%%%%%%%%%%%%%%%%%%%%%%%%%%%%%%%

The separate study of the polarised (longitudinal and transverse) cross sections 
sheds light on the dynamics of the process and on the \qsq\ 
dependence of the total cross section.
Soft physics contributions, related to large transverse dipoles, are predicted
to play a significant role in transverse cross sections, whereas hard features should
be significant in longitudinal amplitudes.
At relatively low values of the scale, $\scaleqsqplmsq\ \lapprox\ 3~\gevsq$, soft, 
``finite size" effects are however expected to also affect longitudinal cross sections.

The extraction of the polarised cross sections presented in this section implies the use
of the measurement of the cross section ratio $R = \sigma_L / \sigma_T$, which 
is performed using angular distributions and is discussed in 
section~\ref{sect:polar_disc_R}.

%%%%%%%%%%%%%%%%%%%%%%%%%%%%%%%%%%%%%
\subsubsection{Cross section measurements}
                                                                                \label{sect:pol-qsq-meas}
%%%%%%%%%%%%%%%%%%%%%%%%%%%%%%%%%%%%%

The total \gstarp\ cross section can be expressed as the sum of the contributions of 
transversely and longitudinally polarised virtual photons:
\begin{equation}
   \sigma_{tot} (\gstarVM) =   \sigma_T + \varepsilon\ \!\sigma_L =
                                     \sigma_T  \ (1 + \varepsilon\ \! R) ,
                                                                                    \label{eq:sigmaT}
\end{equation}
where $\varepsilon$ is the photon polarisation parameter,
$\varepsilon \simeq (1 - y) / (1 - y + y^2/2)$, with $0.91 < \varepsilon < 1.00$
and $\av {\varepsilon} = 0.98$ in the kinematic domain corresponding to the
present measurement.

%-----------------------------------------------------------------------------
\begin{figure}[bp]
\begin{center}
\setlength{\unitlength}{1.0cm}
\begin{picture}(16.0,8.0)   
\put(0.0,0.0){\epsfig{file=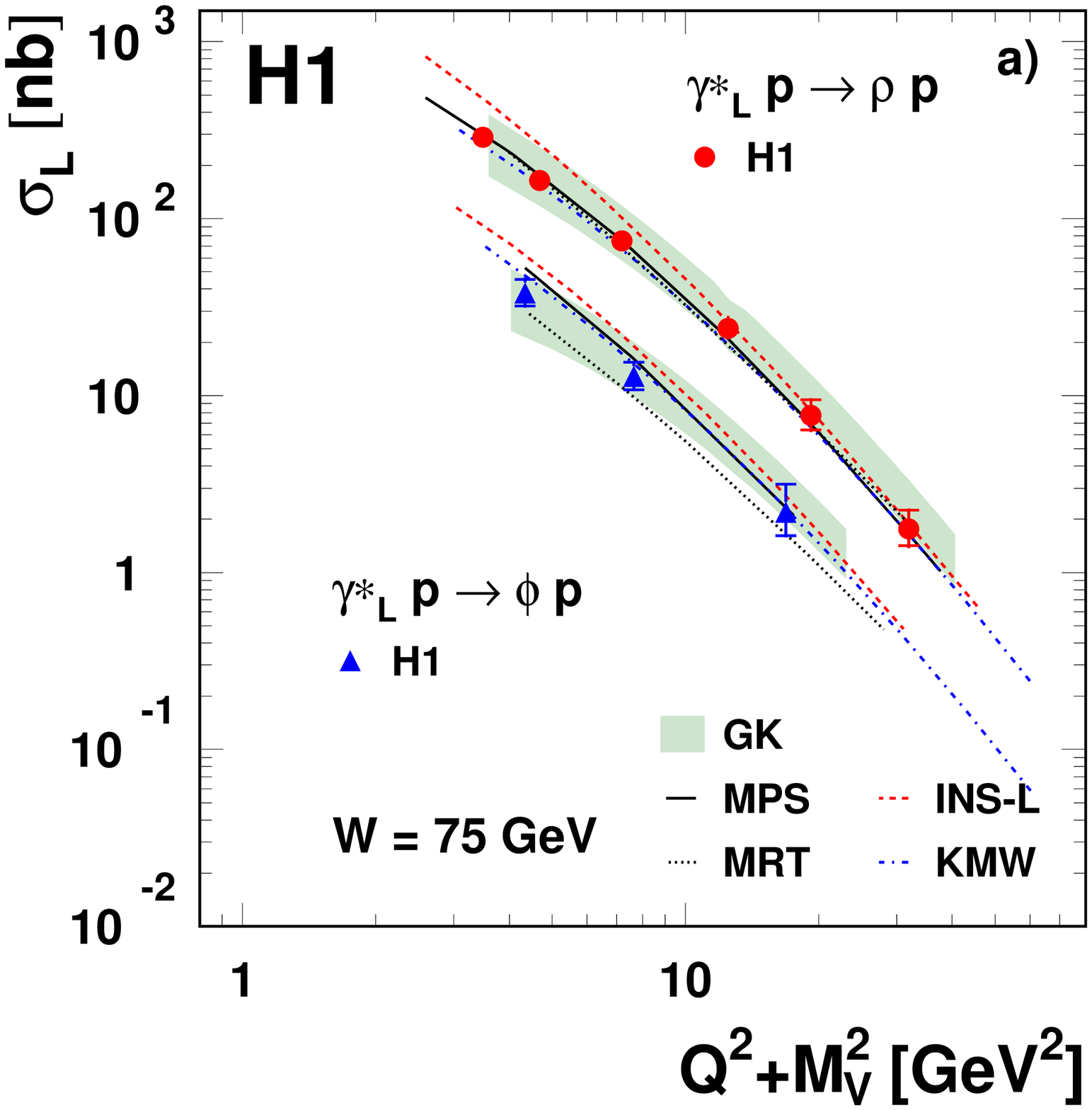,height=8.0cm,width=8.0cm}}
\put(8.0,0.0){\epsfig{file=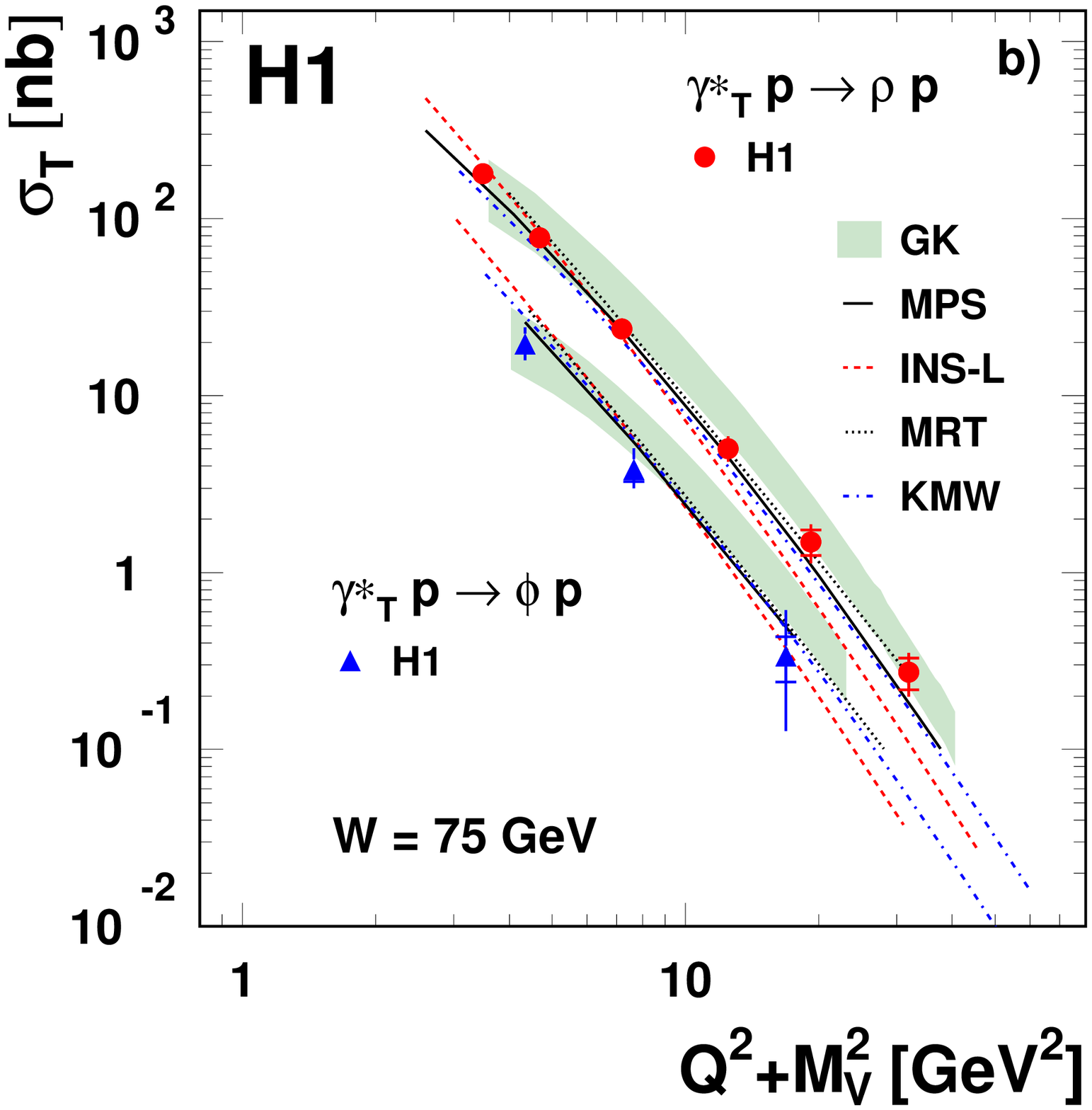,height=8.0cm,width=8.0cm}}
\end{picture}
\caption{\qsqplmsq\ dependence of (a) longitudinal and (b) transverse $\gstarp$ 
cross sections for elastic \rh\ and \ph\ meson production with $W = 75~\gev$.
Overall normalisation errors of~$3.9\%$ for \rh\ and~$4.6\%$ for \ph\ mesons are not included 
in the error bars.
The superimposed curves are model predictions:
GK~\protect\cite{kroll} (shaded bands), 
MPS~\protect\cite{soyez} (solid lines),
INS with large wave function~\protect\cite{ins} (dashed lines),
MRT~\protect\cite{mrt} with CTEQ6.5M PDFs~\protect\cite{cteq} 
and the same normalisation as in Fig.~\ref{fig:sigma_f_qsqplmsq} (dotted lines)
and KMW~\protect\cite{kmw} with GW saturation~\protect\cite{GBW} (dash-dotted lines).
The measurements are given in Tables~\ref{table:rhoel_xslt}-\ref{table:phiel_xslt}.
}
\label{fig:sigma_pol_f_qsqplmsq_models}
\end{center}
\end{figure}
%-----------------------------------------------------------------------------

The polarised cross sections, obtained from the measurements of the total cross sections 
and of $R$, with the value of $\varepsilon$ for the relevant \qsq, are
presented in Fig.~\ref{fig:sigma_pol_f_qsqplmsq_models} 
for elastic \rh\ and \ph\ production, as a function of \qsqplmsq.

%-----------------------------------------------------------------------------
\renewcommand{\arraystretch}{1.15}
\begin{table}[htbp]
\begin{center}
\begin{tabular}{|c|c|c|} 
\multicolumn{2}{c} { }  \\
\hline
                             \multicolumn{2}{|c| } { {\bf  {\boldmath $n$}   constant } } \\
\hline 
\hline
$\sigma_L (\rh)$                          & $\sigma_T (\rh)$              \\
\hline
$2.17 \pm 0.09^{+0.07}_{-0.07}$   &   $2.86 \pm 0.07^{+0.11}_{-0.12}$            \\       
\hline
\hline 
$\sigma_L (\ph)$                       &$\sigma_T (\ph)$               \\
\hline
$2.06 \pm 0.49^{+0.09}_{-0.09}$  & $2.97 \pm 0.52^{+0.14}_{-0.16}$     \\
\hline
\end{tabular} 
\caption{\qsqplmsq\ dependence of the longitudinal and transverse cross sections for \rh\ and
\ph\ meson elastic production, parameterised in the form $1/\qsqplmsq^n$
with $n$ constant.} 
  \label{table:qsq_n_fits_polar}
  \end{center}
  \end{table}
\renewcommand{\arraystretch}{1.}
%-----------------------------------------------------------------------------

Results of power law fits with constant exponents are presented in Table~\ref{table:qsq_n_fits_polar}
(the fit quality does not improve with a \qsqplmsq\ dependent value of $n$).
The fit values differ from the results $n = 3$ for the longitudinal and $n = 4$ 
for the transverse cross sections, obtained from a LO calculation of 
two gluon exchange~\cite{brodsky}.

Model predictions for $\sigma_L$ and $\sigma_T$ are compared to the data in
Fig.~\ref{fig:sigma_pol_f_qsqplmsq_models}.
The GPD predictions of the GK model~\cite{kroll} are slightly too flat, 
both for $\sigma_L$ and for $\sigma_T$, but the global normalisations are within the 
theoretical and experimental errors, which suggests that higher order effects, not 
included in the model, are weak.
The KMW model~\cite{kmw} describes well the shapes of the $\sigma_L$ and $\sigma_T$ 
measurement and the absolute normalisation of $\sigma_L$, whereas the normalisation is 
too low for $\sigma_T$; this is the reflection of the good description of the shape for
$\sigma_{tot}$ and of the prediction for $R$ which is systematically too high
(see Fig.~\ref{fig:rlt-q2} in section~\ref{sect:polar_disc_R}).
The MRT~\cite{mrt} predictions for the \rh\ polarised cross sections are reasonable, 
but for \ph\ production they are too low for $\sigma_L$ and too high for $\sigma_T$,
which reflects the fact that the predictions for $R$ are too low (Fig.~\ref{fig:rlt-q2}).
The INS $k_t$-unintegrated model
with the compact wave function~\cite{ins}  gives predictions which are significantly too high both for 
$\sigma_L$ and for $\sigma_T$, and too steep for $\sigma_T$ (not shown); 
the predictions with the large wave function have better absolute predictions but
are too steep for $\sigma_L$ and for $\sigma_T$.
The MPS dipole saturation model~\cite{soyez} describes the data rather well.

%-----------------------------------------------------------------------------
\begin{figure}[htbp]
\begin{center}
\setlength{\unitlength}{1.0cm}
\begin{picture}(16.0,8.0)   
\put(0.0,0.0){\epsfig{file=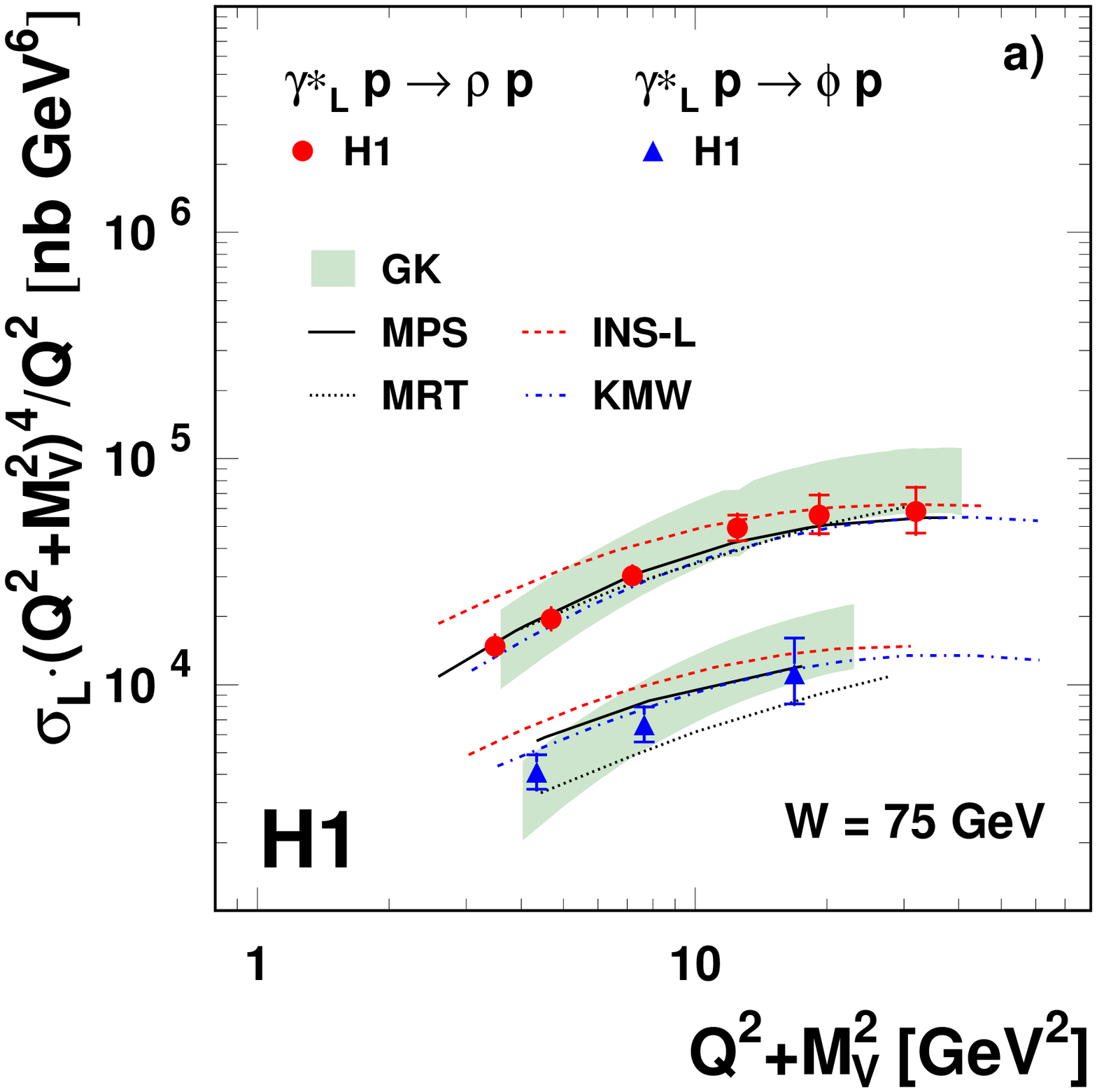,height=8.0cm,width=8.0cm}}
\put(8.0,0.0){\epsfig{file=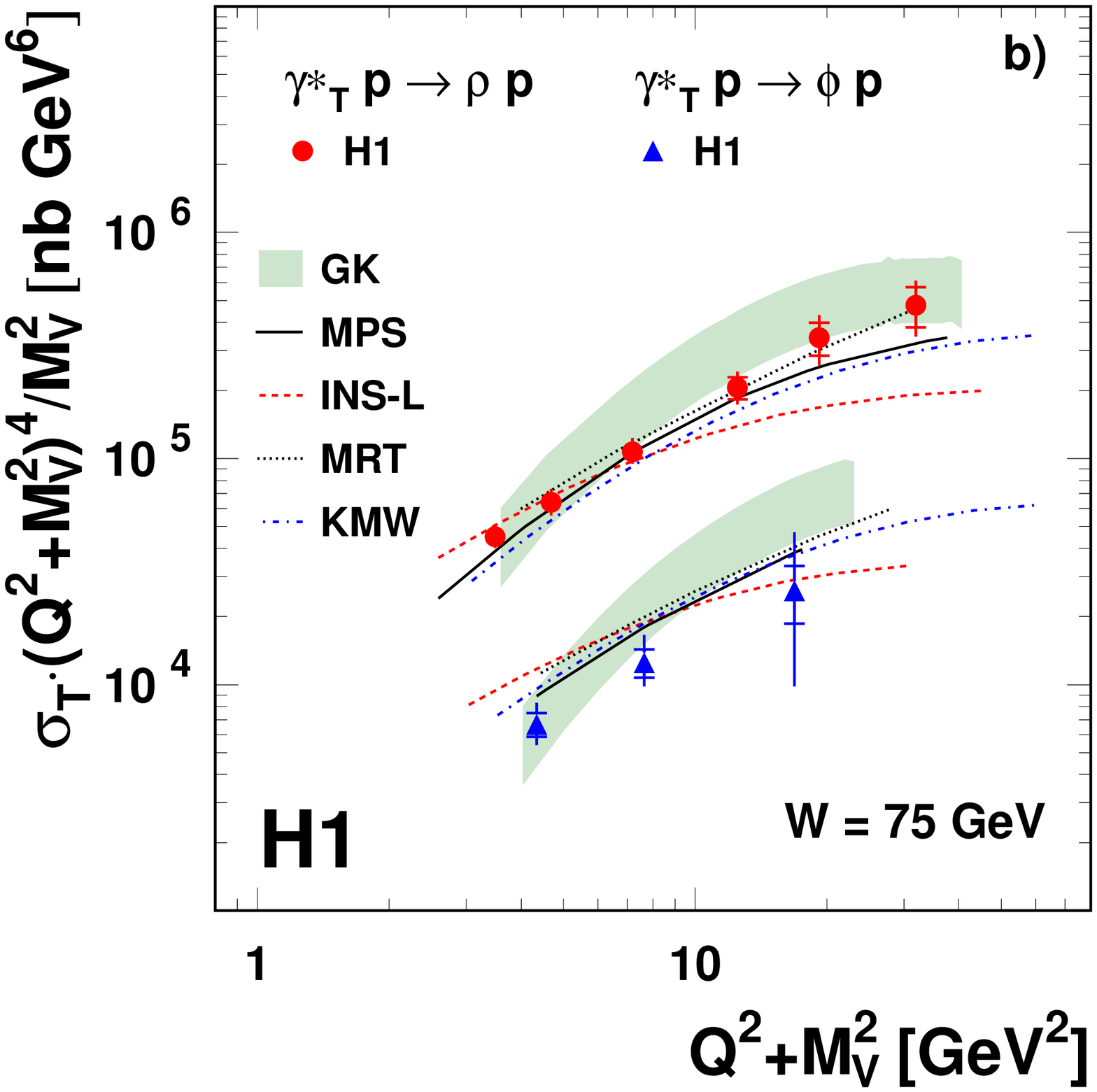,height=8.0cm,width=8.0cm}}
\end{picture}
\caption{\qsqplmsq\ dependences of the $\gstarp$ cross sections 
for \rh\ and \ph\ elastic production with $W = 75~\gev$: 
(a) longitudinal cross sections, multiplied by the scaling 
factor $(\qsq + M_V^2)^4  / \qsq$;
(b) transverse cross sections, multiplied by $(\qsq + M_V^2)^4  / M_V^2$.
The superimposed model predictions are the same as in 
Fig.~\protect\ref{fig:sigma_pol_f_qsqplmsq_models}.}
\label{fig:sigma_pol_f_qsq_scaled}
\end{center}
\end{figure}
%-----------------------------------------------------------------------------

The same data and model predictions are presented in
Fig.~\ref{fig:sigma_pol_f_qsq_scaled}, where
the longitudinal cross sections are divided by \qsq\ and the transverse cross sections  
by \msq, all being in addition multiplied by the scaling factors $ \qsqplmsq^4$ to remove
trivial kinematic dependences~\cite{igor-ivanov}.
The breaking of the formal expectations ($n = 3$, $n = 4$) for the $1/ \qsqplmsq^n$ 
dependence of the longitudinal and transverse cross sections is manifest in this 
presentation. This is expected from the fast increase with \qsq\ of the gluon density 
at small $x$.
Note that the cross sections in Fig.~\ref{fig:sigma_pol_f_qsq_scaled}
are given for a fixed value of $W$ and thus correspond to different values of $x$.
The increase with \qsq\ of the scaled longitudinal cross section is slower than that of
the scaled transverse cross section.
This is reflected in the \qsq\ dependence of the cross section ratio 
$R = \sigma_L / \sigma_T$, which is slower than $Q^2 / M^2_V$ 
(see section~\ref{sect:polar_disc_R}, Figs.~\ref{fig:rlt-q2} and~\ref{fig:rlt-compil}).

%%%%%%%%%%%%%%%%%%%%%%%%%%%%%%%%%%%%%
\subsubsection{Vector meson polarised cross section ratios}
                                                                                \label{sect:sigma_pol_f_qsq_diff-VM}
%%%%%%%%%%%%%%%%%%%%%%%%%%%%%%%%%%%%%

\begin{figure}[htbp]
\begin{center}
\setlength{\unitlength}{1.0cm}
\begin{picture}(16.0,8.0)   
\put(0.0,0.0){\epsfig{file=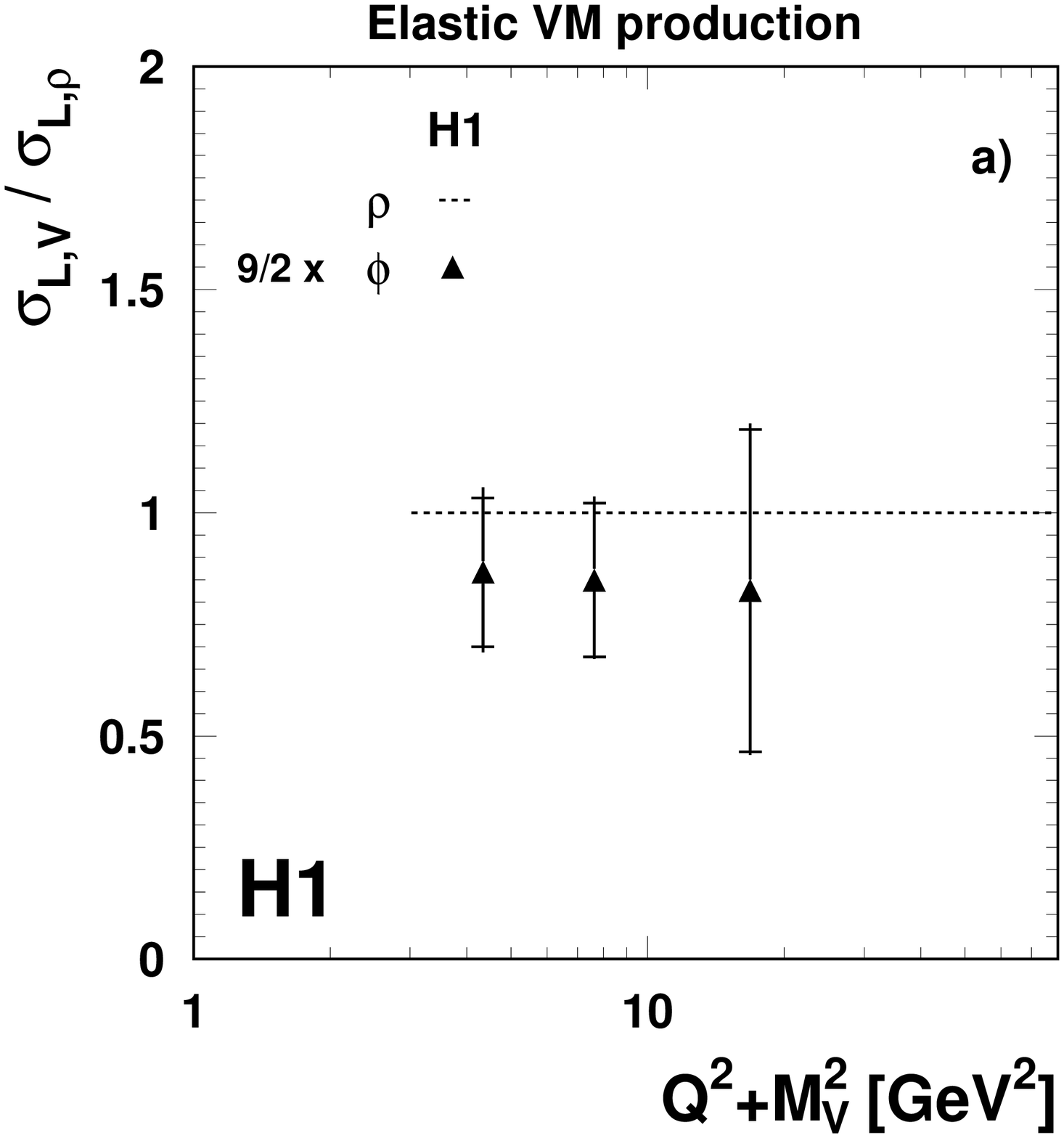,height=8.0cm,width=8.0cm}}
\put(8.0,0.0){\epsfig{file=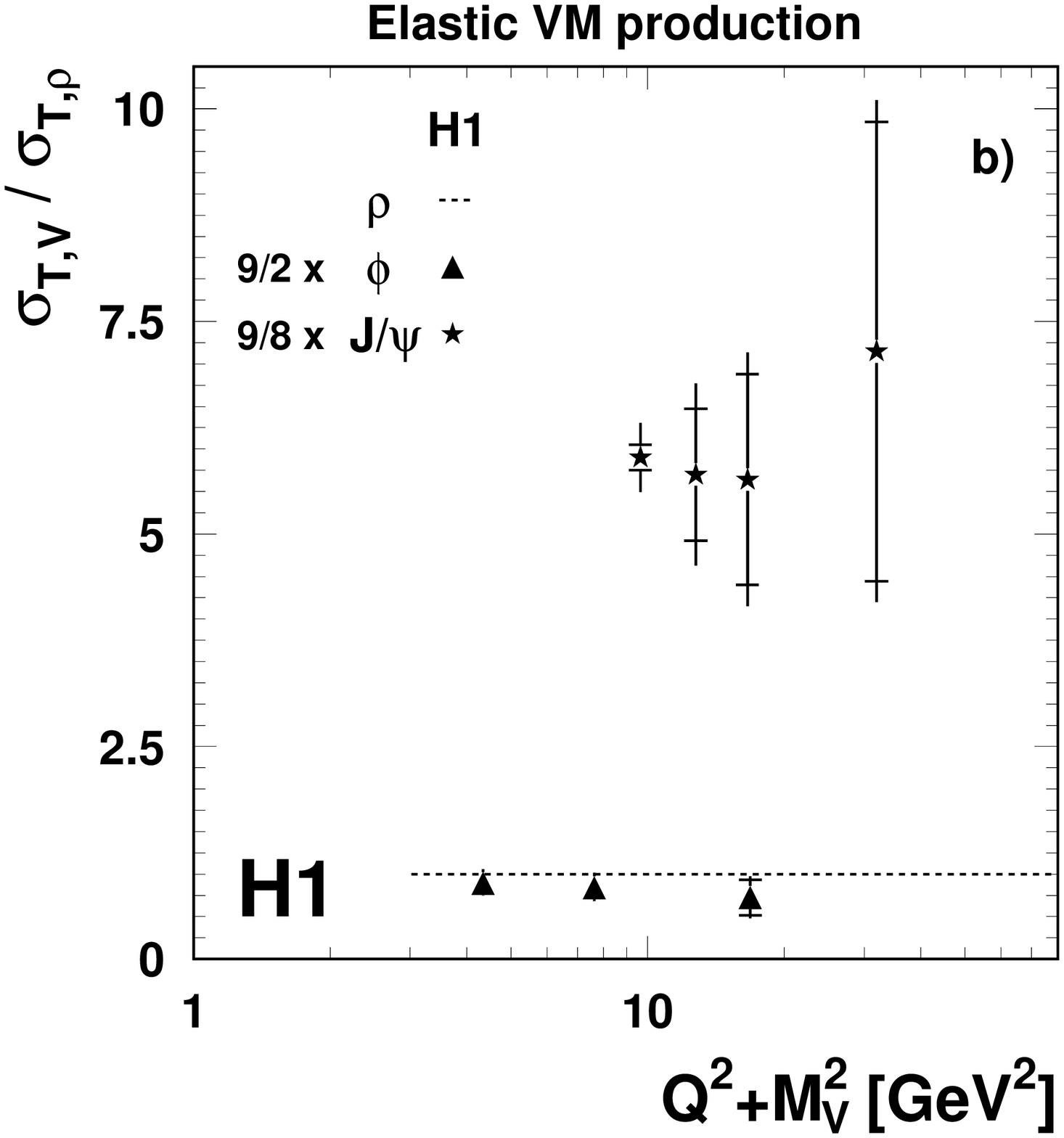,height=8.0cm,width=8.0cm}}
\end{picture}
\caption{Polarised cross sections for the elastic production 
of \ph\ (present measurements) and \jpsi~\protect\cite{h1-jpsi-hera1} mesons,
divided by the parameterisations of the \rh\ elastic polarised cross sections and 
scaled according to the quark charge contents, 
$\rh : \phi : \jpsi  = 9 : 2 : 8$; in
(a) longitudinal; (b) transverse cross sections.}
\label{fig:sigma-pol_ratios_f_qsqplmsq}
\end{center}
\end{figure}

Figure~\ref{fig:sigma-pol_ratios_f_qsqplmsq} shows the  \ph\ to \rh\ and \jpsi\
to \rh\ polarised cross section ratios, scaled according to the quark
charge content of the VM (\jpsi\ longitudinal cross sections are affected by
very large errors due to the measurement errors on $R$ and are not shown).

%-----------------------------------------------------------------------------
%-----------------------------------------------------------------------------

The ratios of the \ph\ to \rh\ polarised production cross sections are within uncertainties 
independent of \qsqplmsq\ and close to the ratio of the total cross sections
(Fig.~\ref{fig:SU5_sigma_f_qsqplmsq}), suggesting little effect of the wave functions.
In contrast, the ratios of the \jpsi\ to \rh\ transverse cross sections are very different 
from~$1$. 
This is because the polarisation states for \rh\ and \ph\ mesons on the one hand
and for \jpsi\ mesons on the other hand
are very different for the same \qsqplmsq\ value, in view of the \qsq\ dependence of $R$.
The fact that the cross section ratios are consistent with being independent of \qsqplmsq\ 
thus indicates that, within the present errors, no large difference is found between 
the small dipoles involved in transverse \jpsi\ production and the dipoles involved in 
transverse \rh\ production, for $\qsqplmsq\ \gapprox\ 10~\gevsq$.

%%%%%%%%%%%%%%%%%%%%%%%%%%%%%%%%%%%%%%%%%%%%
%%%%%%%%%%%%%%%%%%%%%%%%%%%%%%%%%%%%%%%%%%%%
\boldmath
\subsection{$W$ dependences}  
                                                  \label{sect:W_dep}
\unboldmath
%%%%%%%%%%%%%%%%%%%%%%%%%%%%%%%%%%%%%%%%%%%%

%%%%%%%%%%%%%%%%%%%%%%%%%%%%%%%%%%%%%
\subsubsection{Cross section measurements}
                                                                                \label{sect:W-meas}
%%%%%%%%%%%%%%%%%%%%%%%%%%%%%%%%%%%%%

%-----------------------------------------------------------------------------
\begin{figure}[htbp]
\begin{center}
\setlength{\unitlength}{1.0cm}
\begin{picture}(16.0,16.0)   
\put(0.0,8.0){\epsfig{file=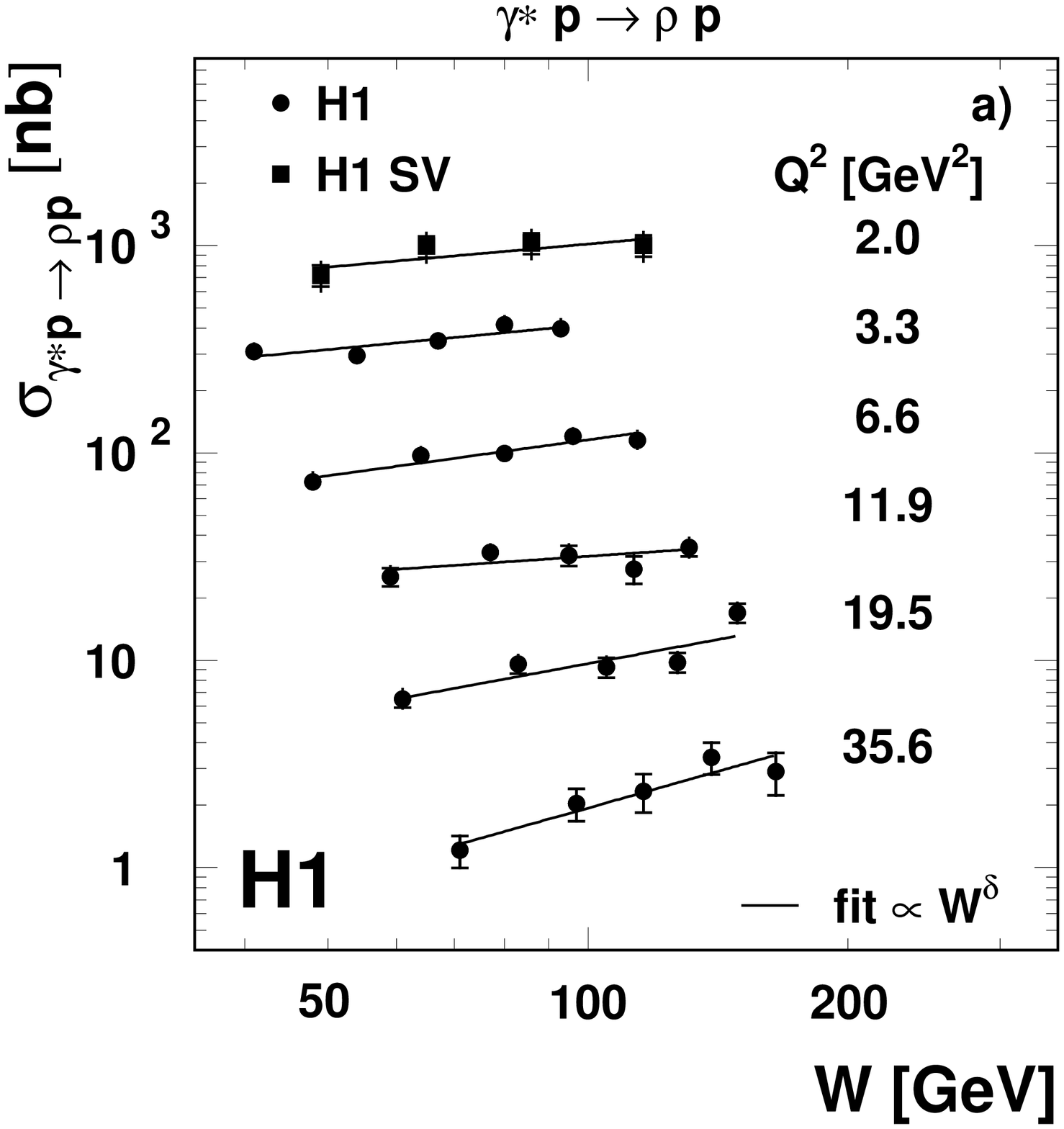,height=8.0cm,width=8.0cm}}
\put(8.0,8.0){\epsfig{file=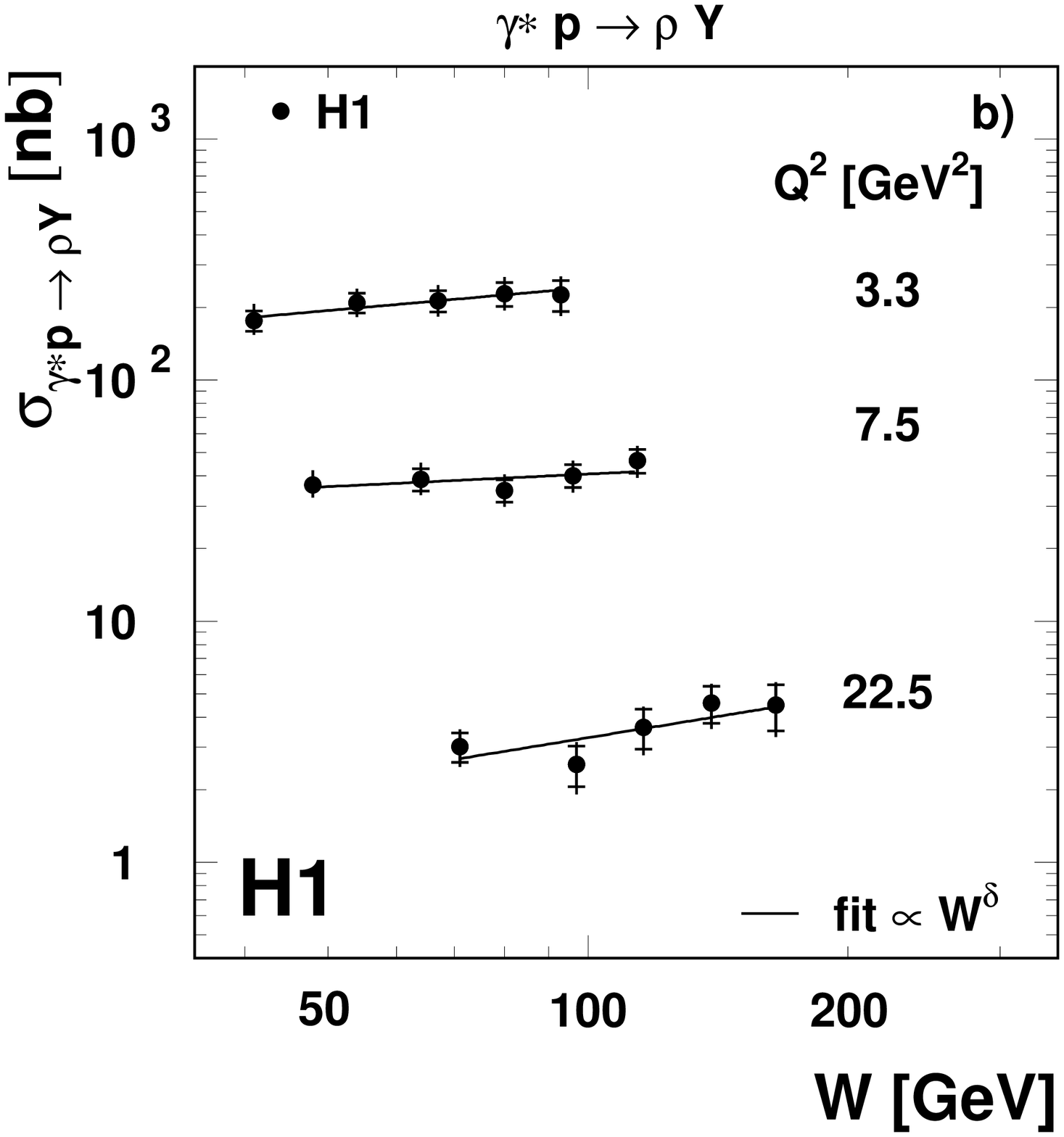,height=8.0cm,width=8.0cm}}
\put(0.0,0.0){\epsfig{file=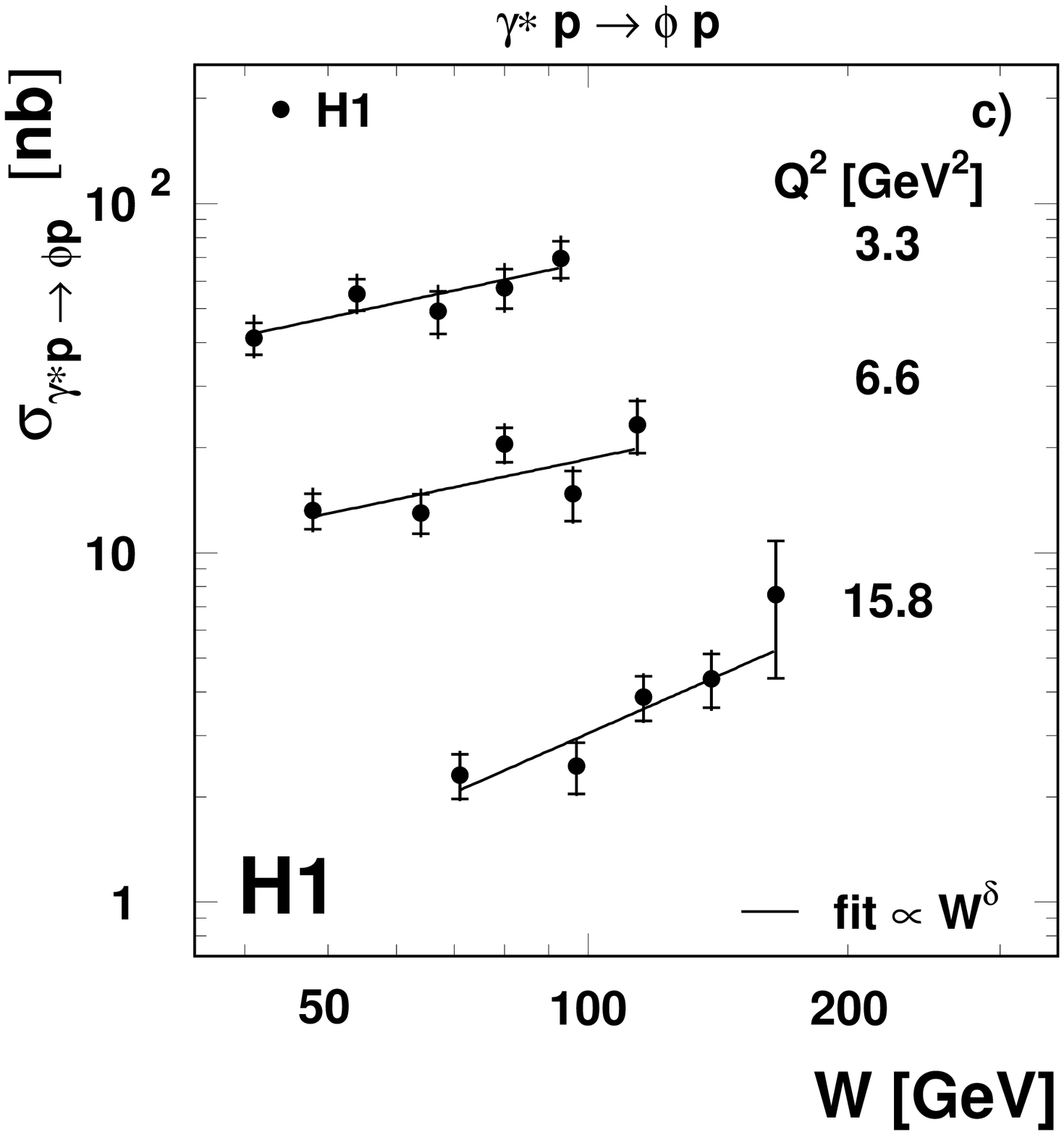,height=8.0cm,width=8.0cm}}
\put(8.0,0.0){\epsfig{file=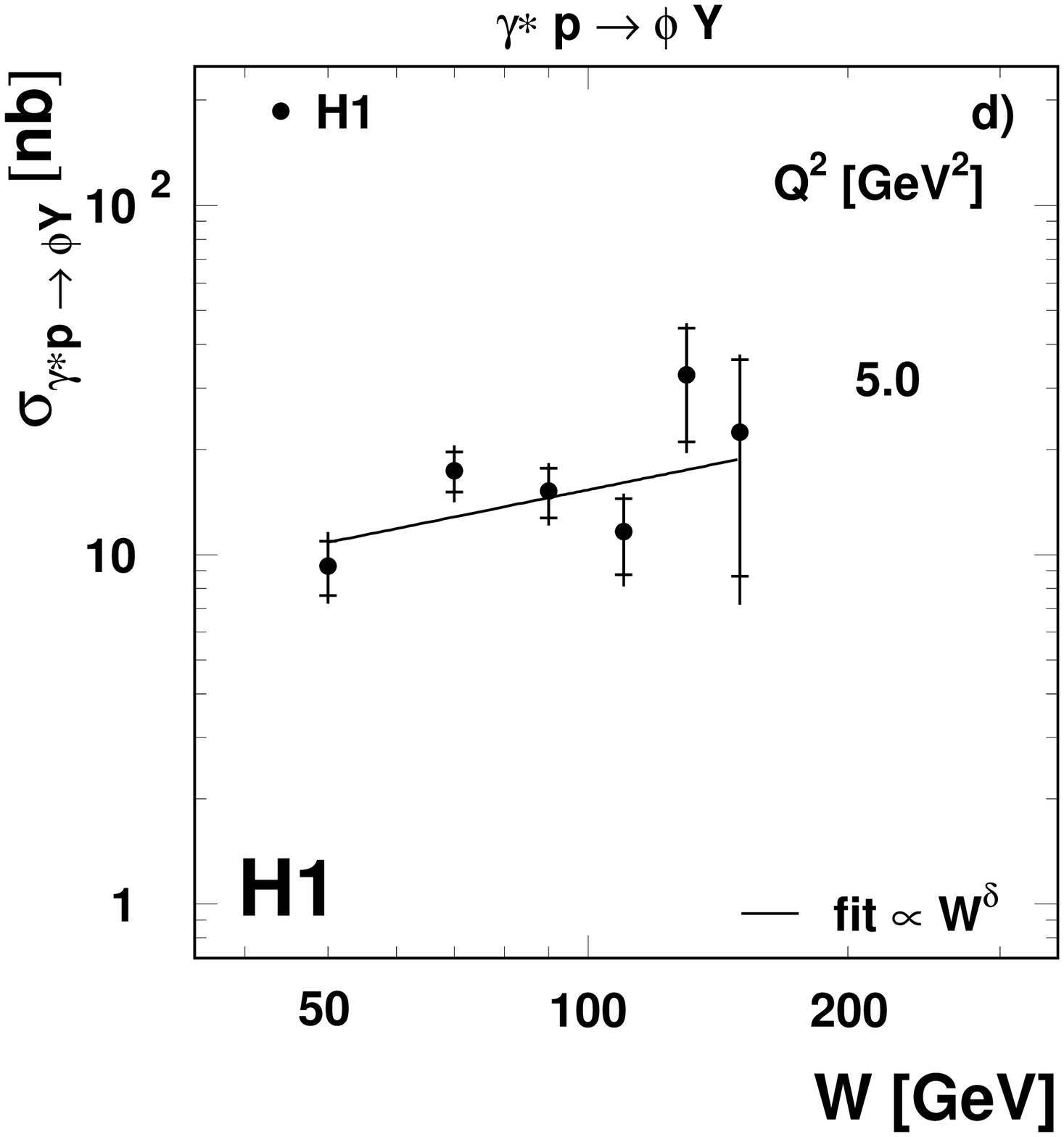,height=8.0cm,width=8.0cm}}
\end{picture}
\caption{$W$ dependence of the $\gstarp$ cross sections for elastic~(a)-(c) and proton 
dissociative~(b)-(d) production for several values of \qsq :
(a)-(b) \rh\ meson production; (c)-(d) \ph\ production.
The overall normalisation errors, not included in the error bars, are the same as in 
Fig.~\ref{fig:sigma_f_qsqplmsq}.
The lines are the results of power law fits.
The present measurements are given in Tables~\ref{table:rhoel_xsq2w}-\ref{table:phipd_xsq2w}.
}
\label{fig:sigma_f_W}
\end{center}
\end{figure}
%-----------------------------------------------------------------------------

Figure~\ref{fig:sigma_f_W} displays the $W$ dependence of  the 
$\gamma^*p$ cross sections for the production of \rh\ and \ph\ mesons, for several 
values of \qsq.
For the first time, measurements are performed for both the elastic and the proton 
dissociative channels.

The $W$ dependence of the cross sections is well described by power laws 
of the form
\begin{equation}
         \sigma (\gstarVM)  \propto W^{\delta},
                                                                                             \label{eq:W-fit}
\end{equation}
represented by the straight lines in Fig.~\ref{fig:sigma_f_W}.
This parameterisation is inspired by the Regge description of hadron interactions at high 
energy, with 
\begin{eqnarray}
    \delta(t) &=& 4 \ ( \alpom(t)  - 1),     \\                                            \label{eq:deltafit}
    \alpom(t) &=& \alpom(0) + \alp \cdot \ t.
                                                                                              \label{eq:traj}
\end{eqnarray}
In hadron interactions, typical values for the intercept and the slope of the pomeron
trajectory are $\alpom(0) = 1.08$ to $1.11$~\cite{dola} and 
$\alp = 0.25~\gevsqm$~\cite{alphaprim}, respectively.

%%%%%%%%%%%%%%%%%%%%%%%%%%%%%%%%%%%%%
\boldmath
\subsubsection{Hardening of the $W$ distributions with \qsq}
                                                                                \label{sect:W-dep_hardening}
\unboldmath
%%%%%%%%%%%%%%%%%%%%%%%%%%%%%%%%%%%%%

%-----------------------------------------------------------------------------
\begin{figure}[tb]
\begin{center}
\setlength{\unitlength}{1.0cm}
\begin{picture}(16.0,8.0)   
\put(0.0,0.0){\epsfig{file=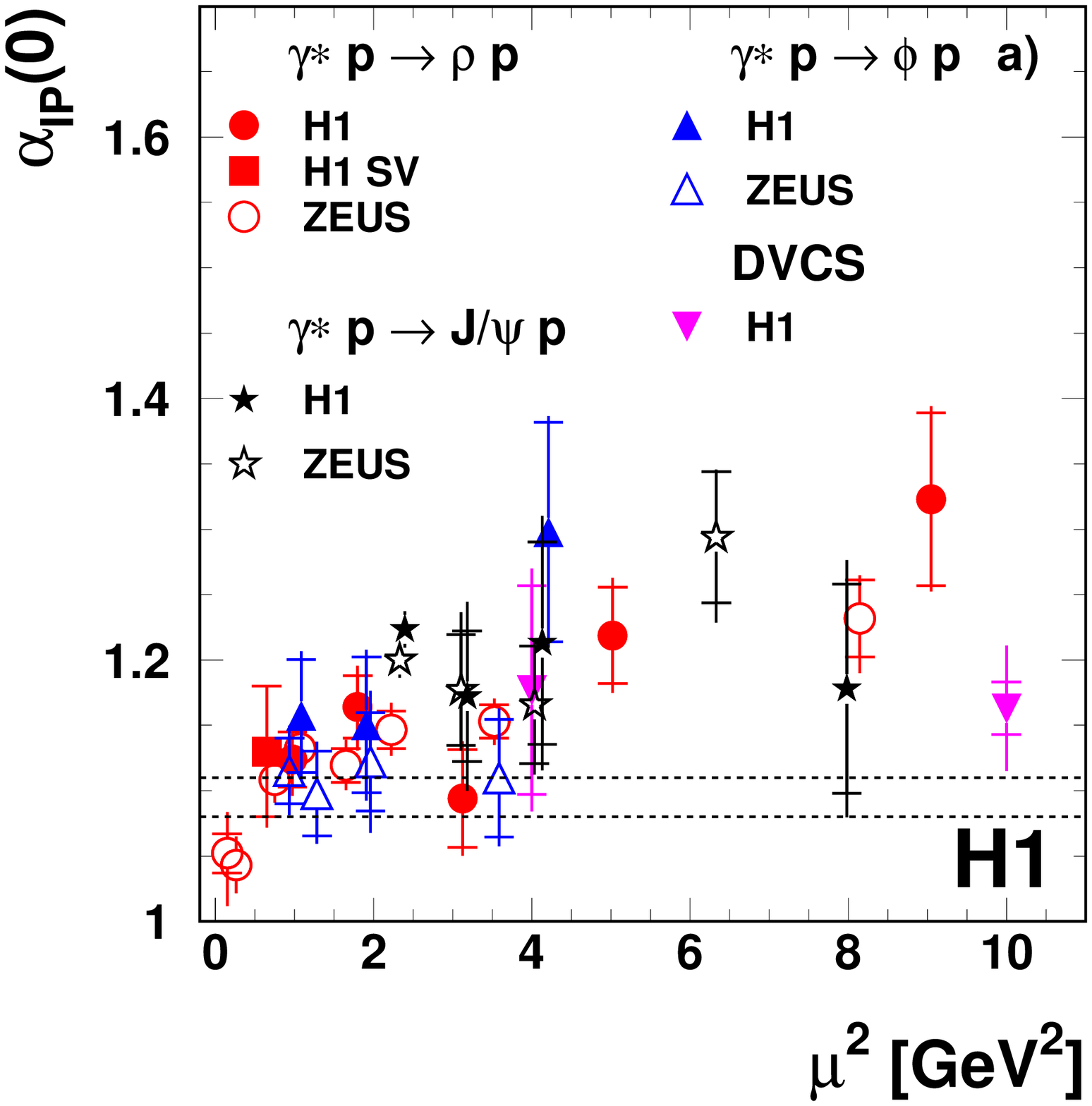,height=8.0cm,width=8.0cm}}
\put(8.0,0.0){\epsfig{file=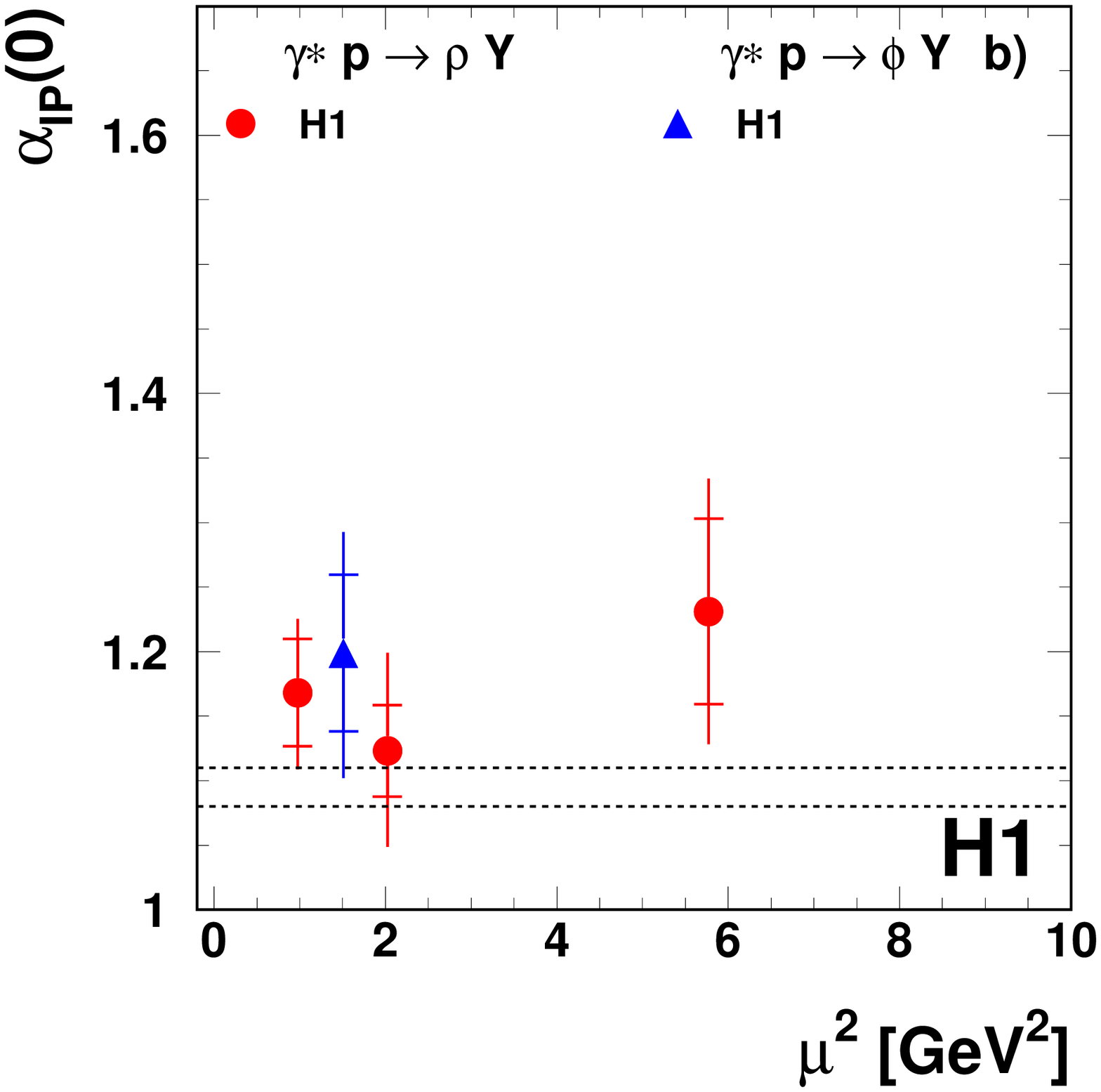,height=8.0cm,width=8.0cm}}
\end{picture}
\caption{Evolution with the scale $\mu^2 = \scaleqsqplmsq$ of the intercept of the 
effective pomeron trajectory, $\alpom(0)$, for \rh\ and \ph\ production: 
(a) elastic production; 
(b) proton dissociation.
H1 measurements of DVCS~\protect\cite{h1-dvcs}
and \jpsi\ production~\protect\cite{h1-jpsi-hera1} 
and ZEUS measurements of 
\rh~\protect\cite{z-rho-photoprod,z-rho}
(for the low \qsq\ points, the value of  \alp\ in~\cite {z-rho-photoprod}
is used ), 
\ph~\protect\cite{z-phi} and 
\jpsi\ production~\protect\cite{z-jpsi-photoprod,z-jpsi-elprod} are also shown.
For DVCS, the scale is taken as $\mu^2 = \qsq$.
The values~$1.08$ and~$1.11$~\cite{dola}, typical for soft diffraction, are indicated by 
the dotted lines.
The present measurements are given in Table~\ref{table:apom}.
}
\label{fig:alphapom0_f_qsq}
\end{center}
\end{figure}
%-----------------------------------------------------------------------------

The $W$ dependence of the cross sections is presented in Fig.~\ref{fig:alphapom0_f_qsq}
in the form of the intercept 
of the effective pomeron trajectory, $\alpom(0)$, to allow comparison between different
channels with different $t$ dependences.
The values of $\alpom(0)$ are calculated for the present \rh\ and \ph\ meson production
from the $W$ dependences following Eqs.~(\ref{eq:W-fit}-\ref{eq:traj}),
using the measured values of $\langle t \rangle$ and the measurements of \alp\ 
for \rh\ production given in Table~\ref{table:shrinkage}; the latter are derived from the evolution 
with $t$ of the $W$ dependence of the cross section.
The measurements of $\alpom(0)$ are presented as a function of the scale 
$\mu^2 = \scaleqsqplmsq $ for \rh, \ph\ and \jpsi\ production, and as a function of 
$\mu^2 = \qsq $ for DVCS, as expected for the LO process.

Up to \scaleqsqplmsq\ values of the order of $3~\gevsq$, 
the $W$ dependence of the elastic cross section for both \rh\ and \ph\ production is
slightly harder than the soft behaviour characteristic of hadron interactions and
photoproduction (Fig.~\ref{fig:alphapom0_f_qsq}(a)).
For the higher \scaleqsqplmsq\ range, higher values of $\alpom(0)$ are reached,
of the order of~$1.2$ to~$1.3$, compatible with \jpsi\ measurements.
This evolution is related to the hardening of the gluon distribution with the 
scale of the interaction.
Consistent results are obtained in the proton dissociative channel, but with larger
uncertainties (Fig.~\ref{fig:alphapom0_f_qsq}(b)).

%%%%%%%%%%%%%%%%%%%%%%%%%%%%%%%%%%%%%
\boldmath
\subsubsection{Comparison with models}
                                                                                \label{sect:W-dep_models}
\unboldmath
%%%%%%%%%%%%%%%%%%%%%%%%%%%%%%%%%%%%%

%-----------------------------------------------------------------------------
\begin{figure}[tb]
\begin{center}
\setlength{\unitlength}{1.0cm}
\begin{picture}(16.0,8.0)   
\put(0.0,0.0){\epsfig{file=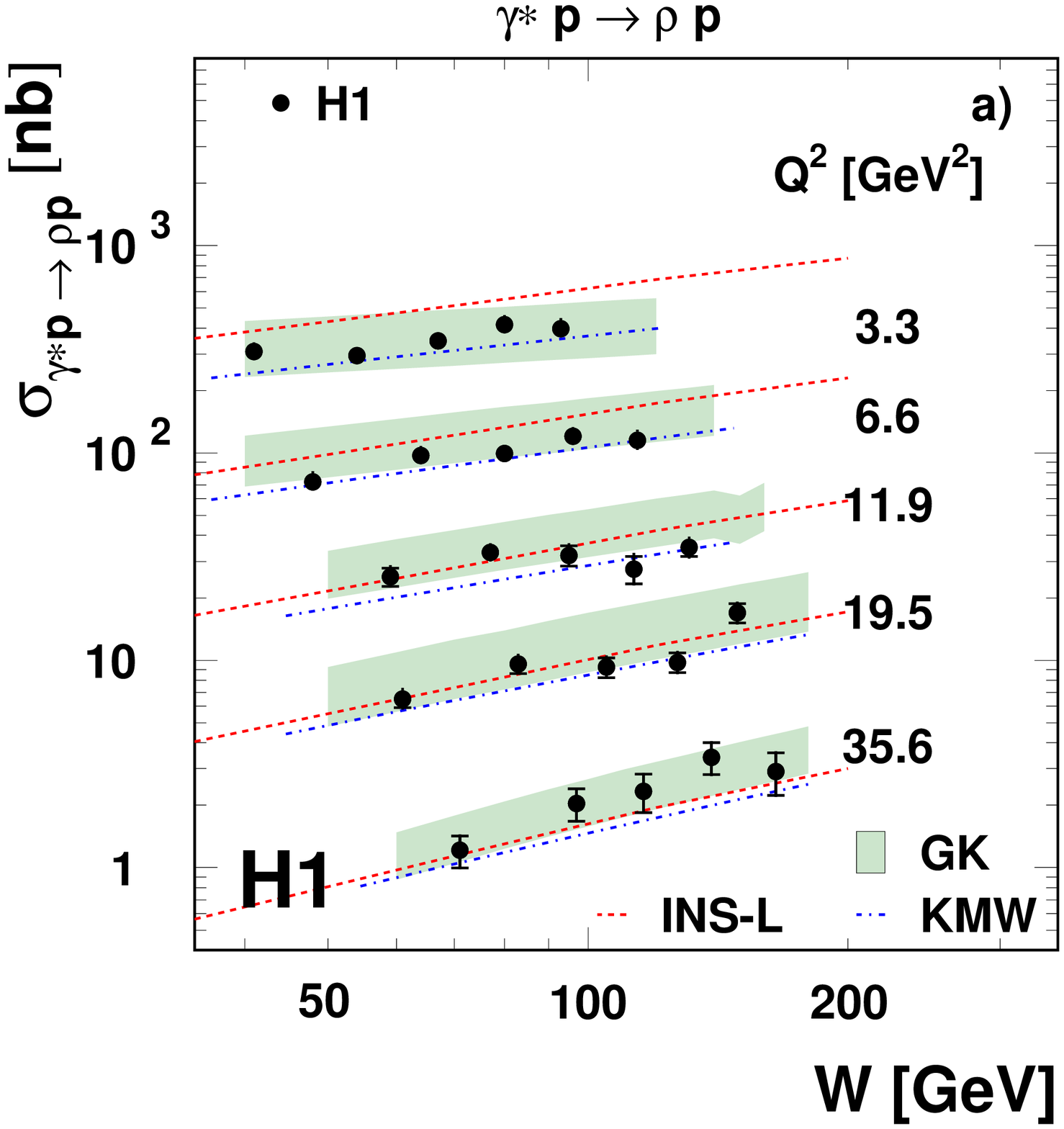,height=8.0cm,width=8.0cm}}
\put(8.0,0.0){\epsfig{file=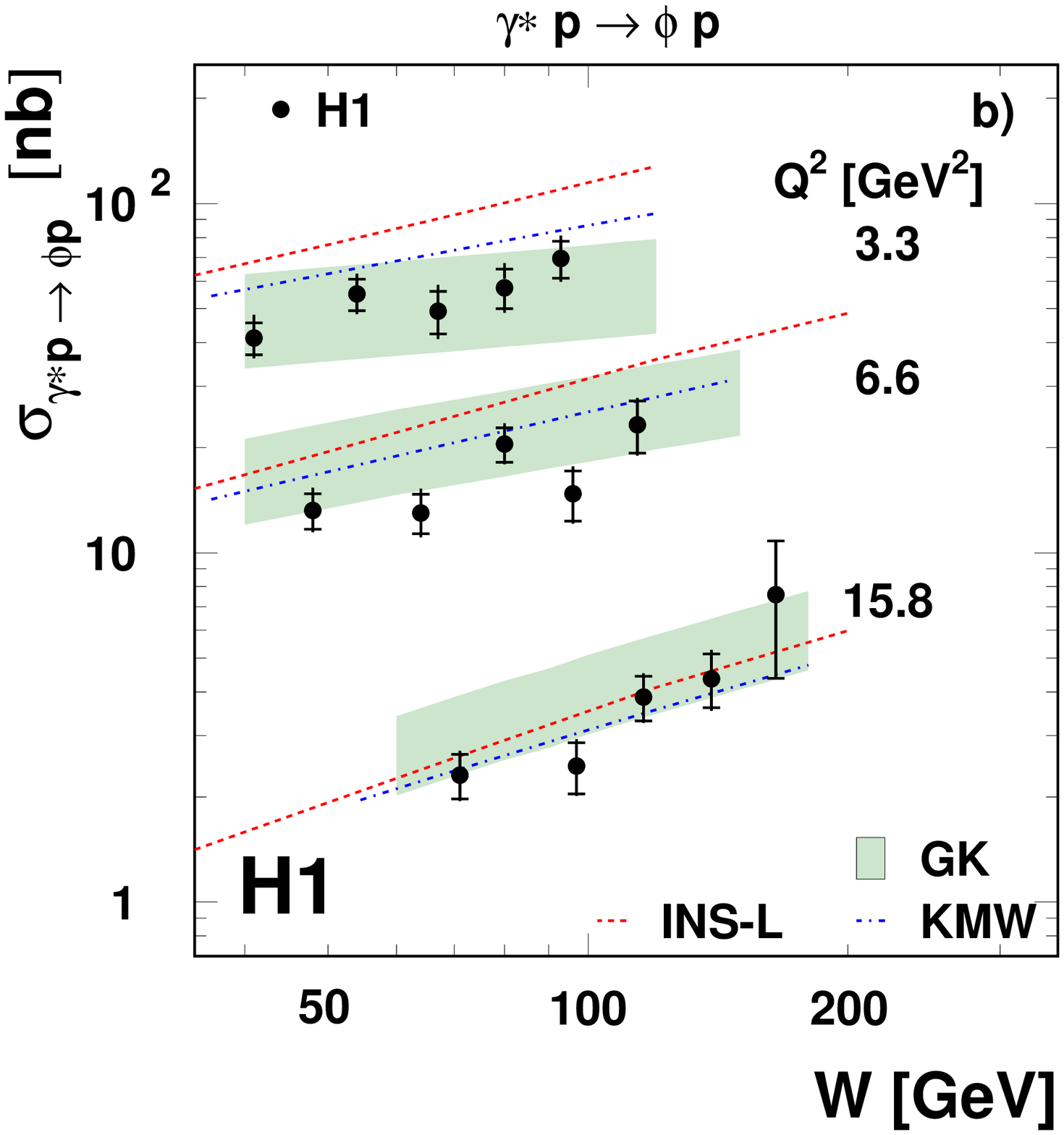,height=8.0cm,width=8.0cm}}
\end{picture}
\caption{Comparison with models of the $W$ dependences of the $\gstarp$ cross 
sections given in Figs.~\protect\ref{fig:sigma_f_W}(a) and~(c), for the elastic production 
of (a) \rh\ mesons; (b) \ph\ mesons.
The superimposed curves are model predictions: GK~\protect\cite{kroll} 
(shaded bands), 
INS with large wave function~\cite{ins} (dashed lines) 
and KMW~\protect\cite{kmw}  with GW saturation~\protect\cite{GBW}
(dash-dotted lines).}
\label{fig:sigma_f_W_models}
\end{center}
\end{figure}
%-----------------------------------------------------------------------------

In principle, the $W$ dependence of VM production can put constraints on gluon 
distributions, including effects like saturation at very low $x$ and
large $W$ values.
All models predict a hardening of the $W$ distribution with increasing \qsq, 
following from the steepening of the gluon distributions.
As examples, predictions are given in Fig.~\ref{fig:sigma_f_W_models} 
for the GK GPD model~\cite{kroll}, 
the INS $k_t$-unintegrated model with the large wave function~\cite{ins}
and the KMW dipole~\cite{kmw} with GW saturation~\cite{GBW}.
The MPS saturation model~\cite{soyez} (not shown) gives predictions for \rh\ 
production nearly identical to those of KMW.
In general, relatively small differences are found between the model predictions 
for the $W$ dependence, and the present data do not provide significant discrimination.
Differences in normalisation between models 
in Fig.~\ref{fig:sigma_f_W_models} reflect differences in the predicted
\qsq\ dependence of the cross sections.

%%%%%%%%%%%%%%%%%%%%%%%%%%%%%%%%%%%%%%%%%%%%
%%%%%%%%%%%%%%%%%%%%%%%%%%%%%%%%%%%%%%%%%%%%
\boldmath
\subsection{$t$ dependences}  
                                                  \label{sect:t_dep}
\unboldmath
%%%%%%%%%%%%%%%%%%%%%%%%%%%%%%%%%%%%%%%%%%%%

%%%%%%%%%%%%%%%%%%%%%%%%%%%%%%%%%%%%%
\subsubsection{Cross section measurements}
                                                                                \label{sect:t-meas}
%%%%%%%%%%%%%%%%%%%%%%%%%%%%%%%%%%%%%

%-----------------------------------------------------------------------------
\begin{figure}[htbp]
\begin{center}
\setlength{\unitlength}{1.0cm}
\begin{picture}(16.0,16.0)   
\put(0.0,8.0){\epsfig{file=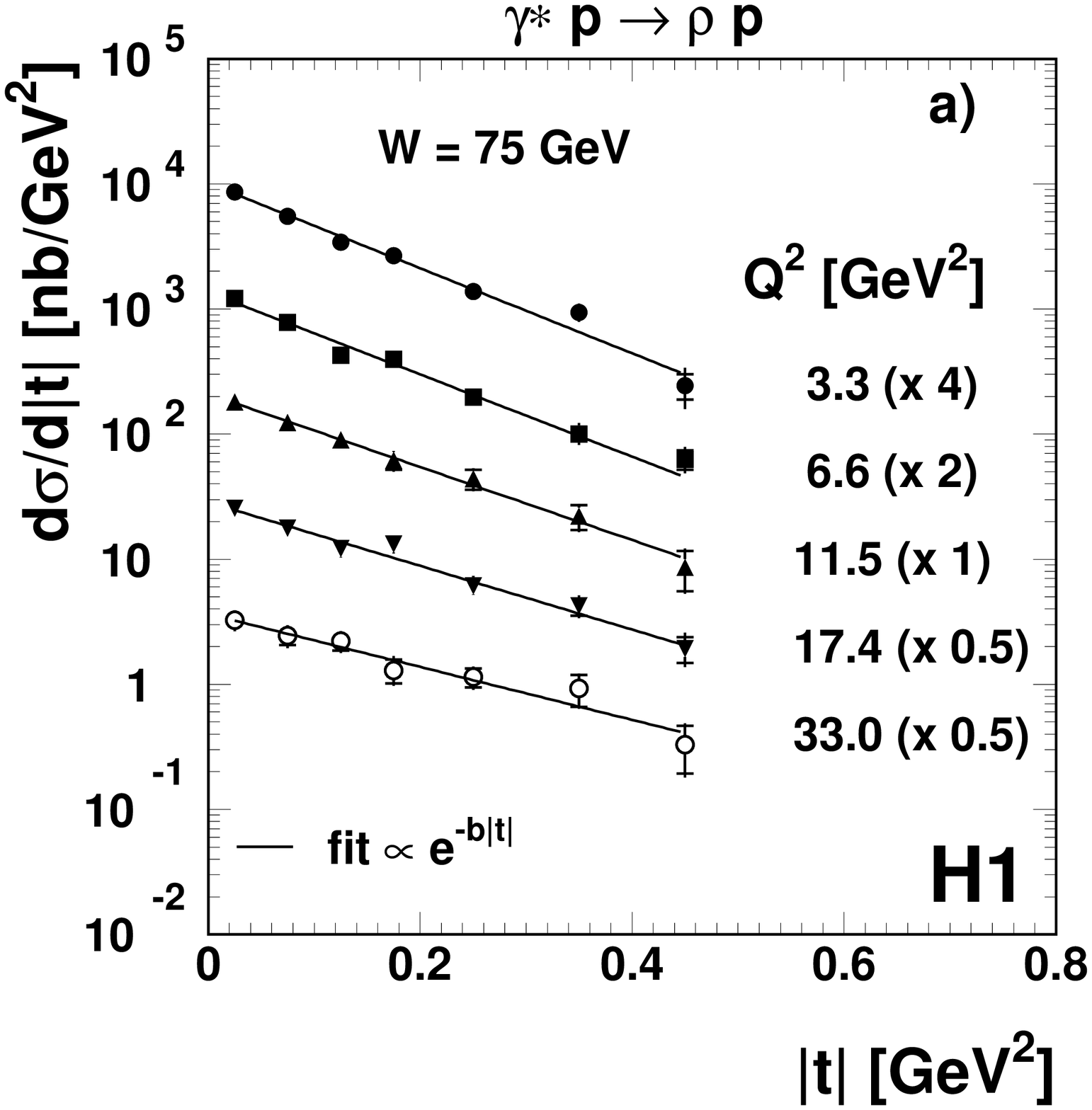,height=8.0cm,width=8.0cm}}
\put(8.0,8.0){\epsfig{file=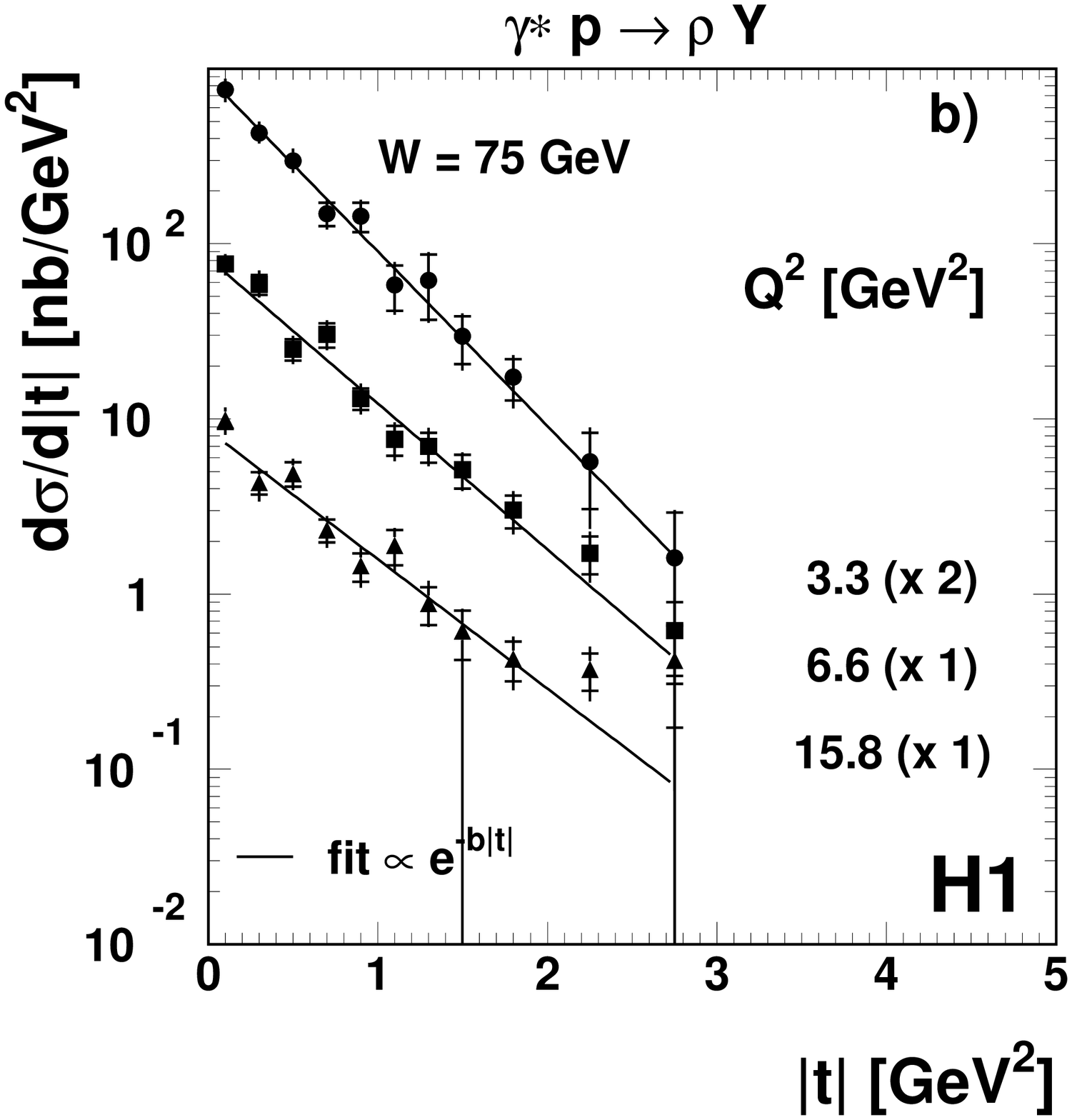,height=8.0cm,width=8.0cm}}
\put(0.0,0.0){\epsfig{file=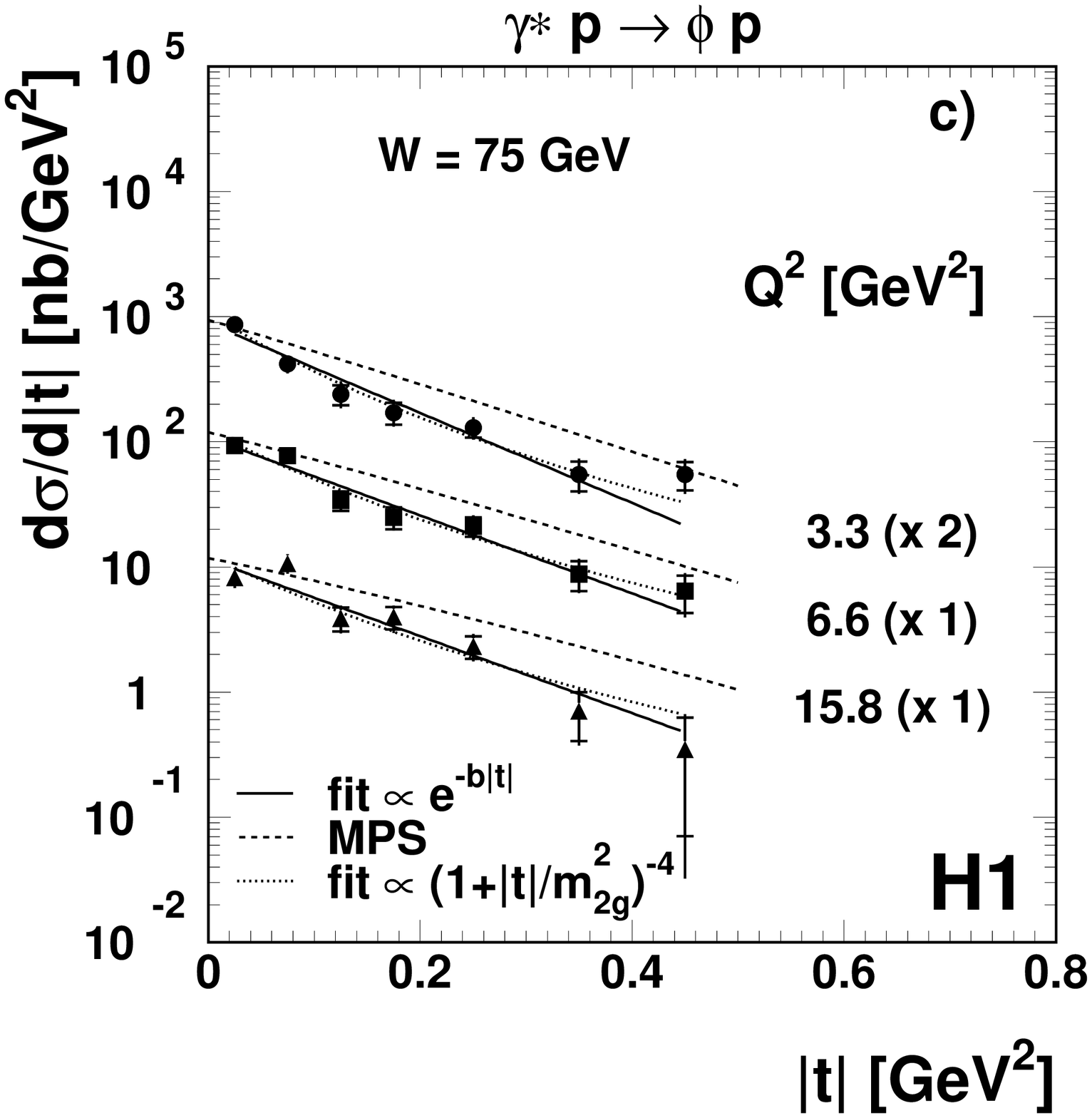,height=8.0cm,width=8.0cm}}
\put(8.0,0.0){\epsfig{file=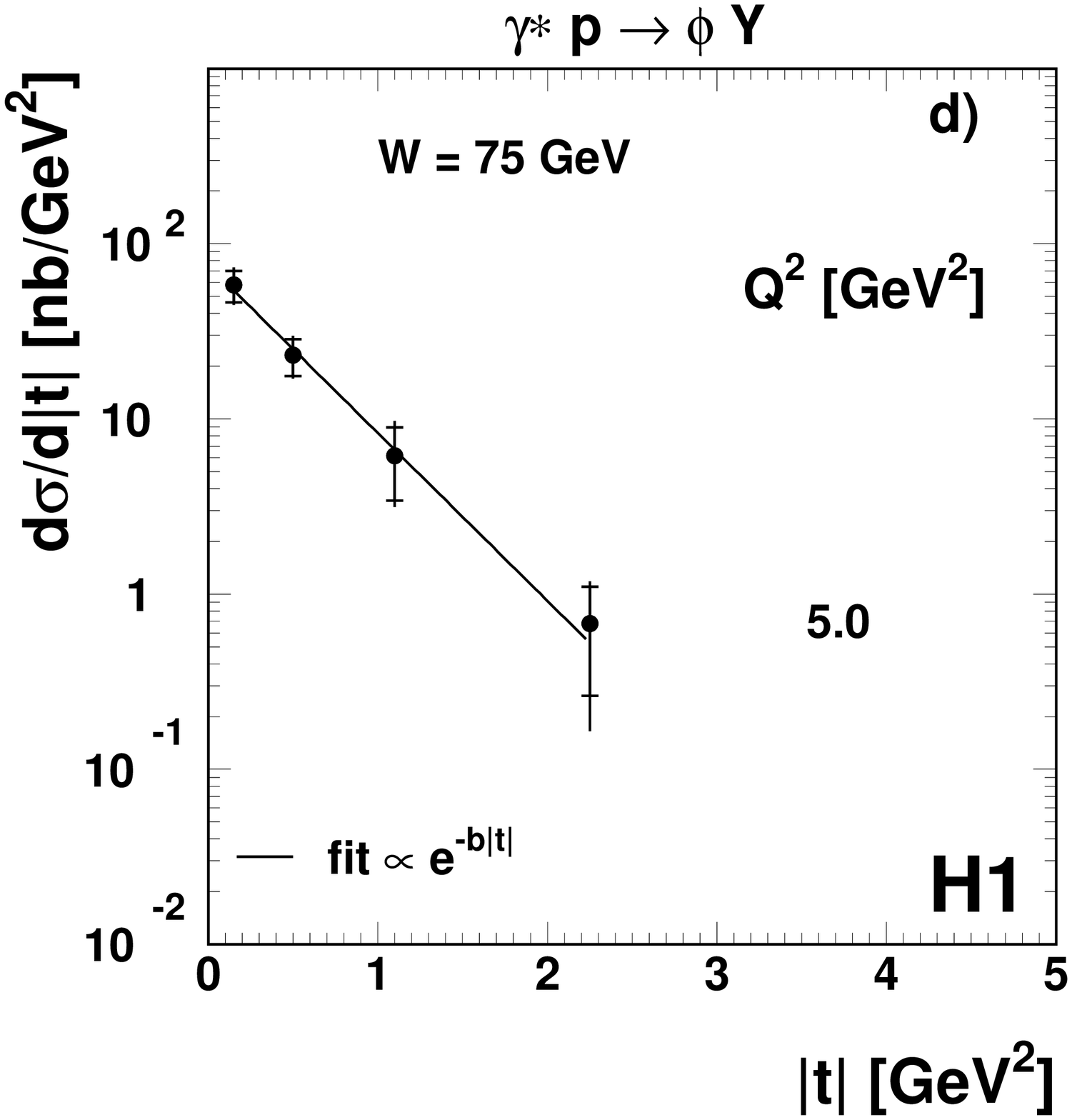,height=8.0cm,width=8.0cm}}
\end{picture}
\caption{$t$ dependence of the $\gstarp$ elastic~(a)-(c) and proton dissociative~(b)-(d) 
production cross sections for several values of \qsq :
(a)-(b) \rh\ production; (c)-(d) \ph\ production.
Some distributions are multiplied by constant factors to improve the readability of the figures.
The overall normalisation errors, not included in the error bars, are the same as in 
Fig.~\ref{fig:sigma_f_qsqplmsq}.
The superimposed curves correspond to exponential fits to the data (solide lines), 
to predictions from the MPS model~\protect\cite{soyez} (dashed lines),
and to fits of Eq.~(\ref{eq:m2g}) parameterising the two-gluon form factor in the FS 
model~\protect\cite{FS-m2g} (dotted lines).
The measurements are given in Tables~\ref{table:rhoel_xsq2t}-\ref{table:phipd_xsq2t}.
}
\label{fig:sigma_f_t}
\end{center}
\end{figure}
%-----------------------------------------------------------------------------

The differential cross sections as a function of \modt\ for \rh\ and \ph\ elastic 
and proton dissociative production are 
presented in Fig.~\ref{fig:sigma_f_t} for different ranges in \qsq.
They are well described by empirical exponential laws of the type 
${\rm d}\sigma / {\rm d}t\  \propto e^{-b~\! |t|}$.

%-----------------------------------------------------------------------------
\begin{figure}[tb]
\begin{center}
\setlength{\unitlength}{1.0cm}
\begin{picture}(16.0,8.0)   
\put(0.0,0.0){\epsfig{file=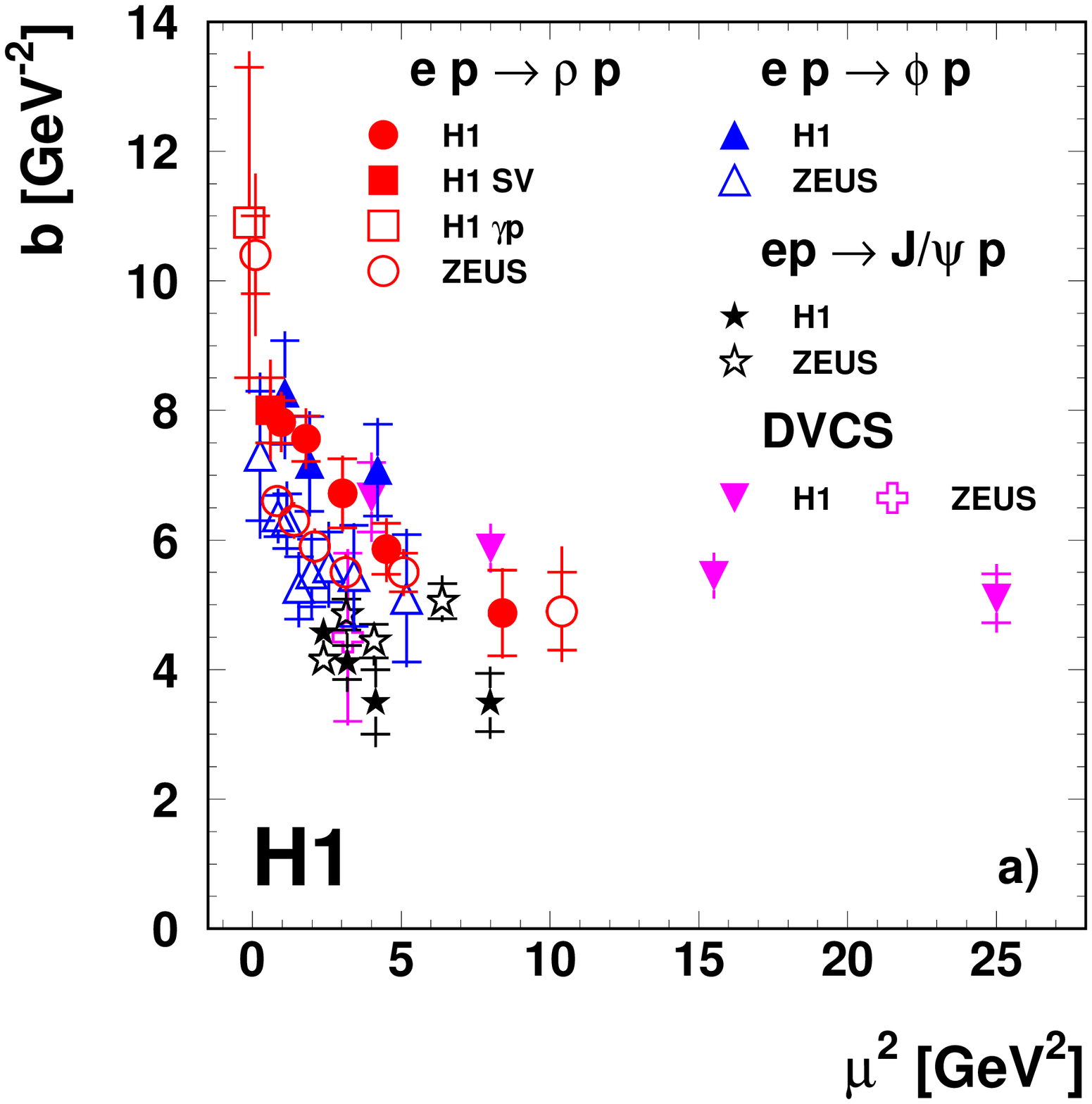,height=8.0cm,width=8.0cm}}
\put(8.0,0.0){\epsfig{file=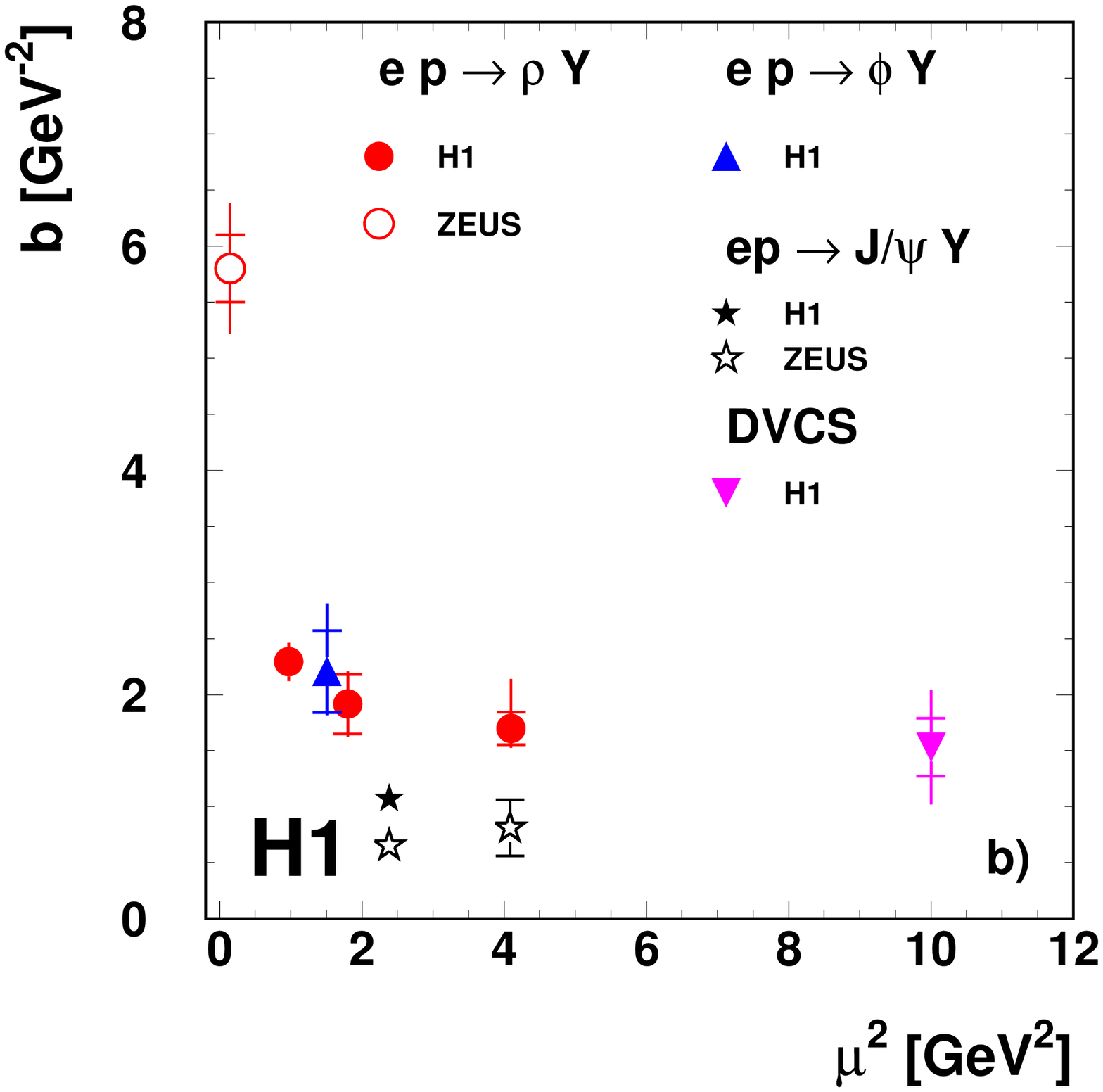,height=8.0cm,width=8.0cm}}
\end{picture}
\caption{Evolution with the scale $\mu^2 = \scaleqsqplmsq $ of the slope parameters 
$b$ of the exponentially falling \modt\ distributions of \rh\ and 
\ph\ electroproduction: (a) elastic scattering; (b) proton dissociation.
H1 data for DVCS~\protect\cite{h1-dvcs}, 
\rh\ photoproduction~\protect\cite{h1-rho-photoprod-94} 
and \jpsi\ production~\protect\cite{h1-jpsi-hera1,h1-psi2s}
and ZEUS data for \rh~\protect\cite{z-rho-photoprod,z-rho}, 
\ph~\protect\cite{z-phi-photoprod,z-phi} 
and \jpsi~\protect\cite{z-jpsi-photoprod,z-jpsi-elprod} production
are also presented.
For DVCS, the scale is taken as $\mu^2 = \qsq$.
The present measurements are given in Table~\ref{table:bslope}.
}
\label{fig:b_f_qsq}
\end{center}
\end{figure}
%-----------------------------------------------------------------------------

The slope parameters $b$ extracted from exponential fits in the range 
$\modt \leq 0.5~\gevsq$ for elastic scattering and $\modt \leq 3~\gevsq$ for proton 
dissociation are presented in Fig.~\ref{fig:b_f_qsq} as a function of the scale 
$\mu^2 = \scaleqsqplmsq$.
The measurements of the proton dissociative slopes are the first precise determination 
at HERA for light VM in electroproduction; they constitute an important ingredient 
for the extraction of the elastic $b$ slopes.
In Fig.~\ref{fig:b_f_qsq}, \rh\ and \ph\ measurements by ZEUS and \jpsi\ measurements
are also presented as a function of \scaleqsqplmsq, together with DVCS measurements 
(with $\mu^2 = \qsq$).

The present measurements of the $b$ slopes for $\scaleqsqplmsq\ \lapprox\ 
5~\gevsq$ are higher than those of ZEUS~\cite{z-rho} and also than 
those of a previous H1 measurement~\cite{h1-rho-95-96}.
Two sources of systematic experimental differences are identified.
The first is related to the estimation of the proton dissociative background, both in size 
and in shape.
The subtraction of a smaller amount of proton dissociative background and the use 
of a steeper proton dissociative slope lead to shallower \modt\ distributions of the
elastic cross section and to smaller $b$ slope measurements.
The use of a central value of $2.5~\gevsqm$ for the proton dissociative slope, 
as assumed in~\cite{h1-rho-95-96}, compared to the values measured here 
(Fig.~\ref{fig:b_f_qsq}(b)), leads to a decrease of the elastic slope determination by
$0.1$~\gevsqm, and
a variation by $\pm 20\%$ of the amount of proton dissociative background induces a 
change of the elastic slope measurement by $\pm 0.2~\gevsqm$ for $\qsq = 5~\gevsq$
and $\pm 0.1~\gevsqm$ for $\qsq = 20~\gevsq$.
The second -- and major -- source of discrepancy, for both VMs, is in the 
treatment of the \om, \ph\ and mostly \rhop\ backgrounds discussed in
section~\ref{sect:rhop_bg}.
Because of the non-detection of the decay photons, these backgrounds exhibit effective 
\modt\ distributions which are much flatter than their genuine distributions and than 
the signal.
Neglecting completely the presence of the \rhop\ background would lead in the present
analysis to a decrease of the measurement of the elastic $b$ slope
by~$0.4~\gevsqm$ for $\qsq = 3~\gevsq$ and~$0.2~\gevsqm$ for $\qsq = 20~\gevsq$.

%%%%%%%%%%%%%%%%%%%%%%%%%%%%%%%%%%%%%%%%%%%%
\subsubsection{Universality of \boldmath{$t$} slopes and hard diffraction}
                                                                                \label{sect:t_slopes_elast}
%%%%%%%%%%%%%%%%%%%%%%%%%%%%%%%%%%%%%%%%%%%%

In an optical model inspired approach, 
the $t$ slopes for DVCS and VM production result from the sum of terms describing  
the form factors due to the transverse sizes of the scattered system $Y$ ($b_Y$), of 
the $q \bar q$ dipole pair ($b_{q \bar q}$) and of the exchange ($b_{\pom}$).
An additional form factor reflecting the VM transverse size may also give a
contribution, $b_V$, to the $t$ slope for light VM production in models where the wave function plays 
an important role in the process, while being negligible for DVCS and for \jpsi.
The value of the slope can thus be decomposed as:
\begin{equation}
b = b_Y + b_{q \bar q} + b_{\pom} + b_{V}.
                                                                                   \label{eq:optical}
\end{equation}

In elastic scattering, the slope $b_Y = b_p$ reflects the colour distribution in the proton.
For baryonic excited states with size larger than that of the proton, larger slopes 
(i.e. steeper $t$ distributions) than for elastic scattering may be expected.
In contrast, when the proton is disrupted in the diffractive scattering, no form factor arises
from the $Y$ system and $b_Y$ is expected to be $\simeq 0$.
The $ b_{\pom} $ contribution of the exchange is generally believed to be small and
independent of \qsq. 
There is indeed a priori 
no relation between \qsq\ and the transverse size of the exchange, at least for 
$\modt \ll \qsq$ and for $\alpha_s$ taken to be constant (LL BFKL).

It is visible in Fig.~\ref{fig:b_f_qsq}  that, already for 
$\scaleqsqplmsq\ \gapprox\ 0.5~\gevsq$, the elastic $b$ slopes for light VM 
electroproduction are significantly lower than in photoproduction,
showing a departure from purely soft diffraction and a decrease of the relevant
$q \bar q$ dipole transverse size. 
Until the scale \scaleqsqplmsq\ reaches values $\gapprox\ 5~\gevsq$, light VM slopes
are however significantly larger than for \jpsi. 
This indicates the presence of dipoles with relatively 
large transverse sizes for light VMs in this \qsq\ domain.
This is expected  in the transverse amplitudes and also in 
longitudinal amplitudes until the fully hard regime is reached (``finite size" effects).
Light VM and DVCS slopes are compatible when plotted as a function of 
the scales \scaleqsqplmsq\ and \qsq, respectively.
For large scale values, they are consistent with the \jpsi\ data, although they may be 
slightly higher.
All these features confirm that the present \qsq\ domain covers the transition from the 
regime where soft diffraction dominates light VM production to the regime where hard 
diffraction dominates.
The comparable values of the slopes for \rh, \ph\ and \jpsi\ production in the harder 
regime suggests that light VM form factors are small.

For proton dissociative diffraction, the $t$ slopes shown in Fig.~\ref{fig:b_f_qsq}(b) have 
significantly smaller values than for elastic scattering.
This is expected for $Y$ systems above the nucleon resonance region, with
vanishing values of $b_Y$.
The proton dissociative slopes for \rh\ and \ph\ mesons are similar 
at the same \scaleqsqplmsq\ value, but remain larger than for \jpsi,
confirming the presence of large dipoles for $\scaleqsqplmsq\ \lapprox\ 5~\gevsq$ 
or, alternatively, leaving room for a light VM form factor.

%%%%%%%%%%%%%%%%%%%%%%%%%%%%%%%%%%%%%%%%%%%%
\boldmath
\subsubsection{Comparison with models}
                                                                                \label{sect:t-dep_models}
\unboldmath
%%%%%%%%%%%%%%%%%%%%%%%%%%%%%%%%%%%%%%%%%%%%

In Figure~\ref{fig:sigma_f_t} predictions of the MPS saturation 
model~\cite{soyez} for the $t$ dependence of the cross sections are shown, superimposed 
on the elastic measurements.
The data fall faster with \modt\ than predicted by the model, especially at small \qsq. 
The discrepancy is particularly significant for \ph\ production.

%-----------------------------------------------------------------------------
\begin{figure}[htbp]
\begin{center}
\setlength{\unitlength}{1.0cm}
\begin{picture}(7.0,7.0)   
\put(0.0,0.0){\epsfig{file=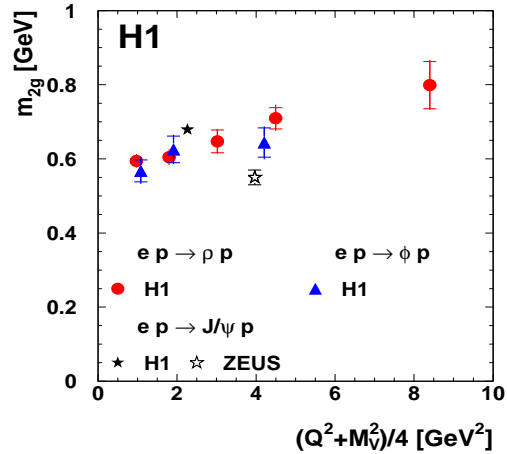  ,height=7.0cm,width=7.0cm}}
\end{picture}
\caption{Dependence on the scale \scaleqsqplmsq\ of the parameter $m_{2g}$ 
of the two-gluon form factor of the FS model~\cite{FS-m2g},
extracted from fits of Eq.~(\ref{eq:m2g}) to the $t$ distributions of \rh\
and \ph\ elastic production cross sections. 
The \jpsi\ measurements by H1 in photoproduction~\cite{h1-jpsi-hera1} and 
by ZEUS in electroproduction~\cite{z-jpsi-elprod} are also presented.
The present measurements are given in Table~\ref{table:m2g}.
}
\label{fig:two-gluon}
\end{center}
\end{figure}
%-----------------------------------------------------------------------------

A dipole function with a $t$ dependent two-gluon form factor has been proposed 
by Frankfurt and Strikman (FS)~\cite{FS-m2g}, with
\begin{equation}
{\rm d}\sigma / {\rm d}t \propto (1 + \modt / m^2_{2g})^{-4},
                                                                                    \label{eq:m2g}
\end{equation}
which tends to $e^{ - b |t| }$ for $ t \to 0$, with $b = 4 / m^2_{2g}$.
Fits of this parameterisation to the data for \rh\ and \ph\ elastic production in several bins
in \qsq\ are shown in Fig.~\ref{fig:sigma_f_t}, superimposed
on the measurements. The fit quality is good, similar to the exponential fits.
Figure~\ref{fig:two-gluon} presents the extracted values of the parameter $m_{2g}$ 
as a function of \scaleqsqplmsq\ for the \rh\ and \ph\ elastic channels.
The parameter increases with \scaleqsqplmsq, from about $0.6~\gevcsq$ at $5~\gevsq$
to about $0.8~\gevcsq$ at $35~\gevsq$.
A measurement in \jpsi\ photoproduction is also shown.
The \qsqplmsq\ dependence of the form factor reflects the \qsq\ dependence of the $t$ 
distributions, as summarised in Fig.~\ref{fig:b_f_qsq}.

%%%%%%%%%%%%%%%%%%%%%%%%%%%%%%%%%%%%%
\boldmath
\subsubsection{Slope of the effective pomeron trajectory}
                                                                                \label{sect:W-dep_shrinkage}
\unboldmath
%%%%%%%%%%%%%%%%%%%%%%%%%%%%%%%%%%%%%

%-----------------------------------------------------------------------------
\begin{figure}[htb]
\begin{center}
\setlength{\unitlength}{1.0cm}
\begin{picture}(16.0,8.0)   
\put(0.0,0.0){\epsfig{file=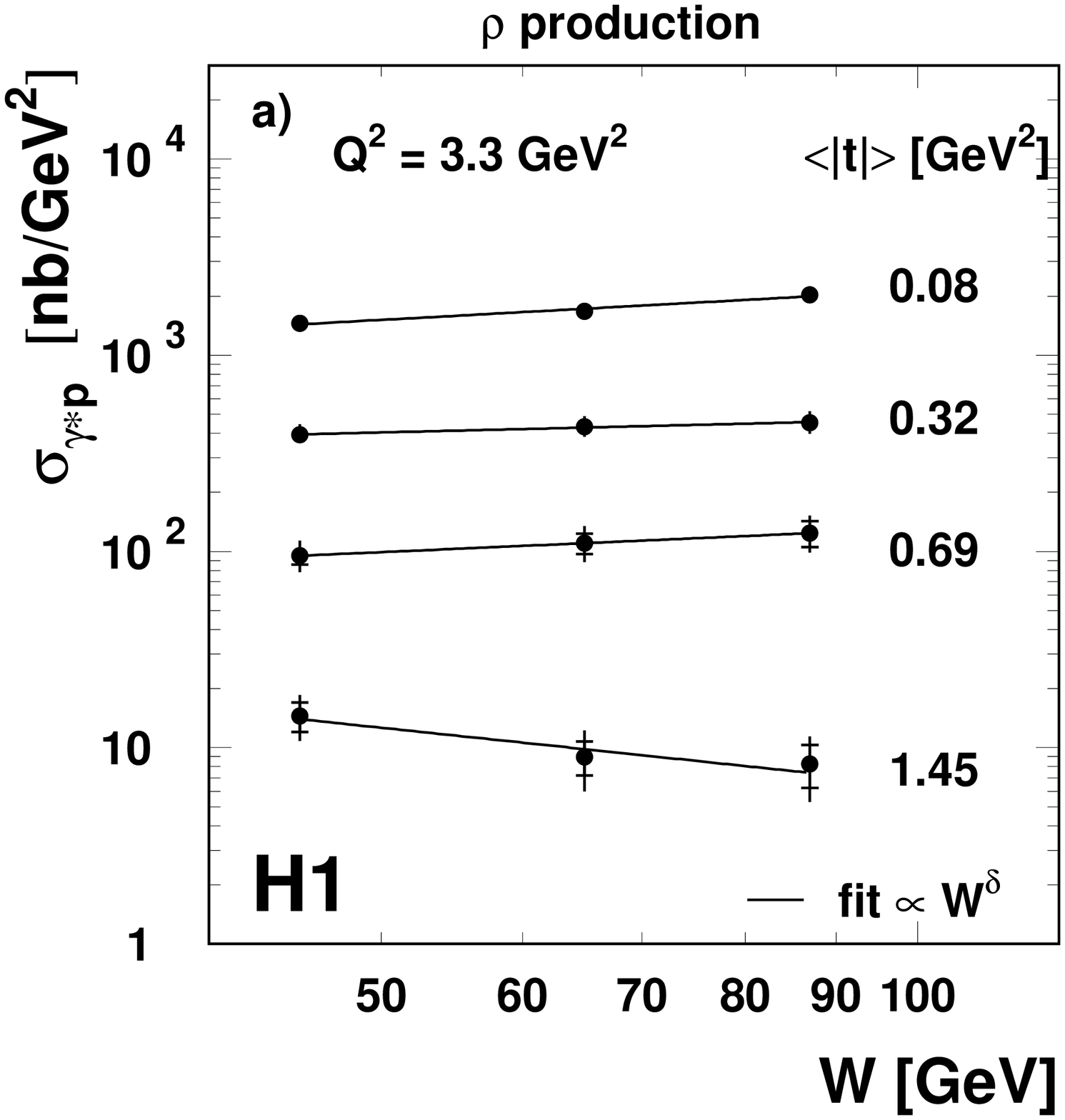,height=8.0cm,width=8.0cm}}
\put(8.0,0.0){\epsfig{file=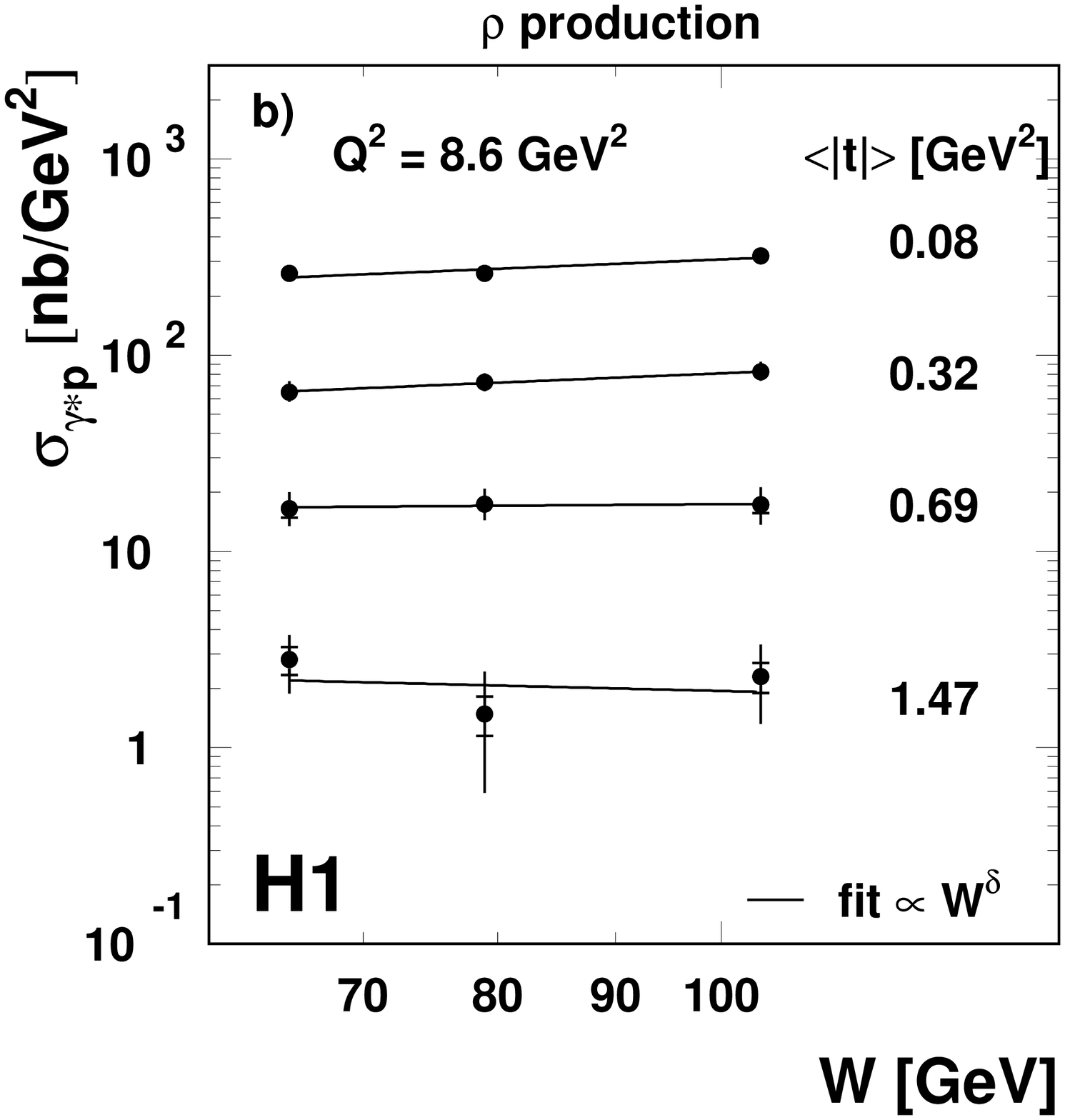,height=8.0cm,width=8.0cm}}
\end{picture}
\caption{$W$ dependence of the \gstarp\ cross sections for \rh\ meson production
in four bins in \modt,
for (a) $\qsq = 3.3~\gevsq$ and (b) $\qsq = 8.6~\gevsq$.
The lines are the results of power law fits.
The notag ($\modt \leq 0.5~\gevsq$) and tag ($\modt \leq 3~\gevsq$) samples are 
combined.
The measurements are given in Table~\ref{table:rho_xstw}.
}
\label{fig:sigma_W_f_t}
\end{center}
\end{figure}
%-----------------------------------------------------------------------------

The $W$ dependences in four bins in \modt\ of the \gstarp\ cross sections for \rh\ 
meson production are presented in Fig.~\ref{fig:sigma_W_f_t} for two values of 
\qsq.
The notag ($\modt \leq 0.5~\gevsq$) and tag ($\modt \leq 3~\gevsq$) samples are 
combined in order to extend the measurement lever arm in \modt.
It was checked that, using only the notag events, compatible values of \alp\ are
obtained, although with much larger errors.
The combination is also supported by the 
fact that the values of $\alpom(0)$ for the elastic and proton 
dissociative processes are compatible (see Fig.~\ref{fig:alphapom0_f_qsq}). 

%-----------------------------------------------------------------------------
\begin{figure}[htb]
\begin{center}
\setlength{\unitlength}{1.0cm}
\begin{picture}(16.0,4.0)   
\put(0.0,0.0){\epsfig{file=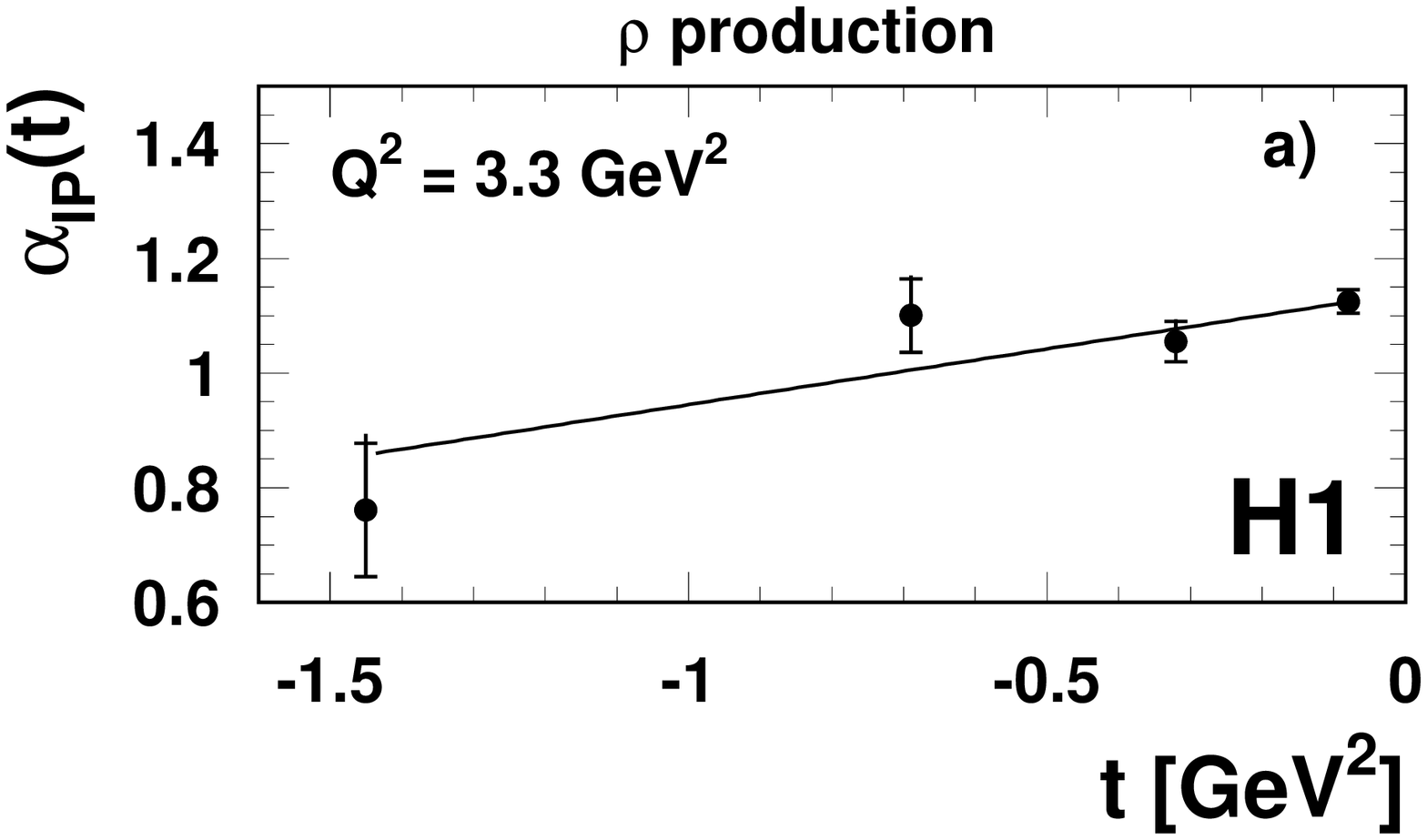,height=4.0cm,width=8.0cm}}
\put(8.0,0.0){\epsfig{file=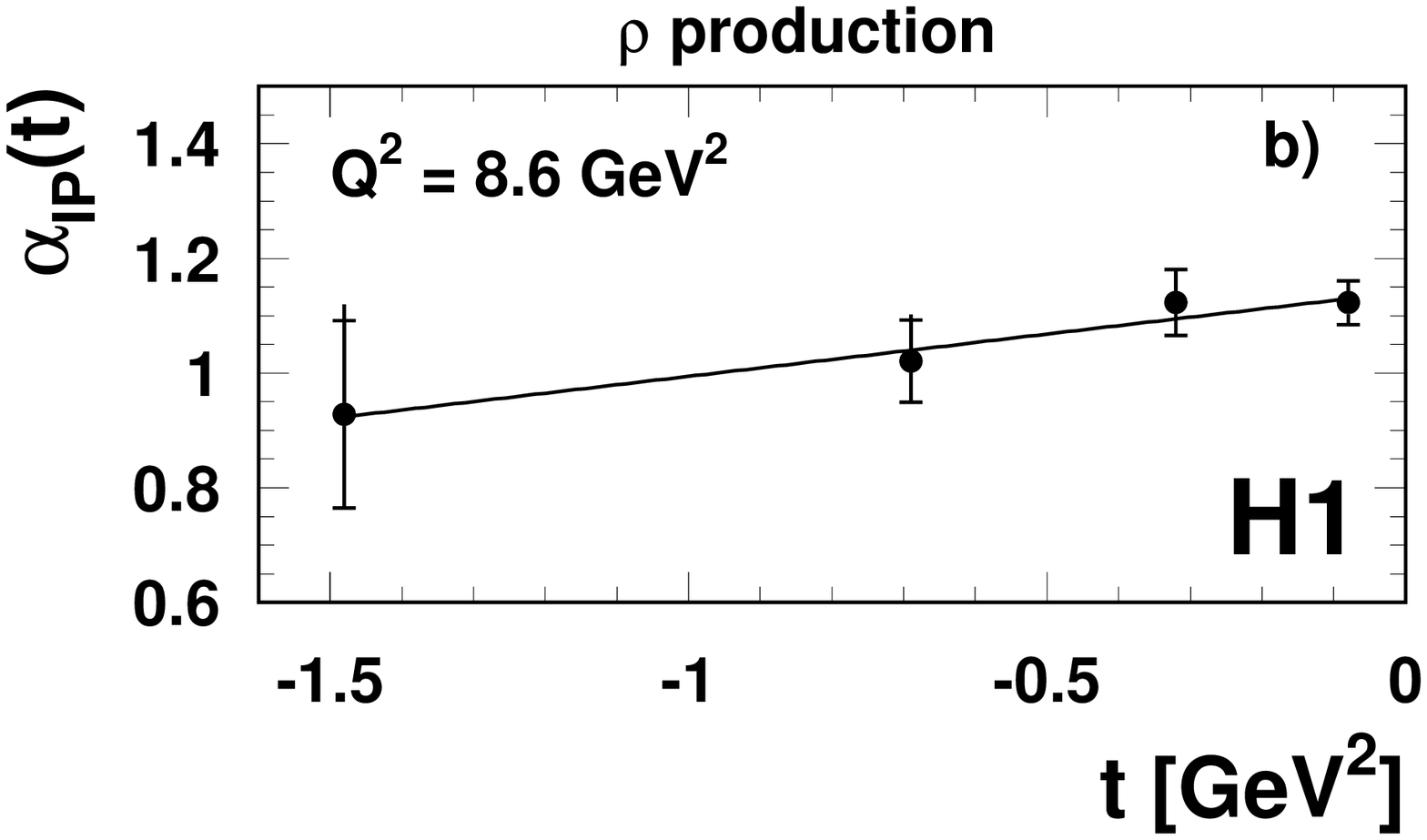,height=4.0cm,width=8.0cm}}
\end{picture}
\caption{$t$ dependence of the values of the $\alpom (t) = \delta(t) /4 +1$ parameters
obtained from the linear fits to the $W$ dependences shown in 
Fig.~\ref{fig:sigma_W_f_t} (\rh\ production), 
for (a) $\qsq = 3.3~\gevsq$; (b) $\qsq = 8.6~\gevsq$.
The lines are the results of linear fits of the form of Eq.~(\ref{eq:traj}).
The measurements are given in Table~\ref{table:apomt}.
} 
\label{fig:delta_f_t}
\end{center}
\end{figure}
%-----------------------------------------------------------------------------

%-----------------------------------------------------------------------------
\renewcommand{\arraystretch}{1.15}
\begin{table}[tbp]
\begin{center}
\begin{tabular}{|c|c|} 
\multicolumn{2}{c} { }  \\
\hline
    \qsq~(\gevsq) &    \multicolumn{1}{c|} {\alp~(\gevsqm)}  \\
\hline
\hline

$3.3$             &   $0.19 \pm 0.07~^{+0.03}_{-0.04}$    \\
$8.6$             &   $0.15 \pm 0.09~^{+0.07}_{-0.06}$     \\
\hline
\end{tabular} 
\caption{Measurement of the slope of the effective pomeron trajectory \alp\ for \rh\ 
production, from the \modt\ evolution of the $W$ dependence of the \rh\
cross section presented in Fig.~\protect\ref{fig:delta_f_t}, using 
Eqs.~(\protect\ref{eq:W-fit}-\protect\ref{eq:traj}), 
for $\qsq = 3.3$ and $8.6~\gevsq$.} 
\label{table:shrinkage}
\end{center}
\end{table}
\renewcommand{\arraystretch}{1.}
%-----------------------------------------------------------------------------

The $W$ dependences, which are observed to depend on \modt, are parameterised 
following the power law of Eq.~(\protect\ref{eq:W-fit}).
The extracted values of  $\alpom (t) = \delta(t) /4 +1$ are presented in 
Fig.~\ref{fig:delta_f_t}.
Linear fits to the $t$ dependence of $\alpom (t)$, following Eq.~(\ref{eq:traj}),
give the measurements of the slope \alp\ of the effective pomeron trajectory
reported in Table~\ref{table:shrinkage}.
Values slightly smaller than~$0.25~\gevsqm$ and higher than $0$ are obtained.

In soft diffraction, the non-zero value of the slope \alprim\ of the pomeron trajectory
($\alprim \simeq 0.25~\gevsqm$) explains the shrinkage of the forward diffractive 
peak with increasing $W$:
\begin{eqnarray}
         \frac { {\rm d} \sigma} { {\rm d} t} (W)
    &=&   \frac { {\rm d} \sigma} { {\rm d} t} (W_0) \ \left( \frac {W} {W_0} \right)^\delta
    \propto e^{b_0  t} \  \left( \frac {W} {W_0} \right)^{4 (\alpha_{I\!\!P}(0) +\alpha^\prime  t - 1 ) }\ ,  
                                                                                                    \nonumber \\
    b &=& b_0 + 4 \ \alp  \ t \ \ln (W / W_0)\ .
                                                                                              \label{eq:shrink}
\end{eqnarray}
The parameter \alp\ can thus in principle also be obtained from the evolution with $W$ of 
the exponential \modt\ slopes for elastic \rh\ production, but this measurement is affected 
by the large errors on $b$ (not shown).

%-----------------------------------------------------------------------------
\begin{figure}[tbp]
\begin{center}
\setlength{\unitlength}{1.0cm}
\begin{picture}(8.0,8.0)   
\put(0.0,0.0){\epsfig{file=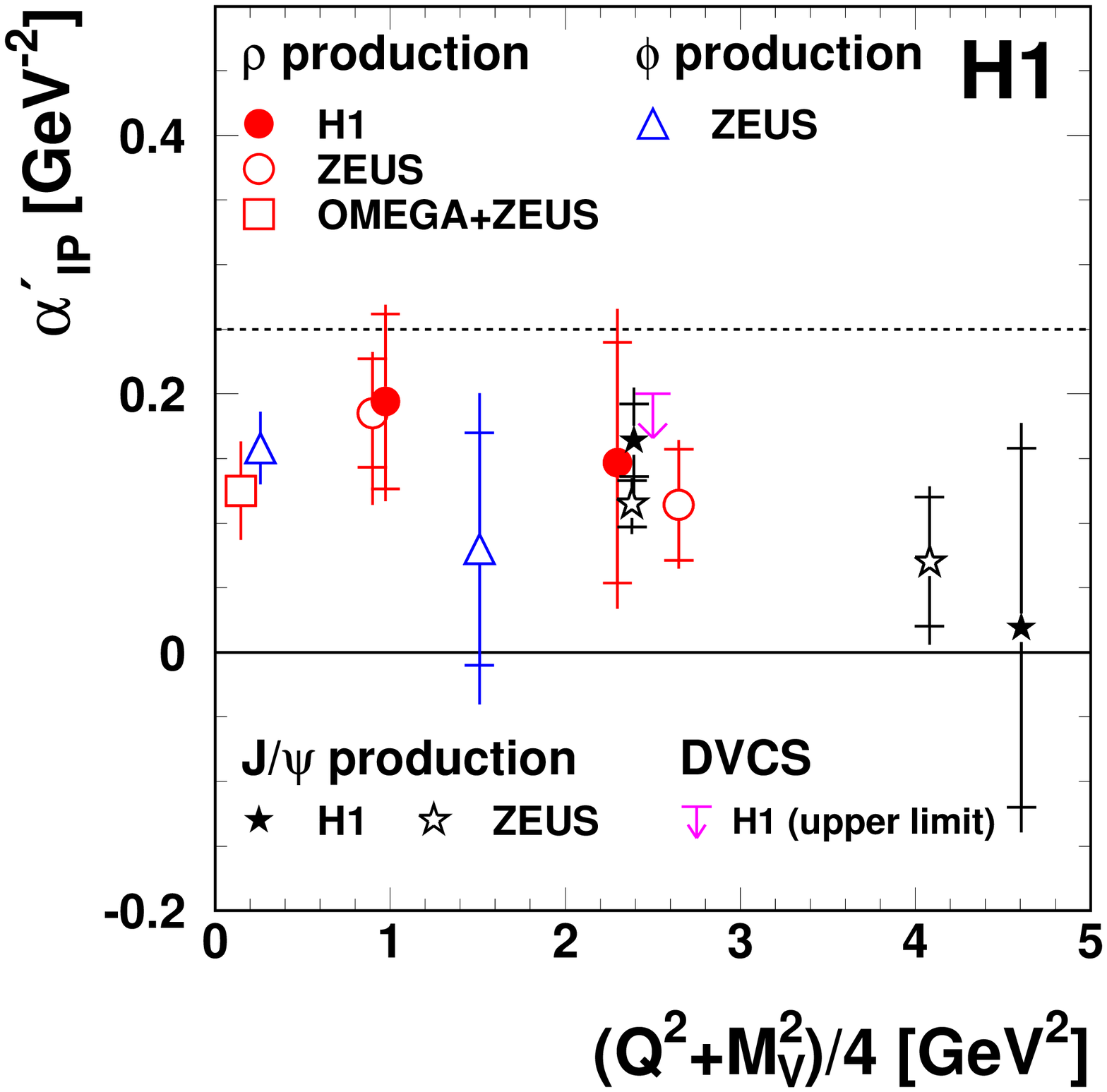  ,height=8.0cm,width=8.0cm}}
\end{picture}
\caption{Slope of the effective pomeron trajectory \alp, presented as a 
function of the scale $\mu^2 = \scaleqsqplmsq$, together with 
measurements by H1~\protect\cite{h1-dvcs,h1-jpsi-hera1}
and ZEUS~\protect\cite{z-rho-photoprod,z-rho,z-phi,z-jpsi-photoprod,z-jpsi-elprod}
for DVCS (upper limit $95\%\ C.L.$, with the scale $\mu^2 = \qsq$) and 
\rh, \ph\ and \jpsi\ in photo- and electroproduction with 
$\modt\ \lsim\ 1.5~\gevsq$.
The line $\alp = 0.25~\gevsqm$ represents a typical value in hadron-hadron 
interactions. }
\label{fig:alprim}
\end{center}
\end{figure}
%-----------------------------------------------------------------------------

Figure~\ref{fig:alprim} summarises \alp\ measurements by H1 
and ZEUS for DVCS and in photo- and electroproduction of \rh, \ph\ and \jpsi\ mesons.
The \alp\ measurement for \rh\ photoproduction~\cite{z-rho-photoprod}, which combines 
the ZEUS data at high energy with OMEGA results~\cite{omega} at low energy, 
is $\alp = 0.12 \pm 0.04~\gevsqm$, which is lower than the value~$0.25$ typical for soft 
hadronic diffraction and is similar, within errors, to values of \alp\ in electroproduction.
Measurements of \alp\ at large \modt\ are consistent with $0$, with small errors on the 
\jpsi\ measurements~\cite{h1-jpsi-hera1,z-jpsi-photoprod,z-jpsi-elprod}.

In the BFKL description of hard scattering, the value of \alp, which reflects the average 
transverse momentum $k_t$ of partons along the diffractive ladder, is expected to be 
small.
In Regge theory, the reggeon trajectories are fixed by the 
resonance positions, and slopes do not depend on \qsq.
Evolutions of the effective pomeron trajectories with \qsq\ or \modt\ are thus an 
indication of additional effects, e.g. multiple exchanges and rescattering processes.

%%%%%%%%%%%%%%%%%%%%%%%%%%%%%%%%%%%%%%%%%%%%
%%%%%%%%%%%%%%%%%%%%%%%%%%%%%%%%%%%%%%%%%%%%
\boldmath
\subsection{Comparison of proton dissociative and elastic cross sections}  
                                                                                       \label{sect:factor}
\unboldmath
%%%%%%%%%%%%%%%%%%%%%%%%%%%%%%%%%%%%%%%%%%%%

This section presents comparisons of the proton dissociative and elastic channels, 
for both \rh\ and \ph\ meson production.
Measurements of the $t$ integrated cross section ratios are first presented,
providing empirical information useful for experimental studies.
The factorisation of VM production amplitudes into photon vertex and proton vertex 
contributions, which can be disentangled by comparing elastic and proton dissociative 
scatterings, is then discussed:
the photon vertex contributions govern the \qsq\ dependence and the relative strength 
of the various helicity amplitudes, whereas proton vertex form factors govern the 
$t$ dependence.
Proton vertex factorisation (``Regge factorisation") has been observed to hold, within 
experimental uncertainties, for inclusive diffraction~\cite{incl-diffr-2006}.
Factorisation is tested here through the study of the \qsq\ independence of
the VM production cross section ratios at $t = 0$ and through 
the measurement of the difference $b_{el.} - b_{p.~diss.}$ between the elastic and 
the proton dissociative exponential $t$ slopes.

%%%%%%%%%%%%%%%%%%%%%%%%%%%%%%%%%%%%%
\subsubsection{ \boldmath {\qsq} dependence of the cross section ratios}
                                                                                \label{sect:pd_on_el_f_qsq}
%%%%%%%%%%%%%%%%%%%%%%%%%%%%%%%%%%%%%

%-----------------------------------------------------------------------------
\begin{figure}[htbp]
\begin{center}
\setlength{\unitlength}{1.0cm}
\begin{picture}(12.0,6.0)   
\put(0.0,0.0){\epsfig{file=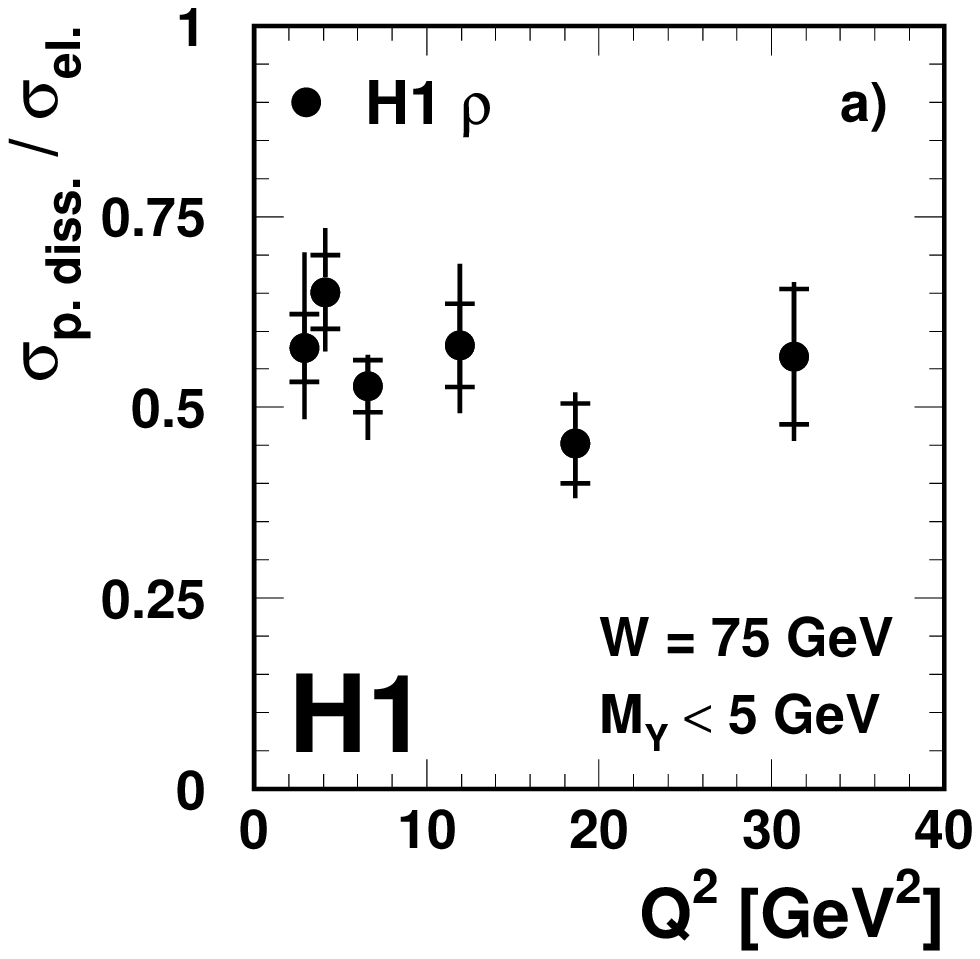,height=6.0cm,width=6.0cm}}
\put(6.0,0.0){\epsfig{file=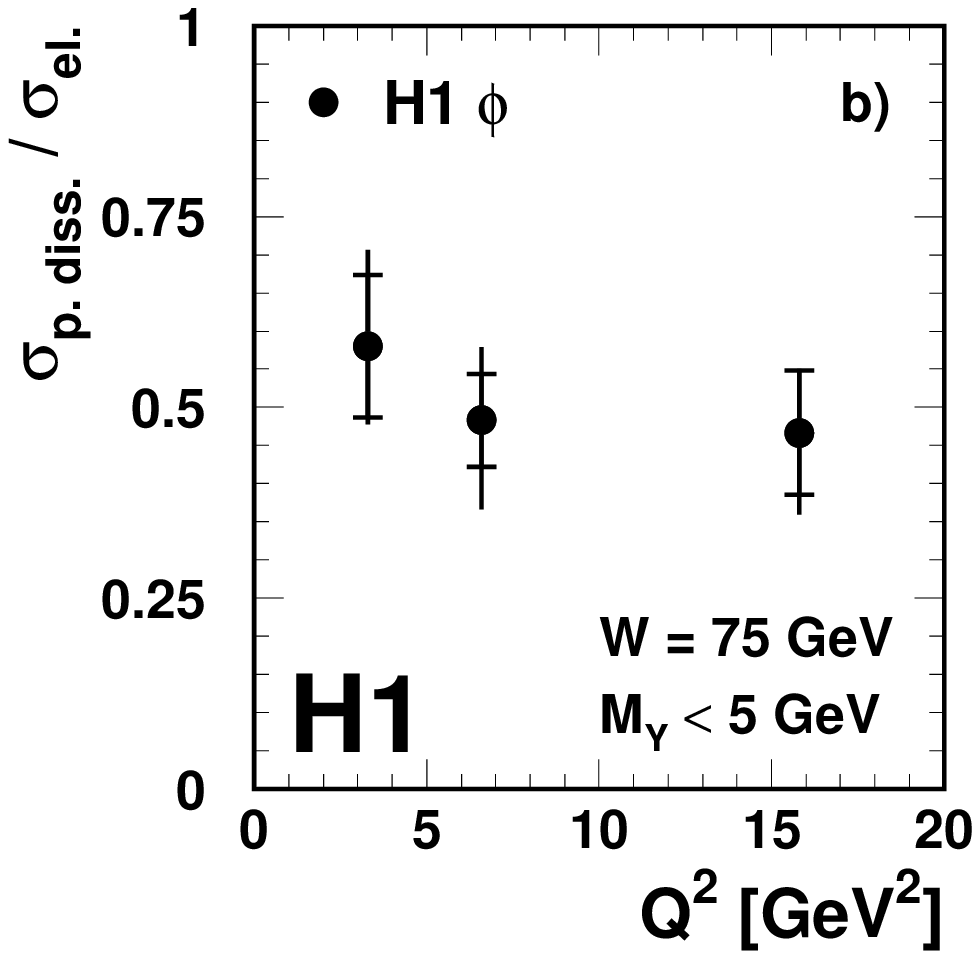,height=6.0cm,width=6.0cm}}
\end{picture}
\caption{\qsq\ dependence of the ratio of proton dissociative ($M_Y < 5~\gevcsq$) to 
elastic \gstarp\ cross sections for $W = 75~\gev$:
(a) \rh\ meson production; (b) \ph\ production.
The overall normalisation error on the ratios, which is not included in the error bars,
is~$2.4\%$.
The measurements are given in Tables~\ref{table:rho_pdel} and \ref{table:phi_pdel}.
} 
\label{fig:pd_on_el_f_qsq}
\end{center}
\end{figure}
%-----------------------------------------------------------------------------

Figure~\ref{fig:pd_on_el_f_qsq} presents, as a function of \qsq, the ratio of the 
proton dissociative to elastic \gstarp\ cross sections, for \rh\ and \ph\ mesons.
In the ratio, several systematic uncertainties cancel, in particular those related to 
meson reconstruction.
No significant dependence of the ratios on \qsq\ is observed.

The average ratios of proton dissociative (with $M_Y < 5~\gevcsq$) to elastic 
cross sections, integrated over $t$, are:
\begin{eqnarray} 
\frac {\sigma_{tot, {\rm p.~diss.}}^{\rm M_Y < 5~{\rm GeV} }} 
      {\sigma_{tot, {\rm el.}}} (\rho) &=&
0.56 \pm 0.02~{\rm (stat.)}~^{+ 0.03}_{-0.05}~{\rm (syst.)}\pm 0.01~{\rm (norm.)}\ ,
         \nonumber \\[0.2cm]
\frac {\sigma_{tot, {\rm p.~diss.}}^{\rm M_Y < 5~{\rm GeV} }} 
      {\sigma_{tot, {\rm el.}}} (\phi) &=&
0.50 \pm 0.04~{\rm (stat.)}~^{+ 0.06}_{-0.08}~{\rm (syst.)}\pm 0.01~{\rm (norm.)}\ . 
                                                                     \label{eq:pd_to_el_ratio}
\end{eqnarray} 
Within uncertainties, the values for the two VMs are compatible.
Using the DIFFVM model to estimate the contributions of proton dissociative
scattering with $M_Y > 5~{\rm GeV}$, the ratio of the proton dissociative 
cross section for the
full $M_Y$ mass range to the elastic cross section is found to be close to~$1$.
This value is used e.g. in~\cite{incl-diffr-2006}.

%%%%%%%%%%%%%%%%%%%%%%%%%%%%%%%%%%%%%%%%%%%%
\boldmath
\subsubsection{Cross section ratios for $t = 0$}  
                                                                                       \label{sect:t0_factor}
\unboldmath
%%%%%%%%%%%%%%%%%%%%%%%%%%%%%%%%%%%%%%%%%%%%

If the same object (e.g. a gluon ladder) is exchanged in proton dissociative and elastic
scattering, proton vertex factorisation should be manifest through the \qsq\ independence 
of the cross section ratio for $t = 0$.

%-----------------------------------------------------------------------------
\begin{figure}[htbp]
\begin{center}
\setlength{\unitlength}{1.0cm}
\begin{picture}(12.0,6.0)   
\put(0.0,0.0){\epsfig{file=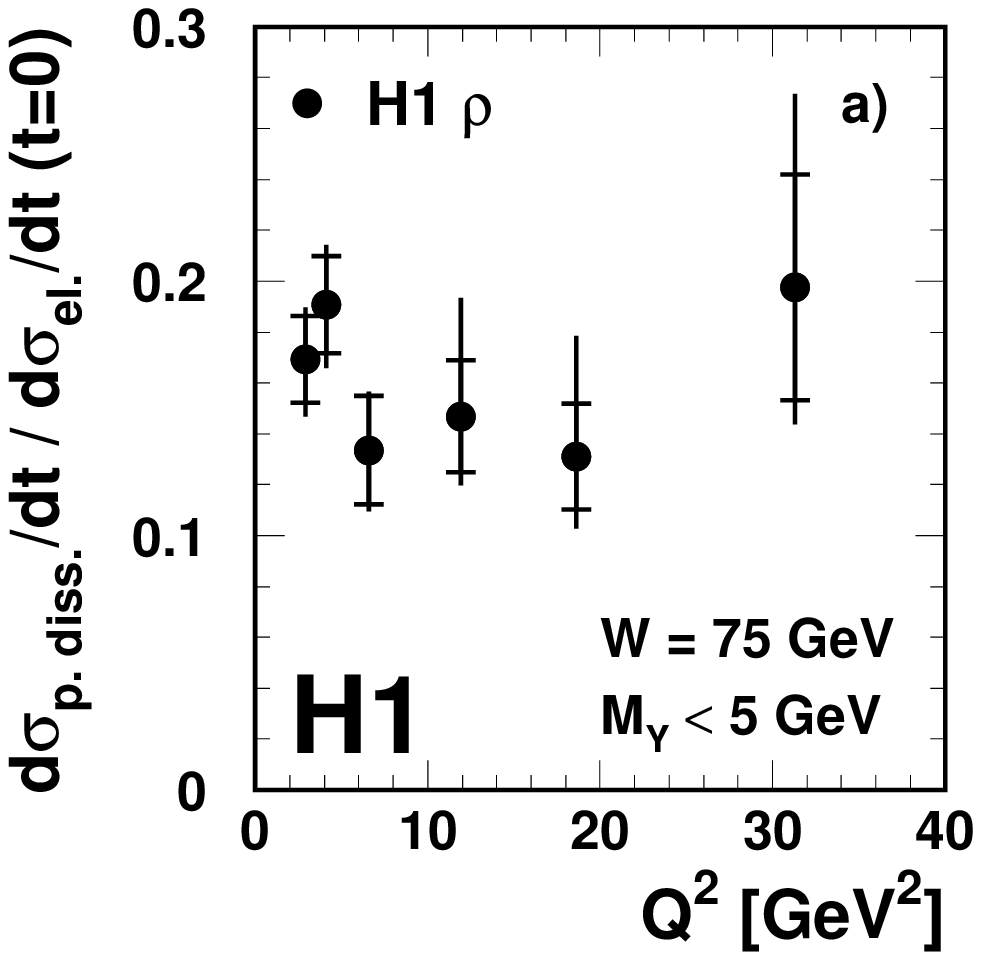,height=6.0cm,width=6.cm}}
\put(6.0,0.0){\epsfig{file=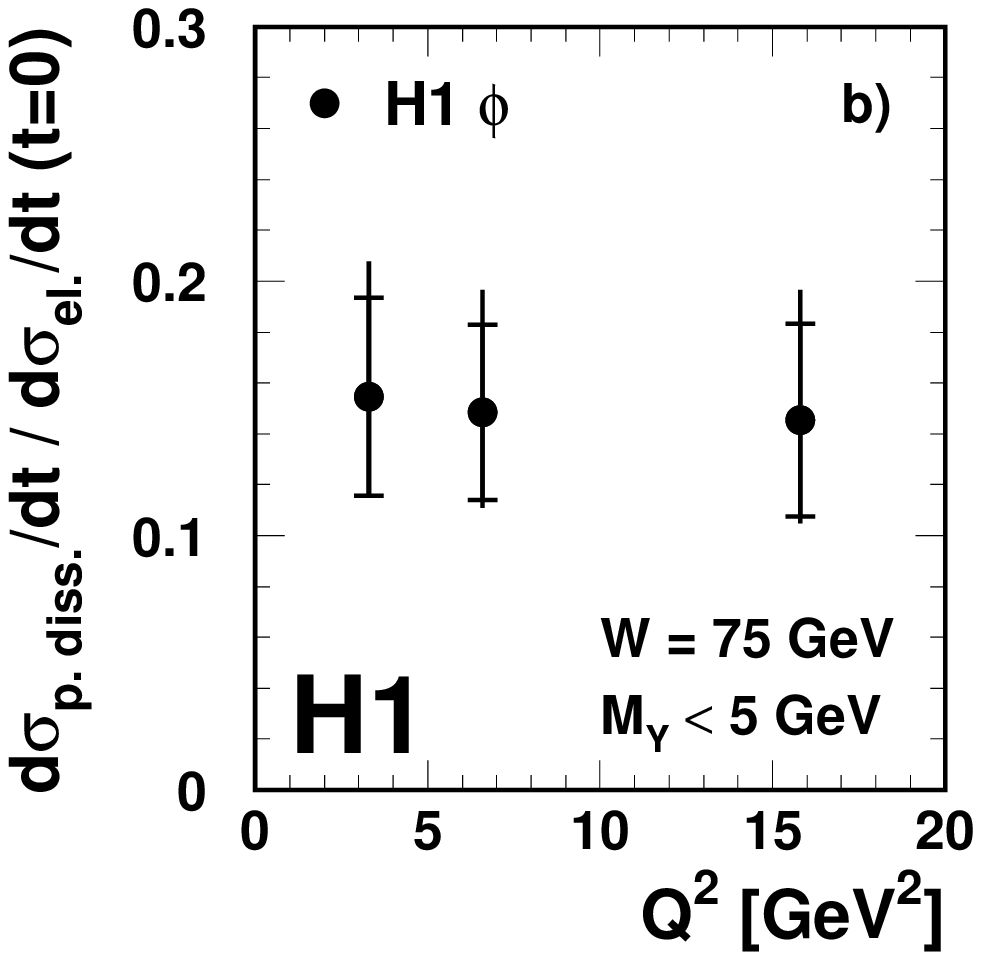,height=6.0cm,width=6.cm}}
\end{picture}
\caption{\qsq\ dependence of the ratio of the proton dissociative (with $M_Y < 5~\gevcsq$)
to the elastic \gstarp\ cross sections at $t = 0$ and $W = 75~\gev$, 
$\frac { {\rm d}\sigma_{{\rm p.~diss.}}  / {\rm d}t } { {\rm d}\sigma_{{\rm el.}} / {\rm d}t } {\rm (t = 0)}$,
for (a) \rh\ meson production; (b) \ph\ production.
The overall normalisation errors, not included in the error bars, are the same as 
in Fig.~\protect\ref{fig:pd_on_el_f_qsq}.
The measurements are given in Tables~\ref{table:rho_pdel}-\ref{table:phi_pdel}.
}
\label{fig:pd_on_el_f_qsq_t0}
\end{center}
\end{figure}
%-----------------------------------------------------------------------------

For exponentially falling $t$ distributions, the cross section ratio at $t = 0$ is obtained from 
the total cross sections and the $b$ slopes as 
\begin{equation}
\frac { {\rm d}\sigma_{{\rm p.~diss.}}  / {\rm d}t } { {\rm d}\sigma_{{\rm el.}} / {\rm d}t } {\rm (t = 0)} =
    \frac { \sigma_{tot, {\rm p.~diss.}}  }  {\sigma_{tot, {\rm el.}} }
    \cdot \frac {b_{\rm {p.~diss.}} } {b_{\rm {el.}} }\ .
                                                                                       \label{eq:t0_ratio}
\end{equation}

Figure~\ref{fig:pd_on_el_f_qsq_t0} presents, as a function of \qsq, the 
cross section ratios at $t = 0$ for \rh\ and \ph\ production, as obtained from 
the total cross section ratios presented in 
Fig.~\ref{fig:pd_on_el_f_qsq} and the $b$ slopes given in Fig.~\ref{fig:b_f_qsq}.

The average ratios for both VMs are measured as:
\begin{eqnarray} 
\frac { {\rm d}\sigma_{{\rm p.~diss.}}^{\rm M_Y < 5~{\rm GeV} } / {\rm d}t } 
       { {\rm d}\sigma_{\rm el.} / {\rm d}t } 
      {\rm (t = 0)} (\rho)                               
&=& 0.159 \pm 0.009~{\rm (stat.)}~^{+ 0.011}_{-0.025}~{\rm (syst.)}\pm 0.004~{\rm (norm.)}\ ,
         \nonumber \\
\frac { {\rm d}\sigma_{{\rm p.~diss.}}^{\rm M_Y < 5~{\rm GeV} } / {\rm d}t } 
       { {\rm d}\sigma_{\rm el.} / {\rm d}t } 
      {\rm (t = 0)} (\phi)                               
&=& 0.149 \pm 0.021~{\rm (stat.)}~^{+ 0.035}_{-0.036}~{\rm (syst.)}\pm 0.003~{\rm (norm.)}\ . 
         \nonumber \\
                                                                     \label{eq:t-0_pd_to_el_ratio}
\end{eqnarray} 

The ratios are observed to be independent of \qsq\ and consistent for the two VMs, 
which supports proton vertex factorisation.

The ratios of the proton dissociative to elastic $b$ slopes are also independent of \qsq, 
with average values of 
\begin{eqnarray} 
b_{p.~diss.} \ / \ b_{el.}  (\rh) &=& 0.28 \pm 0.01~{\rm (stat.)}~^{+ 0.01}_{-0.02}~{\rm (syst.)}\ ,
         \nonumber \\
b_{p.~diss.} \ / \ b_{el.}  (\ph) &=& 0.27 \pm 0.05~{\rm (stat.)}~^{+ 0.06}_{-0.01}~{\rm (syst.)}\ .
                                                                     \label{eq:bpd_to_bel_ratio}
\end{eqnarray} 
This empirical observation is consistent with the \qsq\ independence of the total cross 
section ratios (Fig.~\ref{fig:pd_on_el_f_qsq}) and of the cross section ratios at $t = 0$ 
(Fig.~\ref{fig:pd_on_el_f_qsq_t0}).

%%%%%%%%%%%%%%%%%%%%%%%%%%%%%%%%%%%%%%%%%%%%
\boldmath
\subsubsection{Difference in {\boldmath {$t$}} slope between elastic and proton 
                                                           dissociative scattering}
                                                                                \label{sect:t_slopes_diff}
\unboldmath
%%%%%%%%%%%%%%%%%%%%%%%%%%%%%%%%%%%%%%%%%%%%

In the optical model approach of Eq.~(\ref{eq:optical}), assuming pomeron universality, 
the difference between the elastic and proton dissociative $b$ slopes, $b_{el.} - b_{p.~diss.}$, 
is related only to the proton size and independent of the interaction scale at the photon 
vertex and of the VM species.

%-----------------------------------------------------------------------------
\begin{figure}[htbp]
\begin{center}
\setlength{\unitlength}{1.0cm}
\begin{picture}(6.0,6.0)   
\put(0.0,0.0){\epsfig{file=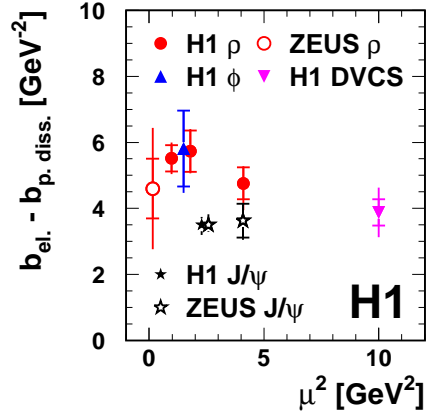  ,height=6.0cm,width=6.0cm}}
\end{picture}
\caption{Slope differences $b_{el.} - b_{p.~diss.}$ between elastic and proton dissociative 
scattering for \rh\ and \ph\ meson production, as a function of \scaleqsqplmsq.
Results of H1 
for DCVS~\protect\cite{h1-dvcs} 
and \jpsi\ photoproduction~\protect\cite{h1-jpsi-hera1,h1-psi2s}
and of ZEUS 
for \rh~\protect\cite{z-rho-photoprod} 
and \jpsi\ \protect\cite{z-jpsi-photoprod,z-jpsi-elprod} photo- and
electroproduction
are also shown.
The present measurements are given in Table~\ref{table:rho_pdel_bslp}.
}
\label{fig:b_el-minus-pd}
\end{center}
\end{figure}
%-----------------------------------------------------------------------------

Figure~\ref{fig:b_el-minus-pd} presents the slope difference $b_{el.} - b_{p.~diss.}$ 
for \rh\ and \ph\ meson production, as a function of \scaleqsqplmsq.
Within errors, \qsq\ independent values for the slope differences are found, with 
consistent average values of
\begin{eqnarray} 
b_{el.} - b_{p.~diss.} (\rh) &=& 5.31 \pm 0.28~{\rm (stat.)}~^{+ 0.29}_{-0.24}~{\rm (syst.)}\ ,
         \nonumber \\
b_{el.} - b_{p.~diss.} (\ph) &=& 5.81 \pm 1.14~{\rm (stat.)}~^{+ 0.14}_{-0.74}~{\rm (syst.)}\ .
                                                                     \label{eq:bel-bpd}
\end{eqnarray} 
These observations support proton vertex factorisation, with a proton form factor 
contribution of about $5.5~\gevsqm$.

Measurements of \jpsi\ photo- and electroproduction are also presented 
in Fig.~\ref{fig:b_el-minus-pd}. 
They are consistent with \qsq\ independence, with 
$b_{el.} - b_{p.~diss.} = 3.50 \pm 0.07~\gevsqm$, a value significantly 
smaller than for \rh\ and \ph\ production;
for DVCS~\cite{h1-dvcs}, the measurement is $3.88 \pm 0.61~\gevsqm$. 
The difference observed between light and heavy VMs is difficult to understand in the 
optical model, since the contributions to the slopes of the $q \bar{q}$ dipole form factors 
and of possible VM form factors should cancel in the difference.
It may indicate that the hard regime is not reached for \rh\ and \ph\ mesons in the 
present kinematic domain.

%%%%%%%%%%%%%%%%%%%%%%%%%%%%%%%%%%%%%%%%%%%%
%%%%%%%%%%%%%%%%%%%%%%%%%%%%%%%%%%%%%%%%%%%%
%%%%%%%%%%%%%%%%%%%%%%%%%%%%%%%%%%%%%%%%%%%%
%%%%%%%%%%%%%%%%%%%%%%%%%%%%%%%%%%%%%%%%%%%%

%\newpage

%%%%%%%%%%%%%%%%%%%%%%%%%%%%%%%%%%%%%%%%%%%%
\section{Polarisation Measurements}  
                                                  \label{sect:polarisation}
%%%%%%%%%%%%%%%%%%%%%%%%%%%%%%%%%%%%%%%%%%%%

Information on the spin and parity properties of the exchange and on the 
contribution of the various polarisation amplitudes are accessed in diffractive VM production 
through the distributions of the angles $\theta$, $\phib$ and $\phi$ defined in 
Fig.~\ref{fig:dec_ang}.
The present section presents, successively, the  measurements of the spin density matrix 
elements, a discussion of the nature of the exchange, measurements
of the longitudinal over transverse cross section ratio $R$, and measurements of the ratios
and relative phases of the helicity amplitudes.
The results are compared with QCD models.

%%%%%%%%%%%%%%%%%%%%%%%%%%%%%%%%%%%%%%%%%%%%
%%%%%%%%%%%%%%%%%%%%%%%%%%%%%%%%%%%%%%%%%%%%
\subsection{Spin density matrix elements}  
                                                                              \label{sect:matrix-elements}
%%%%%%%%%%%%%%%%%%%%%%%%%%%%%%%%%%%%%%%%%%%%

%%%%%%%%%%%%%%%%%%%%%%%%%%%%%%%%%%%%%%%%%%%%
\subsubsection{Measurement procedure}  
                                                                              \label{sect:matrix-elements_procedure}
%%%%%%%%%%%%%%%%%%%%%%%%%%%%%%%%%%%%%%%%%%%%

In the formalism of Schilling and Wolf~\cite{sch-w}, summarised in the Appendix, 
the angular distributions allow the measurement of spin density matrix 
elements given in the form $r^{i}_{jk}$, which are normalised bilinear combinations 
of the complex helicity amplitudes 
$T_{\lambda_V \lambda_{N'}, \lambda_{\gamma} \lambda_{N}}$, 
$\lambda_{\gamma}$ and  $\lambda_{V}$ being the helicities of the virtual photon and 
of the VM, respectively, and $\lambda_{N}$ and $\lambda_{N'}$  those
of the incoming proton and of the outgoing baryonic system $Y$.

At HERA, the proton beam is not polarised and the helicity of the outgoing baryonic system 
$Y$ is not measured;
the helicities $\lambda_{N}$ and $\lambda_{N'}$ are thus integrated over.
For the electron beam, transverse polarisation builds up progressively over the running period
through the Sokolov-Ternov effect but the related matrix elements are measurable only for 
$\qsq \approx m_e^2$, where $m_e$ is the electron mass, and are not accessible in 
electroproduction.
The electron beam is thus treated here as unpolarised.

In these conditions, a total of 15 independent components of the spin density 
matrix remain accessible to measurement.
Under natural parity exchange (NPE) in the $t$ channel\footnote{NPE trajectories
are defined as containing for $t > 0$ poles with $P = (-1)^J$, $P$ and $J$ being the 
particle parity and spin, respectively.}, 
five $T_{\lambda_V \lambda_{\gamma}}$ amplitudes are independent:
two helicity conserving amplitudes ($ T_{00}$ and $T_{11}$), 
two single helicity flip amplitudes ($T_{01}$ and $T_{10}$) 
and one double flip amplitude ($ T_{-11}$).

The 15 matrix elements enter the normalised angular distribution 
$W(\theta, \phib, \phi)$ which is given in Eq.~(\ref{eq:W}) of the Appendix.
They are measured as projections of the $W(\theta, \phib, \phi)$ 
distribution onto 15 orthogonal functions of the $\theta$, \phib\ and \ph\ angles,
listed in Appendix~C of~\cite{sch-w}.
In practice, each matrix element is given by the average value of the corresponding
($\theta$, \phib, \ph) function, calculated over the relevant data sample.
For \rh\ production, the $\omega$, \ph\ and \rhop\ background contributions to the 
angular distributions 
are subtracted following the results of the Monte Carlo simulations;
no correction is performed for the interfering non-resonant $\pi \pi$ channel
but this is expected to have a small effect since the interference contribution is
small, decreases with \qsq\ and changes sign at the resonance mass value, so 
that it largely cancels when integrated over the selected mass range 
(see Fig.~\ref{fig:rh_mass_sod}).
For \ph\ production, the $\omega$, \rhop\ and dipion backgrounds are subtracted. 
Kinematic and angular distributions are corrected for detector acceptance and migration
effects.
The systematic errors on the measurements are estimated by varying the MC simulations 
according to the list given in Table~\ref{table:syst}.
In addition, a systematic error related to the binning is assigned to the acceptance 
correction used for
determining the average value of the projection functions; it is quantified by varying the 
number of bins in the $\theta$, \phib\ and \ph\ angular variables.

For both \rh\ and \ph\ mesons, the matrix element measurements for the elastic and proton 
dissociative channels are found to be compatible within experimental errors.
In order to improve the statistical significance of the measurements and to reach higher
\modt\ values, the notag and tag samples with $\modt \leq 0.5~\gevsq$ and 
$\modt \leq 3~\gevsq$, respectively, are combined. The large \modt\ notag sample 
is not used because of the large \rhop\ background, as shown in 
Figs.~\ref{fig:mass_distrib-mpp-NT}(c)-(d).

%%%%%%%%%%%%%%%%%%%%%%%%%%%%%%%%%%%%%%%%%%%%
\subsubsection{Matrix element measurements}  
                                                                              \label{sect:matrix-elements_meas}
%%%%%%%%%%%%%%%%%%%%%%%%%%%%%%%%%%%%%%%%%%%%

%-----------------------------------------------------------------------------
\begin{figure}[htbp]
\begin{center}
\setlength{\unitlength}{1.0cm}
\begin{picture}(16.0,16.0)   
\put(0.0,0.0){\epsfig{file=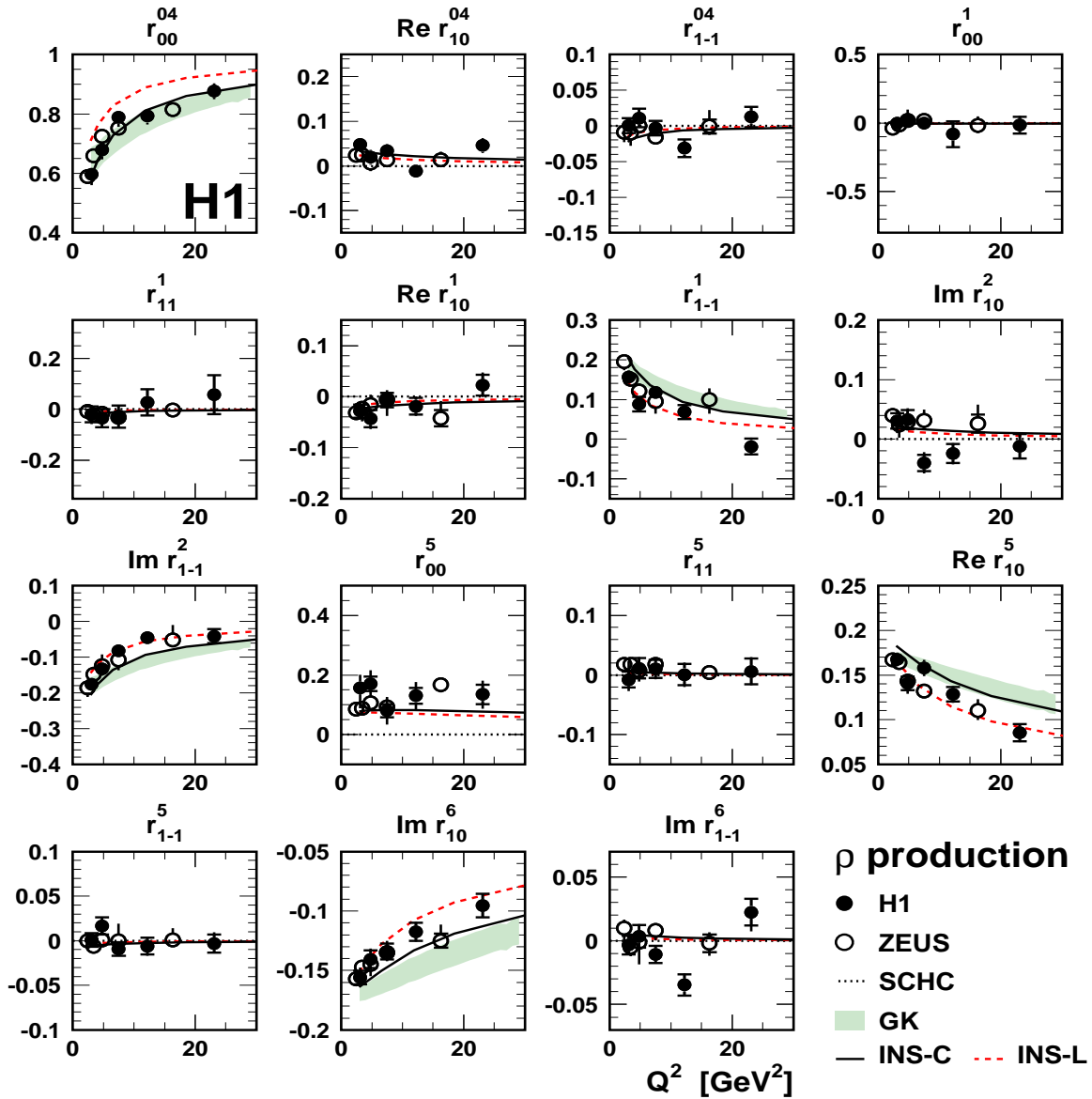  ,height=16.0cm,width=16.0cm}}
\end{picture}
\caption{Spin density matrix elements for the diffractive electroproduction of \rh\ mesons, 
as a function of \qsq.
The notag ($\modt \leq 0.5~\gevsq$) and tag ($\modt \leq 3~\gevsq$) samples are combined.
ZEUS results~\cite{z-rho} are also shown.
Where appropriate, the dotted lines show the expected vanishing values of the matrix 
elements if only the SCHC amplitudes are non-zero.
The shaded bands are predictions of the GK GPD model~\cite{kroll} for the
elements which are non-zero in the SCHC approximation;
the curves are predictions of the INS $k_t$-unintegrated model~\cite{ins}
for the compact (solid lines) and large (dashed lines)  wave functions, respectively.
The present measurements are given in Table~\ref{table:matelem_f_qsq_rho}.
}
\label{fig:matelem_f_qsq_rho}
\end{center}
\end{figure}
%-----------------------------------------------------------------------------

%-----------------------------------------------------------------------------
\begin{figure}[htbp]
\begin{center}
\setlength{\unitlength}{1.0cm}
\begin{picture}(16.0,16.0)   
\put(0.0,0.0){\epsfig{file=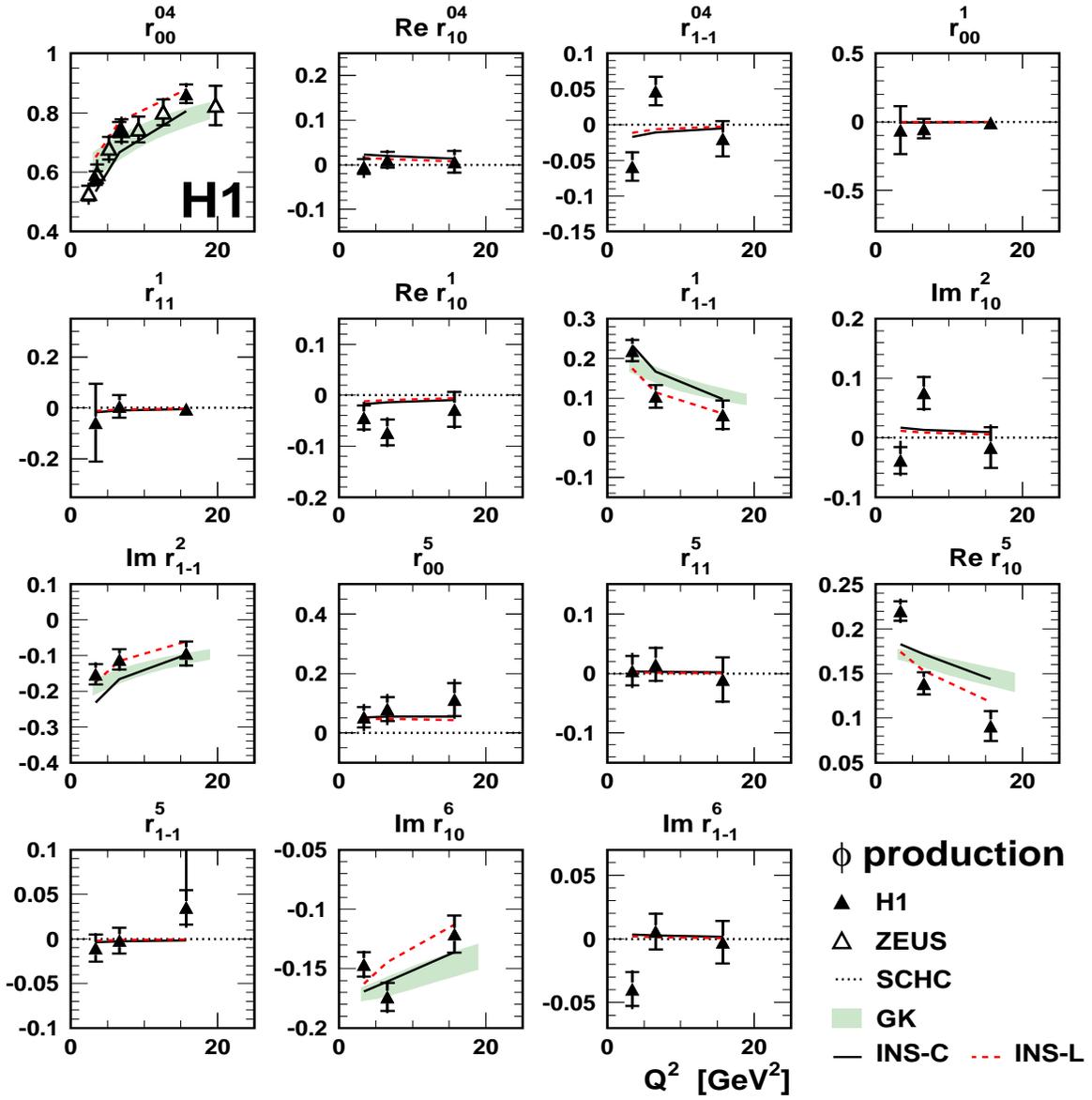  ,height=16.0cm,width=16.0cm}}
\end{picture}
\caption{Same as Fig.~\ref{fig:matelem_f_qsq_rho}, for \ph\ mesons.
ZEUS results~\protect\cite{z-phi} for the \rzqzz\ matrix element are also shown.
The present measurements are given in Table~\ref{table:matelem_f_qsq_phi}.
}
\label{fig:matelem_f_qsq_phi}
\end{center}
\end{figure}
%-----------------------------------------------------------------------------

%-----------------------------------------------------------------------------
\begin{figure}[htbp]
\begin{center}
\setlength{\unitlength}{1.0cm}
\begin{picture}(16.0,16.0)   
\put(0.0,0.0){\epsfig{file=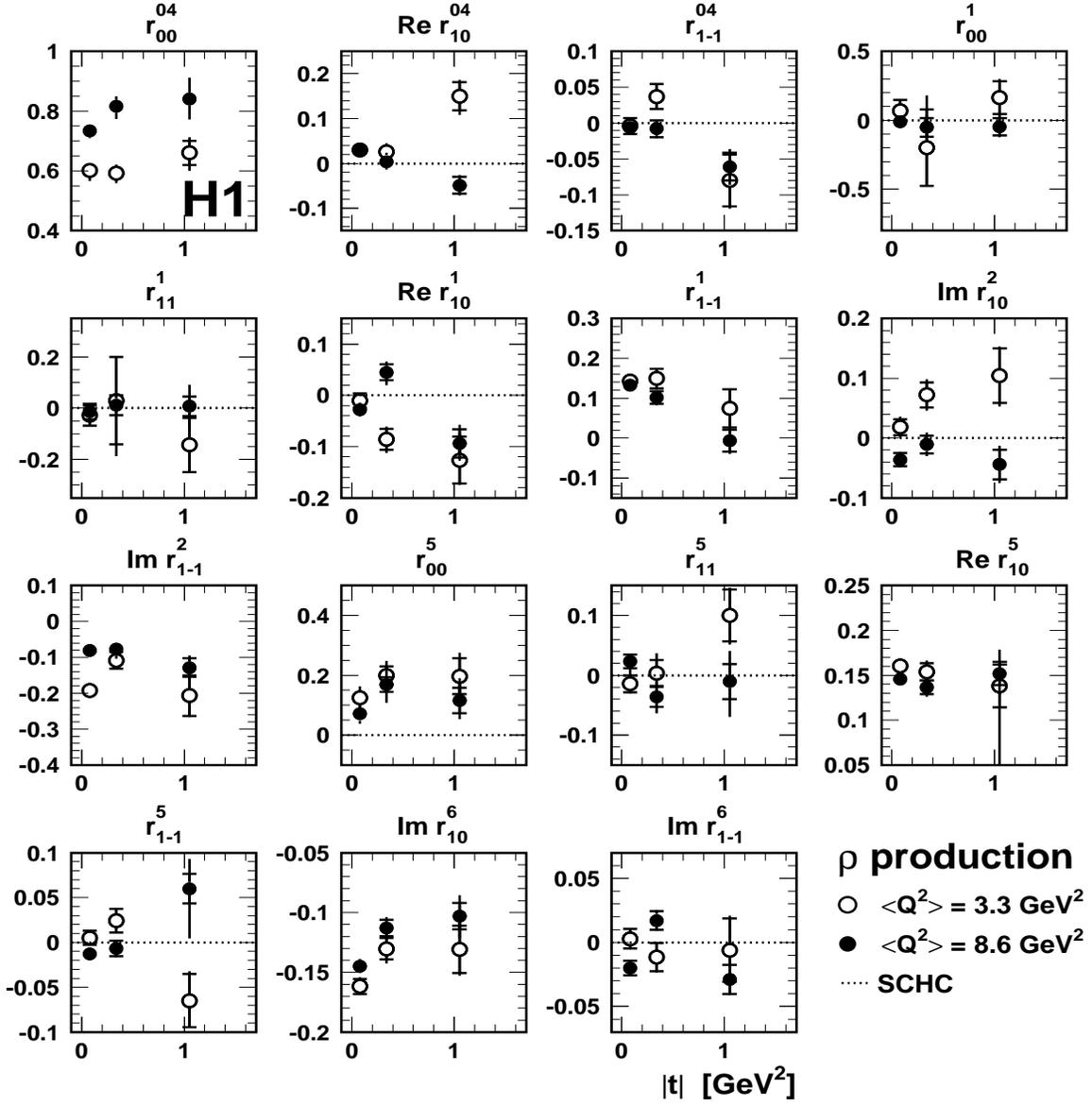  ,height=16.0cm,width=16.0cm}}
\end{picture}
\caption{Spin density matrix elements for the diffractive electroproduction of \rh\ mesons, 
as a function of \modt, for two intervals in \qsq: $2.5 \leq \qsq < 5~\gevsq$ and 
$5 \leq \qsq  \leq 60~\gevsq$.
Where appropriate, the dotted lines show the expected vanishing values of the matrix 
elements if only the SCHC amplitudes are non-zero.
The measurements are given in Table~\ref{table:matelem_f_t_two_qsq_rho}.
}
\label{fig:matelem_f_t_two_qsq_rho}
\end{center}
\end{figure}
%-----------------------------------------------------------------------------

%-----------------------------------------------------------------------------
\begin{figure}[htbp]
\begin{center}
\setlength{\unitlength}{1.0cm}
\begin{picture}(16.0,16.0)   
\put(0.0,0.0){\epsfig{file=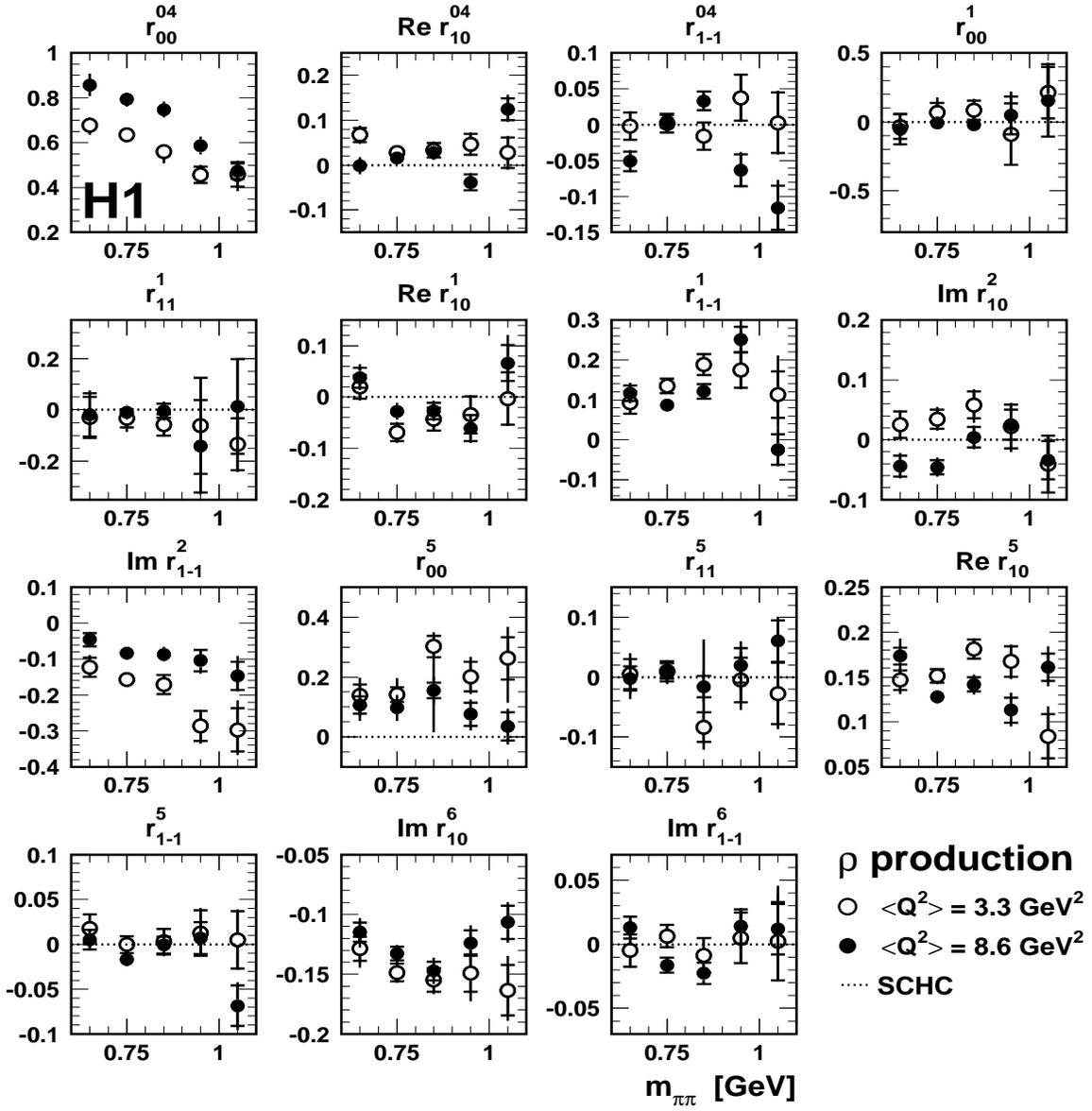  ,height=16.0cm,width=16.0cm}}
\end{picture}
\caption{Same as Fig.~\ref{fig:matelem_f_t_two_qsq_rho}, as a function
of the mass \mpp.
The measurements are given in Tables~\ref{table:matelem_f_m_two_qsq_rho}
and~\ref{table:matelem_f_m_two_qsq_rho_2}.
}
\label{fig:matelem_f_m_two_qsq_rho}
\end{center}
\end{figure}
%-----------------------------------------------------------------------------

The matrix element measurements are presented 
as a function of \qsq\ for \rh\ and \ph\ production 
in Figs.~\ref{fig:matelem_f_qsq_rho} and~\ref{fig:matelem_f_qsq_phi}, 
and as a function of \modt\ and the mass \mpp\ for \rh\ production 
in two intervals of \qsq, in Figs.~\ref{fig:matelem_f_t_two_qsq_rho}
and~\ref{fig:matelem_f_m_two_qsq_rho}.

%-----------------------------------------------------------------------------
\begin{figure}[htbp]
\begin{center}
\setlength{\unitlength}{1.0cm}
\begin{picture}(16.5,4.5)   
\put(0.0,0.0){\epsfig{file=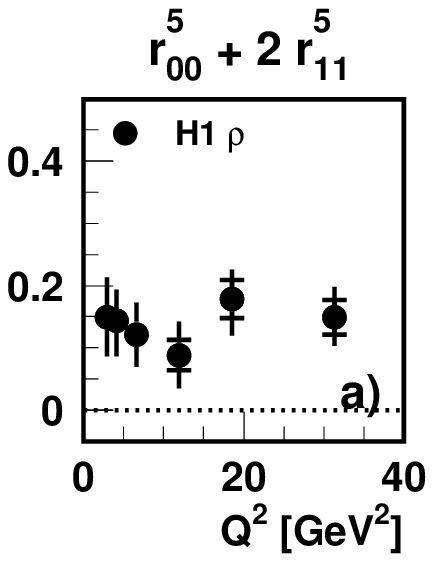,height=4.5cm,width=4.5cm}}
\put(4.0,0.0){\epsfig{file=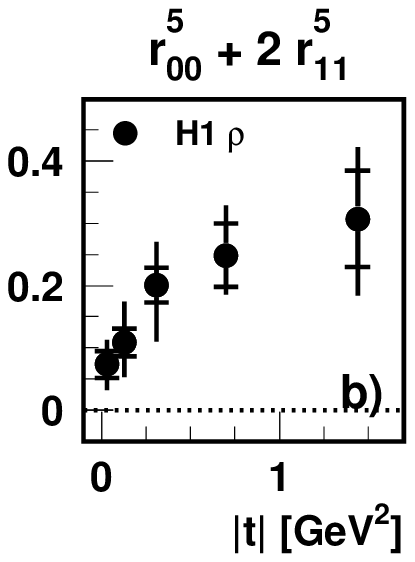,height=4.5cm,width=4.5cm}}
\put(8.0,0.0){\epsfig{file=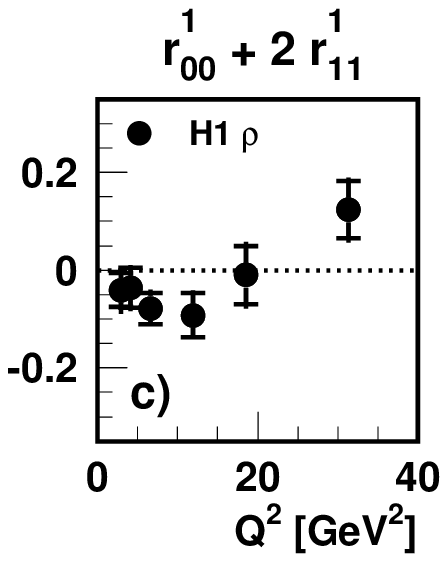,height=4.5cm,width=4.5cm}}
\put(12.,0.0){\epsfig{file=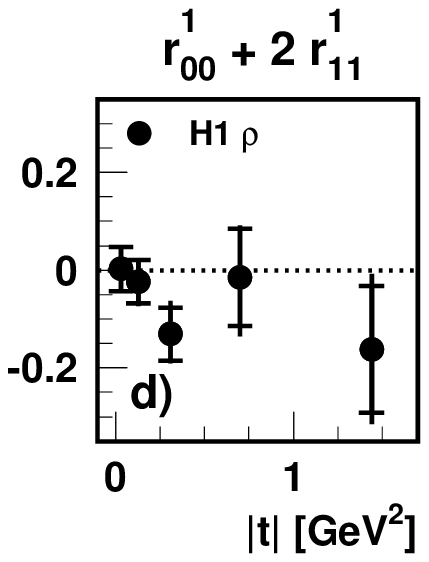,height=4.5cm,width=4.5cm}}
\end{picture}
\caption{Measurements, as a function of \qsq\ and \modt, of the \rh\ matrix element 
combinations \rfivecomb\ and \ronecomb, obtained from fits of Eq.~(\ref{eq:angle_phi}) 
to the $\phi$ angle distributions.
The notag ($\modt \leq 0.5~\gevsq$) and tag ($\modt \leq 3~\gevsq$) samples are combined.
The dotted lines show the expected vanishing values of the matrix elements if only 
the SCHC amplitudes are non-zero.
The measurements are given in Table~\ref{table:r5comb}.
}
\label{fig:rfive-one-comb}
\end{center}
\end{figure}
%-----------------------------------------------------------------------------

The present measurements as a function of \qsq\ and \modt\ confirm with increased 
precision the previous H1 results~\cite{h1-rho-95-96, h1-phi-95-96, h1-rho-large-t-97}
and they are globally compatible with ZEUS measurements as a function of 
\qsq~\cite{z-rho,z-phi}. 
No significant dependence of the matrix elements with $W$ is observed within the present
data.
Measurements (not shown) of the matrix elements \rzzzz\ and \rzqumu, obtained 
from fits to the $ \cos \theta$ and \phib\ distributions as given 
by Eqs.~(\ref{eq:cosths}-\ref{eq:angle_varphi}) of the Appendix, are in agreement with those 
presented in Figs.~\ref{fig:matelem_f_qsq_rho} to~\ref{fig:matelem_f_m_two_qsq_rho}.
For the combinations \rfivecomb\ and \ronecomb\ for \rh\ mesons, measurements from fits 
of Eq.~(\ref{eq:angle_phi}) to the $\phi$ distribution, which give smaller errors than 
the projection method, are presented  in Fig.~\ref{fig:rfive-one-comb}.

%%%%%%%%%%%%%%%%%%%%%%%%%%%%%%%%%%%%%%%%%%%%
\subsubsection{Comparison with models}  
                                                                              \label{sect:matrix-elements_models}
%%%%%%%%%%%%%%%%%%%%%%%%%%%%%%%%%%%%%%%%%%%%

Figures~\ref{fig:matelem_f_qsq_rho} and~\ref{fig:matelem_f_qsq_phi} present,
superimposed on the \rh\ and \ph\ measurements, predictions of the GK GPD 
model~\cite{kroll} and of the INS $k_t$-unintegrated model~\cite{ins} for two 
different wave functions;
for the GK model, the SCHC approximation is used and only non-zero elements are 
shown.

For \rh\ production (Fig.~\ref{fig:matelem_f_qsq_rho}), taking into account the experimental 
and theoretical uncertainties and the use of the SCHC approximation, the GK 
model~\cite{kroll} gives a description of the data which is reasonable in shape 
but does not describe the normalisation well.
The INS model~\cite{ins} reproduces the gross features of the \qsq\ 
evolution but there are problems in the details.
The model with the compact wave function describes the \rzqzz\ matrix element evolution, but it fails 
for the other elements which are non-zero under SCHC 
(\ruumu, Im \rdumu, Re \rcuz, Im \rsuz);
on the other hand, the model with the large wave function gives a rather good description of these four elements, 
but fails badly for \rzqzz.
In addition, both wave functions predict too low values for \rczz, also in the regime 
with $\qsq > 10~\gevsq$.

For \ph\ mesons (Fig.~\ref{fig:matelem_f_qsq_phi}) with less statistics, the picture 
is slightly different
for the INS model~\cite{ins}, where the use of a large wave function gives a better description of all matrix 
elements, including \rzqzz, than the compact wave function.

%%%%%%%%%%%%%%%%%%%%%%%%%%%%%%%%%%%%%%%%%%%%
%%%%%%%%%%%%%%%%%%%%%%%%%%%%%%%%%%%%%%%%%%%%
%%%%%%%%%%%%%%%%%%%%%%%%%%%%%%%%%%%%%%%%%%%%
\subsection{Nature of the exchange}  
                                                                                  \label{sect:polar_discussion}
%%%%%%%%%%%%%%%%%%%%%%%%%%%%%%%%%%%%%%%%%%%%

%%%%%%%%%%%%%%%%%%%%%%%%%%%%%%%%%%%%%%%%%%%%
\subsubsection{Natural parity exchange}  
                                                                                  \label{sect:polar_disc_NPE}
%%%%%%%%%%%%%%%%%%%%%%%%%%%%%%%%%%%%%%%%%%%%

%-----------------------------------------------------------------------------
\begin{figure}[htbp]
\begin{center}
\setlength{\unitlength}{1.0cm}
\begin{picture}(16.5,4.5)   
\put(0.0,0.0){\epsfig{file=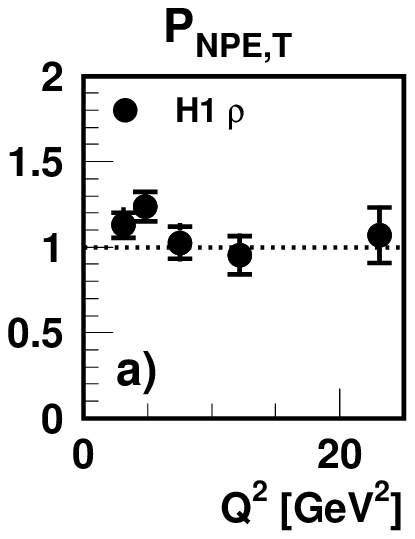,height=4.5cm,width=4.5cm}}
\put(4.0,0.0){\epsfig{file=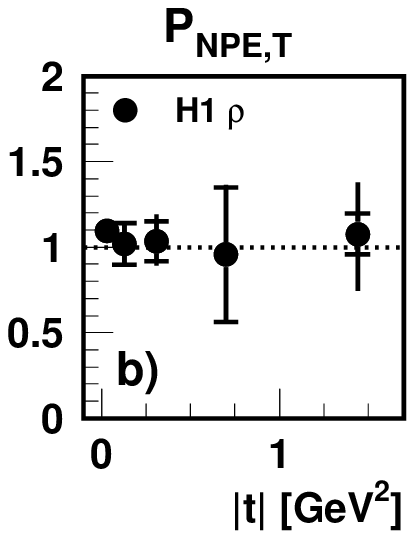,height=4.5cm,width=4.5cm}}
\put(8.0,0.0){\epsfig{file=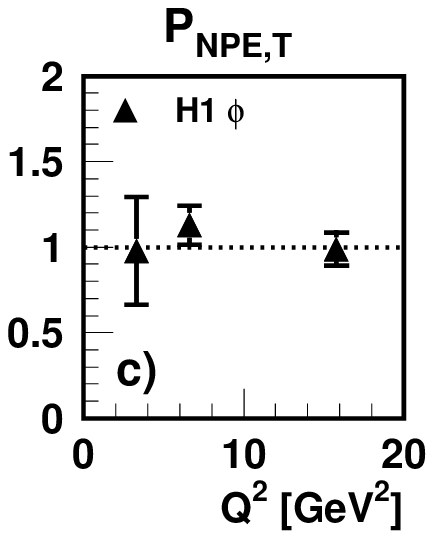,height=4.5cm,width=4.5cm}}
\put(12.,0.0){\epsfig{file=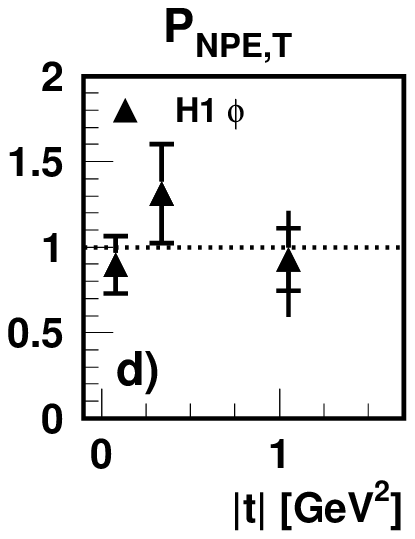,height=4.5cm,width=4.5cm}}
\end{picture}
\caption{Asymmetry $P_{\text{NPE,T}}$ between natural and unnatural 
parity exchange for transverse photons:
(a)-(b) \rh\ mesons, as a function of \qsq\ and \modt;
(c)-(d) \ph\ mesons.
The dotted lines indicate the value~$1$ expected for NPE.
The measurements are given in Table~\ref{table:npe}.
}
\label{fig:NPE-asym}
\end{center}
\end{figure}
%-----------------------------------------------------------------------------
 
The observation at low energy~\cite{bauer, joos,chio} of dominant 
natural parity exchange (NPE) supports the 
attribution of the vacuum quantum numbers ($J^{PC} = 0^{++}$) to the pomeron;
the recent observation by the HERMES collaboration~\cite{hermes} of the
presence at low energy of a small contribution (about $6\%$) of unnatural parity 
exchange is attributed to quark exchange ($\pi$, $a_1$ or $b_1$ exchange).
At high energy, the modeling of diffraction as two gluon exchange implies a NPE
character, in particular in the GK GPD model~\cite{kroll}.

With unpolarised beams and for a single value of the beam energies, the only accessible 
information about the parity of the exchange is the asymmetry 
$P_{\text{NPE,T}} = (\sigma^N_T - \sigma^U_T) \ / \ (\sigma^N_T + \sigma^U_T)$ between 
natural ($\sigma^N_T$) and unnatural  ($\sigma^U_T$) parity exchange for transverse 
photons, using Eq.~(\ref{eq:NPE-asym}) of the Appendix.
Measurements of $P_{\text{NPE,T}}$ as a function of \qsq\ and \modt\
for \rh\ and \ph\ mesons are presented in Fig.~\ref{fig:NPE-asym}.
They are globally compatible with~$1$,
which supports NPE for transverse photons.
Natural parity exchange is assumed in the following.

%%%%%%%%%%%%%%%%%%%%%%%%%%%%%%%%%%%%%%%%%%%%
\boldmath
\subsubsection{Helicity conserving amplitudes; SCHC approximation}  
                                                                                  \label{sect:polar_disc_schc}
\unboldmath
%%%%%%%%%%%%%%%%%%%%%%%%%%%%%%%%%%%%%%%%%%%%

Inspection of Figs.~\ref{fig:matelem_f_qsq_rho} and~\ref{fig:matelem_f_qsq_phi} 
shows that, for both \rh\ and \ph\ meson electroproduction, the five matrix elements
listed in Eq.~(\ref{eq:non-0-schc}) of the Appendix 
(\rzqzz, \ruumu, Im~\rdumu, Re~\rcuz , Im~\rsuz), which 
contain products of the two helicity conserving amplitudes, $T_{00}$ and $T_{11}$, 
are significantly different from zero, with the SCHC relations of Eq.~(\ref{eq:schc_pairing}) 
being approximately satisfied.
In addition, with the significant exception of \rczz, the other matrix elements are
small or consistent with~$0$.

In the present kinematic domain, SCHC is thus a reasonable
approximation, which can be used to obtain information on the transition amplitudes.
In order to decrease the sensitivity to the SCHC violating amplitudes, 
which increase with \modt\ (see sections~\ref{sect:polar_disc_hel_flip} 
and~\ref{sect:polar_disc_ampl_ratios}), only events with $\modt \leq 0.5~\gevsq$ are used 
in the rest of this section.

%%%%%%%%%%%%%%%
\paragraph{{\boldmath $\psi$} distributions; phase {\boldmath $\delta$} between the SCHC 
                  amplitudes}
%%%%%%%%%%%%%%%

%-----------------------------------------------------------------------------
\begin{figure}[htbp]
\begin{center}
\setlength{\unitlength}{1.0cm}
\begin{picture}(4.5,4.5)   
\put(0.0,0.0){\epsfig{file=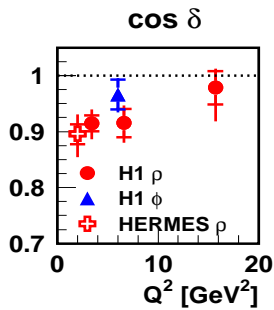  ,height=4.50cm,width=4.50cm}}
\end{picture}
\caption{Cosine of the phase $\delta$ between the $T_{00}$ and $T_{11}$ helicity 
conserving amplitudes for \rh\ and \ph\ production with $\modt \leq 0.5~\gevsq$, 
measured as a function of \qsq\ from two-dimensional fits of Eq.~(\ref{eq:Wcosdelta}), 
in the SCHC approximation.
The HERMES~\protect\cite{hermes} measurement on protons is also shown.
The dotted line indicates the value~$1$ which corresponds to amplitudes in phase.
The present measurements are given in Table~\ref{table:cosdelta}.
}
\label{fig:cosdelta}
\end{center}
\end{figure}
%-----------------------------------------------------------------------------

Under SCHC, the angular distribution 
$W(\theta, \phib, \phi)$ reduces to a function of the angles $\theta$ and 
$\psi = \phi - \phib$, Eq.~(\ref{eq:Wcosdelta}), which allows the extraction in this 
approximation of the cross section ratio $R = \sigma_L / \sigma_T$ and of the 
phase $\delta$ between the $T_{00}$ and $T_{11}$ amplitudes.

Measurements of $\cos \delta$ obtained from two-dimensional fits of
Eq.~(\ref{eq:Wcosdelta}) with $R$ left free are presented in 
Fig.~\ref{fig:cosdelta} as a function of \qsq\ for \rh\ and \ph\ production 
($\modt \leq 0.5~\gevsq$).
They are in agreement with the measurements obtained with $R$ fixed to 
the values measured in the SCHC approximation using the \rzqzz\ matrix element and 
Eq.~(\ref{eq:R_schc}).

The measurements of $\cos\delta$ are close to~$1$, indicating that the transverse and 
longitudinal amplitudes are nearly in phase.
For \rh\ production with $\qsq < 10~\gevsq$, $\cos\delta$ differs however significantly 
from~$1$, as is also observed for \qsq\ around $2~\gevsq$ in the low energy measurement 
by HERMES~\cite{hermes}.
An indication of an increase of $\cos\delta$ toward~$1$ at high \qsq\ may be present
in the data.
An interpretation of a value of  $\cos \delta$ different from~$1$ at high energy in terms
of a $W$ dependence of $\sigma_L / \sigma_T$ will be given
in section~\ref{sect:polar_disc_ampl_phases}.

%%%%%%%%%%%%%%%%%%%%%%%%%%%%%%%%%%%%%%%%%%%%
\boldmath
\subsubsection{Helicity flip amplitudes}  
                                                                                  \label{sect:polar_disc_hel_flip}
\unboldmath
%%%%%%%%%%%%%%%%%%%%%%%%%%%%%%%%%%%%%%%%%%%%

A significant violation of SCHC is observed in Figs.~\ref{fig:matelem_f_qsq_rho} 
and~\ref{fig:matelem_f_qsq_phi} through the non-zero value of the \rczz\ matrix element,
for \rh\ and for \ph\ mesons (see also Fig.~\ref{fig:rfive-one-comb} for the \rfivecomb\ 
combination measurement for \rh\ mesons).
The \rczz\ matrix element is proportional to the product ${\rm Re} \ (T_{00}T^\dagger_{01})$  
of $T_{00}$, the leading SCHC amplitude, and $T_{01}$, the helicity flip amplitude
describing the transition from a transverse photon to a longitudinal VM.
In Figs.~\ref{fig:matelem_f_t_two_qsq_rho} and~\ref{fig:matelem_f_m_two_qsq_rho},
non-zero values, with \qsq\ dependent strengths, are also observed in \rh\ production 
for the matrix elements Re~\rzquz, Re~\ruuz\ and Im~\rduz, which contain the product 
${\rm Re} \ (T_{11}T^\dagger_{01})$ of $T_{01}$ and the second SCHC amplitude 
$T_{11}$.
The data tend to support the relation ${\rm Im}~\rduz =  - {\rm Re}~\ruuz$ 
of Eq.~(\ref{eq:T01-non-schc_pairing}).
Other matrix elements are, within errors, consistent with~$0$ when integrated over $t$.

These findings confirm the previous H1 observation~\cite{h1-rho-95-96, h1-phi-95-96}
in \rh\ production
that the \tzu\ helicity flip amplitude is significantly different from~$0$ in the present \qsq\ 
domain and is dominant among the SCHC violating amplitudes, supporting the hierarchy
(see for instance~\cite{ik})
\begin{equation}
|T_{00}| > |T_{11}| > |T_{01}| > |T_{10}| \ , \ |T_{-11}|.
                                                                                               \label{eq:hierarchy2}
\end{equation}
Note that helicity violation as such is not a signature for hard processes.
When integrated over \modt, the \tzu\ amplitude in the present kinematic domain is larger 
for low \qsq\ than for large \qsq, as shown by the \rczz\ matrix element measurement in 
Fig.~\ref{fig:matelem_f_m_two_qsq_rho}.
At low energy and for $\av{\qsq}$ around $0.5~\gevsq$, the \tzu\ amplitude is 
non-zero, with  $|T_{01}| \ / \ \sqrt{|T_{00}|^2 + |T_{11}|^2} = 15$ to~$20\% $ for 
$W$ about $2.5~\gev$~\cite{joos} and $11$ to 
$14\%$ for $10 \leq W \leq 16~\gev$~\cite{chio}.

The \rczz\ matrix element increases with $|t|$, as observed in 
Fig.~\ref{fig:matelem_f_t_two_qsq_rho} (see also Fig.~\ref{fig:rfive-one-comb}).
This is expected on quite general grounds for helicity flip amplitudes, as will be discussed 
in section~\ref{sect:polar_disc_ampl_ratios}.

%%%%%%%%%%%%%%%%%%%%%%%%%%%%%%%%%%%%%%%%%%%%
%%%%%%%%%%%%%%%%%%%%%%%%%%%%%%%%%%%%%%%%%%%%
%%%%%%%%%%%%%%%%%%%%%%%%%%%%%%%%%%%%%%%%%%%%
\boldmath
\subsection{Cross section ratio $R = \sigma_L / \sigma_T$}  
                                                                                  \label{sect:polar_disc_R}
\unboldmath
%%%%%%%%%%%%%%%%%%%%%%%%%%%%%%%%%%%%%%%%%%%%

The cross section ratio $R = \sigma_L / \sigma_T$ is one of the most 
important observables in the study of light VM production since 
it is sensitive to the interaction dynamics, including effects related
to the interacting dipole size or depending on the VM wave function.

In the SCHC approximation, $R$ can be calculated from the \rzqzz\ matrix element:
\begin{equation}
R_{SCHC} = \frac {T_{00}^2} {T_{11}^2} = 
                                                 \frac{1}{\varepsilon} \  \frac{\rzzzz}{1-\rzzzz}. 
                                                                                         \label{eq:R_schc}
\end{equation}

In view of the observed violation of SCHC, a better approximation takes into
account the dominant helicity flip amplitude $T_{01}$ and uses in addition
the measurement of \rczz:
\begin{equation}
\rapproch = \frac {T_{00}^2} {T_{11}^2 + T_{01}^2} =
\frac{1}{\varepsilon} \ 
      \frac{\rzqzz - \varepsilon (\rczz)^2 + \sqrt{(\rzqzz)^2 - 2 \varepsilon (\rczz)^2}}
             {2 - 2 \rzqzz + \varepsilon (\rczz)^2},
                                                                                         \label{eq:R_T01}
\end{equation}
where NPE is assumed and the amplitudes are taken to be in phase.
As expected, the effect of this improved approximation is mostly significant at 
large \modt\ values, in view of the increase with \modt\ of the helicity flip
amplitudes: 
the corresponding measurement of $R$ is lower than that obtained in the SCHC
approximation by about $0.05$ for $\modt = 0.1~\gevsq$ and about $0.30$ for 
$\modt = 1~\gevsq$, independently of \qsq.
Integrated over $t$, this makes a $7\%$ difference.
Measurements of $R$ are presented in the following using the improved 
approximation of Eq.~(\ref{eq:R_T01}).
The general features of the kinematic variable dependences discussed below are 
similar when the SCHC approximation of Eq.~(\ref{eq:R_schc}) is used.

%%%%%%%%%%%%%%%%%%%%%%%%%%%%%%%%%%%%%%%%%%%%
\subsubsection{\boldmath {\qsq} dependence}
                                                                                  \label{sect:polar_disc_R-qsq}
%%%%%%%%%%%%%%%%%%%%%%%%%%%%%%%%%%%%%%%%%%%%

%-----------------------------------------------------------------------------
\begin{figure}[htbp]
\begin{center}
\setlength{\unitlength}{1.0cm}
\begin{picture}(14.0,7.0)   
\put(0.0,0.0){\epsfig{file=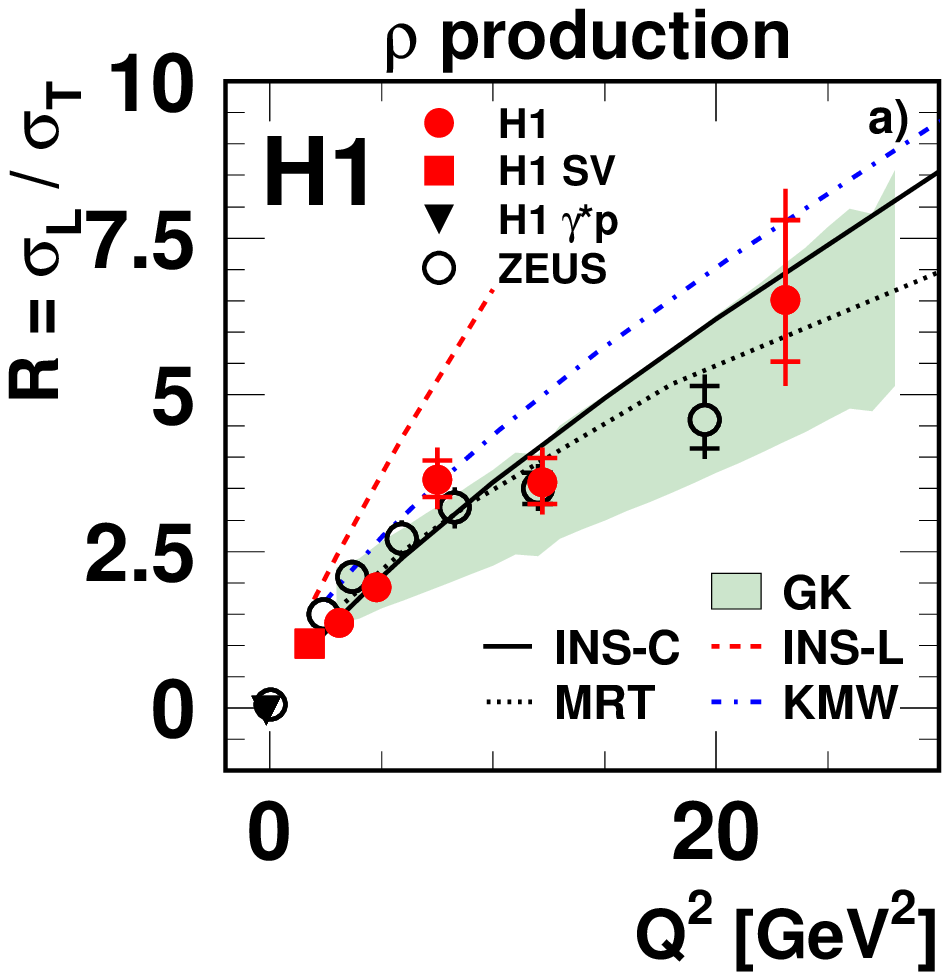,height=7.0cm,width=7.0cm}}
\put(7.0,0.0){\epsfig{file=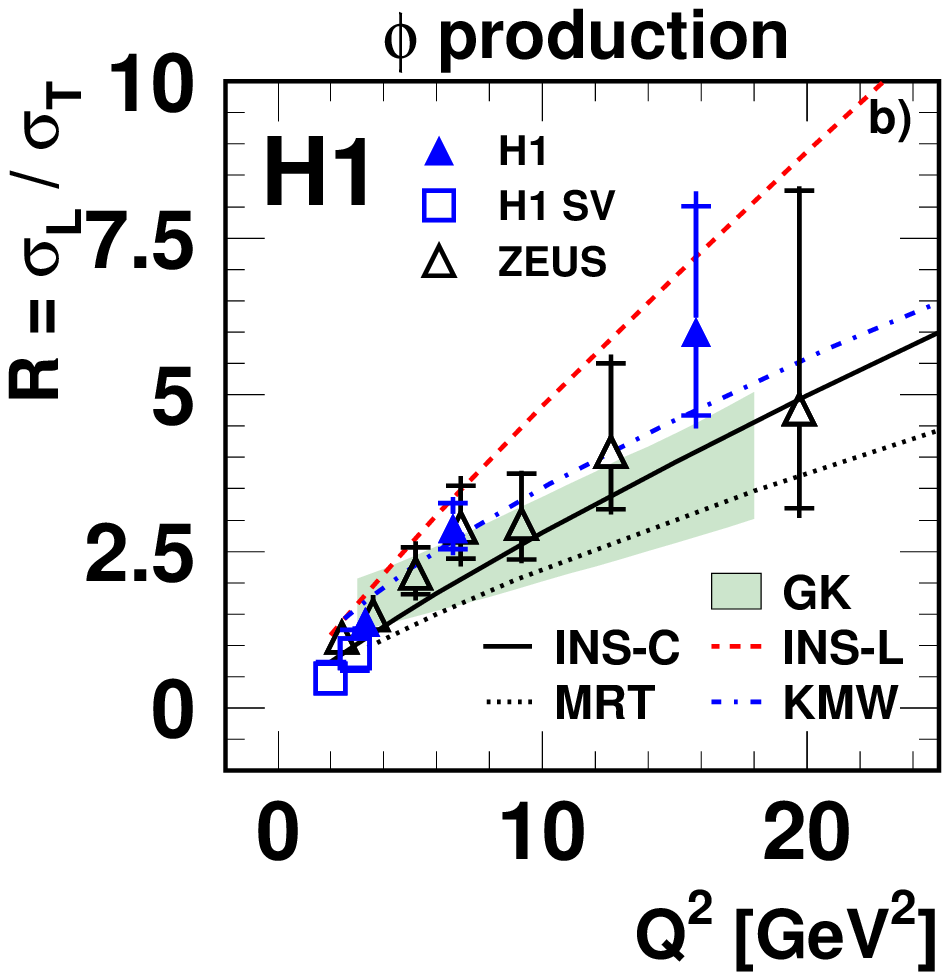,height=7.0cm,width=7.0cm}}
\end{picture}
\caption{\qsq\ dependence of the ratio $R = \sigma_L / \sigma_T$ of the longitudinal 
to transverse cross sections measured using Eq.~(\ref{eq:R_T01})
for (a) \rh\ meson production; (b) \ph\ production.
Measurements of $R$ in the SCHC approximation, for \rh\ photoproduction by 
H1~\protect\cite{h1-rho-jpsi-94} and 
ZEUS~\protect\cite{z-rho-photoprod} and for \rh\ and \ph\ electroproduction
by ZEUS~\protect\cite{z-rho-photoprod,z-rho,z-phi} are also shown.
The superimposed curves are predictions of the models of GK~\cite{kroll} (shaded bands), 
INS~\cite{ins} with the compact (solid lines) and the large (dashed lines) wave functions, 
MRT with the CTEQ6.5M PDF parameterisation~\cite{mrt} 
(dotted lines) and KMW~\cite{kmw} (dash-dotted lines).
The present measurements are given in Table~\ref{table:rlt-q2}.
}
\label{fig:rlt-q2}
\end{center}
\end{figure}
%-----------------------------------------------------------------------------

The measurements of $R$ presented in Fig.~\ref{fig:rlt-q2} show a strong increase 
with \qsq, which is tamed at large \qsq, a feature already noted in previous 
H1~\cite{h1-rho-95-96} and ZEUS~\cite{z-rho} publications.

For \rh\ production,
the GK GPD model~\cite{kroll}, the MRT model~\cite{mrt} and the INS model~\cite{ins} 
with the compact wave function give a good description of the measurements,  whereas the 
KMW~\cite{kmw} predictions are too high and the INS model
with the large wave function is ruled out.
The predictions of the MPS model~\cite{soyez} (not shown) are very similar to those 
of KMW up to $10~\gevsq$, and then slightly lower.
The \qsq\ dependence of the IK~\cite{ik} model (not shown) is
similar to that of the MRT model, since it is derived in a similar
way.
For \ph\ production, the KMW model gives a good description
while the MRT predictions are too low; within the quoted uncertainty,
the GK model describes the data; for the INS model, the large wave function gives a slightly better 
description than the compact wave function; the predictions of the MPS model (not shown) are again
similar to those of KMW, although slightly higher at low \qsq.

%-----------------------------------------------------------------------------
\begin{figure}[htb]
\begin{center}
\setlength{\unitlength}{1.0cm}
\begin{picture}(10.0,10.0)   
\put(0.0,0.0){\epsfig{file=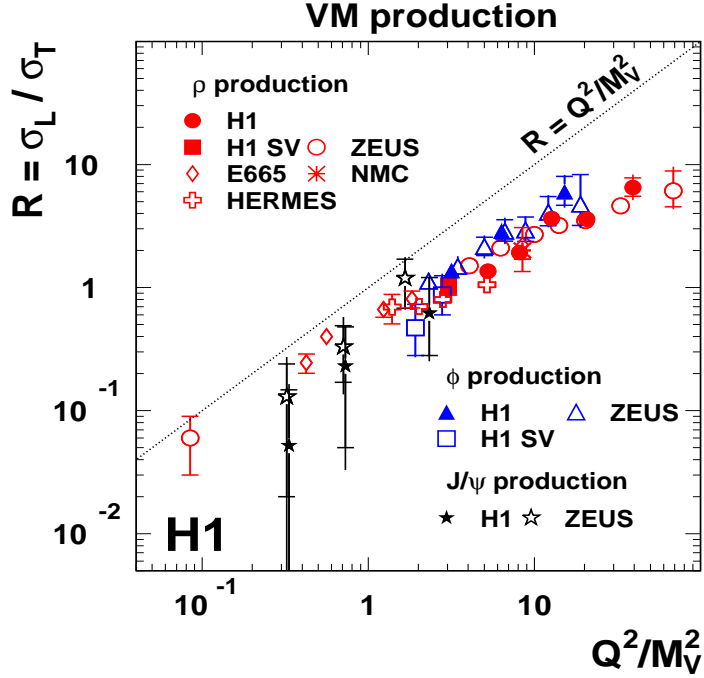  ,height=10.0cm,width=10.0cm}}
\end{picture}
\caption{Ratio $R = \sigma_L / \sigma_T$ as a function of the variable 
$Q^2\ / M_{V}^2$.
Electroproduction measurements of \rh\ mesons by fixed target experiments 
(NMC~\protect\cite{NMC}, E665~\protect\cite{E665} and 
HERMES~\protect\cite{hermes}), of \rh\ and \ph\
mesons by ZEUS~\protect\cite{z-rho,z-phi} and of \jpsi\ mesons by
H1 and ZEUS~\protect\cite{h1-jpsi-hera1,z-jpsi-elprod} are also shown.
The dotted line represents the scaling behaviour $R = Q^2 / M_{V}^2$.}
\label{fig:rlt-compil}
\end{center}
\end{figure}
%-----------------------------------------------------------------------------

$R$ measurements for \rh, \ph\ and \jpsi\ mesons are presented as a function of the
scaling variable {\mbox{$Q^2 / M_{V}^2$}} in Fig.~\ref{fig:rlt-compil}.
The improved approximation, Eq.~(\ref{eq:R_T01}), is used for the present data
whereas the SCHC approximation is used for the other data, which 
makes little difference for the $t$ integrated measurements.
A smooth and common behaviour is observed for the three VMs over the full 
$Q^2  / M_{V}^2$ range and the full energy range, from the fixed target experiments
 to the HERA collider measurements.

The data are close to a law $R = Q^2 / M_{V}^2$, represented by the dotted line, 
but they lie systematically below the line, with a slower increase of $R$ with 
increasing \qsq.
These features are easily understood in the MRT~\cite{mrt} and IK~\cite{ik} models 
%where, in the SCHC approximation $R \simeq |T_{00}|^2 / |T_{11}|^2$, the
where the formal $Q^2 / M_{V}^2$ evolution is damped by a factor 
$ \gamma^2 / (1 + \gamma)^2$ and the taming of the $R$ evolution results from the 
decrease of $\gamma$ with increasing \qsq.

%%%%%%%%%%%%%%%%%%%%%%%%%%%%%%%%%%%%%%%%%%%%
\subsubsection{\boldmath {$W$} dependence}
                                                                                  \label{sect:polar_disc_R-W}
%%%%%%%%%%%%%%%%%%%%%%%%%%%%%%%%%%%%%%%%%%%%

%-----------------------------------------------------------------------------
\begin{figure}[htbp]
\begin{center}
\setlength{\unitlength}{1.0cm}
\begin{picture}(14.0,14.0)   
\put(0.0,7.0){\epsfig{file=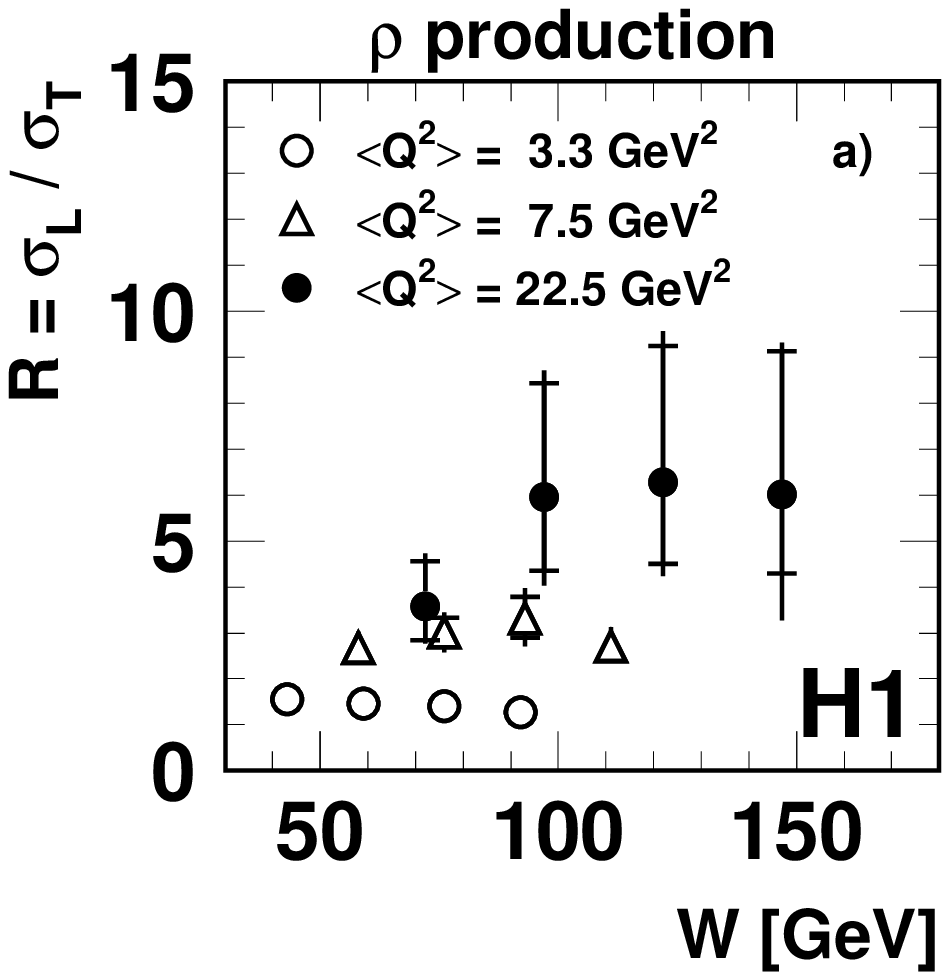,height=7.0cm,width=7.0cm}}
\put(7.0,7.0){\epsfig{file=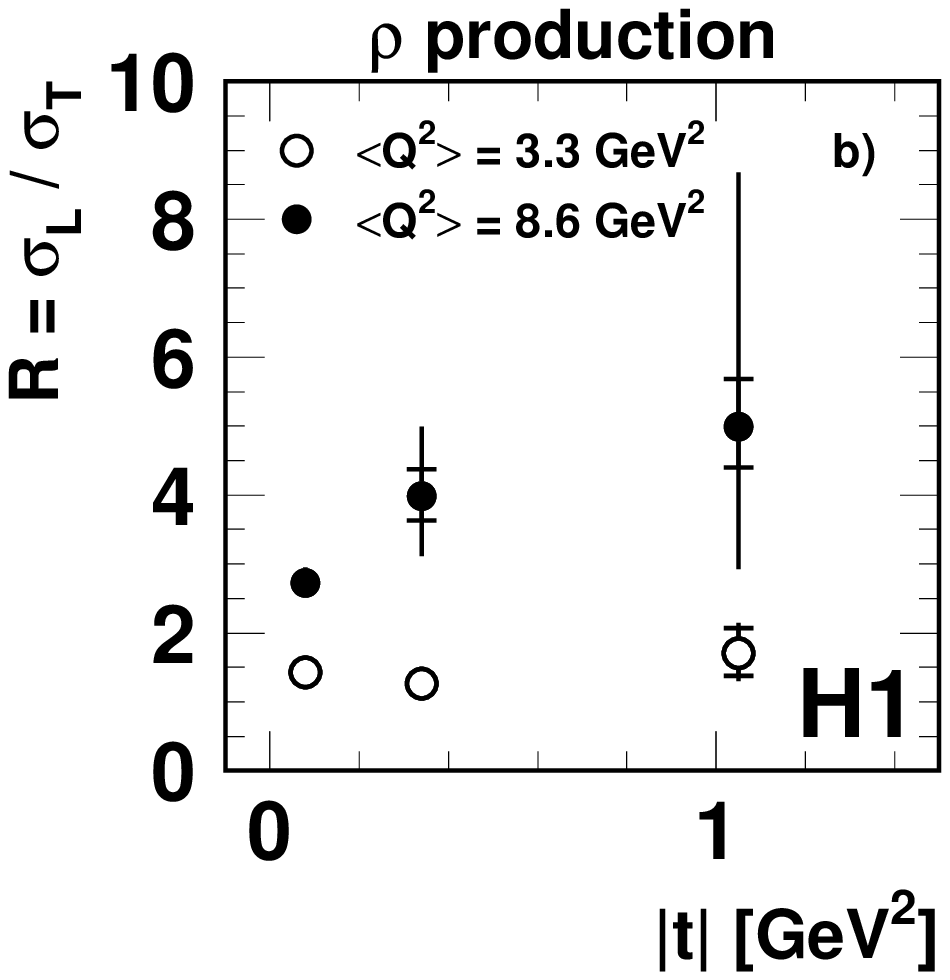,height=7.0cm,width=7.0cm}}
\put(3.5,0.0){\epsfig{file=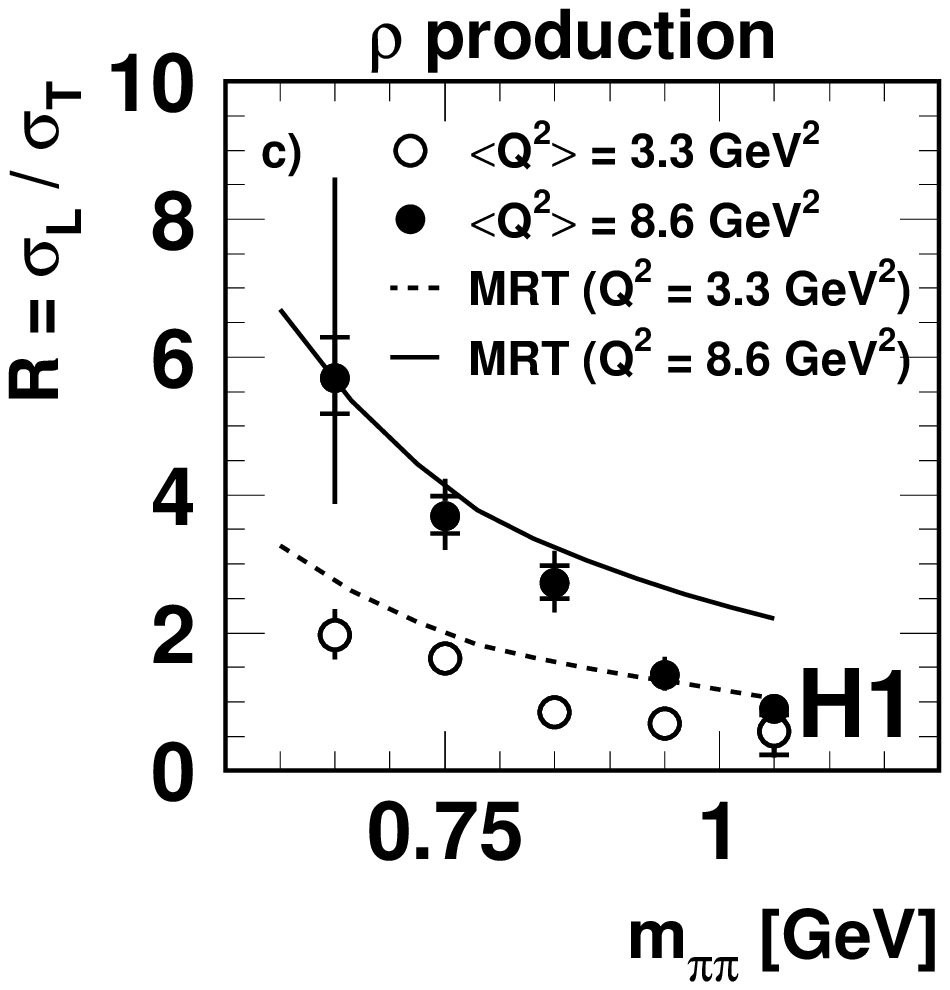,height=7.0cm,width=7.0cm}}
\end{picture}
\caption{Dependence, for \rh\ meson production, of the ratio 
$R = \sigma_L / \sigma_T$ of the longitudinal 
to the transverse cross sections, determined using Eq.~(\ref{eq:R_T01}), on
(a) $W$; (b) \modt; (c) \mpp, 
separately for $2.5 \leq \qsq < 5~\gevsq$ and for $5 \leq \qsq  \leq 60~\gevsq$;
for $W$, the latter bin is divided into $5 \leq \qsq  < 15.5$ and 
$15.5 \leq \qsq   \leq 60~\gevsq$.
The curves in (c) are from the MRT model~\protect\cite{teubner}.
The measurements are given in Tables~\ref{table:rlt-w}-\ref{table:rlt-m}.
}
\label{fig:rlt-wtm}
\end{center}
\end{figure}
%-----------------------------------------------------------------------------

The $W$ dependence of $R$ is presented for \rh\ meson production in Fig.~\ref{fig:rlt-wtm}(a) 
for three intervals in \qsq.
Because of the strong correlation in detector acceptance between $W$ and \qsq, the lever 
arm in $W$ for each domain in \qsq\ is rather limited.
As discussed in section~\ref{sect:th_context}, the onset 
of hard diffraction, characterised by a strong $W$ dependence, is expected to be delayed 
for transverse amplitudes compared to longitudinal amplitudes.
A harder $W$ dependence is thus expected for $\sigma_L$ than 
for $\sigma_T$, resulting in an increase of $R$ with $W$.
In view of the limited precision, no significant conclusion can be drawn from the present 
measurements.

%%%%%%%%%%%%%%%%%%%%%%%%%%%%%%%%%%%%%%%%%%%%
\subsubsection{\boldmath {$t$} dependence; \boldmath {$b_L - b_T$} slope difference }
                                                                                  \label{sect:polar_disc_R-t}
%%%%%%%%%%%%%%%%%%%%%%%%%%%%%%%%%%%%%%%%%%%%

Figure~\ref{fig:rlt-wtm}(b) presents the measurement of $R$ as a function of \modt\ for 
\rh\ mesons, in two 
bins in \qsq.
For exponentially falling $t$ distributions, this can be translated into a measurement of the 
difference between the longitudinal and transverse $t$ slopes, through the relation
$R(t) = \sigma_L(t) / \sigma_T(t) \propto e^{- (b_L - b_T) |t|}$.
Measurements of the slope difference $b_L - b_T$ extracted from a fit of the $t$ 
dependence of $R$ are given in Table~\ref{tab:bL-bT} (for completeness, the
result for \ph\ production in one bin in $t$ is also given in spite of the large errors).
The errors are dominated by the systematic uncertainty on the \rhop\ background 
subtraction.
A slight indication ($1.5 \sigma$) is found for a negative value of $b_L - b_T$ in the higher 
bin in \qsq.
The use of the SCHC approximation of Eq.~(\ref{eq:R_schc}) instead of the improved 
approximation of  Eq.~(\ref{eq:R_T01}) for the measurement of $R$ does not affect the 
measurements of $b_L - b_T$.

%-----------------------------------------------------------------------------
\renewcommand{\arraystretch}{1.15}
\begin{table}[htbp]
\begin{center}
\begin{tabular}{|c|c|} 
\multicolumn{2}{c} { }  \\
\hline
    \av{\qsq}  (\gevsq)        &        $b_L - b_T$    (\gevsqm)         \\ 
\hline
\hline
    \multicolumn{2}{|c|} {\rh\ production }  \\
\hline
$3.3$                             &   $-0.03 \pm 0.27~_{-0.17}^{+0.19}$   \\
$8.6$                             &   $-0.65 \pm 0.14~_{-0.51}^{+0.41}$   \\
\hline
\hline
    \multicolumn{2}{|c|} {\ph\ production }  \\
\hline
$5.3$                             &   $-0.16 \pm 0.56_{-1.10}^{+0.46}$   \\
\hline
\end{tabular} 
\caption{Difference between the longitudinal and transverse slopes,
$b_L - b_T$, of the $t$
distributions for \rh\ (two bins in \qsq) and \ph\ meson production, calculated from 
the $t$ dependence of the cross section 
ratio $R = \sigma_L / \sigma_T$ obtained using Eq.~(\ref{eq:R_T01}). }
\label{tab:bL-bT}
\end{center}
\end{table}
\renewcommand{\arraystretch}{1.}
%-----------------------------------------------------------------------------

A difference between the $b$ slopes is expected to indicate a difference between 
the transverse size of the dominant dipoles for longitudinal and transverse amplitudes 
(see e.g.~\cite{ins}).
The indication for a negative value of $b_L - b_T$ in the higher bin in \qsq\ is consistent 
with the expectation that $\sigma_L$ reaches a harder QCD regime than $\sigma_T$.
Conversely, the absence of a \modt\ dependence of $R$ in the lower \qsq\ range is 
consistent with the interpretation of  $b$ slope measurements in 
section~\ref{sect:t_slopes_elast}, suggesting that
large dipoles may be present in longitudinal amplitudes (``finite size" effects) 
for moderate values of the scale $\scaleqsqplmsq$.

%%%%%%%%%%%%%%%%%%%%%%%%%%%%%%%%%%%%%%%%%%%%
\subsubsection{\boldmath {\mpp} dependences}
                                                                                  \label{sect:polar_disc_R-m}
%%%%%%%%%%%%%%%%%%%%%%%%%%%%%%%%%%%%%%%%%%%%

%-----------------------------------------------------------------------------
\begin{figure}[htbp]
\begin{center}
\setlength{\unitlength}{1.0cm}
\begin{picture}(6.0,6.0)   
\put(0.0,0.0){\epsfig{file=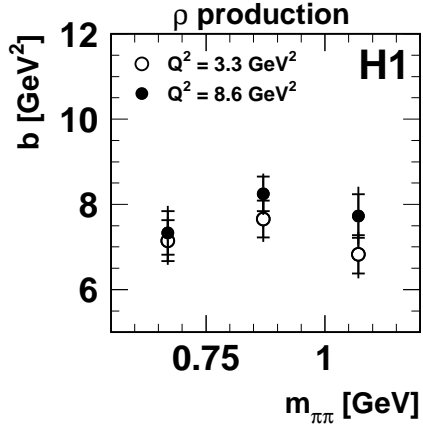  ,height=6.0cm,width=6.0cm}}
\end{picture}
\caption{Dependence of the exponential $t$ slope for \rh\ elastic production as a function 
of the mass \mpp, for $\qsq = 3.3$ and $8.6~\gevsq$.
The measurements are given in Table~\ref{table:rho_masst_bslp}.
}
\label{fig:bfmq2}
\end{center}
\end{figure}
%-----------------------------------------------------------------------------
 
A striking decrease of the cross section ratio $R$ with the increase of the \mpp\ mass, 
which was also reported by ZEUS~\cite{z-rho}, is observed in Fig.~\ref{fig:rlt-wtm}(c).
This strong effect is not expected in calculations where the \rh\ meson is treated as
a particle with well defined mass and wave function.
A simple interpretation of the \mpp\ dependence follows from  the formal
$\qsq / M^2$ dependence of the cross section ratio, if the mass $M$ is understood as 
the dipion mass rather than the nominal resonance mass.
Such an interpretation is in line with the open quark approach of the MRT parton-hadron duality
model~\cite{mrt}, and is qualitatively supported by the calculations superimposed to
the data in Fig.~\ref{fig:rlt-wtm}(c)~\cite{teubner}.
The mass dependence of $R$ expected from the interference of resonant \rh\ and non-resonant 
$\pi \pi$ production, discussed in~\cite{ryskin-shab-skewing}, is small compared to that 
observed here and should decrease with \qsq.

The $b$ slopes of the \modt\ distributions do not show any significant dependence
on the mass (see Fig.~\ref{fig:bfmq2}), which indicates that the \mpp\ dependence 
of $R$ can not be explained by an hypothetic kinematic selection of dipoles with specific  
size, related either to transverse or longitudinal amplitudes.
All this suggests that the VM wave function plays a limited role in the description of VM diffractive
production.

%%%%%%%%%%%%%%%%%%%%%%%%%%%%%%%%%%%%%%%%%%%%
%%%%%%%%%%%%%%%%%%%%%%%%%%%%%%%%%%%%%%%%%%%%
%%%%%%%%%%%%%%%%%%%%%%%%%%%%%%%%%%%%%%%%%%%%
\subsection{Helicity amplitude ratios and relative phases}  
                                                                                   \label{sect:polar_disc_ampl_ratios}
%%%%%%%%%%%%%%%%%%%%%%%%%%%%%%%%%%%%%%%%%%%%

The measurements of the spin density matrix elements presented in 
Figs.~\ref{fig:matelem_f_qsq_rho} to~\ref{fig:matelem_f_m_two_qsq_rho} give access to the 
ratios and relative phases of the helicity amplitudes.
Following the IK analysis~\cite{ik}, four amplitude ratios, taken relative to the dominant 
$T_{00}$ amplitude, are measured from global fits to the 15 matrix element measurements, 
assuming NPE and taking all amplitudes as purely imaginary;
negative values correspond to opposite phases.
The measurements are presented in the following sections for \rh\ and \ph\ 
mesons as a function of \qsq\ and \modt, 
and additionally for \rh\ mesons as a function of the  \mpp\ invariant mass.
The relative phases are then discussed.

%%%%%%%%%%%%%%%%%%%%%%%%%%%%%%%%%%%%%%%%%%%%
%%%%%%%%%%%%%%%%%%%%%%%%%%%%%%%%%%%%%%%%%%%%
\subsubsection{\boldmath {\qsq} dependences}
                                                                            \label{sect:polar_qsq_ampl_ratios}
%%%%%%%%%%%%%%%%%%%%%%%%%%%%%%%%%%%%%%%%%%%%

%-----------------------------------------------------------------------------
\begin{figure}[tb]
\begin{center}
\setlength{\unitlength}{1.0cm}
\begin{picture}(16.5,9.0)   
\put(0.0,4.5){\epsfig{file=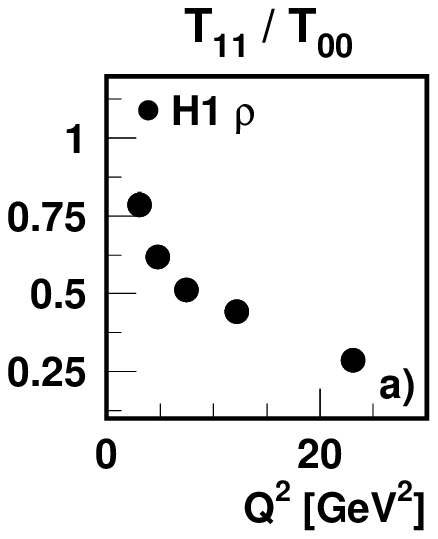,height=4.5cm,width=4.5cm}}
\put(4.0,4.5){\epsfig{file=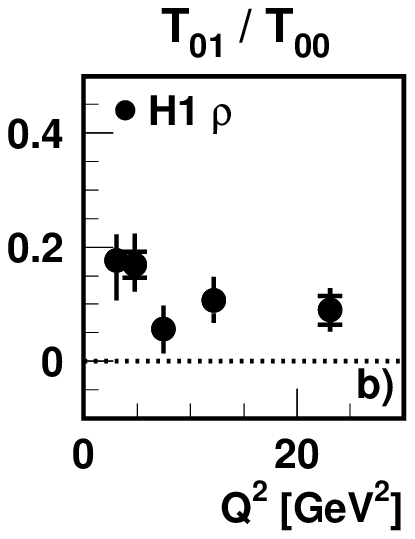,height=4.5cm,width=4.5cm}}
\put(8.0,4.5){\epsfig{file=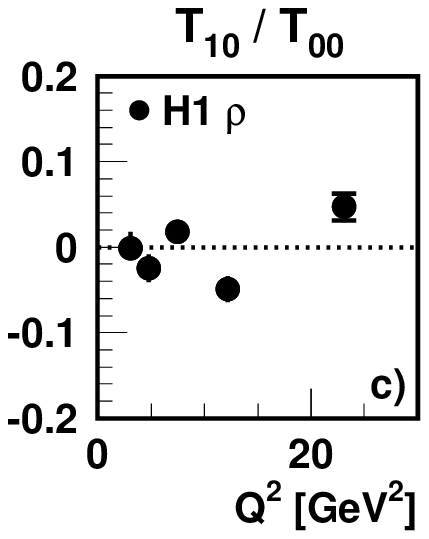,height=4.5cm,width=4.5cm}}
\put(12.,4.5){\epsfig{file=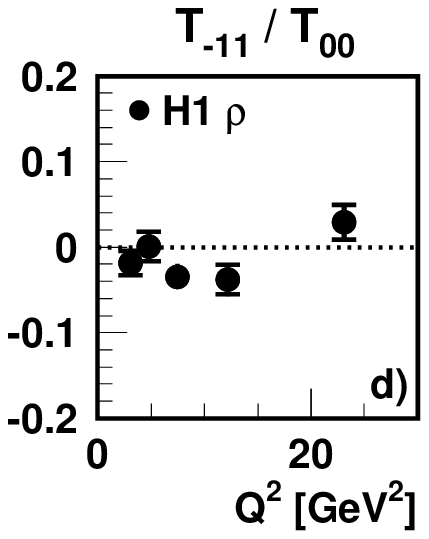,height=4.5cm,width=4.5cm}}
\put(0.0,0.0){\epsfig{file=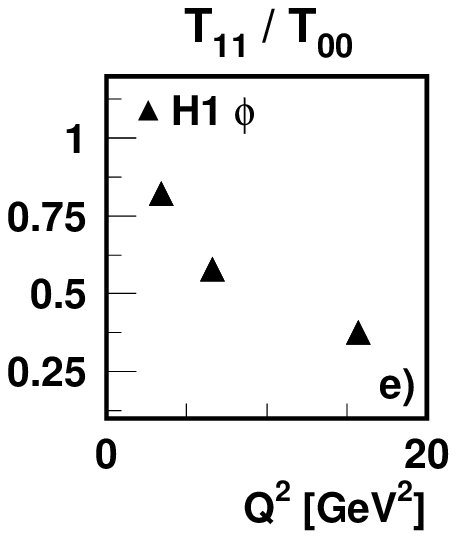,height=4.5cm,width=4.5cm}}
\put(4.0,0.0){\epsfig{file=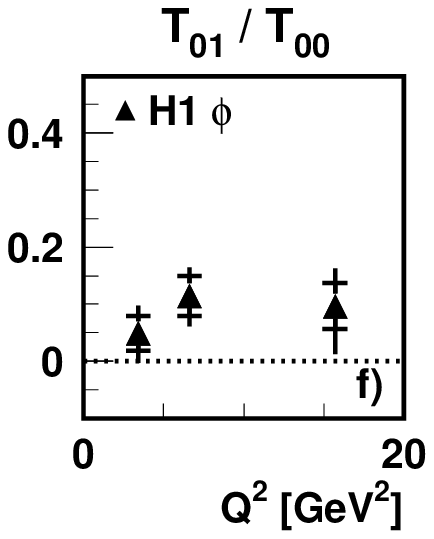,height=4.5cm,width=4.5cm}}
\put(8.0,0.0){\epsfig{file=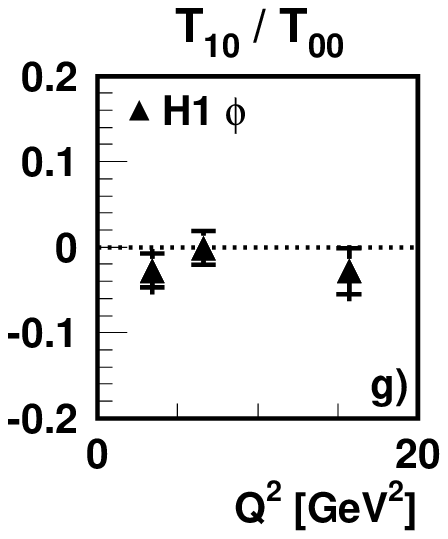,height=4.5cm,width=4.5cm}}
\put(12.,0.0){\epsfig{file=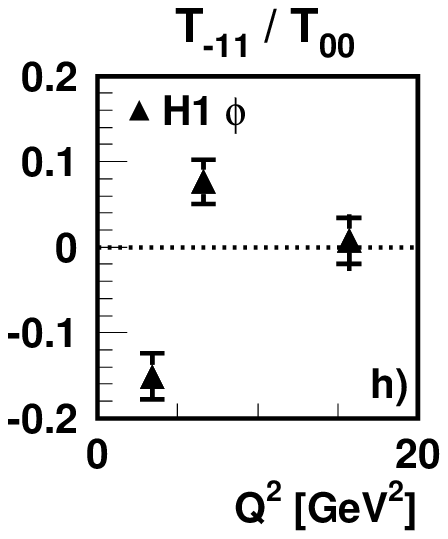,height=4.5cm,width=4.5cm}}
\end{picture}
\caption{Ratios of the helicity amplitudes, calculated from global fits to the measurements 
of  the 15 spin density matrix elements, as a function of \qsq:
(a)-(d) \rh\ meson production; (e)-(h) \ph\ production. 
NPE is assumed and all amplitudes are taken as purely imaginary.
Where appropriate, the dotted lines show the expected null value of the ratio
if the non-SCHC amplitudes are vanishing.
The measurements are given in Table~\ref{table:ampl-ratios-f-qsq}.
}
\label{fig:ampl-ratios-f-qsq}
\end{center}
\end{figure}
%-----------------------------------------------------------------------------

The \qsq\ dependence of the four amplitude ratios for \rh\ and \ph\ meson production
are presented in Fig.~\ref{fig:ampl-ratios-f-qsq}.
The strong decrease with \qsq\ of the amplitude ratio \ralpha\ for both VMs,
which is consistent with a linear increase with $1/Q$,
is related to the increase of the cross section ratio $R$ 
through the dominance of the SCHC amplitudes.
%(section~\ref{sect:polar_disc_R-qsq}, Fig.~\ref{fig:rlt-q2}). 
%The evolution of $R$ is governed by the formal $ Q^2 / M_{V}^2$ dependence, with a 
%modulation due to the \qsq\ evolution of the gluon distribution function.
For the first time, a \qsq\ dependence of the amplitude ratio \rbeta\ 
is also observed, for \rh\ meson production.
This dependence is also visible in the comparison of the two \qsq\ 
ranges in Figs.~\ref{fig:ampl-ratios-f-t} and~\ref{fig:ampl-ratios-f-m}.
No significant \qsq\ dependence is observed for the amplitude ratios \rdelta and
\reta.

In the IK~\cite{ik} model, the amplitude ratio \ralpha\ is given by
\begin{equation}
\ralpha =   \frac {M} {Q} \ \frac {1 + \gamma} {\gamma},       
%  \ralpha = M / Q \cdot (1 + \gamma) / \gamma,
                                                                                          \label{eq:ik:alpha}     
\end{equation}
where the decrease with \qsq\ of the anomalous dimension $\gamma$ slows down
the \qsq\ evolution, 
and the amplitude ratio \rbeta\ is given by
\begin{equation}
\rbeta   =   \frac {\sqrt{\modt}} {Q} \ \frac {1} {\sqrt{2} \gamma}
%$ \rbeta   =  \sqrt{\modt} / Q}  \cdot 1/ \sqrt{2} \gamma.$
                                                                                          \label{eq:ik:beta}     
\end{equation}
The model describes the \ralpha\ evolution well for values of
$M = 0.6~\gev < m_{\rho}$ and $\gamma = 0.7$, or 
$M = m_{\rho}$ and $\gamma = 1.1$ (not shown).
The latter is preferred for the description of \rbeta, though the physical interpretation 
of this high value for the parameter $\gamma$ is unclear.

%%%%%%%%%%%%%%%%%%%%%%%%%%%%%%%%%%%%%%%%%%%%
%%%%%%%%%%%%%%%%%%%%%%%%%%%%%%%%%%%%%%%%%%%%
%%%%%%%%%%%%%%%%%%%%%%%%%%%%%%%%%%%%%%%%%%%%
\subsubsection{\boldmath {\modt} dependences}
                                                                            \label{sect:polar_t_ampl_ratios}
%%%%%%%%%%%%%%%%%%%%%%%%%%%%%%%%%%%%%%%%%%%%
%%%%%%%%%%%%%%%%%%%%%%%%%%%%%%%%%%%%%%%%%%%%

The $t$ dependence of the amplitudes, empirically parameterised as exponentially falling,
is mainly determined by the proton and VM form factors.
It is a reasonable assumption that these form factors affect in a similar way all 
amplitudes, and that their effects cancel in matrix elements and in amplitude 
ratios~\cite{ik}.
The study of the $t$ dependence of the amplitude ratios thus gives access, in the reaction
dynamics, to features specific to the different amplitudes.
Note, however, that this line of reasoning neglects the different 
$t$ dependences for transverse and longitudinal amplitudes, related to different dipole 
sizes.

%-----------------------------------------------------------------------------
\begin{figure}[p]
\begin{center}
\setlength{\unitlength}{1.0cm}
\begin{picture}(16.5,9.0)   
\put(0.0,4.5){\epsfig{file=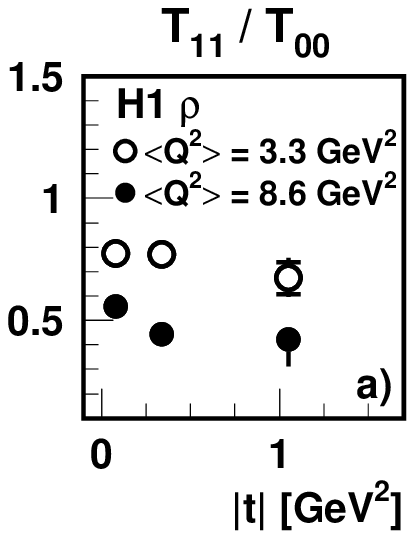,height=4.5cm,width=4.5cm}}
\put(4.0,4.5){\epsfig{file=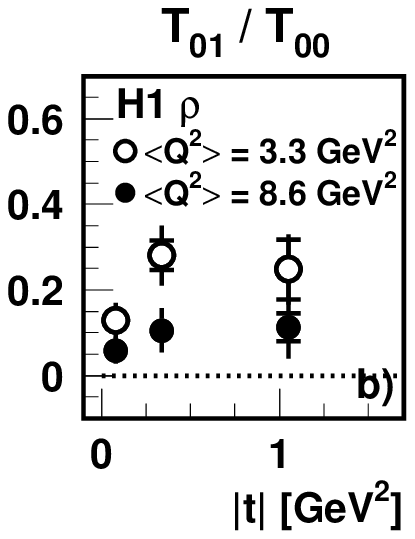,height=4.5cm,width=4.5cm}}
\put(8.0,4.5){\epsfig{file=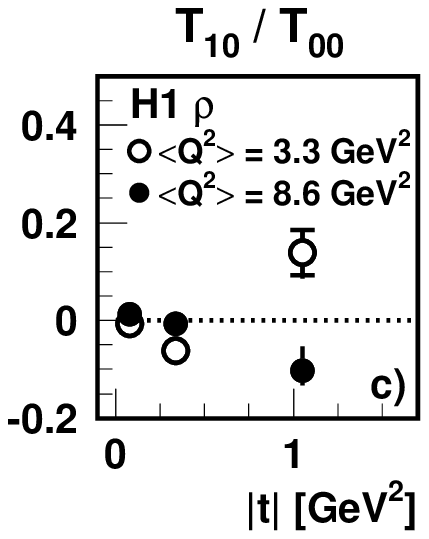,height=4.5cm,width=4.5cm}}
\put(12.,4.5){\epsfig{file=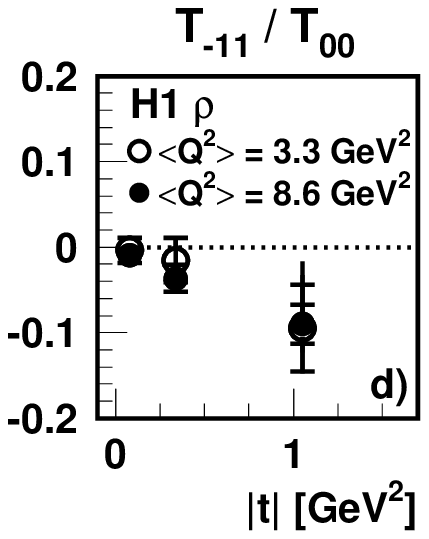,height=4.5cm,width=4.5cm}}
\put(0.0,0.0){\epsfig{file=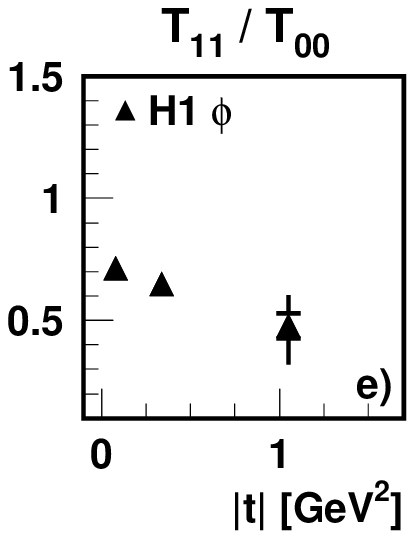,height=4.5cm,width=4.5cm}}
\put(4.0,0.0){\epsfig{file=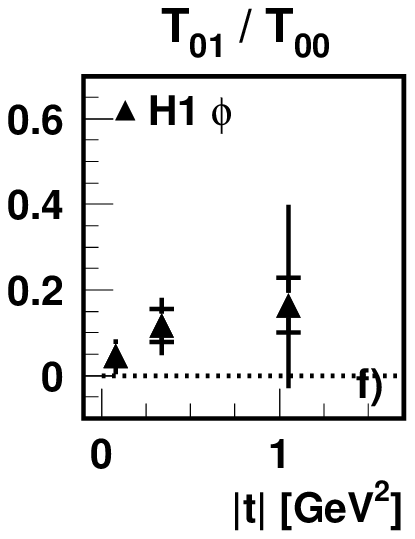,height=4.5cm,width=4.5cm}}
\put(8.0,0.0){\epsfig{file=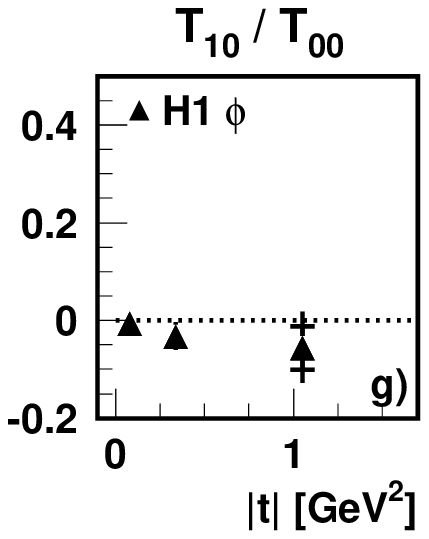,height=4.5cm,width=4.5cm}}
\put(12.,0.0){\epsfig{file=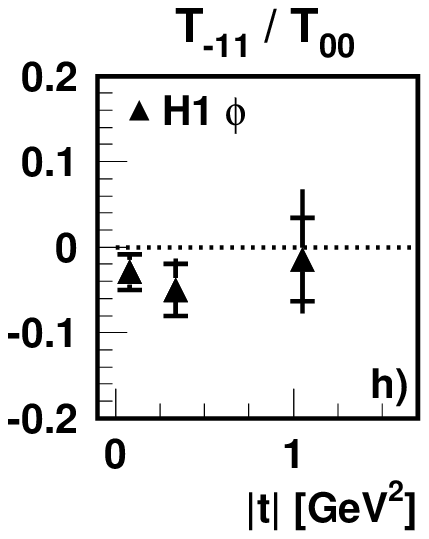,height=4.5cm,width=4.5cm}}
\end{picture}
\caption{Same as Fig.~\ref{fig:ampl-ratios-f-qsq}, as a function of \modt:
(a)-(d) \rh\ meson production, for two bins in \qsq: $2.5  \leq \qsq < 5~\gevsq$ 
(open circles) and $5  \leq \qsq  \leq 60~\gevsq$
(closed  circles); 
(e)-(h) \ph\ production.
The measurements are given in Table~\ref{table:ampl-ratios-f-t}.
}
\label{fig:ampl-ratios-f-t}
\end{center}
\end{figure}
%-----------------------------------------------------------------------------

Figure~\ref{fig:ampl-ratios-f-t} shows, for both VMs, the \modt\ dependences of the four 
amplitude ratios.
For the first time, a decrease with \modt\ of the ratio of amplitudes \ralpha\ is observed, 
both for \rh\ and for \ph\ production (Figs.~\ref{fig:ampl-ratios-f-t}(a) and~(e)).
The increase with \modt\ of the normalised \tzu\ helicity flip amplitudes, 
which could be deduced from the behaviour of the \rczz\ matrix element, is confirmed
in Figs.~\ref{fig:ampl-ratios-f-t}(b) and~(f).
For the second single flip amplitude, \tuz, negative values with increased strength 
relative to \tzz\ are observed in Fig.~\ref{fig:ampl-ratios-f-t}(c) at large 
\qsq.
Finally, non-zero values are found in Fig.~\ref{fig:ampl-ratios-f-t}(d) for the ratio of the 
double flip \tmuu\ to the \tzz\ amplitude, with negative values of the ratio and intensity 
increasing with \modt\ for both bins in \qsq.

The \modt\ dependence of the \tuu\ to \tzz\ amplitude ratio, which is not predicted in the
IK model, Eq.~\ref{eq:ik:alpha}, may be understood as 
an indication of different transverse dipole sizes in transverse and longitudinal
photon scattering, as discussed in section~\ref{sect:polar_disc_R-t} for the $t$
dependence of the cross section ratio $R$.
This is substantiated by the calculation of the cross section ratio using 
the helicity amplitude ratios,
the cross section ratio $R = \sigma_L / \sigma_T$ being given by:
\begin{equation}
R = \frac {1 + 2 \ (\rdelta)^2} {(\ralpha)^2 + (\rbeta)^2 + (\reta)^2}.
                                                                                         \label{eq:R_ampl}
\end{equation}
Following the procedure of section~\ref{sect:polar_disc_R-t}, the difference
between the longitudinal and transverse slopes are extracted from 
the $t$ dependence of $R$.
The results are given in Table~\ref{tab:bL-bT_ampl}.
For \rh\ production, the same effect is observed as in Table~\ref{tab:bL-bT}, where 
the value of $R$ was obtained only from the measurements of
the \rzqzz\ and \rczz\ matrix elements using Eq.~(\ref{eq:R_T01}):
a value of $b_L-b_T$ consistent with~$0$ for $\qsq < 5~\gevsq$, and a negative value
for $\qsq > 5~\gevsq$.
Errors are reduced due to the use of all amplitude ratios in the global fits, 
and the value of $b_L-b_T$ in the \qsq\ range with $\qsq > 5~\gevsq$ is $3 \sigma$ away
from~$0$.
For \ph\ production, the limited statistics do not allow to measure separately
the slope difference in two bins in \qsq.

%-----------------------------------------------------------------------------
\renewcommand{\arraystretch}{1.15}
\begin{table}[htbp]
\begin{center}
\begin{tabular}{|c|c|} 
\multicolumn{2}{c} { }  \\
\hline
    \av{\qsq}  (\gevsq)        &        $b_L - b_T$    (\gevsqm)         \\ 
\hline
\hline
    \multicolumn{2}{|c|} {\rh\ production}    \\ 
\hline
$3.3$                             &   $-0.06 \pm 0.22~_{-0.11}^{+0.24}$   \\
$8.6$                             &   $-0.53 \pm 0.10~_{-0.57}^{+0.14}$   \\
\hline
\hline
    \multicolumn{2}{|c|} {\ph\ production}    \\ 
\hline
$5.3$                             &   $-0.70 \pm 0.23~_{-0.63}^{+0.58}$   \\
\hline
\end{tabular} 
\caption{Difference between the longitudinal and transverse slopes of the $t$
distributions for \rh\ (two bins in \qsq) and for \ph\ meson production, calculated from
the $t$ dependence of the cross section ratio $R = \sigma_L / \sigma_T$
obtained using fits to the amplitude ratios, Eq.~(\ref{eq:R_ampl}).}
\label{tab:bL-bT_ampl}
\end{center}
\end{table}
\renewcommand{\arraystretch}{1.}
%-----------------------------------------------------------------------------

The $t$ dependence of the helicity flip amplitudes for light quarks can be explained 
as follows.
In the case of the $T_{01}$ amplitude, the virtual photon with transverse polarisation 
fluctuates into a quark and an antiquark which, given their opposite helicities, must be in 
an orbital momentum state with projection~$1$ onto the photon direction.
During the hard interaction, 
the dipole size and the quark and antiquark helicities are unchanged, but 
a transverse momentum $k_t \simeq \sqrt{|t|}$  is transferred to the dipole, which 
modifies its line of flight and thus allows a change of the orbital momentum projection.
The $T_{01}$ amplitude, which describes the production of a longitudinal meson 
from a transverse photon, is thus proportional to $\sqrt{|t|}$.
Similar reasons explain the $t$ dependence of the $T_{10}$ amplitude.
Note that, at variance with the case of light VMs, for heavy VMs with a non-relativistic 
wave function ($z \simeq 1-z \simeq 1/2$),
the exchange of orbital momentum cannot take place, thus implying SCHC.

In the IK model the \modt\ dependence of the single-flip to no-flip amplitude ratio 
\rbeta\ is given by Eq.~(\ref{eq:ik:beta}), and that of \rdelta\ by
\begin{equation}
\rdelta   = - \frac {M \ \sqrt{\modt}} {Q^2} \ \frac {\sqrt{2}} {\gamma},
                                                                                          \label{eq:ik:delta}     
\end{equation}
respectively, where the negative value of the ratio is consistent
with the \rh\ data in the higher \qsq\ domain, 
Fig.~\ref{fig:ampl-ratios-f-t}(c).

In the two-gluon exchange picture of diffraction for the double flip $T_{-11}$ amplitude, 
the change by two units from the photon to the VM
helicities requires in addition spin transfer by the exchanged gluons.
The observation of a non-zero value for this amplitude may thus provide important 
information concerning gluon polarisation in the proton~\cite{dima-ivanov-pc}.
The prediction of the IK model for \reta\ is
\begin{eqnarray}
\reta     &=& \eta^0 + \eta^1,       
                                                                                          \label{eq:ik:eta}     \\
\ \ \ \ \ \ \ \ \eta^0 &=& - \frac {{\bar \alpha}^2_S \modt \ M} {\pi \ \alpha_S \ Q \ m_\rho^2} 
       \ \frac {1} {4^\gamma \ \frac {\Gamma^2(\gamma+1)}{\Gamma(2\gamma+2)}
                  \ xG(x,\qsq/4)} \\
                                                                                          \label{eq:ik:eta0}
\ \ \ \ \ \ \ \ \eta^1 &=& \frac {M \ \modt} {Q^3} \ \frac {2 (\gamma+2)} {\gamma}
                                                                                          \label{eq:ik:eta1},
\end{eqnarray}
with a dependence proportional to \modt.
The model describes the \modt\ dependence of the data, but the negative sign 
of \reta, both for $\qsq < 5~\gevsq$ and $\qsq > 5~\gevsq$, is at variance with the 
model expectation; this is attributed to the strong approximations involved in the
parameterisations~\cite{dima-ivanov-pc}.

%%%%%%%%%%%%%%%%%%%%%%%%%%%%%%%%%%%%%%%%%%%%
%%%%%%%%%%%%%%%%%%%%%%%%%%%%%%%%%%%%%%%%%%%%
%%%%%%%%%%%%%%%%%%%%%%%%%%%%%%%%%%%%%%%%%%%%
\subsubsection{\boldmath{$W$} and \boldmath{\mpp} dependences}
                                                                            \label{sect:polar_W-m_ampl_ratios}
%%%%%%%%%%%%%%%%%%%%%%%%%%%%%%%%%%%%%%%%%%%%

%-----------------------------------------------------------------------------
\begin{figure}[p]
\begin{center}
\setlength{\unitlength}{1.0cm}
\begin{picture}(16.5,4.)   
\put(0.0,0.0){\epsfig{file=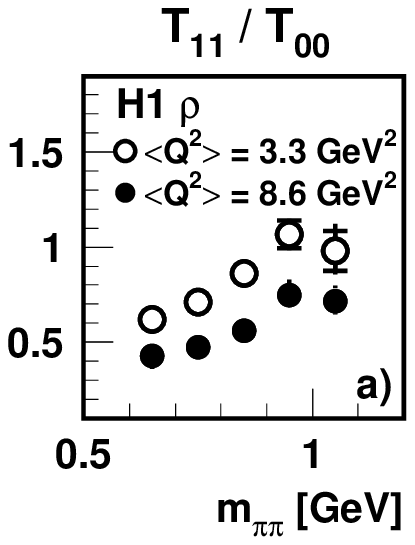,height=4.5cm,width=4.5cm}}
\put(4.0,0.0){\epsfig{file=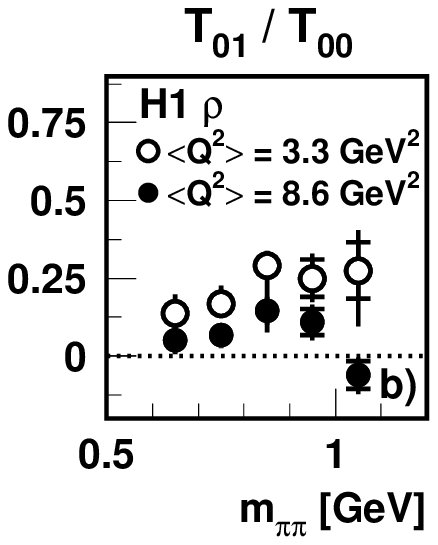,height=4.5cm,width=4.5cm}}
\put(8.0,0.0){\epsfig{file=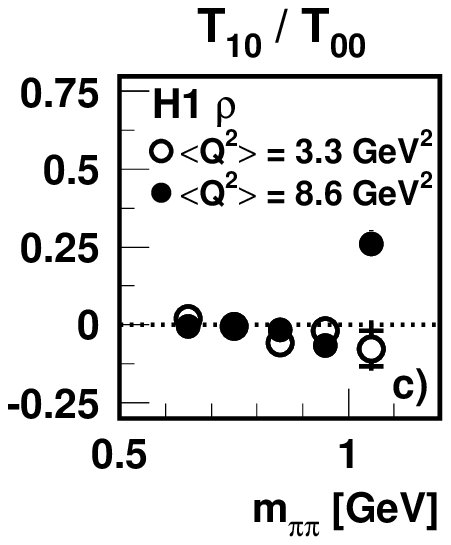,height=4.5cm,width=4.5cm}}
\put(12.,0.0){\epsfig{file=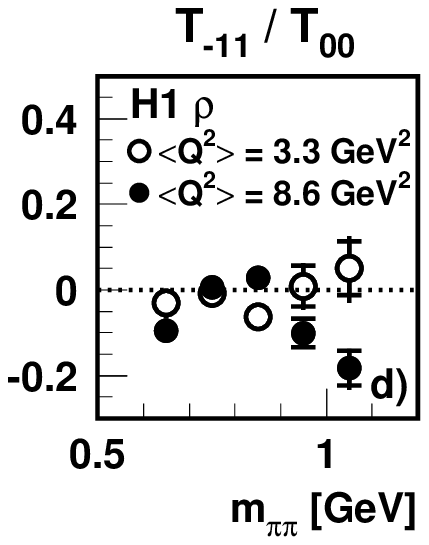,height=4.5cm,width=4.5cm}}
\end{picture}
\caption{Same as Fig.~\ref{fig:ampl-ratios-f-t} for \rh\ meson production, as a function 
of the mass \mpp.
The measurements are given in Table~\ref{table:ampl-ratios-f-m}.
}
\label{fig:ampl-ratios-f-m}
\end{center}
\end{figure}
%-----------------------------------------------------------------------------

No significant $W$ dependence of the amplitude ratios is observed (not shown),
which follows from the absence of a $W$ dependence of the matrix elements.
The strong \mpp\ dependence of the $\sigma_L / \sigma_T$ cross section ratio observed in 
Fig.~\ref{fig:rlt-wtm}(c) is confirmed in the ratio \ralpha\ of the dominant SCHC amplitudes,
as seen in Fig.~\ref{fig:ampl-ratios-f-m}, with a similar hint for \rbeta. 
As suggested in section~\ref{sect:polar_disc_R-m}, these features may be related to
the $M/Q$ dependences in Eqs.~(\ref{eq:ik:alpha}) and~(\ref{eq:ik:beta}).

%%%%%%%%%%%%%%%%%%%%%%%%%%%%%%%%%%%%%%%%%%%%
\subsubsection{Amplitude relative phases}  
                                                                                  \label{sect:polar_disc_ampl_phases}
%%%%%%%%%%%%%%%%%%%%%%%%%%%%%%%%%%%%%%%%%%%%

In an extension of the fits performed in the previous sections, the phases between the
amplitudes can be left free. 
To ensure proper convergence, the number of fitted quantities has to be reduced.
In view of their small values, the approximation is made to put to~$0$ the 
amplitudes $T_{10}$ and $T_{-11}$.
When the phase difference $\cos (\phi_{01} - \phi_{00})$ is left free, it is 
pushed to the bound~$1$; it is therefore fixed to this value in the fit\footnote{The observation 
that $\cos (\phi_{01} - \phi_{00})$ is close to~$1$ is at variance with 
calculations in~\cite{kroll-paperI}, where an attempt was made in a GPD approach to 
estimate the size of the $T_{01}$ amplitude within the handbag approach.
The prediction in~\cite{kroll-paperI} that the amplitudes should be out of phase depends
in fact strongly on a number of assumptions~\cite{kroll-pc}.}.

The fitted phase difference $\cos (\phi_{11} - \phi_{00})$ is found to be systematically lower 
than 1, with the amplitude ratios \ralpha\ and \rbeta\ being compatible with those presented in 
the previous section.
The average value of the phase difference for \rh\ mesons is 
\begin{equation}
\cos (\phi_{11} - \phi_{00}) = 0.936 \pm 0.016~{\rm (stat.)}~^{+ 0.025}_{-0.038}~{\rm (syst.)}, 
\end{equation}
which confirms the 
result of section~\ref{sect:polar_disc_schc} under the SCHC approximation, that the 
dominant longitudinal and transverse amplitudes are nearly but not completely in phase.

%%%%%%%%%%%%%%%%%%%%%%%%%%%%%%%%%%%%%%%%%%%%
%%%%%%%%%%%%%%%%%%%%%%%%%%%%%%%%%%%%%%%%%%%%
%%%%%%%%%%%%%%%%%%%%%%%%%%%%%%%%%%%%%%%%%%%%
%%%%%%%%%%%%%%%%%%%%%%%%%%%%%%%%%%%%%%%%%%%%

%\newpage

%%%%%%%%%%%%%%%%%%%%%%%%%%%%%%%%%%%%%%%%%%%%
\section{Summary and Conclusions}  
                                                                                            \label{sect:conclusions}
%%%%%%%%%%%%%%%%%%%%%%%%%%%%%%%%%%%%%%%%%%%%

This paper reports on the measurement of diffractive \rh\ and \ph\ meson
electroproduction at high energy, both in the elastic and proton dissociative channels.
The data were taken in the years 1996 to 2000 with the H1 detector at the $ep$ collider HERA,
in the kinematic domain $2.5  \leq \qsq  \leq 60~\gevsq$, $35 \leq W \leq 180~\gev$, 
$|t| \leq 3~\gevsq$ and $M_Y < 5~\gevcsq$.

The total, longitudinal and transverse \gstarp\ cross sections are measured as a function 
of the scaling variable $\qsq + \msq$.
They roughly follow power laws, and are well described by empirical 
parameterisations allowing the power to linearly depend on $\qsq + \msq$.
The \ph\ to \rh\ total cross section ratios are found to be independent of $\qsq + \msq$
and consistent for elastic and proton dissociative scattering, with a
value close to but slightly lower than the ratio expected from quark charge counting,
$\phi : \rh  = 2 : 9$.
The measurements significantly differ from the formal predictions 
$n = 3$ and $n = 4$ for the $1/ \qsqplmsq^n$ 
dependence of the longitudinal and transverse cross sections, respectively, which
is attributed mainly to the increase with \qsq\ of the gluon density at small $x$.

The \gstarp\ cross sections increase with the photon--proton centre of mass energy 
$W$, which is parameterised in the Regge inspired form $\propto W^\delta$, 
where $\delta$ increases significantly with \qsq.
This ``hardening" of the $W$ distribution
is described in terms of the intercept $\alpom(0)$ of the effective Regge trajectory.
For values of the scale $\mu^2 = \scaleqsqplmsq$ up to about $3~\gevsq$, 
the $W$ dependence of \rh\ and \ph\ production is slightly harder than the 
soft behaviour characteristic of hadron interactions and photoproduction,
$\alpom(0) = 1.08$ to~$1.11$.
For the higher \scaleqsqplmsq\ range, values of $\alpom(0)$ of the order of~$1.2$ 
to~$1.3$ are reached, compatible with \jpsi\ measurements.
DVCS measurements show a similar behaviour as a function of the scale  $\mu^2 = \qsq$.

The $t$ dependences of the cross sections are well described as exponentially falling
distributions $\propto e^{-b~\! |t|}$, up to \modt\ values of $0.5~\gevsq$ for elastic production 
and~$3~\gevsq$ for proton dissociation.
The $t$ slopes are measured for all four channels, providing the first precise determination 
at HERA of the proton dissociative slopes for light VM electroproduction.
The values of the $t$ slopes are lower than those in photoproduction
and they decrease with
increasing scale, in a way which is common to light VMs and DVCS.
Values of the $t$ slopes comparable to those for \jpsi\ production, or slightly larger, are 
reached for a scale \scaleqsqplmsq\ $\gapprox\ 5~\gevsq$, which suggests that light 
VM form factors are small and confirms that the 
dominant longitudinal amplitudes approach a perturbative behaviour for \scaleqsqplmsq\ 
around~$3$ to $5~\gevsq$.
The correlation between the $W$ and $t$ dependences of the cross sections is
parameterised in the form of the slope \alp\ of the effective pomeron trajectory.
For \rh\ meson production, this slope is smaller than that in soft hadron-hadron 
interactions, albeit with large errors.

The ratio of the proton dissociative to elastic cross sections for $\modt = 0$ and 
the difference between the elastic and proton dissociative slopes are measured
to be independent of \qsq.
These observations support the relevance of the factorisation of the process into a hard 
scattering contribution at the photon vertex and a soft diffractive scattering at the proton 
vertex (``Regge factorisation").
The value measured for \rh\ and \ph\ production for the slope difference, 
$b_{el.} - b_{p.~diss.} \simeq 5.5~\gevsqm$, however, is
larger than for \jpsi\ production.

Polarisation effects are studied through the measurement of $15$ spin density 
matrix elements, which are normalised bilinear combinations of the complex helicity 
amplitudes $T_{\lambda_{V} \lambda_{\gamma}}$.
The dependence on the kinematic variables and, for \rh\, mesons, on the dipion mass is
measured.
The main feature in the present domain is the dominance of the $s$-channel helicity 
conserving (SCHC) amplitudes, $T_{00}$ and $T_{11}$, with $T_{00} > T_{11}$.
In addition, a significant breaking of SCHC is manifest through 
the non-zero value of the \rczz\ matrix element, especially at large \modt\ values.

The ratio $R = \sigma_L / \sigma_T$ of the longitudinal to transverse cross sections
increases strongly with \qsq, as predicted in pQCD, with a scaling behaviour as a 
function of $Q^2  / M_{V}^2$ for the different VMs.
The linear dependence $R = \qsq / \msq$ predicted at LO, however, is damped for 
large values of \qsq.
No $W$ dependence of $R$ is observed within errors.
For $t$, an indication of the dependence of $R$ is found for \rh\ meson production
with $\qsq > 5~\gevsq$.
This can be interpreted as a difference between the longitudinal and transverse 
$t$ slopes, $b_L - b_T$, which differs from zero by $1.5 \sigma$, with dominant systematic 
errors.
A strong \mpp\ dependence of $R$ is observed for \rh\ meson production, both for
\qsq\ smaller and larger than $5$~\gevsq.
This behaviour may be interpreted as following from the 
general $\qsq / M^2$ dependence of VM production, if the mass $M$ is understood as 
the dipion mass rather than the nominal resonance mass.

The ratio of the helicity amplitudes is measured from global fits to the 15 matrix 
elements.
Several features expected in pQCD are observed for the first time. 
A decrease with increasing \qsq\ is found for the amplitude ratio \rbeta, which supports 
the higher twist nature of the helicity flip amplitudes.
The amplitude ratio \ralpha\ is observed to decrease with increasing \modt,
which may be related to different transverse sizes of transverse and longitudinal 
dipoles. This is substantiated by the non-zero value of 
the slope difference $b_L - b_T$ obtained from the measurement of $R$ 
from global fits of the helicity amplitudes, with a $3 \sigma$ significance.
At large \qsq, the amplitude ratio \rdelta\ which involves the second 
single flip amplitude is found to exhibit a \modt\ dependence.
Finally, a non-zero value at large \modt\ is found for the ratio \reta\ which involves the 
double flip amplitude, an observation which may provide information on gluon polarisation 
in the proton.
The phase between the $T_{00}$ and $T_{11}$ amplitudes is measured to be non-zero, 
which may suggest different $W$ dependences of the transverse and longitudinal 
amplitudes. 

The general features of the kinematic dependences of the cross sections and of the spin 
density matrix elements are understood qualitatively in QCD.
In particular, the $W$ and $t$ dependences indicate that ``hard", perturbative QCD features 
become dominant in the longitudinal cross section in the present kinematic domain,
for $\scaleqsqplmsq\ \gapprox\ 3-5~\gevsq$.
The measurements are globally described by models using GPDs or a dipole approach, 
which differ in detail but agree on the gross features.

The study of VM production at HERA thus provides new insights for the understanding
of QCD and the interplay of soft and hard diffraction.

%%%%%%%%%%%%%%%%%%%%%%%%%%%%%%%%%%%%%%%%%%%%
%%%%%%%%%%%%%%%%%%%%%%%%%%%%%%%%%%%%%%%%%%%%
%%%%%%%%%%%%%%%%%%%%%%%%%%%%%%%%%%%%%%%%%%%%
%%%%%%%%%%%%%%%%%%%%%%%%%%%%%%%%%%%%%%%%%%%%

%%%%%%%%%%%%%%%%%%%%%%%%%%%%%%%%%%%%%%%%%%%%
\section*{Acknowledgements}
%%%%%%%%%%%%%%%%%%%%%%%%%%%%%%%%%%%%%%%%%%%%

We are grateful to the HERA machine group whose outstanding
efforts have made this experiment possible. 
We thank the engineers and technicians for their work in 
constructing and maintaining the H1 detector, our funding 
agencies for financial support, the DESY technical staff for 
continual assistance, and the DESY directorate for the hospitality 
which they extend to the non-DESY members of the collaboration.
We thank A.~Bruni, J.-R.~Cudell, M.~Diehl, S.V.~Goloskokov, D.Yu.~Ivanov, I.~Ivanov, 
R.~Kirschner, P.~Kroll, G.~Soyez, M.~Strikman, T.~Teubner and G.~Watt
for useful discussions and 
for providing us with the predictions of their calculations.

%%%%%%%%%%%%%%%%%%%%%%%%%%%%%%%%%%%%%%%%%%%%
%%%%%%%%%%%%%%%%%%%%%%%%%%%%%%%%%%%%%%%%%%%%
%%%%%%%%%%%%%%%%%%%%%%%%%%%%%%%%%%%%%%%%%%%%
%%%%%%%%%%%%%%%%%%%%%%%%%%%%%%%%%%%%%%%%%%%%

\newpage

\section*{Appendix}

\appendix

\def\thesubsection{\Alph{subsection}}

%%%%%%%%%%%%%%%
\paragraph {Matrix elements}                                                  \label{sect:polar:ang_distri}
%%%%%%%%%%%%%%%

In the formalism of~\cite{sch-w}, 
the spin density matrix elements are normalised sums of products of two helicity
amplitudes $T_{\lambda_{\rho} \lambda_{N'}, \lambda_{\gamma} \lambda_{N}}$.
They are given in the form $r^{i}_{jk}$, where the notation~$^{(04)}$ of the upper index~$^{(i)}$ 
denotes the combination of unpolarised transverse and longitudinal photons\footnote{The
separation of the~$^{(0)}$ and~$^{(4)}$ components is only possible through measurements 
with different polarisation parameters $\varepsilon$, i.e. with different beam energies in the 
same detector configuration. In this case, 18 matrix elements in total can be measured.},
the notations~$^{(1)}$ and~$^{(2)}$ are used for VM production by transverse photons with 
orthogonal linear polarisations, and~$^{(5)}$ and~$^{(6)}$ for the interference between VM 
production by transverse and longitudinal photons.
The lower indices~$_{(j, k)}$ refer to the VM helicities $\lambda_{V}$ of the pair
of amplitudes. 

%%%%%%%%%%%%%%%
\paragraph  {Angular distribution}
%%%%%%%%%%%%%%%

In the absence of longitudinal beam polarisation, 15 independent components of the spin 
density matrix can be measured (8 additional matrix elements are accessible with
a longitudinally polarised lepton beam).
They enter in the normalised angular distribution $W(\theta, \phib, \phi)$:
\begin{eqnarray}
 W(\theta, \phib, \phi) &=& \frac{3}{4\pi}\ \; \left\{ \ \
  \frac{1}{2} (1 - \rzqzz) + \frac{1}{2} (3 \ \rzqzz -1) \
  \cos^2\theta \right. \nonumber \\
  &-& \sqrt{2}\ {\rm Re} \ \rzquz\ \sin 2\theta  \cos\phib
    - \rzqumu\ \sin^2\theta \cos 2\phib \nonumber \\
  &-& \varepsilon\ \cos2\phi  \left(
      \ruuu\ \sin^2\theta + \ruzz\ \cos^2\theta
      - \sqrt{2}\ {\rm Re} \ \ruuz\ \sin2\theta \cos\phib \right.
\nonumber \\
& &    \ \ \ \ \ \ \ \ \ \ \ \ \ \ \ \ \ \
      - \left. \ruumu\ \sin^2\theta \cos2\phib \frac{}{} \right) \nonumber \\
    &-& \varepsilon\ \sin2\phi  \left( \sqrt{2}\ {\rm Im}\ \rduz\
\sin2\theta
\sin\phib
      +  {\rm Im}\ \rdumu\ \sin^2\theta \sin2\phib \right) \nonumber \\
  &+& \sqrt{2\varepsilon\ (1+\varepsilon)}\ \cos\phi\ \left(  \frac{ }{ } \rcuu\
\sin^2\theta
      + \rczz\ \cos^2\theta \right. \nonumber \\
& &    \ \ \ \ \ \ \ \ \ \ \ \ \ \ \ \ \ \ - \sqrt{2}\ {\rm Re} \ \rcuz\
\sin 2\theta \cos\phib
       - \left. \rcumu\ \sin^2\theta \cos2\phib \frac{ }{ } \right) \nonumber \\
  &+& \sqrt{2\varepsilon\ (1+\varepsilon)}\ \sin\phi\ \left( \sqrt{2}\
{\rm Im}
\
          \rsuz\ \sin2\theta \sin\phib  \right. \nonumber \\
& &    \ \ \ \ \ \ \ \ \ \ \ \ \ \ \ \ \ \ \left.
        \left. + {\rm Im}\ \rsumu\ \sin^2\theta \sin2\phib \frac{ }{ } \right)\
\right\} .
                                                                                                    \label{eq:W}
\end{eqnarray}

%%%%%%%%%%%%%%%
\paragraph {Measurement of the matrix elements}
%%%%%%%%%%%%%%%

The matrix elements are measured as projections of the normalised angular distribution, 
Eq.~(\ref{eq:W}), onto orthogonal functions of the $\theta$, \phib\ and \ph\ angles, with
one specific function corresponding to each matrix element (see Appendix~C 
of~\cite{sch-w}).
In practice, each matrix element is measured as the average value of the corresponding
function, taken over all events in the data sample.

Alternatively, fits to the projections of the angular distribution $W(\theta, \phib, \phi)$ 
onto each of the three angles provide measurements of the matrix elements
 \rzzzz\ and \rzqumu\ and of the combinations (\rfivecomb) and (\ronecomb):
\begin{eqnarray}
W(\theta) &\propto&   1 - \rzzzz + (3 \ \rzzzz -1) \ \cos^2\theta            \label{eq:cosths} \\
W(\varphi) &\propto&   1  - 2 \rzqumu \ \cos 2 \varphi                        \label{eq:angle_varphi} \\
W(\phi) &\propto&   1  + \sqrt{2\varepsilon (1+\varepsilon)} \cos \phi\ (\rfivecomb) 
                            - \varepsilon\ \cos2\phi\ (\ronecomb).               \label{eq:angle_phi}
\end{eqnarray}

%%%%%%%%%%%%%%%
\paragraph{Natural parity exchange}
                                                                                          \label{sect:polar:NPE}
%%%%%%%%%%%%%%%

Natural parity exchange (NPE) in the $t$ channel implies the following relations between 
amplitudes\footnote{More precisely, 
Eq.~(\ref{eq:NPE}) implies that, for $\modt = |t|_{min}$, the trajectory exchanged 
in the $t$ channel has natural parity.}:
\begin{equation}
  T_{-\lambda_V \lambda_{N'}, -\lambda_{\gamma} \lambda_{N}} =
  (-1)^{\lambda_V-\lambda_{\gamma}} \
  T_{\lambda_V \lambda_{N'}, \lambda_{\gamma} \lambda_{N}}  .
                                                                                                \label{eq:NPE}
\end{equation}  
For unnatural parity exchange, an additional factor $(-1)$ appears in the right hand
side of Eq.~(\ref{eq:NPE}).

Under NPE and integrating over the nucleon polarisations,
the number of independent $T_{\lambda_{V} \lambda_{\gamma}}$ amplitudes is 
reduced from 9 to 5:
two helicity conserving amplitudes ($ T_{00}$ and $T_{11} = T_{-1-1}$), 
two single helicity flip amplitudes ($T_{01} = -T_{0-1}$ and $T_{10} = -T_{-10}$) 
and one double flip amplitude ($ T_{-11} =  T_{1-1}$).

In general, longitudinally polarised lepton beams are required to separate natural and unnatural
parity exchange process.
However, unpolarised beams allow the measurement of the asymmetry $P_{NPE,T}$ between 
natural ($\sigma^N_T$) and unnatural  ($\sigma^U_T$) parity exchange for transverse photons:
\begin{equation}
P_{NPE,T} = \frac {\sigma^N_T - \sigma^U_T} {\sigma^N_T + \sigma^U_T}
            = 2 - \rzqzz + 2 \rzqumu - 2 \ruuu - 2 \ruumu.
                                                                                          \label{eq:NPE-asym}
\end{equation}
The measurement of the corresponding asymmetry for longitudinal photons requires 
different values of $\varepsilon$, i.e. different beam energies.

%%%%%%%%%%%%%%%
\paragraph{\boldmath {$s$}-channel helicity conservation}
%                                                                                          \label{sect:polar:schc}
%%%%%%%%%%%%%%%

In the approximation of $s$-channel helicity conservation (SCHC)~\cite{gilman}, the helicity
of the virtual photon (measured in the helicity frame defined in section~\ref{sect:variables}) 
is retained by the final state VM (with the nucleon helicity also remaining unchanged).
Single and double helicity flip amplitudes thus vanish 
($T_{01} = T_{10} =  T_{-11} = 0$) and only five matrix elements are non-zero:
\begin{equation}
 \rzqzz, \ \ \ \ruumu, \ \ \ {\rm Im} \ \rdumu, \ \ \ {\rm Re} \ \rcuz, \ \ \ {\rm Im} \ \rsuz;
                                                                                         \label{eq:non-0-schc}
\end{equation}
Under SCHC and NPE, the following relations hold between these elements:
\begin{eqnarray}
       \ruumu  =  - {\rm Im} \ \rdumu = \frac {1} {2} \ (1 -  \rzqzz),  
       \ \ \ {\rm Re} \ \rcuz  =  -{\rm Im} \ \rsuz.
                                                                                      \label{eq:schc_pairing}
\end{eqnarray}

In the case of SCHC, only two independent parameters are left, conveniently chosen as 
the cross section ratio $R = \sigma_L / \sigma_T$ and the phase $\delta$ between 
the $T_{00}$ and $T_{11}$ amplitudes, with 
\begin{equation}
T_{00} \ T_{11}^* = |T_{00}| \ |T_{11}| \ e^{-i\delta}.
                                                                                    \label{eq:def_cosdelta}
\end{equation}
The angular distribution $W(\theta, \phib, \phi)$ then reduces to a function of 
$\theta$ and $\psi = \phi - \phib$, the angle between the electron scattering plane and the 
\rh\ meson decay plane, in the \gstarp\ frame:
\begin{eqnarray}
 W(\costh, \psi) &=& \frac{3}{8\pi}\ \frac{1}{1+\varepsilon\ R}\ 
  \left\{ \ \sin^2\theta \ (1+ \varepsilon\ \cos2\psi)  \frac{ } { }  
  \right. \nonumber \\
    &+& \left. 2\ \varepsilon\ R\ \cos^2\theta
  - \sqrt{2\varepsilon\ (1+\varepsilon)\ R}\ \cos\delta \sin2\theta
 \cos\psi \right\}.
                                                                                                   \label{eq:Wcosdelta}
\end{eqnarray}

In the SCHC approximation, the cross section ratio $R$ is obtained 
from the measurement of the matrix element \rzqzz, as given by Eq.~(\ref{eq:R_schc}).

%%%%%%%%%%%%%%%
\paragraph{Dominant helicity flip amplitude {\boldmath $T_{01}$} }
                                                                                          \label{sect:polar:T01}
%%%%%%%%%%%%%%%

The precision of measurements performed in the SCHC approximation, especially at 
large \modt, can be improved by retaining the dominant helicity flip amplitude $T_{01}$.
Five additional matrix elements 
are then non-zero, supplementing the five elements given in Eq.~(\ref{eq:non-0-schc}):
\begin{equation}
  {\rm Re}~\rzquz, \ \ \ \ruzz, \ \ \ {\rm Re}~\ruuz, \ \ \ {\rm Im}~\rduz, \ \ \ \rczz.
                                                                                         \label{eq:non-0-T01}
\end{equation}
Under NPE, the following relations hold in addition to the SCHC
relations~(\ref{eq:schc_pairing}):
\begin{equation}
{\rm Re}~\rzquz = - {\rm Re}~\ruuz = {\rm Im}~\rduz.
                                                                                      \label{eq:T01-non-schc_pairing}
\end{equation}

Assuming that the amplitudes are in phase, an improved approximation of the cross 
section ratio $R$ is given by Eq.~(\ref{eq:R_T01}), which uses the matrix elements
\rzqzz\ and \rczz.

%%%%%%%%%%%%%%%%%%%%%%%%%%%%%%%%%%%%%%%%%%%%
%%%%%%%%%%%%%%%%%%%%%%%%%%%%%%%%%%%%%%%%%%%%
%%%%%%%%%%%%%%%%%%%%%%%%%%%%%%%%%%%%%%%%%%%%
%%%%%%%%%%%%%%%%%%%%%%%%%%%%%%%%%%%%%%%%%%%%

% \include{pap-hera1-biblio}
\clearpage

%%%%%%%%%%%%%%%%%%%%%%%%%%%%%%%%%%%%%%%%%%%%

%\bibitem{olsson}
%H1 Collab., 
%``A new measurement of exclusive $\rho^{0}$ photoproduction at HERA", 
%J. Olsson, 
%XIV Int. Workshop on DIS,
%Tsukuba, Japan, 2006;
%R. Weber, PhD Thesis, Swiss Federal Institute of Technology Zurich (ETH) (2006),
%http://www-h1.desy.de/psfiles/theses/h1th-440.pdf.

% \include{tables}
\renewcommand{\arraystretch}{1.2}

 \begin{table}[H]
 \begin{center}
\begin{tabular}{|d|lll|lll|} 
 \hline
 \multicolumn{1}{|c|}{\qsq\ (\gevsq)} &\multicolumn{3}{c|}{$f_I$   } &\multicolumn{3}{c|}{$n$ } \\
 \hline
  3.3 & $-0.470$ & $\pm\ 0.049$ & $^{+ 0.018}_{- 0.017}$ & $1.73$ & $\pm\ 0.21$ & $^{+0.07}_{-0.07}$ \\
  6.5 & $-0.270$ & $\pm\ 0.059$ & $^{+ 0.019}_{- 0.015}$ & $0.91$ & $\pm\ 0.24$ & $^{+0.06}_{-0.08}$ \\
 11.9 & $-0.351$ & $\pm\ 0.095$ & $^{+ 0.017}_{- 0.020}$ & $1.27$ & $\pm\ 0.39$ & $^{+0.09}_{-0.07}$ \\
 23.0 & $-0.100$ & $\pm\ 0.090$ & $^{+ 0.009}_{- 0.018}$ & $0.20$ & $\pm\ 0.37$ & $^{+0.08}_{-0.04}$ \\
 \hline
 \end{tabular}
 \caption{
  \qsq\ dependence, for elastic \rh\ meson production, of the                                                                          
  S\"{o}ding skewing parameter $f_I$ defined in Eq.~(\ref{eq:sod}) and         
  of the Ross-Stodolsky parameter $n$ defined in Eq.~(\ref{eq:rs}).            
 }
 \label{table:skew}
 \end{center}
 \end{table}

 \begin{table}[H]
 \begin{center}
 \begin{tabular}{|d|lll|} 
 \hline
 \multicolumn{1}{|c|}{\qsq\ (\gevsq)} & \multicolumn{3}{c|}{$\sigma (\gsrel) $ ($ {\rm nb} $)} \\ 
 \hline
 2.65 & $563 $ & $\pm\    32$ & $^{+ 59}_{- 45}$ \\
 2.95 & $423 $ & $\pm\    24$ & $^{+ 49}_{- 34}$ \\
 3.30 & $383 $ & $\pm\    20$ & $^{+ 43}_{- 30}$ \\
 3.70 & $295 $ & $\pm\    17$ & $^{+ 33}_{- 25}$ \\
 4.15 & $232 $ & $\pm\    16$ & $^{+ 27}_{- 17}$ \\
 4.65 & $198 $ & $\pm\    13$ & $^{+ 20}_{- 19}$ \\
 5.20 & $154 $ & $\pm\     9$ & $^{+ 17}_{-  8}$ \\
 5.85 & $131 $ & $\pm\     9$ & $^{+ 12}_{-  7}$ \\
 6.55 & $102 $ & $\pm\     8$ & $^{+  8}_{-  7}$ \\
 7.35 & $79.1$ & $\pm\   6.4$ & $^{+ 6.6}_{- 3.6}$ \\
 8.20 & $56.5$ & $\pm\   4.3$ & $^{+ 5.5}_{- 2.2}$ \\
 9.20 & $53.0$ & $\pm\   3.9$ & $^{+ 4.2}_{- 3.7}$ \\
 10.3 & $39.3$ & $\pm\   3.9$ & $^{+ 4.2}_{- 2.8}$ \\
 11.5 & $30.8$ & $\pm\   2.9$ & $^{+ 2.8}_{- 2.5}$ \\
 12.9 & $25.8$ & $\pm\   2.2$ & $^{+ 2.7}_{- 1.7}$ \\
 14.5 & $15.7$ & $\pm\   1.3$ & $^{+ 1.7}_{- 1.2}$ \\
 16.5 & $12.5$ & $\pm\   1.0$ & $^{+ 1.3}_{- 1.1}$ \\
 18.8 & $9.22$ & $\pm\  0.82$ & $^{+1.11}_{-0.80}$ \\
 21.7 & $5.99$ & $\pm\  0.58$ & $^{+0.67}_{-0.55}$ \\
 25.0 & $3.54$ & $\pm\  0.42$ & $^{+0.39}_{-0.31}$ \\
 29.3 & $2.24$ & $\pm\  0.31$ & $^{+0.31}_{-0.28}$ \\
 35.0 & $1.68$ & $\pm\  0.27$ & $^{+0.24}_{-0.21}$ \\
 46.0 &$0.742$ & $\pm\ 0.105$ & $^{+0.101}_{-0.088}$ \\
\hline
 \end{tabular}
 \caption{
  \qsq\ dependence of the \gstarp\ cross section for elastic \rh\              
  meson production                                                             
  for $W = 75~\gev$.                                                           
  The overall normalisation uncertainty of $3.9\%$ is not included in the              
  systematic errors.                                                           
 }
 \label{table:rhoel_xsq2}
 \end{center}
 \end{table}

 \begin{table}[H]
 \begin{center}
 \begin{tabular}{|d|lll|} 
 \hline
 \multicolumn{1}{|c|}{\qsq\ (\gevsq)} & \multicolumn{3}{c|}{$\sigma (\gsrpd) $ ($ {\rm nb} $)} \\ 
 \hline
 2.71 & $314 $ & $\pm\   32$ & $^{+ 45}_{- 35}$ \\
 3.21 & $217 $ & $\pm\   20$ & $^{+ 29}_{- 24}$ \\
 3.82 & $182 $ & $\pm\   16$ & $^{+ 26}_{- 17}$ \\
 4.52 & $131 $ & $\pm\   13$ & $^{+ 17}_{- 16}$ \\
 5.36 & $57.7$ & $\pm\  6.2$ & $^{+ 6.8}_{- 6.0}$ \\
 6.35 & $50.5$ & $\pm\  5.7$ & $^{+ 6.5}_{- 4.9}$ \\
 7.60 & $40.4$ & $\pm\  4.6$ & $^{+ 5.1}_{- 3.7}$ \\
 9.30 & $25.9$ & $\pm\  2.3$ & $^{+ 3.0}_{- 2.3}$ \\
12.00 & $17.0$ & $\pm\  1.8$ & $^{+ 2.3}_{- 1.7}$ \\
14.85 & $10.8$ & $\pm\  1.4$ & $^{+ 1.2}_{- 1.2}$ \\
19.20 & $3.39$ & $\pm\ 0.45$ & $^{+0.53}_{-0.56}$ \\
32.15 & $1.01$ & $\pm\ 0.12$ & $^{+0.16}_{-0.16}$ \\
 \hline
 \end{tabular}
 \caption{
  \qsq\ dependence of the \gstarp\ cross section for proton                    
  dissociative \rh\                                                            
  meson production for $W = 75~\gev$.                                          
  The overall normalisation uncertainty of $4.6\%$ is not included in the              
  systematic errors.                                                           
 }
 \label{table:rhopd_xsq2}
 \end{center}
 \end{table}

 \begin{table}[H]
 \begin{center}
 \begin{tabular}{|d|lll|} 
 \hline
 \multicolumn{1}{|c|}{\qsq\ (\gevsq)} & \multicolumn{3}{c|}{$\sigma (\gspel) $ ($ {\rm nb} $)} \\ 
 \hline
 2.71 & $72.6 $ & $\pm\  8.5$ & $^{+10.6}_{- 7.6}$ \\
 3.21 & $64.5 $ & $\pm\  6.2$ & $^{+ 8.3}_{- 6.0}$ \\
 3.82 & $46.4 $ & $\pm\  4.5$ & $^{+ 5.2}_{- 3.6}$ \\
 4.52 & $35.0 $ & $\pm\  4.0$ & $^{+ 4.1}_{- 3.4}$ \\
 5.36 & $25.1 $ & $\pm\  2.7$ & $^{+ 2.6}_{- 1.4}$ \\
 6.35 & $18.2 $ & $\pm\  2.0$ & $^{+ 1.9}_{- 1.5}$ \\
 7.60 & $12.6 $ & $\pm\  1.5$ & $^{+ 1.2}_{- 0.7}$ \\
 9.30 & $7.04 $ & $\pm\ 0.90$ & $^{+0.78}_{-0.52}$ \\
12.00 & $5.34 $ & $\pm\ 0.63$ & $^{+0.56}_{-0.37}$ \\
14.85 & $2.25 $ & $\pm\ 0.38$ & $^{+0.30}_{-0.21}$ \\
19.20 & $1.28 $ & $\pm\ 0.24$ & $^{+0.17}_{-0.13}$ \\
32.15 & $0.371$ & $\pm\ 0.076$ & $^{+0.049}_{-0.042}$ \\
 \hline
 \end{tabular}
 \caption{
  \qsq\ dependence of the  \gstarp\ cross section for elastic \ph\ meson       
  production for $W = 75~\gev$.                                                
  The overall normalisation uncertainty of $4.7\%$ is not included in the              
  systematic errors.                                                           
 }
 \label{table:phiel_xsq2}
 \end{center}
 \end{table}

 \begin{table}[H]
 \begin{center}
 \begin{tabular}{|d|lll|} 
 \hline
 \multicolumn{1}{|c|}{\qsq\ (\gevsq)} & \multicolumn{3}{c|}{$\sigma (\gsppd) $ ($ {\rm nb} $)} \\ 
 \hline
 3.3 &$ 33.4$ & $\pm\  5.1$ & $^{+ 6.9}_{- 4.2}$ \\
 6.6 &$ 8.04$ & $\pm\ 0.89$ & $^{+0.93}_{-1.04}$ \\
11.9 &$ 2.66$ & $\pm\ 0.51$ & $^{+0.31}_{-0.33}$ \\
18.6 &$0.779$ & $\pm\ 0.216$ & $^{+0.115}_{-0.141}$ \\
31.3 &$0.273$ & $\pm\ 0.090$ & $^{+0.034}_{-0.038}$ \\
 \hline
 \end{tabular}
 \caption{
  \qsq\ dependence of the \gstarp\ cross section for proton                    
  dissociative \rh\                                                            
  meson production for $W = 75~\gev$.                                          
  The overall normalisation uncertainty of $5.3\%$ is not included in the              
  systematic errors.                                                           
 }
 \label{table:phipd_xsq2}
 \end{center}
 \end{table}

 \begin{table}[H]
 \begin{center}
 \begin{tabular}{|d|lll||d|lll|} 
 \hline
 \multicolumn{1}{|c|}{\qsq\ (\gevsq)} & \multicolumn{3}{c||}
 {$\sigma (\gspel) / \sigma (\gsrel) $ } &  
 \multicolumn{1}{c|}{$\qsq+M^2$ (\gevsq)} & \multicolumn{3}{c|}
 {$\sigma (\gspel) / \sigma (\gsrel) $ } \\ 
 \hline
   2.9 & $0.148 $ & $\pm\ 0.012$ & $^{+0.011}_{-0.009}$ &  3.94 & $0.186$ & $\pm\ 0.015$ & $^{+0.014}_{-0.012}$ \\
   4.1 & $0.171 $ & $\pm\ 0.014$ & $^{+0.009}_{-0.007}$ &  5.14 & $0.205$ & $\pm\ 0.017$ & $^{+0.011}_{-0.008}$ \\
   6.6 & $0.169 $ & $\pm\ 0.011$ & $^{+0.007}_{-0.005}$ &  7.64 & $0.192$ & $\pm\ 0.013$ & $^{+0.008}_{-0.005}$ \\
  11.9 & $0.171 $ & $\pm\ 0.018$ & $^{+0.007}_{-0.005}$ & 12.94 & $0.185$ & $\pm\ 0.019$ & $^{+0.007}_{-0.005}$ \\
  18.6 & $0.146 $ & $\pm\ 0.025$ & $^{+0.009}_{-0.006}$ & 19.64 & $0.154$ & $\pm\ 0.027$ & $^{+0.010}_{-0.007}$ \\
  31.3 & $0.187 $ & $\pm\ 0.047$ & $^{+0.010}_{-0.009}$ & 32.34 & $0.195$ & $\pm\ 0.049$ & $^{+0.010}_{-0.009}$ \\
 \hline
 \end{tabular}
 \caption{
  Ratio of the \ph\ to \rh\ elastic production cross sections                  
  for $W = 75~\gev$,                                                           
  as a function of \qsq\ and of \qsqplmsq.                                     
  The overall normalisation uncertainty of $4.0\%$ is not included in the              
  systematic errors.                                                           
 }
 \label{table:p2r}
 \end{center}
 \end{table}

 \begin{table}[H]
 \begin{center}
 \begin{tabular}{|d|lll|lll|} 
 \hline
 \multicolumn{1}{|c|}{\qsq\ (\gevsq)} & \multicolumn{3}{c|}{$\sigma_T (\gsrel) $ ($ {\rm nb} $)}     
 & \multicolumn{3}{c|}{$\sigma_L (\gsrel) $ ($ {\rm nb} $)} \\  
 \hline
 2.9 & $180 $ & $\pm\	11$ &	$^{+ 23}_{- 17}$ & $288 $ & $\pm\   26$ & $^{+ 29}_{- 20}$ \\
 4.1 & $78.1$ & $\pm\  5.5$ & $^{+10.2}_{- 8.2}$ & $165 $ & $\pm\   17$ & $^{+ 17}_{- 12}$ \\
 6.6 & $23.7$ & $\pm\  1.6$ & $^{+ 3.1}_{- 2.1}$ & $74.7$ & $\pm\  6.9$ & $^{+ 5.6}_{- 3.2}$ \\
11.9 & $ 5.0$ & $\pm\  0.6$ & $^{+ 0.7}_{- 0.5}$ & $24.0$ & $\pm\  3.4$ & $^{+ 2.3}_{- 1.6}$ \\
18.6 & $1.49$ & $\pm\ 0.25$ & $^{+0.30}_{-0.27}$ & $ 7.7$ & $\pm\  1.8$ & $^{+ 0.9}_{- 0.7}$ \\
31.3 & $0.27$ & $\pm\ 0.06$ & $^{+0.05}_{-0.05}$ & $1.76$ & $\pm\ 0.49$ & $^{+0.20}_{-0.17}$ \\
 \hline
 \end{tabular}
 \caption{
  \qsq\ dependence of the transverse and longitudinal \gstarp\                 
  cross sections for elastic \rh\ meson production with $W = 75~\gev$.         
  The overall normalisation uncertainty of $3.9\%$ is not included in the              
  systematic errors.                                                           
 }
 \label{table:rhoel_xslt}
 \end{center}
 \end{table}

 \begin{table}[H]
 \begin{center}
 \begin{tabular}{|d|lll|lll|} 
 \hline
 \multicolumn{1}{|c|}{\qsq\ (\gevsq)} & \multicolumn{3}{c|}{$\sigma_T (\gspel) $ ($ {\rm nb} $)}     
 & \multicolumn{3}{c|}{$\sigma_L (\gspel) $ ($ {\rm nb} $)} \\  
 \hline
 3.3 & $19.6$ & $\pm\  2.4$ & $^{+ 2.9}_{- 2.1}$ & $38.1$ & $\pm\  7.3$ & $^{+ 4.1}_{- 2.9}$ \\
 6.6 & $ 3.8$ & $\pm\  0.5$ & $^{+ 0.6}_{- 0.5}$ & $12.8$ & $\pm\  2.6$ & $^{+ 1.1}_{- 0.6}$ \\
15.8 & $0.34$ & $\pm\ 0.10$ & $^{+0.07}_{-0.06}$ & $ 2.2$ & $\pm\  1.0$ & $^{+ 0.3}_{- 0.2}$ \\
 \hline
 \end{tabular}
 \caption{
  \qsq\ dependence of the transverse and longitudinal \gstarp\                 
  cross sections for elastic \ph\ meson production with $W = 75~\gev$.         
  The overall normalisation uncertainty of $4.7\%$ is not included in the              
  systematic errors.                                                           
 }
 \label{table:phiel_xslt}
 \end{center}
 \end{table}

 \begin{table}[H]
 \begin{center}
 \begin{tabular}{|d|c|lll|} 
 \hline
 \multicolumn{1}{|c|}{\qsq\ (\gevsq)} & $W$ (\mbox{${\rm GeV}$}) & \multicolumn{3}{c|}{$\sigma (\gsrel) $ ($ {\rm nb} $)} \\ 
 \hline
\hline
 3.3 &   $\phantom{0}41$ & $308$ & $\pm\  14$ & $^{+ 30}_{- 16}$ \\
 3.3 &   $\phantom{0}54$ & $294$ & $\pm\  16$ & $^{+ 26}_{- 17}$ \\
 3.3 &   $\phantom{0}67$ & $346$ & $\pm\  19$ & $^{+ 30}_{- 21}$ \\
 3.3 &   $\phantom{0}80$ & $416$ & $\pm\  23$ & $^{+ 40}_{- 27}$ \\
 3.3 &   $\phantom{0}93$ & $397$ & $\pm\  27$ & $^{+ 40}_{- 25}$ \\
\hline
 6.6 &  $ \phantom{0}48$ & $72.6$ & $\pm\  4.3$ & $^{+ 7.8}_{- 4.2}$ \\
 6.6 &  $ \phantom{0}64$ & $97.2$ & $\pm\  5.9$ & $^{+ 9.4}_{- 5.5}$ \\
 6.6 &  $ \phantom{0}80$ & $99.3$ & $\pm\  6.2$ & $^{+ 7.2}_{- 5.0}$ \\
 6.6 &  $ \phantom{0}96$ & $120 $ & $\pm\   9 $ & $^{+ 10}_{-  6}$ \\
 6.6 &  $114$ & $115 $ & $\pm\  10 $ & $^{+ 10}_{-  6}$ \\
\hline
11.9 &  $ \phantom{0}59$ & $25.3$ & $\pm\  2.6$ & $^{+ 2.4}_{- 1.3}$ \\
11.9 &  $ \phantom{0}77$ & $33.1$ & $\pm\  2.7$ & $^{+ 2.4}_{- 1.5}$ \\
11.9 &  $ \phantom{0}95$ & $32.1$ & $\pm\  3.5$ & $^{+ 3.0}_{- 1.8}$ \\
11.9 &  $113$ & $27.5$ & $\pm\  4.1$ & $^{+ 3.5}_{- 1.9}$ \\
11.9 &  $131$ & $34.9$ & $\pm\  3.2$ & $^{+ 3.1}_{- 2.3}$ \\
\hline
19.5 &  $ \phantom{0}61$ & $ 6.5$ & $\pm\  0.6$ & $^{+ 0.6}_{- 0.4}$ \\
19.5 &  $ \phantom{0}83$ & $ 9.6$ & $\pm\  0.9$ & $^{+ 0.7}_{- 0.5}$ \\
19.5 &  $105$ & $ 9.3$ & $\pm\  1.0$ & $^{+ 0.8}_{- 0.5}$ \\
19.5 &  $127$ & $ 9.8$ & $\pm\  1.1$ & $^{+ 0.8}_{- 0.6}$ \\
19.5 &  $149$ & $16.9$ & $\pm\  1.8$ & $^{+ 1.4}_{- 1.0}$ \\
\hline
35.6 &  $ \phantom{0}71$ &  $1.2$ & $\pm\  0.2$ & $^{+ 0.1}_{- 0.1}$ \\
35.6 &  $ \phantom{0}97$ &  $2.0$ & $\pm\  0.4$ & $^{+ 0.2}_{- 0.1}$ \\
35.6 &  $116$ &  $2.3$ & $\pm\  0.5$ & $^{+ 0.2}_{- 0.1}$ \\
35.6 &  $139$ &  $3.4$ & $\pm\  0.6$ & $^{+ 0.3}_{- 0.2}$ \\
35.6 &  $165$ &  $2.9$ & $\pm\  0.7$ & $^{+ 0.3}_{- 0.2}$ \\
 \hline
 \end{tabular}
 \caption{
  $W$ dependence of the \gstarp\ cross section for elastic \rh\ meson          
  production, for several values of \qsq.                                      
  The overall normalisation uncertainty of $3.9\%$ is not included in the              
  systematic errors.                                                           
 }
 \label{table:rhoel_xsq2w}
 \end{center}
 \end{table}

 \begin{table}[H]
 \begin{center}
 \begin{tabular}{|d|c|lll|} 
 \hline
 \multicolumn{1}{|c|}{\qsq\ (\gevsq)} & $W$ (\mbox{${\rm GeV}$}) & \multicolumn{3}{c|}{$\sigma (\gsrpd) $ ($ {\rm nb} $)} \\ 
 \hline
\hline
 3.3 &  $ \phantom{0}41$ & $177$ & $\pm\  17$ & $^{+ 25}_{- 14}$ \\
 3.3 &  $ \phantom{0}54$ & $209$ & $\pm\  20$ & $^{+ 23}_{- 17}$ \\
 3.3 &  $ \phantom{0}67$ & $213$ & $\pm\  21$ & $^{+ 27}_{- 20}$ \\
 3.3 &  $ \phantom{0}80$ & $228$ & $\pm\  26$ & $^{+ 29}_{- 22}$ \\
 3.3 &  $ \phantom{0}93$ & $226$ & $\pm\  33$ & $^{+ 29}_{- 25}$ \\
\hline
 7.5 &  $ \phantom{0}48$ & $36.6$ & $\pm\  2.8$ & $^{+ 4.9}_{- 3.0}$ \\
 7.5 &  $ \phantom{0}64$ & $38.8$ & $\pm\  4.2$ & $^{+ 5.1}_{- 3.6}$ \\
 7.5 &  $ \phantom{0}80$ & $34.9$ & $\pm\  3.7$ & $^{+ 4.3}_{- 3.4}$ \\
 7.5 &  $ \phantom{0}96$ & $40.2$ & $\pm\  4.3$ & $^{+ 4.4}_{- 4.2}$ \\
 7.5 &  $114$ & $46.2$ & $\pm\  5.2$ & $^{+ 5.0}_{- 4.5}$ \\
\hline
22.5 &  $ \phantom{0}71$ &  $3.0$ & $\pm\  0.4$ & $^{+ 0.3}_{- 0.3}$ \\
22.5 &  $ \phantom{0}97$ &  $2.5$ & $\pm\  0.5$ & $^{+ 0.4}_{- 0.4}$ \\
22.5 &  $116$ &  $3.6$ & $\pm\  0.7$ & $^{+ 0.4}_{- 0.5}$ \\
22.5 &  $139$ &  $4.6$ & $\pm\  0.8$ & $^{+ 0.5}_{- 0.6}$ \\
22.5 &  $165$ &  $4.5$ & $\pm\  1.0$ & $^{+ 0.5}_{- 0.7}$ \\
 \hline
 \end{tabular}
 \caption{
  $W$ dependence of the \gstarp\ cross section for proton dissociative \rh\    
  meson production, for several values of \qsq.                                
  The overall normalisation uncertainty of $4.6\%$ is not included in the              
  systematic errors.                                                           
 }
 \label{table:rhopd_xsq2w}
 \end{center}
 \end{table}

 \begin{table}[H]
 \begin{center}
 \begin{tabular}{|d|c|lll|} 
 \hline
 \multicolumn{1}{|c|}{\qsq\ (\gevsq)} & $W$ (\mbox{${\rm GeV}$}) & \multicolumn{3}{c|}{$\sigma (\gspel) $ ($ {\rm nb} $)} \\ 
 \hline
\hline
 3.3 &  $ \phantom{0}41$ & $41.2$ & $\pm\  4.3$ & $^{+ 5.3}_{- 2.9}$ \\
 3.3 &  $ \phantom{0}54$ & $55.1$ & $\pm\  5.7$ & $^{+ 5.4}_{- 3.7}$ \\
 3.3 &  $ \phantom{0}67$ & $49.2$ & $\pm\  6.9$ & $^{+ 6.2}_{- 4.4}$ \\
 3.3 &  $ \phantom{0}80$ & $57.5$ & $\pm\  7.4$ & $^{+ 6.8}_{- 4.6}$ \\
 3.3 &  $ \phantom{0}93$ & $69.6$ & $\pm\  8.4$ & $^{+ 7.5}_{- 5.2}$ \\
\hline
 6.6 &  $ \phantom{0}48$ & $13.2$ & $\pm\  1.5$ & $^{+ 1.5}_{- 0.9}$ \\
 6.6 &  $ \phantom{0}64$ & $13.0$ & $\pm\  1.7$ & $^{+ 1.5}_{- 1.0}$ \\
 6.6 &  $ \phantom{0}80$ & $20.5$ & $\pm\  2.3$ & $^{+ 1.9}_{- 1.1}$ \\
 6.6 &  $ \phantom{0}96$ & $14.7$ & $\pm\  2.4$ & $^{+ 1.7}_{- 1.0}$ \\
 6.6 &  $114$ & $23.3$ & $\pm\  4.0$ & $^{+ 2.2}_{- 1.6}$ \\
\hline
15.8 &  $ \phantom{0}71$ & $ 2.3$ & $\pm\  0.3$ & $^{+ 0.2}_{- 0.1}$ \\
15.8 &  $ \phantom{0}97$ & $ 2.5$ & $\pm\  0.4$ & $^{+ 0.2}_{- 0.2}$ \\
15.8 &  $116$ & $ 3.9$ & $\pm\  0.6$ & $^{+ 0.3}_{- 0.2}$ \\
15.8 &  $139$ & $ 4.4$ & $\pm\  0.8$ & $^{+ 0.5}_{- 0.3}$ \\
15.8 &  $165$ & $ 7.6$ & $\pm\  3.2$ & $^{+ 0.6}_{- 0.4}$ \\
 \hline
 \end{tabular}
 \caption{
  $W$ dependence of the \gstarp\ cross section for elastic \ph\ meson          
  production for several values of \qsq.                                       
  The overall normalisation uncertainty of $4.7\%$ is not included in the              
  systematic errors.                                                           
 }
 \label{table:phiel_xsq2w}
 \end{center}
 \end{table}

 \begin{table}[H]
 \begin{center}
 \begin{tabular}{|d|c|lll|} 
 \hline
 \multicolumn{1}{|c|}{\qsq\ (\gevsq)} & $W$ (\mbox{${\rm GeV}$}) & \multicolumn{3}{c|}{$\sigma (\gsppd) $ ($ {\rm nb} $)} \\ 
 \hline
 5.0 &  $ \phantom{0}50$ & $ 9.3$ & $\pm\  1.6$ & $^{+ 1.6}_{- 1.2}$ \\
 5.0 &  $ \phantom{0}70$ & $17.4$ & $\pm\  2.3$ & $^{+ 2.2}_{- 2.3}$ \\
 5.0 &  $ \phantom{0}90$ & $15.2$ & $\pm\  2.5$ & $^{+ 1.9}_{- 1.9}$ \\
 5.0 &  $110$ & $11.6$ & $\pm\  2.9$ & $^{+ 1.7}_{- 2.0}$ \\
 5.0 &  $130$ & $32.7$ & $\pm\ 11.7$ & $^{+ 6.3}_{- 6.1}$ \\
 5.0 &  $150$ & $22.4$ & $\pm\ 13.7$ & $^{+ 6.1}_{- 6.5}$ \\
 \hline
 \end{tabular}
 \caption{
  $W$ dependence of the \gstarp\ cross section for proton dissociative         
  \ph\ meson production for $\qsq = 5$ \gevsq.                                 
  The overall normalisation uncertainty of $5.3\%$ is not included in the              
  systematic errors.                                                           
 }
 \label{table:phipd_xsq2w}
 \end{center}
 \end{table}

  \begin{table}[H]
  \begin{center}
 \begin{tabular}{|d|lll|lll|} 
 \hline
 \multicolumn{1}{|c|}{\qsq\ (\gevsq)} &\multicolumn{3}{c|}{$\delta$} &
 \multicolumn{3}{c|}{$\alpha_{\pom}(0)$} \\
 \hline
\hline
 \multicolumn{7}{|c|}{$\gsrel$}\\
 \hline
 3.3 & $0.40$ & $\pm\ 0.08$ & $^{+0.06}_{-0.06}$ & $1.10$ & $\pm\ 0.02$ & $^{+0.02}_{-0.02}$ \\
 6.6 & $0.57$ & $\pm\ 0.10$ & $^{+0.05}_{-0.07}$ & $1.14$ & $\pm\ 0.02$ & $^{+0.01}_{-0.02}$ \\
11.9 & $0.28$ & $\pm\ 0.15$ & $^{+0.05}_{-0.05}$ & $1.07$ & $\pm\ 0.04$ & $^{+0.01}_{-0.01}$ \\
19.5 & $0.77$ & $\pm\ 0.15$ & $^{+0.05}_{-0.05}$ & $1.19$ & $\pm\ 0.04$ & $^{+0.01}_{-0.01}$ \\
35.6 & $1.17$ & $\pm\ 0.26$ & $^{+0.04}_{-0.04}$ & $1.29$ & $\pm\ 0.07$ & $^{+0.01}_{-0.01}$ \\
 \hline
 \hline 
 \multicolumn{7}{|c|}{$\gspel$}\\
 \hline
 3.3 & $0.53$ & $\pm\ 0.17$ & $^{+0.09}_{-0.09}$ & $1.13$ & $\pm\ 0.04$ & $^{+0.02}_{-0.02}$ \\
 6.6 & $0.52$ & $\pm\ 0.21$ & $^{+0.07}_{-0.08}$ & $1.13$ & $\pm\ 0.05$ & $^{+0.02}_{-0.02}$ \\
15.8 & $1.09$ & $\pm\ 0.34$ & $^{+0.08}_{-0.08}$ & $1.27$ & $\pm\ 0.08$ & $^{+0.02}_{-0.02}$ \\
 \hline
 \hline 
 \multicolumn{7}{|c|}{$\gsrpd$}\\
 \hline
 3.3 & $0.32$ & $\pm\ 0.17$ & $^{+0.08}_{-0.09}$ & $1.17$ & $\pm\ 0.04$ & $^{+0.04}_{-0.04}$ \\
 7.5 & $0.17$ & $\pm\ 0.14$ & $^{+0.07}_{-0.09}$ & $1.12$ & $\pm\ 0.04$ & $^{+0.07}_{-0.07}$ \\
22.5 & $0.58$ & $\pm\ 0.29$ & $^{+0.10}_{-0.13}$ & $1.23$ & $\pm\ 0.07$ & $^{+0.07}_{-0.07}$ \\
 \hline
 \hline 
 \multicolumn{7}{|c|}{$\gsppd$}\\
 \hline
 5.0 & $0.50$ & $\pm\ 0.24$ & $^{+0.16}_{-0.20}$ & $1.20$ & $\pm\ 0.06$ & $^{+0.07}_{-0.08}$ \\
 \hline
 \end{tabular}
 \caption{
  \qsq\ dependence of the parameters $\delta$ and $\alpom(0)$,                 
  for elastic and proton dissociative                                          
  \rh\ and \ph\ meson production,                                              
  computed from the $W$ dependence of the cross section using                  
  Eqs.~(\ref{eq:W-fit}-\ref{eq:traj}).                                         
  The values of $\alpom(0)$ are obtained                                       
  using the measured values of $\langle t \rangle$ 
  and the measurements of \alp\ for \rh\ production                          
  given in Table~\ref{table:shrinkage}.      
 }
 \label{table:apom}
 \end{center}
 \end{table}

 \clearpage

 \begin{table}[H]
 \begin{center}
 \begin{tabular}{|d|c| l l l |} 
 \hline
 \multicolumn{1}{|c|}{\qsq\ (\gevsq)} & $|t|$ (\gevsq) & \multicolumn{3}{c|}
 {${\rm{d}\sigma}/{\rm{d}t}~(\gsrel) $ (${\rm nb} / \gevsq$)} \\ 
 \hline
 \hline
 3.3 & $0.025$ &  $2156$ & $\pm\    82$ & $^{+  202}_{-  141}$ \\
 3.3 & $0.075$ &  $1379$ & $\pm\    73$ & $^{+  132}_{-   90}$ \\
 3.3 & $0.125$ &  $ 858$ & $\pm\    54$ & $^{+   96}_{-   65}$ \\
 3.3 & $0.175$ &  $ 665$ & $\pm\    48$ & $^{+   72}_{-   60}$ \\
 3.3 & $0.250$ &  $ 346$ & $\pm\    27$ & $^{+   46}_{-   38}$ \\
 3.3 & $0.350$ &  $ 234$ & $\pm\    25$ & $^{+   35}_{-   28}$ \\
 3.3 & $0.450$ &  $61.2$ & $\pm\  13.9$ & $^{+  18.8}_{-  16.4}$ \\
\hline
 \end{tabular}
 \caption{
  $t$ dependences of the \gstarp\ cross section for elastic \rh\ meson         
  production for several values of \qsq.                                       
  The overall normalisation uncertainty of $3.9\%$ is not included in the              
  systematic errors.                                                           
 }
 \label{table:rhoel_xsq2t}
 \end{center}
 \end{table}

 \begin{table}[H]
 \begin{center}
 \begin{tabular}{|d|c|lll|} 
 \hline
 \multicolumn{1}{|c|}{\qsq\ (\gevsq)} & $|t|$ (\gevsq) & \multicolumn{3}{c|}
 {${\rm{d}\sigma}/{\rm{d}t}~(\gsrpd) $ (${\rm nb} / \gevsq$)} \\ 
 \hline
\hline
 3.3 & $0.100$ &  $  379$ & $\pm\     32$ & $^{+   47}_{-   48}$ \\
 3.3 & $0.300$ &  $  214$ & $\pm\     19$ & $^{+   28}_{-   19}$ \\
 3.3 & $0.500$ &  $  149$ & $\pm\     16$ & $^{+   21}_{-   15}$ \\
 3.3 & $0.700$ &  $ 74.1$ & $\pm\   11.3$ & $^{+  11.1}_{-   8.0}$ \\
 3.3 & $0.900$ &  $ 71.8$ & $\pm\   13.7$ & $^{+  11.0}_{-   8.0}$ \\
 3.3 & $1.100$ &  $ 29.1$ & $\pm\    8.4$ & $^{+   6.0}_{-   4.3}$ \\
 3.3 & $1.300$ &  $ 30.8$ & $\pm\   12.4$ & $^{+   6.6}_{-   4.0}$ \\
\hline
 \end{tabular}
 \caption{
  $t$ dependences of the \gstarp\ cross section for proton dissociative        
  \rh\ meson production for several values of \qsq.                            
  The overall normalisation uncertainty of $4.6\%$ is not included in the              
  systematic errors.                                                           
 }
 \label{table:rhopd_xsq2t}
 \end{center}
 \end{table}

 \begin{table}[H]
 \begin{center}
 \begin{tabular}{|d|c|lll|} 
 \hline
 \multicolumn{1}{|c|}{\qsq\ (\gevsq)} & $|t|$ (\gevsq) & \multicolumn{3}{c|}
 {${\rm{d}\sigma}/{\rm{d}t}~(\gspel) $ (${\rm nb} / \gevsq$)} \\ 
 \hline
\hline
 3.3 & $0.025$ &   $ 431$ & $\pm\    34$ & $^{+   40}_{-   25}$ \\
 3.3 & $0.075$ &   $ 209$ & $\pm\    24$ & $^{+   24}_{-   22}$ \\
 3.3 & $0.125$ &   $ 120$ & $\pm\    21$ & $^{+   24}_{-   16}$ \\
 3.3 & $0.175$ &   $85.4$ & $\pm\  17.1$ & $^{+  15.2}_{-   9.6}$ \\
 3.3 & $0.250$ &   $64.6$ & $\pm\  10.5$ & $^{+   9.0}_{-   7.4}$ \\
 3.3 & $0.350$ &   $27.5$ & $\pm\   7.4$ & $^{+   5.8}_{-   3.6}$ \\
 3.3 & $0.450$ &   $27.4$ & $\pm\   7.1$ & $^{+   5.4}_{-   4.5}$ \\
\hline
 6.6 & $0.025$ &  $ 93.1$ & $\pm\   10.4$ & $^{+   7.8}_{-   3.9}$ \\
 6.6 & $0.075$ &  $ 77.7$ & $\pm\    9.1$ & $^{+   6.8}_{-   3.4}$ \\
 6.6 & $0.125$ &  $ 34.1$ & $\pm\    6.1$ & $^{+   3.1}_{-   2.6}$ \\
 6.6 & $0.175$ &  $ 24.9$ & $\pm\    5.0$ & $^{+   2.8}_{-   1.7}$ \\
 6.6 & $0.250$ &  $ 21.2$ & $\pm\    3.8$ & $^{+   2.7}_{-   2.7}$ \\
 6.6 & $0.350$ &  $ 8.77$ & $\pm\   2.37$ & $^{+  1.54}_{-  1.00}$ \\
 6.6 & $0.450$ &  $ 6.41$ & $\pm\   2.12$ & $^{+  1.29}_{-  1.26}$ \\
\hline
15.8 & $0.025$ &  $ 8.24$ & $\pm\   1.14$ & $^{+  1.09}_{-  0.89}$ \\
15.8 & $0.075$ &  $ 10.7$ & $\pm\    1.4$ & $^{+   1.3}_{-   1.1}$ \\
15.8 & $0.125$ &  $ 3.89$ & $\pm\   0.85$ & $^{+  0.59}_{-  0.42}$ \\
15.8 & $0.175$ &  $ 3.96$ & $\pm\   0.80$ & $^{+  0.48}_{-  0.56}$ \\
15.8 & $0.250$ &  $ 2.32$ & $\pm\   0.47$ & $^{+  0.39}_{-  0.29}$ \\
15.8 & $0.350$ &  $0.702$ & $\pm\  0.296$ & $^{+ 0.188}_{- 0.140}$ \\
15.8 & $0.450$ &  $0.349$ & $\pm\  0.278$ & $^{+ 0.160}_{- 0.150}$ \\
 \hline
 \end{tabular}
 \caption{
  $t$ dependences of the \gstarp\ cross section for elastic \ph\ meson         
  production for several values of \qsq.                                       
  The overall normalisation uncertainty of $4.7\%$ is not included in the              
  systematic errors.                                                           
 }
 \label{table:phiel_xsq2t}
 \end{center}
 \end{table}

 \begin{table}[H]
 \begin{center}
 \begin{tabular}{|d|c|lll|} 
 \hline
 \multicolumn{1}{|c|}{\qsq\ (\gevsq)} & $|t|$ (\gevsq) & \multicolumn{3}{c|}
 {${\rm{d}\sigma}/{\rm{d}t}~(\gsppd) $ (${\rm nb} / \gevsq$)} \\ 
 \hline
 5.0 & $0.150$ &  $ 58.2$ & $\pm\   11.8$ & $^{+   9.0}_{-   6.1}$ \\
 5.0 & $0.500$ &  $ 23.1$ & $\pm\    5.5$ & $^{+   4.1}_{-   2.6}$ \\
 5.0 & $1.100$ &  $ 6.17$ & $\pm\   2.76$ & $^{+  2.28}_{-  1.25}$ \\
 5.0 & $2.250$ &  $0.681$ & $\pm\  0.418$ & $^{+ 0.285}_{- 0.301}$ \\
 \hline
 \end{tabular}
 \caption{
  $t$ dependence of the \gstarp\ cross section for proton dissociative         
  \ph\ meson production for $\qsq = 5$ \gevsq.                                 
  The overall normalisation uncertainty of $5.3\%$ is not included in the              
  systematic errors.                                                           
 }
 \label{table:phipd_xsq2t}
 \end{center}
 \end{table}

  \begin{table}[H]
  \begin{center}
 \begin{tabular}{|d|lll|} 
 \hline
 \multicolumn{1}{|c|}{\qsq\ (\gevsq)} &\multicolumn{3}{c|}{$b$ (\mbox{${\rm GeV}^{-2}$})} \\
 \hline
\hline
 \multicolumn{4}{|c|}{$\gsrel$}\\
 \hline
 3.3 & $7.82$ & $\pm\ 0.33$ & $^{+0.33}_{-0.33}$ \\
 6.6 & $7.57$ & $\pm\ 0.35$ & $^{+0.30}_{-0.31}$ \\
11.5 & $6.72$ & $\pm\ 0.53$ & $^{+0.23}_{-0.25}$ \\
17.4 & $5.86$ & $\pm\ 0.40$ & $^{+0.26}_{-0.33}$ \\
33.0 & $4.87$ & $\pm\ 0.66$ & $^{+0.21}_{-0.22}$ \\
 \hline
\hline
  \multicolumn{4}{|c|}{$\gspel$}\\
 \hline
 3.3 & $8.28$ & $\pm\ 0.80$ & $^{+0.49}_{-0.65}$ \\
 6.6 & $7.17$ & $\pm\ 0.73$ & $^{+0.36}_{-0.34}$ \\
15.8 & $7.08$ & $\pm\ 0.71$ & $^{+0.38}_{-0.34}$ \\
 \hline
 \hline
 \multicolumn{4}{|c|}{$\gsrpd$}\\
 \hline
 3.3 & $2.29$ & $\pm\ 0.12$ & $^{+0.12}_{-0.12}$ \\
 6.6 & $1.91$ & $\pm\ 0.26$ & $^{+0.13}_{-0.13}$ \\
15.8 & $1.70$ & $\pm\ 0.15$ & $^{+0.42}_{-0.10}$ \\
 \hline
\hline
 \multicolumn{4}{|c|}{$\gsppd$}\\
 \hline
 5.0 & 2.21 & $\pm$ 0.37 & $^{+0.48}_{-0.13}$ \\
 \hline
 \end{tabular}
 \caption{
  \qsq\ dependence of the $b$ slope parameters of the exponentially            
  falling \modt\ distributions of \rh\ and \ph\                                
  elastic and proton dissociative production.                                  
 }
 \label{table:bslope}
 \end{center}
 \end{table}

  \begin{table}[H]
  \begin{center}
 \begin{tabular}{|d|lll|} 
 \hline
 \multicolumn{1}{|c|}{\qsq\ (\gevsq)} &\multicolumn{3}{c|}{$m_{2g}$ (\mbox{${\rm GeV}$})} \\
 \hline
\hline
 \multicolumn{4}{|c|}{$\gsrel$}\\
 \hline
 3.3 & $0.59$ & $\pm\ 0.01$ & $^{+0.01}_{-0.01}$ \\
 6.6 & $0.60$ & $\pm\ 0.02$ & $^{+0.01}_{-0.01}$ \\
11.5 & $0.65$ & $\pm\ 0.03$ & $^{+0.01}_{-0.01}$ \\
17.4 & $0.71$ & $\pm\ 0.03$ & $^{+0.02}_{-0.02}$ \\
33.0 & $0.80$ & $\pm\ 0.06$ & $^{+0.02}_{-0.02}$ \\
\hline
\hline
 \multicolumn{4}{|c|}{$\gspel$}\\
 \hline
 3.3 & $0.57$ & $\pm\ 0.03$ & $^{+0.02}_{-0.02}$ \\
 6.6 & $0.63$ & $\pm\ 0.04$ & $^{+0.02}_{-0.02}$ \\
15.8 & $0.64$ & $\pm\ 0.04$ & $^{+0.02}_{-0.02}$ \\
 \hline
 \end{tabular}
 \caption{
  Parameter $m_{2g}$ of the two-gluon form factor of the FS model~\cite{FS-m2g},                                        
  extracted from fits of Eq.~(\ref{eq:m2g}) to the $t$ distributions of \rh\   
  and \ph\ elastic production cross sections.                                  
 }
 \label{table:m2g}
 \end{center}
 \end{table}

 \begin{table}[H]
 \begin{center}
 \begin{tabular}{|d|c|lll|} 
 \hline
 \multicolumn{1}{|c|}{\modt\ (\gevsq)} & $W$ (\mbox{${\rm GeV}$}) & \multicolumn{3}{c|}{$\sigma (\gsrpd) $ ($ {\rm nb} $)} \\ 
 \hline
 \hline
\multicolumn{5}{|c|}{$\qsq = 3.3$ \gevsq }\\
 \hline
 0.08 &   $45$ & $1451$ & $\pm\  51$ & $^{+124}_{- 56}$ \\
 0.08 &   $65$ & $1677$ & $\pm\  66$ & $^{+123}_{- 75}$ \\
 0.08 &   $87$ & $2030$ & $\pm\  82$ & $^{+173}_{-102}$ \\
\hline
 0.32 &  $ 45$ & $393$ & $\pm\  24$ & $^{+ 45}_{- 29}$ \\
 0.32 &  $ 65$ & $433$ & $\pm\  30$ & $^{+ 48}_{- 37}$ \\
 0.32 &  $ 87$ & $453$ & $\pm\  33$ & $^{+ 55}_{- 44}$ \\
\hline
 0.69 &   $45$ & $95.1$ & $\pm\  9.4$ & $^{+16.1}_{-13.5}$ \\
 0.69 &   $65$ & $110 $ & $\pm\  13 $ & $^{+ 20}_{- 17}$ \\
 0.69 &   $87$ & $124 $ & $\pm\  19 $ & $^{+ 22}_{- 17}$ \\
\hline
 1.45 &   $45$ & $14.5$ & $\pm\  2.5$ & $^{+ 3.2}_{- 2.7}$ \\
 1.45 &   $65$ & $ 9.0$ & $\pm\  1.8$ & $^{+ 2.7}_{- 2.4}$ \\
 1.45 &   $87$ & $ 8.3$ & $\pm\  2.1$ & $^{+ 2.3}_{- 2.2}$ \\
\hline
 \hline
 \multicolumn{5}{|c|}{$\qsq = 8.6$ \gevsq }\\
 \hline
 0.08 &  $ \phantom{0}65$ & $261$ & $\pm\  14$ & $^{+ 23}_{- 11}$ \\
 0.08 &  $ \phantom{0}79$ & $261$ & $\pm\  11$ & $^{+ 17}_{- 10}$ \\
 0.08 &  $104$ & $320$ & $\pm\  14$ & $^{+ 22}_{- 13}$ \\
\hline
 0.32 &  $ \phantom{0}65$ & $64.8$ & $\pm\  5.9$ & $^{+ 6.4}_{- 3.8}$ \\
 0.32 &  $ \phantom{0}79$ & $72.8$ & $\pm\  5.3$ & $^{+ 6.0}_{- 5.3}$ \\
 0.32 &  $104$ & $82.1$ & $\pm\  5.6$ & $^{+ 8.7}_{- 5.9}$ \\
\hline
 0.69 &  $ \phantom{0}65$ & $16.5$ & $\pm\  1.7$ & $^{+ 3.1}_{- 2.6}$ \\
 0.69 &  $ \phantom{0}79$ & $17.5$ & $\pm\  1.6$ & $^{+ 3.0}_{- 2.5}$ \\
 0.69 &  $104$ & $17.3$ & $\pm\  1.7$ & $^{+ 3.6}_{- 3.2}$ \\
\hline
 1.47 &  $ \phantom{0}65$ & $ 2.8$ & $\pm\  0.5$ & $^{+ 0.8}_{- 0.8}$ \\
 1.47 &  $ \phantom{0}79$ & $ 1.5$ & $\pm\  0.3$ & $^{+ 0.9}_{- 0.8}$ \\
 1.47 &  $104$ & $ 2.3$ & $\pm\  0.4$ & $^{+ 1.0}_{- 0.9}$ \\
 \hline
 \end{tabular}
 \caption{
$W$ dependence of the \gstarp\ cross sections for \rh\ meson production
in four bins in \modt, for $\qsq = 3.3~\gevsq$ and $\qsq = 8.6~\gevsq$.
The notag ($\modt \leq 0.5~\gevsq$) and tag ($\modt \leq 3~\gevsq$) samples are 
combined.
The overall normalisation uncertainty of~$4\%$ is not included in the systematic errors.
 } 
 \label{table:rho_xstw}
 \end{center}
 \end{table}

 \begin{table}[H]
 \begin{center}
 \begin{tabular}{|d|lll|} 
 \hline
 \multicolumn{1}{|c|}{$t$ (\gevsq)} &\multicolumn{3}{c|}{$\alpha_{\pom}(t)$} \\
 \hline
\hline
 \multicolumn{4}{|c|}{\qsq = $3.3$ \gevsq } \\
 \hline
-0.08 & $1.12$ & $\pm\  0.02$ & $^{+0.01}_{-0.01}$ \\
-0.32 & $1.05$ & $\pm\  0.04$ & $^{+0.02}_{-0.02}$ \\
-0.69 & $1.10$ & $\pm\  0.06$ & $^{+0.03}_{-0.02}$ \\
-1.45 & $0.76$ & $\pm\  0.12$ & $^{+0.06}_{-0.07}$ \\
 \hline
\hline
 \multicolumn{4}{|c|}{\qsq = $8.6$ \gevsq } \\
 \hline
-0.08 & $1.12$ & $\pm\  0.04$ & $^{+0.01}_{-0.01}$ \\
-0.32 & $1.12$ & $\pm\  0.06$ & $^{+0.01}_{-0.01}$ \\
-0.69 & $1.02$ & $\pm\  0.07$ & $^{+0.04}_{-0.04}$ \\
-1.48 & $0.93$ & $\pm\  0.16$ & $^{+0.10}_{-0.15}$ \\
 \hline
 \end{tabular}
 \caption{
  $t$ dependence of $\alpha_{\pom}(t)$ for \rh\ meson production,             
  for two values of \qsq.                                                      
 }
 \label{table:apomt}
 \clearpage 
 \end{center}
 \end{table}

 \renewcommand{\arraystretch}{1.25}
  
 \begin{table}[H]
 \begin{center}
 \begin{tabular}{|d|lll|lll|} 
 \hline
 \multicolumn{1}{|c|}{\qsq\ (\gevsq)} & \multicolumn{3}{c|}
{$ \frac {\sigma_{tot,\rm{p.~diss.} }} {\sigma_{tot,\rm{el.}}} (\rho) $} &
 \multicolumn{3}{c|}
{$ \frac { {\rm d}\sigma_{{\rm p.~diss.}}  / {\rm d}t } 
       { {\rm d}\sigma_{{\rm el.}} / {\rm d}t } 
      {\rm (t = 0)} (\rho) $} \\
 \hline
 2.9 & $0.58$ & $\pm\ 0.04$ & $^{+0.12}_{-0.08}$ & $0.169$ & $\pm\ 0.017$ & $^{+0.011}_{-0.015}$ \\
 4.1 & $0.65$ & $\pm\ 0.05$ & $^{+0.07}_{-0.06}$ & $0.191$ & $\pm\ 0.019$ & $^{+0.014}_{-0.016}$ \\
 6.6 & $0.53$ & $\pm\ 0.03$ & $^{+0.02}_{-0.06}$ & $0.133$ & $\pm\ 0.021$ & $^{+0.009}_{-0.011}$ \\
11.9 & $0.58$ & $\pm\ 0.05$ & $^{+0.09}_{-0.07}$ & $0.147$ & $\pm\ 0.022$ & $^{+0.041}_{-0.016}$ \\
18.6 & $0.45$ & $\pm\ 0.05$ & $^{+0.04}_{-0.05}$ & $0.131$ & $\pm\ 0.021$ & $^{+0.042}_{-0.019}$ \\
31.3 & $0.57$ & $\pm\ 0.09$ & $^{+0.04}_{-0.06}$ & $0.198$ & $\pm\ 0.044$ & $^{+0.062}_{-0.030}$ \\
 \hline
 \end{tabular}
 \caption{
  \qsq\ dependences,  for                          
  $W = 75~\gev$, of the ratios of proton dissociative ($M_Y < 5~\gevcsq$)     
  to elastic \rh\ meson production cross sections integrated over $t$ and for
  $t = 0$.                                                               
  The overall normalisation uncertainty of $2.4\%$ is not included in the              
  systematic errors.                                                           
 }
 \label{table:rho_pdel}
 \end{center}
 \end{table}

 \begin{table}[H]
 \begin{center}
 \begin{tabular}{|d|lll|lll|} 
 \hline
 \multicolumn{1}{|c|}{\qsq\ (\gevsq)} & \multicolumn{3}{c|}
{$ \frac {\sigma_{tot,\rm{p.~diss. }}} {\sigma_{tot,\rm{el.}}} (\phi) $} &
 \multicolumn{3}{c|}
{$ \frac { {\rm d}\sigma_{{\rm p.~diss.}}  / {\rm d}t } 
       { {\rm d}\sigma_{{\rm el.}} / {\rm d}t } 
      {\rm (t = 0)} (\phi) $} \\
 \hline
 3.3 & $0.58$ & $\pm\  0.09$ & $^{+0.09}_{-0.04}$ & $0.155$ & $\pm\ 0.039$ & $^{+0.037}_{-0.006}$ \\
 6.6 & $0.48$ & $\pm\  0.06$ & $^{+0.07}_{-0.10}$ & $0.148$ & $\pm\ 0.034$ & $^{+0.034}_{-0.015}$ \\
15.8 & $0.47$ & $\pm\  0.08$ & $^{+0.02}_{-0.07}$ & $0.146$ & $\pm\ 0.038$ & $^{+0.034}_{-0.015}$ \\
 \hline
 \end{tabular}
 \caption{
  \qsq\ dependences,  for                          
  $W = 75~\gev$, of the ratios of proton dissociative ($M_Y < 5~\gevcsq$)     
  to elastic \ph\ meson production cross sections integrated over $t$ and for
  $t = 0$.                                                               
  The overall normalisation uncertainty of $2.4\%$ is not included in the              
  systematic errors.                                                           
 }
 \label{table:phi_pdel}
 \end{center}
 \end{table} 
  
  \renewcommand{\arraystretch}{1.15}

 \begin{table}[H]
 \begin{center}
 \begin{tabular}{|d|lll|} 
 \hline
 \multicolumn{1}{|c|}{\qsq\ (\gevsq)} & \multicolumn{3}{c|}{ $b_{el.} - b_{p.~diss.}$ (\mbox{${\rm GeV}^{-2}$})} \\ 
 \hline
 \hline
 \multicolumn{4}{|c|}{$\rho$} \\
 \hline
 3.3 & $5.52$ & $\pm\ 0.40$ & $^{+0.26}_{-0.26}$ \\
 6.6 & $5.74$ & $\pm\ 0.62$ & $^{+0.22}_{-0.25}$ \\
15.8 & $4.76$ & $\pm\ 0.48$ & $^{+0.19}_{-0.65}$ \\
 \hline
 \hline
 \multicolumn{4}{|c|}{$\phi$} \\
 \hline
 5.0 & $5.81$ & $\pm\ 1.15$ & $^{+0.16}_{-0.70}$ \\
 \hline
 \end{tabular}
 \caption{
Slope differences $b_{el.} - b_{p.~diss.}$ between elastic and proton dissociative 
scattering for \rh\ and \ph\ meson production as a function of \qsq.     
 }
 \label{table:rho_pdel_bslp}
 \end{center}
 \end{table}

  \renewcommand{\arraystretch}{1.25}

  \clearpage
  \begin{sidewaystable}[p]
  \begin{center}
 \begin{tabular}{|c|r|r|r|r|r|}
 \hline
 $\langle\qsq\rangle$ (\gevsq) &
\multicolumn{1}{c|}{$ 3.1$}
  &
\multicolumn{1}{c|}{$ 4.8$}
  &
\multicolumn{1}{c|}{$ 7.5$}
  &
\multicolumn{1}{c|}{$12.2$}
  &
\multicolumn{1}{c|}{$23.1$}
  \\
 \hline
 \rzqzz   &
 $0.597$ $\pm\ 0.013$ $^{+ 0.014}_{- 0.034}$ 
   & 
 $0.680$ $\pm\ 0.017$ $^{+ 0.017}_{- 0.028}$ 
   & 
 $0.789$ $\pm\ 0.013$ $^{+ 0.018}_{- 0.029}$ 
   & 
 $0.793$ $\pm\ 0.016$ $^{+ 0.018}_{- 0.023}$ 
   & 
 $0.877$ $\pm\ 0.019$ $^{+ 0.019}_{- 0.020}$ 
  \\
 Re \rzquz&
 $0.049$ $\pm\ 0.009$ $^{+ 0.010}_{- 0.010}$ 
   & 
 $0.019$ $\pm\ 0.012$ $^{+ 0.012}_{- 0.013}$ 
   & 
 $0.034$ $\pm\ 0.010$ $^{+ 0.011}_{- 0.010}$ 
   & 
$-0.011$ $\pm\ 0.012$ $^{+ 0.009}_{- 0.009}$ 
   & 
 $0.046$ $\pm\ 0.014$ $^{+ 0.008}_{- 0.009}$ 
  \\
 \rzqumu  &
$ 0.000$ $\pm\ 0.011$ $^{+ 0.004}_{- 0.004}$ 
   & 
$ 0.011$ $\pm\ 0.013$ $^{+ 0.002}_{- 0.003}$ 
   & 
$-0.003$ $\pm\ 0.010$ $^{+ 0.002}_{- 0.002}$ 
   & 
$-0.031$ $\pm\ 0.013$ $^{+ 0.003}_{- 0.004}$ 
   & 
$ 0.012$ $\pm\ 0.014$ $^{+ 0.002}_{- 0.004}$ 
  \\
 \ruzz    &
$-0.001$ $\pm\ 0.029$ $^{+ 0.035}_{- 0.012}$ 
   & 
$ 0.021$ $\pm\ 0.024$ $^{+ 0.015}_{- 0.028}$ 
   & 
$ 0.001$ $\pm\ 0.039$ $^{+ 0.020}_{- 0.013}$ 
   & 
$-0.081$ $\pm\ 0.095$ $^{+ 0.016}_{- 0.018}$ 
   & 
$-0.015$ $\pm\ 0.061$ $^{+ 0.026}_{- 0.011}$ 
  \\
 \ruuu    &
$-0.019$ $\pm\ 0.031$ $^{+ 0.006}_{- 0.021}$ 
   & 
$-0.034$ $\pm\ 0.036$ $^{+ 0.010}_{- 0.005}$ 
   & 
$-0.028$ $\pm\ 0.044$ $^{+ 0.011}_{- 0.011}$ 
   & 
$ 0.027$ $\pm\ 0.051$ $^{+ 0.015}_{- 0.013}$ 
   & 
$ 0.058$ $\pm\ 0.077$ $^{+ 0.014}_{- 0.018}$ 
  \\
 Re \ruuz &
$-0.029$ $\pm\ 0.013$ $^{+ 0.014}_{- 0.010}$ 
   & 
$-0.043$ $\pm\ 0.017$ $^{+ 0.011}_{- 0.012}$ 
   & 
$-0.007$ $\pm\ 0.014$ $^{+ 0.013}_{- 0.010}$ 
   & 
$-0.019$ $\pm\ 0.017$ $^{+ 0.010}_{- 0.009}$ 
   & 
$ 0.023$ $\pm\ 0.021$ $^{+ 0.013}_{- 0.011}$ 
  \\
\hline
 \end{tabular}
 \caption{
  Spin density matrix elements for the diffractive electroproduction of \rh\   
  mesons, as a function of \qsq.                                               
  The notag ($\modt \leq  0.5~\gevsq$) and tag ($\modt \leq  3~\gevsq$)        
  samples are combined.                                                        
 }
\label{table:matelem_f_qsq_rho}
 \end{center}
 \end{sidewaystable}

  \clearpage
  \begin{table}[p]
  \begin{center}
 \begin{tabular}{|c|r|r|r|}
 \hline
 $\langle \qsq \rangle $ (\gevsq) &
\multicolumn{1}{c|}{$ 3.3$}
  &
\multicolumn{1}{c|}{$ 6.6$}
  &
\multicolumn{1}{c|}{$15.8$}
  \\
 \hline
 \rzqzz   &
$ 0.581$ $\pm\ 0.023$ $^{+ 0.014}_{- 0.015}$ 
   & 
$ 0.746$ $\pm\ 0.024$ $^{+ 0.017}_{- 0.016}$ 
   & 
$ 0.864$ $\pm\ 0.031$ $^{+ 0.019}_{- 0.016}$ 
   \\
  Re \rzquz&
$-0.004$ $\pm\ 0.017$ $^{+ 0.018}_{- 0.020}$ 
   & 
$ 0.011$ $\pm\ 0.018$ $^{+ 0.012}_{- 0.013}$ 
   & 
$ 0.007$ $\pm\ 0.024$ $^{+ 0.014}_{- 0.011}$ 
   \\
  \rzqumu  &
$-0.059$ $\pm\ 0.020$ $^{+ 0.008}_{- 0.004}$ 
   & 
$ 0.047$ $\pm\ 0.020$ $^{+ 0.004}_{- 0.003}$ 
   & 
$-0.020$ $\pm\ 0.025$ $^{+ 0.006}_{- 0.002}$ 
   \\
  \ruzz     &
$-0.060$ $\pm\ 0.174$ $^{+ 0.011}_{- 0.013}$ 
   & 
$-0.049$ $\pm\ 0.070$ $^{+ 0.018}_{- 0.018}$ 
   & 
$-0.008$ $\pm\ 0.018$ $^{+ 0.020}_{- 0.026}$ 
  \\
 \ruuu    &
$-0.059$ $\pm\ 0.153$ $^{+ 0.008}_{- 0.010}$ 
   & 
$ 0.006$ $\pm\ 0.044$ $^{+ 0.012}_{- 0.009}$ 
   & 
$-0.004$ $\pm\ 0.014$ $^{+ 0.015}_{- 0.011}$ 
   \\
  Re \ruuz &
$-0.044$ $\pm\ 0.023$ $^{+ 0.021}_{- 0.015}$ 
   & 
$-0.073$ $\pm\ 0.025$ $^{+ 0.017}_{- 0.013}$ 
   & 
$-0.028$ $\pm\ 0.034$ $^{+ 0.013}_{- 0.018}$ 
   \\
  \ruumu   &
$ 0.220$ $\pm\ 0.027$ $^{+ 0.018}_{- 0.013}$ 
   & 
$ 0.104$ $\pm\ 0.029$ $^{+ 0.009}_{- 0.007}$ 
   & 
$ 0.058$ $\pm\ 0.036$ $^{+ 0.010}_{- 0.017}$ 
   \\
\hline
 \end{tabular}
 \caption{
  Spin density matrix elements for the diffractive electroproduction of \ph\   
  mesons, as a function of \qsq.                                               
  The notag ($\modt \leq  0.5~\gevsq$) and tag ($\modt \leq  3~\gevsq$)        
  samples are combined.                                                        
 }
\label{table:matelem_f_qsq_phi}
 \end{center}
 \end{table}

  \clearpage
  \begin{sidewaystable}[p]
  \begin{center}
 \begin{tabular}{|c|r|r|r|r|}
 \hline
 \multicolumn{5}{|c|}{$\langle\qsq\rangle$ = 3.3 \gevsq } \\
 \hline
 $\langle W \rangle$ (\mbox{${\rm GeV}$}) &
\multicolumn{1}{c|}{$ 43$}
  &
\multicolumn{1}{c|}{$ 59$}
  &
\multicolumn{1}{c|}{$ 76$}
  &
\multicolumn{1}{c|}{$ 92$}
  \\
 \hline
 \rzqzz   &
$ 0.619$ $\pm\ 0.021$ $^{+ 0.012}_{- 0.048}$ 
   & 
$ 0.600$ $\pm\ 0.021$ $^{+ 0.012}_{- 0.020}$ 
   & 
$ 0.612$ $\pm\ 0.023$ $^{+ 0.014}_{- 0.029}$ 
   & 
$ 0.580$ $\pm\ 0.028$ $^{+ 0.014}_{- 0.027}$ 
   \\
  Re \rzquz&  
$ 0.580$ $\pm\ 0.028$ $^{+ 0.014}_{- 0.027}$ 
   & 
$ 0.729$ $\pm\ 0.019$ $^{+ 0.016}_{- 0.020}$ 
   & 
$ 0.755$ $\pm\ 0.019$ $^{+ 0.020}_{- 0.024}$ 
   & 
$ 0.795$ $\pm\ 0.021$ $^{+ 0.023}_{- 0.045}$ 
   \\
  \rzqumu  &  
$ 0.795$ $\pm\ 0.021$ $^{+ 0.023}_{- 0.045}$ 
   & 
$ 0.731$ $\pm\ 0.021$ $^{+ 0.021}_{- 0.021}$ 
   & 
$ 0.798$ $\pm\ 0.037$ $^{+ 0.024}_{- 0.018}$ 
   & 
$ 0.878$ $\pm\ 0.036$ $^{+ 0.018}_{- 0.027}$ 
   \\
  \ruzz     & 
$ 0.878$ $\pm\ 0.036$ $^{+ 0.018}_{- 0.027}$ 
   & 
$ 0.872$ $\pm\ 0.040$ $^{+ 0.023}_{- 0.032}$ 
   & 
$ 0.856$ $\pm\ 0.045$ $^{+ 0.041}_{- 0.059}$ 
   & 
$ 0.047$ $\pm\ 0.014$ $^{+ 0.012}_{- 0.013}$ 
  \\
 \ruuu    &
$ 0.047$ $\pm\ 0.014$ $^{+ 0.012}_{- 0.013}$ 
   & 
$ 0.056$ $\pm\ 0.015$ $^{+ 0.014}_{- 0.010}$ 
   & 
$ 0.029$ $\pm\ 0.016$ $^{+ 0.010}_{- 0.010}$ 
   & 
$ 0.028$ $\pm\ 0.019$ $^{+ 0.010}_{- 0.011}$ 
   \\
  Re \ruuz &  
$ 0.028$ $\pm\ 0.019$ $^{+ 0.010}_{- 0.011}$ 
   & 
$ 0.043$ $\pm\ 0.014$ $^{+ 0.011}_{- 0.011}$ 
   & 
$ 0.028$ $\pm\ 0.014$ $^{+ 0.012}_{- 0.012}$ 
   & 
$-0.001$ $\pm\ 0.015$ $^{+ 0.012}_{- 0.010}$ 
   \\
\hline
 \end{tabular}
 \caption{
  Spin density matrix elements for the diffractive electroproduction of \rh\   
  mesons as a function of $W$, for  $2.5 \leq  Q^2 < 5$,                       
  $5 \leq  \qsq < 15.5$ and $15.5 \leq  Q^2 \leq  60~\gevsq$.                  
  The notag ($\modt \leq  0.5~\gevsq$) and tag ($\modt \leq  3~\gevsq$)        
  samples are combined.                                                        
 }
 \label{table:matelem_f_w_three_qsq_rho}
 \end{center}
 \end{sidewaystable}

 \begin{sidewaystable}[p]
 \begin{center}
 \begin{tabular}{|c|r|r|r|r|}
 \hline
 \multicolumn{5}{|c|}{$\langle\qsq\rangle$ = 7.5 \gevsq } \\
 \hline
 $\langle W \rangle$ (\mbox{${\rm GeV}$}) &
\multicolumn{1}{c|}{$ 58$}
  &
\multicolumn{1}{c|}{$ 76$}
  &
\multicolumn{1}{c|}{$ 93$}
  &
\multicolumn{1}{c|}{$111$}
  \\
 \hline
 \rzqzz   &
$ 0.729$ $\pm\ 0.019$ $^{+ 0.016}_{- 0.020}$ 
   & 
$ 0.755$ $\pm\ 0.019$ $^{+ 0.020}_{- 0.024}$ 
   & 
$ 0.795$ $\pm\ 0.021$ $^{+ 0.023}_{- 0.045}$ 
   & 
$ 0.731$ $\pm\ 0.021$ $^{+ 0.021}_{- 0.021}$ 
   \\
  Re \rzquz& 
$ 0.731$ $\pm\ 0.021$ $^{+ 0.021}_{- 0.021}$ 
   & 
$ 0.798$ $\pm\ 0.037$ $^{+ 0.024}_{- 0.018}$ 
   & 
$ 0.878$ $\pm\ 0.036$ $^{+ 0.018}_{- 0.027}$ 
   & 
$ 0.872$ $\pm\ 0.040$ $^{+ 0.023}_{- 0.032}$ 
   \\
  \rzqumu  & 
$ 0.872$ $\pm\ 0.040$ $^{+ 0.023}_{- 0.032}$ 
   & 
$ 0.856$ $\pm\ 0.045$ $^{+ 0.041}_{- 0.059}$ 
   & 
$ 0.047$ $\pm\ 0.014$ $^{+ 0.012}_{- 0.013}$ 
   & 
$ 0.056$ $\pm\ 0.015$ $^{+ 0.014}_{- 0.010}$ 
   \\
  \ruzz     & 
$ 0.056$ $\pm\ 0.015$ $^{+ 0.014}_{- 0.010}$ 
   & 
$ 0.029$ $\pm\ 0.016$ $^{+ 0.010}_{- 0.010}$ 
   & 
$ 0.028$ $\pm\ 0.019$ $^{+ 0.010}_{- 0.011}$ 
   & 
$ 0.043$ $\pm\ 0.014$ $^{+ 0.011}_{- 0.011}$ 
   \\
  \ruuu     &
$ 0.043$ $\pm\ 0.014$ $^{+ 0.011}_{- 0.011}$ 
   & 
$ 0.028$ $\pm\ 0.014$ $^{+ 0.012}_{- 0.012}$ 
   & 
$-0.001$ $\pm\ 0.015$ $^{+ 0.012}_{- 0.010}$ 
   & 
$ 0.006$ $\pm\ 0.015$ $^{+ 0.008}_{- 0.007}$ 
   \\
  Re \ruuz & 
$ 0.006$ $\pm\ 0.015$ $^{+ 0.008}_{- 0.007}$ 
   & 
$ 0.047$ $\pm\ 0.027$ $^{+ 0.009}_{- 0.010}$ 
   & 
$ 0.047$ $\pm\ 0.029$ $^{+ 0.009}_{- 0.011}$ 
   & 
$ 0.033$ $\pm\ 0.028$ $^{+ 0.008}_{- 0.015}$ 
   \\
\hline
 \end{tabular}
 \caption{
  Spin density matrix elements for the diffractive electroproduction of \rh\   
  mesons as a function of $W$,  for  $2.5 \leq  Q^2 < 5$,                       
  $5 \leq  \qsq < 15.5$ and $15.5 \leq  Q^2 \leq  60~\gevsq$, continued from 
  Table~\ref{table:matelem_f_w_three_qsq_rho}.                                                    
 }
 \label{table:matelem_f_w_three_qsq_rho_2}
 \end{center}
 \end{sidewaystable}

 \begin{sidewaystable}[p]
 \begin{center}
 \begin{tabular}{|c|r|r|r|r|}
 \hline
 \multicolumn{5}{|c|}{$\langle\qsq\rangle$ = 22.5 \gevsq } \\
 \hline
 $\langle W \rangle$ (\mbox{${\rm GeV}$}) &
\multicolumn{1}{c|}{$ 72$}
  &
\multicolumn{1}{c|}{$ 97$}
  &
\multicolumn{1}{c|}{$122$}
  &
\multicolumn{1}{c|}{$147$}
  \\
 \hline
 \rzqzz   &
$ 0.798$ $\pm\ 0.037$ $^{+ 0.024}_{- 0.018}$ 
   & 
$ 0.878$ $\pm\ 0.036$ $^{+ 0.018}_{- 0.027}$ 
   & 
$ 0.872$ $\pm\ 0.040$ $^{+ 0.023}_{- 0.032}$ 
   & 
$ 0.856$ $\pm\ 0.045$ $^{+ 0.041}_{- 0.059}$ 
   \\
  Re \rzquz& 
$ 0.856$ $\pm\ 0.045$ $^{+ 0.041}_{- 0.059}$ 
   & 
$ 0.047$ $\pm\ 0.014$ $^{+ 0.012}_{- 0.013}$ 
   & 
$ 0.056$ $\pm\ 0.015$ $^{+ 0.014}_{- 0.010}$ 
   & 
$ 0.029$ $\pm\ 0.016$ $^{+ 0.010}_{- 0.010}$ 
   \\
  \rzqumu  & 
$ 0.029$ $\pm\ 0.016$ $^{+ 0.010}_{- 0.010}$ 
   & 
$ 0.028$ $\pm\ 0.019$ $^{+ 0.010}_{- 0.011}$ 
   & 
$ 0.043$ $\pm\ 0.014$ $^{+ 0.011}_{- 0.011}$ 
   & 
$ 0.028$ $\pm\ 0.014$ $^{+ 0.012}_{- 0.012}$ 
   \\
  \ruzz     & 
$ 0.028$ $\pm\ 0.014$ $^{+ 0.012}_{- 0.012}$ 
   & 
$-0.001$ $\pm\ 0.015$ $^{+ 0.012}_{- 0.010}$ 
   & 
$ 0.006$ $\pm\ 0.015$ $^{+ 0.008}_{- 0.007}$ 
   & 
$ 0.047$ $\pm\ 0.027$ $^{+ 0.009}_{- 0.010}$ 
   \\
  \ruuu     &
$ 0.047$ $\pm\ 0.027$ $^{+ 0.009}_{- 0.010}$ 
   & 
$ 0.047$ $\pm\ 0.029$ $^{+ 0.009}_{- 0.011}$ 
   & 
$ 0.033$ $\pm\ 0.028$ $^{+ 0.008}_{- 0.015}$ 
   & 
$ 0.004$ $\pm\ 0.032$ $^{+ 0.013}_{- 0.012}$ 
   \\
  Re \ruuz & 
$ 0.004$ $\pm\ 0.032$ $^{+ 0.013}_{- 0.012}$ 
   & 
$ 0.006$ $\pm\ 0.016$ $^{+ 0.003}_{- 0.009}$ 
   & 
$-0.011$ $\pm\ 0.017$ $^{+ 0.005}_{- 0.002}$ 
   & 
$ 0.024$ $\pm\ 0.019$ $^{+ 0.004}_{- 0.002}$ 
   \\
\hline
 \end{tabular}
 \caption{
  Spin density matrix elements for the diffractive electroproduction of \rh\   
  mesons as a function of $W$, for  $2.5 \leq  Q^2 < 5$,                       
  $5 \leq  \qsq < 15.5$ and $15.5 \leq  Q^2 \leq  60~\gevsq$, continued from 
  Table~\ref{table:matelem_f_w_three_qsq_rho}.                  
 }
 \label{table:matelem_f_w_three_qsq_rho_3}
\end{center}
 \end{sidewaystable}

  \clearpage
  \begin{table}[p]
  \begin{center}
 \begin{tabular}{|c|r|r|r|}
 \hline
 $\langle |t| \rangle$ (\gevsq) &
\multicolumn{1}{c|}{$0.08$}
  &
\multicolumn{1}{c|}{$0.34$}
  &
\multicolumn{1}{c|}{$1.05$}
  \\
 \hline
\hline
 \multicolumn{4}{|c|}{$\langle\qsq\rangle = 3.3$ \gevsq } \\
 \hline
 \rzqzz   &
$ 0.602$ $\pm\ 0.014$ $^{+ 0.011}_{- 0.031}$ 
   & 
$ 0.593$ $\pm\ 0.020$ $^{+ 0.019}_{- 0.029}$ 
   & 
$ 0.660$ $\pm\ 0.040$ $^{+ 0.033}_{- 0.045}$ 
   \\
  Re \rzquz&
$ 0.030$ $\pm\ 0.009$ $^{+ 0.011}_{- 0.012}$ 
   & 
$ 0.026$ $\pm\ 0.014$ $^{+ 0.013}_{- 0.009}$ 
   & 
$ 0.150$ $\pm\ 0.032$ $^{+ 0.017}_{- 0.028}$ 
   \\
  \rzqumu  &
$-0.004$ $\pm\ 0.011$ $^{+ 0.002}_{- 0.004}$ 
   & 
$ 0.037$ $\pm\ 0.017$ $^{+ 0.006}_{- 0.002}$ 
   & 
$-0.080$ $\pm\ 0.036$ $^{+ 0.010}_{- 0.015}$ 
   \\
  \ruzz     &
$ 0.070$ $\pm\ 0.077$ $^{+ 0.020}_{- 0.014}$ 
   & 
$-0.198$ $\pm\ 0.277$ $^{+ 0.259}_{- 0.025}$ 
   & 
$ 0.163$ $\pm\ 0.119$ $^{+ 0.067}_{- 0.076}$ 
   \\
  \ruuu     &
$-0.028$ $\pm\ 0.040$ $^{+ 0.006}_{- 0.011}$ 
   & 
$ 0.029$ $\pm\ 0.171$ $^{+ 0.025}_{- 0.132}$ 
   & 
$-0.144$ $\pm\ 0.106$ $^{+ 0.015}_{- 0.015}$ 
   \\
  Re \ruuz &
$-0.010$ $\pm\ 0.014$ $^{+ 0.008}_{- 0.006}$ 
   & 
$-0.086$ $\pm\ 0.020$ $^{+ 0.015}_{- 0.017}$ 
   & 
$-0.126$ $\pm\ 0.045$ $^{+ 0.028}_{- 0.016}$ 
   \\
\hline
 \end{tabular}
 \caption{
  Spin density matrix elements for the diffractive electroproduction of \rh\   
  mesons as a function of \modt, for  $2.5 \leq  Q^2 < 5$ and                  
  $5 \leq  Q^2 \leq  60~\gevsq$.                                               
  The notag ($\modt \leq  0.5~\gevsq$) and tag ($\modt \leq  3~\gevsq$)        
  samples are combined.                                                        
 }
\label{table:matelem_f_t_two_qsq_rho}
 \end{center}
 \end{table}

  \clearpage
  \begin{table}[p]
  \begin{center}
 \begin{tabular}{|c|r|r|r|}
 \hline
 $\langle |t| \rangle $ (\gevsq) &
\multicolumn{1}{c|}{$0.08$}
  &
\multicolumn{1}{c|}{$0.34$}
  &
\multicolumn{1}{c|}{$1.05$}
  \\
 \hline
 \rzqzz   &
$ 0.667$ $\pm\ 0.019$ $^{+ 0.012}_{- 0.011}$ 
   & 
$ 0.641$ $\pm\ 0.027$ $^{+ 0.021}_{- 0.021}$ 
   & 
$ 0.830$ $\pm\ 0.039$ $^{+ 0.087}_{- 0.088}$ 
   \\
  Re \rzquz&
$ 0.002$ $\pm\ 0.014$ $^{+ 0.010}_{- 0.012}$ 
   & 
$ 0.008$ $\pm\ 0.020$ $^{+ 0.020}_{- 0.021}$ 
   & 
$-0.010$ $\pm\ 0.041$ $^{+ 0.033}_{- 0.026}$ 
   \\
  \rzqumu  &
$-0.014$ $\pm\ 0.016$ $^{+ 0.004}_{- 0.002}$ 
   & 
$-0.003$ $\pm\ 0.023$ $^{+ 0.010}_{- 0.006}$ 
   & 
$-0.036$ $\pm\ 0.044$ $^{+ 0.012}_{- 0.024}$ 
   \\
  \ruzz     &
$-0.058$ $\pm\ 0.108$ $^{+ 0.005}_{- 0.007}$ 
   & 
$-0.069$ $\pm\ 0.166$ $^{+ 0.037}_{- 0.020}$ 
   & 
$ 0.007$ $\pm\ 0.032$ $^{+ 0.140}_{- 0.127}$ 
   \\
  \ruuu     &
$-0.012$ $\pm\ 0.079$ $^{+ 0.005}_{- 0.002}$ 
   & 
$-0.044$ $\pm\ 0.138$ $^{+ 0.020}_{- 0.015}$ 
   & 
$-0.029$ $\pm\ 0.040$ $^{+ 0.068}_{- 0.065}$ 
   \\
  Re \ruuz &
$-0.028$ $\pm\ 0.019$ $^{+ 0.016}_{- 0.009}$ 
   & 
$-0.071$ $\pm\ 0.027$ $^{+ 0.016}_{- 0.020}$ 
   & 
$-0.139$ $\pm\ 0.059$ $^{+ 0.026}_{- 0.025}$ 
   \\
  \ruumu   &
$ 0.215$ $\pm\ 0.021$ $^{+ 0.009}_{- 0.007}$ 
   & 
$ 0.064$ $\pm\ 0.032$ $^{+ 0.010}_{- 0.013}$ 
   & 
$ 0.114$ $\pm\ 0.067$ $^{+ 0.060}_{- 0.031}$ 
   \\
\hline
 \end{tabular}
 \caption{
  Spin density matrix elements for the diffractive electroproduction of \ph\   
  mesons as a function of \modt.                                               
  The notag ($\modt \leq  0.5~\gevsq$) and tag ($\modt \leq  3~\gevsq$)        
  samples are combined.                                                        
 }
\label{table:matelem_f_t_phi}
 \end{center}
 \end{table}

  \clearpage
  \begin{sidewaystable}[p]
  \begin{center}
 \begin{tabular}{|c|r|r|r|r|r|}
 \hline
 \multicolumn{6}{|c|}{$\langle\qsq\rangle = 3.3$ \gevsq } \\
 \hline
 $\langle \mpipi \rangle$ (\mbox{${\rm GeV}$}) &
\multicolumn{1}{c|}{$0.65$}
  &
\multicolumn{1}{c|}{$0.75$}
  &
\multicolumn{1}{c|}{$0.85$}
  &
\multicolumn{1}{c|}{$0.95$}
  &
\multicolumn{1}{c|}{$1.05$}
  \\
\hline
 \rzqzz   &
$ 0.677$ $\pm\ 0.021$ $^{+ 0.029}_{- 0.037}$ 
   & 
$ 0.635$ $\pm\ 0.017$ $^{+ 0.012}_{- 0.021}$ 
   & 
$ 0.559$ $\pm\ 0.023$ $^{+ 0.013}_{- 0.046}$ 
   & 
$ 0.456$ $\pm\ 0.037$ $^{+ 0.024}_{- 0.012}$ 
   & 
$ 0.458$ $\pm\ 0.055$ $^{+ 0.027}_{- 0.051}$ 
   \\
  Re \rzquz& 
$ 0.067$ $\pm\ 0.016$ $^{+ 0.013}_{- 0.012}$ 
   & 
$ 0.028$ $\pm\ 0.012$ $^{+ 0.011}_{- 0.013}$ 
   & 
$ 0.033$ $\pm\ 0.016$ $^{+ 0.011}_{- 0.012}$ 
   & 
$ 0.046$ $\pm\ 0.023$ $^{+ 0.013}_{- 0.010}$ 
   & 
$ 0.027$ $\pm\ 0.034$ $^{+ 0.014}_{- 0.016}$ 
   \\
  \rzqumu  &  
$-0.002$ $\pm\ 0.019$ $^{+ 0.009}_{- 0.003}$ 
   & 
$ 0.002$ $\pm\ 0.013$ $^{+ 0.002}_{- 0.002}$ 
   & 
$-0.016$ $\pm\ 0.019$ $^{+ 0.005}_{- 0.007}$ 
   & 
$ 0.038$ $\pm\ 0.032$ $^{+ 0.008}_{- 0.010}$ 
   & 
$ 0.003$ $\pm\ 0.042$ $^{+ 0.014}_{- 0.014}$ 
   \\
  \ruzz     & 
$-0.033$ $\pm\ 0.091$ $^{+ 0.026}_{- 0.033}$ 
   & 
$ 0.068$ $\pm\ 0.070$ $^{+ 0.084}_{- 0.017}$ 
   & 
$ 0.084$ $\pm\ 0.070$ $^{+ 0.045}_{- 0.054}$ 
   & 
$-0.089$ $\pm\ 0.223$ $^{+ 0.061}_{- 0.029}$ 
   & 
$ 0.212$ $\pm\ 0.186$ $^{+ 0.060}_{- 0.077}$ 
   \\
  \ruuu     & 
$-0.029$ $\pm\ 0.079$ $^{+ 0.027}_{- 0.027}$ 
   & 
$-0.033$ $\pm\ 0.035$ $^{+ 0.011}_{- 0.041}$ 
   & 
$-0.058$ $\pm\ 0.042$ $^{+ 0.025}_{- 0.019}$ 
   & 
$-0.062$ $\pm\ 0.187$ $^{+ 0.015}_{- 0.034}$ 
   & 
$-0.135$ $\pm\ 0.101$ $^{+ 0.030}_{- 0.037}$ 
   \\
  Re \ruuz &  
$ 0.020$ $\pm\ 0.023$ $^{+ 0.009}_{- 0.013}$ 
   & 
$-0.069$ $\pm\ 0.017$ $^{+ 0.011}_{- 0.011}$ 
   & 
$-0.043$ $\pm\ 0.022$ $^{+ 0.009}_{- 0.009}$ 
   & 
$-0.035$ $\pm\ 0.036$ $^{+ 0.014}_{- 0.014}$ 
   & 
$-0.003$ $\pm\ 0.051$ $^{+ 0.038}_{- 0.018}$ 
   \\
\hline
 \end{tabular}
 \caption{
  Spin density matrix elements for the diffractive electroproduction of \rh\   
  mesons as a function of \mpp, for  $2.5 \leq  Q^2 < 5$ and                   
  $5 \leq  Q^2 \leq  60~\gevsq$.                                               
  The notag ($\modt \leq  0.5~\gevsq$) and tag ($\modt \leq  3~\gevsq$)        
  samples are combined.                                                        
 }
\label{table:matelem_f_m_two_qsq_rho}
 \end{center}
 \end{sidewaystable}

 \begin{sidewaystable}[p]
 \begin{center}
 \begin{tabular}{|c|r|r|r|r|r|}
 \hline
 \multicolumn{6}{|c|}{$\langle\qsq\rangle = 8.6$ \gevsq } \\
 \hline
 $\langle \mpipi \rangle$ (\mbox{${\rm GeV}$}) &
\multicolumn{1}{c|}{$0.65$}
  &
\multicolumn{1}{c|}{$0.75$}
  &
\multicolumn{1}{c|}{$0.85$}
  &
\multicolumn{1}{c|}{$0.95$}
  &
\multicolumn{1}{c|}{$1.05$}
  \\
\hline
 \rzqzz   &
$ 0.857$ $\pm\ 0.012$ $^{+ 0.048}_{- 0.049}$ 
   & 
$ 0.793$ $\pm\ 0.012$ $^{+ 0.019}_{- 0.030}$ 
   & 
$ 0.747$ $\pm\ 0.016$ $^{+ 0.032}_{- 0.032}$ 
   & 
$ 0.586$ $\pm\ 0.027$ $^{+ 0.031}_{- 0.028}$ 
   & 
$ 0.473$ $\pm\ 0.035$ $^{+ 0.021}_{- 0.026}$ 
   \\
  Re \rzquz&
$-0.002$ $\pm\ 0.013$ $^{+ 0.015}_{- 0.014}$ 
   & 
$ 0.016$ $\pm\ 0.009$ $^{+ 0.011}_{- 0.010}$ 
   & 
$ 0.028$ $\pm\ 0.012$ $^{+ 0.009}_{- 0.012}$ 
   & 
$-0.039$ $\pm\ 0.018$ $^{+ 0.012}_{- 0.009}$ 
   & 
$ 0.125$ $\pm\ 0.024$ $^{+ 0.021}_{- 0.015}$ 
   \\
  \rzqumu  & 	     
$-0.051$ $\pm\ 0.014$ $^{+ 0.007}_{- 0.008}$ 
   & 
$ 0.005$ $\pm\ 0.009$ $^{+ 0.002}_{- 0.001}$ 
   & 
$ 0.033$ $\pm\ 0.013$ $^{+ 0.002}_{- 0.004}$ 
   & 
$-0.063$ $\pm\ 0.022$ $^{+ 0.012}_{- 0.010}$ 
   & 
$-0.116$ $\pm\ 0.031$ $^{+ 0.024}_{- 0.025}$ 
   \\
  \ruzz     & 	     
$-0.053$ $\pm\ 0.109$ $^{+ 0.022}_{- 0.036}$ 
   & 
$-0.008$ $\pm\ 0.025$ $^{+ 0.026}_{- 0.010}$ 
   & 
$-0.020$ $\pm\ 0.037$ $^{+ 0.031}_{- 0.022}$ 
   & 
$ 0.047$ $\pm\ 0.137$ $^{+ 0.105}_{- 0.033}$ 
   & 
$ 0.156$ $\pm\ 0.263$ $^{+ 0.074}_{- 0.052}$ 
   \\
  \ruuu     & 	     
$-0.019$ $\pm\ 0.084$ $^{+ 0.042}_{- 0.032}$ 
   & 
$-0.009$ $\pm\ 0.022$ $^{+ 0.016}_{- 0.018}$ 
   & 
$-0.004$ $\pm\ 0.028$ $^{+ 0.024}_{- 0.026}$ 
   & 
$-0.141$ $\pm\ 0.180$ $^{+ 0.025}_{- 0.054}$ 
   & 
$ 0.013$ $\pm\ 0.184$ $^{+ 0.041}_{- 0.040}$ 
   \\
  Re \ruuz & 	     
$ 0.038$ $\pm\ 0.020$ $^{+ 0.017}_{- 0.014}$ 
   & 
$-0.028$ $\pm\ 0.012$ $^{+ 0.012}_{- 0.010}$ 
   & 
$-0.027$ $\pm\ 0.016$ $^{+ 0.012}_{- 0.010}$ 
   & 
$-0.061$ $\pm\ 0.025$ $^{+ 0.012}_{- 0.014}$ 
   & 
$ 0.066$ $\pm\ 0.035$ $^{+ 0.042}_{- 0.009}$ 
   \\
\hline
 \end{tabular}
 \caption{
  Spin density matrix elements for the diffractive electroproduction of \rh\   
  mesons as a function of \mpp, for  $2.5 \leq  Q^2 < 5$ and                   
  $5 \leq  Q^2 \leq  60~\gevsq$, continued from 
  Table~\ref{table:matelem_f_m_two_qsq_rho}.                                               
 }
\label{table:matelem_f_m_two_qsq_rho_2}
 \end{center}
 \end{sidewaystable}

  \begin{table}[H]
  \begin{center}
 \begin{tabular}{|d|lll|lll|} 
 \hline
 \multicolumn{1}{|c|}{$\langle\qsq\rangle$ (\gevsq)} &\multicolumn{3}{c|}{\rfivecomb} &
 \multicolumn{3}{c|}{\ronecomb} \\
 \hline
  2.9 &  $0.149$ & $\pm\  0.018$ & $^{+ 0.062}_{- 0.060}$ & $-0.041$ & $\pm\  0.034$ & $^{+ 0.029}_{- 0.034}$ \\
  4.1 &  $0.144$ & $\pm\  0.019$ & $^{+ 0.046}_{- 0.055}$ & $-0.036$ & $\pm\  0.041$ & $^{+ 0.023}_{- 0.025}$ \\
  6.6 &  $0.122$ & $\pm\  0.016$ & $^{+ 0.049}_{- 0.049}$ & $-0.078$ & $\pm\  0.032$ & $^{+ 0.023}_{- 0.014}$ \\
 11.9 &  $0.088$ & $\pm\  0.024$ & $^{+ 0.049}_{- 0.047}$ & $-0.093$ & $\pm\  0.045$ & $^{+ 0.024}_{- 0.018}$ \\
 18.5 &  $0.178$ & $\pm\  0.031$ & $^{+ 0.037}_{- 0.049}$ & $-0.010$ & $\pm\  0.059$ & $^{+ 0.034}_{- 0.034}$ \\
 31.3 &  $0.149$ & $\pm\  0.028$ & $^{+ 0.041}_{- 0.037}$ & $ 0.123$ & $\pm\  0.059$ & $^{+ 0.030}_{- 0.032}$ \\
\hline
 \hline
 \multicolumn{1}{|c|}{$\langle |t| \rangle$ (\gevsq)} &\multicolumn{3}{c|}{\rfivecomb} &
 \multicolumn{3}{c|}{\ronecomb} \\
 \hline
 0.03 &  $0.074$ & $\pm\  0.022$ & $^{+ 0.032}_{- 0.036}$ & $ 0.003$ & $\pm\  0.045$ & $^{+ 0.004}_{- 0.013}$ \\
 0.13 &  $0.109$ & $\pm\  0.022$ & $^{+ 0.062}_{- 0.052}$ & $-0.024$ & $\pm\  0.044$ & $^{+ 0.017}_{- 0.024}$ \\
 0.31 &  $0.200$ & $\pm\  0.028$ & $^{+ 0.064}_{- 0.086}$ & $-0.131$ & $\pm\  0.054$ & $^{+ 0.041}_{- 0.028}$ \\
 0.70 &  $0.249$ & $\pm\  0.051$ & $^{+ 0.063}_{- 0.038}$ & $-0.015$ & $\pm\  0.099$ & $^{+ 0.040}_{- 0.067}$ \\
 1.44 &  $0.308$ & $\pm\  0.077$ & $^{+ 0.086}_{- 0.096}$ & $-0.162$ & $\pm\  0.129$ & $^{+ 0.086}_{- 0.080}$ \\
 \hline
 \end{tabular}
 \caption{
  \qsq\ and \modt\ dependences of the matrix element                           
  combinations \rfivecomb\ and \ronecomb, obtained from fits                   
  of Eq.~(\ref{eq:angle_phi}) to the $\phi$ distribution,                      
  for \rh\ meson electroproduction.                                            
  The notag ($\modt \leq 0.5~\gevsq$) and tag ($\modt \leq 3~\gevsq$)          
  samples are combined.                                                        
 }
 \label{table:r5comb}
 \end{center}
 \end{table}

  \begin{table}[H]
  \begin{center}
 \begin{tabular}{|d|lll||d|lll|} 
 \hline
 \multicolumn{1}{|c|}{$\langle\qsq\rangle$  (\gevsq)} &\multicolumn{3}{c||}{$P_{\rm NPE,T}$} & 
 \multicolumn{1}{c|}{$\langle |t| \rangle$ (\gevsq)} &\multicolumn{3}{c|}{$P_{\rm NPE,T}$} \\
\hline
 \hline
 \multicolumn{8}{|c|}{$\rho$ production}\\
 \hline
  3.1 & $1.13$ & $\pm\ 0.07$ & $^{+0.07}_{-0.03}$ &  0.03 & $1.10$ & $\pm\ 0.06$ & $^{+0.05}_{-0.02}$ \\
  4.8 & $1.24$ & $\pm\ 0.09$ & $^{+0.04}_{-0.03}$ &  0.13 & $1.02$ & $\pm\ 0.12$ & $^{+0.03}_{-0.03}$ \\
  7.5 & $1.03$ & $\pm\ 0.10$ & $^{+0.06}_{-0.05}$ &  0.31 & $1.03$ & $\pm\ 0.12$ & $^{+0.11}_{-0.08}$ \\
 12.2 & $0.95$ & $\pm\ 0.11$ & $^{+0.05}_{-0.05}$ &  0.70 & $0.96$ & $\pm\ 0.39$ & $^{+0.11}_{-0.09}$ \\
 23.1 & $1.07$ & $\pm\ 0.16$ & $^{+0.06}_{-0.05}$ &  1.44 & $1.08$ & $\pm\ 0.12$ & $^{+0.28}_{-0.31}$ \\
 \hline
\hline
 \multicolumn{8}{|c|}{$\phi$ production}\\
 \hline
  3.3 & $0.98$ & $\pm\ 0.31$ & $^{+0.04}_{-0.05}$ &  0.08 & $0.90$ & $\pm\ 0.17$ & $^{+0.02}_{-0.02}$ \\
  6.6 & $1.13$ & $\pm\ 0.11$ & $^{+0.04}_{-0.05}$ &  0.34 & $1.31$ & $\pm\ 0.29$ & $^{+0.06}_{-0.05}$ \\
 15.8 & $0.99$ & $\pm\ 0.10$ & $^{+0.06}_{-0.05}$ &  1.05 & $0.93$ & $\pm\ 0.18$ & $^{+0.22}_{-0.28}$ \\
 \hline
 \end{tabular}
 \caption{
  Asymmetry $P_{\rm NPE,T}$ between natural and unnatural parity exchange for      
  transverse photons, as a function of \qsq\ and \modt, for \rh\ and           
  \ph\ meson production.                                                       
 }
 \label{table:npe}
 \end{center}
 \end{table}

 \begin{table}[H]
 \begin{center}
 \begin{tabular}{|d|lll|} 
 \hline
 \multicolumn{1}{|c|}{$\langle\qsq\rangle$ (\gevsq)} &\multicolumn{3}{c|}{$\cos\delta$} \\
 \hline
\hline
 \multicolumn{4}{|c|}{$\rho$ production} \\
 \hline
  3.3 &  $0.914$ & $\pm\  0.014$ & $^{+ 0.021}_{- 0.024}$ \\
  6.6 &  $0.915$ & $\pm\  0.026$ & $^{+ 0.018}_{- 0.024}$ \\
 15.8 &  $0.978$ & $\pm\  0.030$ & $^{+ 0.016}_{- 0.052}$ \\
 \hline
\hline
 \multicolumn{4}{|c|}{$\phi$ production} \\
 \hline
  5.3 &  $0.966$ & $\pm\  0.027$ & $^{+ 0.012}_{- 0.018}$ \\
 \hline
 \end{tabular}
 \caption{
  Cosine of the phase $\delta$ between the $T_{00}$ and $T_{11}$               
  helicity conserving amplitudes for \rh\ and \ph\ meson production,           
  measured as a function of \qsq\ from                                         
  two-dimensional fits to Eq.~(\ref{eq:Wcosdelta}), in the SCHC                
  approximation ($\modt \leq  0.5~\gevsq$)                                     
 }
\label{table:cosdelta}
 \end{center}
 \end{table}

 \begin{table}[H]
 \begin{center}
 \begin{tabular}{|d|lll|} 
 \hline
 \multicolumn{1}{|c|}{$\langle\qsq\rangle$ (\gevsq)} &
 \multicolumn{3}{c|}{$R = \sigma_L / \sigma_T$} \\
 \hline
\hline
 \multicolumn{4}{|c|}{$\rho$ production} \\
 \hline
  3.1 &   $1.36$ & $^{+  0.08}_{-  0.08}$  & $^{+  0.11}_{-  0.13}$ \\
  4.8 &   $1.92$ & $^{+  0.16}_{-  0.15}$  & $^{+  0.15}_{-  0.12}$ \\
  7.5 &   $3.65$ & $^{+  0.31}_{-  0.27}$  & $^{+  0.41}_{-  0.38}$ \\
 12.2 &   $3.60$ & $^{+  0.38}_{-  0.34}$  & $^{+  0.39}_{-  0.37}$ \\
 23.1 &   $6.52$ & $^{+  1.27}_{-  0.98}$  & $^{+  1.23}_{-  0.96}$ \\
\hline
 \hline
 \multicolumn{4}{|c|}{$\phi$ production} \\
 \hline
  3.3 &   $1.37$ & $^{+  0.14}_{-  0.13}$  & $^{+  0.09}_{-  0.09}$ \\
  6.6 &   $2.87$ & $^{+  0.40}_{-  0.34}$  & $^{+  0.28}_{-  0.26}$ \\
 15.8 &   $6.01$ & $^{+  2.00}_{-  1.33}$  & $^{+  1.00}_{-  0.78}$ \\
 \hline
 \end{tabular}
 \caption{
  \qsq\ dependence of the ratio $R = \sigma_L / \sigma_T$ of the            
  longitudinal to transverse cross sections, for \rh\ and \ph\ meson           
  production.                                                                  
 }
\label{table:rlt-q2}
 \end{center}
 \end{table}

 \begin{table}[H]
 \begin{center}
 \begin{tabular}{|c|lll|} 
 \hline
 $\langle W \rangle$ (\mbox{${\rm GeV}$}) &\multicolumn{3}{c|}{$R = \sigma_L / \sigma_T$} \\
\hline
 \hline
 \multicolumn{4}{|c|}{$\langle\qsq\rangle = 3.3$ \gevsq} \\
 \hline
  $43$ &   $1.56$ & $^{+  0.15}_{-  0.14}$  & $^{+  0.11}_{-  0.08}$ \\
  $59$ &   $1.46$ & $^{+  0.14}_{-  0.13}$  & $^{+  0.09}_{-  0.08}$ \\
  $76$ &   $1.40$ & $^{+  0.16}_{-  0.15}$  & $^{+  0.10}_{-  0.08}$ \\
  $92$ &   $1.27$ & $^{+  0.17}_{-  0.15}$  & $^{+  0.10}_{-  0.11}$ \\
 \hline
\hline
 \multicolumn{4}{|c|}{$\langle\qsq\rangle = 7.5$ \gevsq} \\
 \hline
 $ \phantom{0}58$ &   $2.67$ & $^{+  0.28}_{-  0.24}$  & $^{+  0.22}_{-  0.18}$ \\
 $ \phantom{0}76$ &   $3.00$ & $^{+  0.34}_{-  0.29}$  & $^{+  0.29}_{-  0.30}$ \\
 $ \phantom{0}93$ &   $3.31$ & $^{+  0.47}_{-  0.41}$  & $^{+  0.46}_{-  0.45}$ \\
 $111$ &   $2.72$ & $^{+  0.32}_{-  0.28}$  & $^{+  0.26}_{-  0.22}$ \\
 \hline
\hline
 \multicolumn{4}{|c|}{$\langle\qsq\rangle = 22.5$ \gevsq} \\
 \hline
 $ \phantom{0}72$ &   $3.59$ & $^{+  0.98}_{-  0.74}$  & $^{+  0.59}_{-  0.37}$ \\
 $ \phantom{0}97$ &   $5.96$ & $^{+  2.48}_{-  1.60}$  & $^{+  1.19}_{-  1.05}$ \\
 $122$ &   $6.27$ & $^{+  2.98}_{-  1.76}$  & $^{+  1.41}_{-  1.02}$ \\
 $147$ &   $6.01$ & $^{+  3.13}_{-  1.71}$  & $^{+  1.08}_{-  2.14}$ \\
 \hline
 \end{tabular}
 \caption{
  $W$ dependence of the ratio $R = \sigma_L / \sigma_T$ of the              
  longitudinal to transverse cross sections for \rh\ meson production,         
  for  $2.5 \leq  Q^2 < 5~\gevsq$, $5 \leq  \qsq < 15.5~\gevsq$ and            
  $15.5 \leq  Q^2 \leq  60~\gevsq$.                                            
 }
\label{table:rlt-w}
 \end{center}
 \end{table}

 \begin{table}[H]
 \begin{center}
 \begin{tabular}{|c|lll|} 
 \hline
 $\langle |t| \rangle$ (\gevsq) &
 \multicolumn{3}{c|}{$R = \sigma_L / \sigma_T$} \\
 \hline
\hline
 \multicolumn{4}{|c|}{$\langle\qsq\rangle = 3.3$ \gevsq} \\
 \hline
 $0.08$ &  $ 1.43$ & $^{+  0.09}_{-  0.09}$  & $^{+  0.08}_{-  0.08}$ \\
 $0.34$ &  $ 1.26$ & $^{+  0.13}_{-  0.12}$  & $^{+  0.16}_{-  0.16}$ \\
 $1.05$ &  $ 1.70$ & $^{+  0.37}_{-  0.32}$  & $^{+  0.24}_{-  0.23}$ \\
 \hline
\hline
 \multicolumn{4}{|c|}{$\langle\qsq\rangle = 8.6$ \gevsq} \\
 \hline
 $0.08$ &   $2.73$ & $^{+  0.16}_{-  0.15}$  & $^{+  0.15}_{-  0.15}$ \\
 $0.34$ &   $3.98$ & $^{+  0.40}_{-  0.35}$  & $^{+  0.92}_{-  0.79}$ \\
 $1.05$ &   $4.99$ & $^{+  0.69}_{-  0.59}$  & $^{+  3.63}_{-  1.98}$ \\
 \hline
 \end{tabular}
 \caption{
  \modt\ dependence of the ratio $R = \sigma_L / \sigma_T$ of the           
  longitudinal to transverse cross sections for \rh\ meson production,         
  for  $2.5 \leq  Q^2 < 5~\gevsq$ and $5 \leq  Q^2 \leq  60~\gevsq$.           
 }
\label{table:rlt-t}
 \end{center}
 \end{table}

 \begin{table}[H]
 \begin{center}
 \begin{tabular}{|c|lll|} 
 \hline
 $\langle \mpipi \rangle$ (\mbox{${\rm GeV}$}) &
 \multicolumn{3}{c|}{$R = \sigma_L / \sigma_T$} \\
\hline
 \hline
 \multicolumn{4}{|c|}{$\langle\qsq\rangle = 3.3$ \gevsq} \\
 \hline
 $0.65$ &   $1.97$ & $^{+  0.21}_{-  0.19}$  & $^{+  0.32}_{-  0.30}$ \\
 $0.75$ &   $1.62$ & $^{+  0.13}_{-  0.12}$  & $^{+  0.11}_{-  0.13}$ \\
 $0.85$ &   $0.85$ & $^{+  0.14}_{-  0.15}$  & $^{+  0.12}_{-  0.09}$ \\
 $0.95$ &   $0.69$ & $^{+  0.14}_{-  0.15}$  & $^{+  0.05}_{-  0.10}$ \\
 $1.05$ &   $0.57$ & $^{+  0.24}_{-  0.34}$  & $^{+  0.14}_{-  0.19}$ \\
 \hline
\hline
 \multicolumn{4}{|c|}{$\langle\qsq\rangle = 8.6$ \gevsq} \\
 \hline
 $0.65$ &   $5.70$ & $^{+  0.59}_{-  0.52}$  & $^{+  2.85}_{-  1.75}$ \\
 $0.75$ &   $3.70$ & $^{+  0.28}_{-  0.26}$  & $^{+  0.46}_{-  0.43}$ \\
 $0.85$ &   $2.72$ & $^{+  0.25}_{-  0.23}$  & $^{+  0.40}_{-  0.36}$ \\
 $0.95$ &   $1.39$ & $^{+  0.17}_{-  0.15}$  & $^{+  0.20}_{-  0.17}$ \\
 $1.05$ &   $0.90$ & $^{+  0.14}_{-  0.12}$  & $^{+  0.08}_{-  0.06}$ \\
 \hline
 \end{tabular}
 \caption{
  \mpp\ dependence of the ratio $R = \sigma_L / \sigma_T$ of the            
  longitudinal to transverse cross sections for \rh\ meson production,         
  for $2.5 \leq  Q^2 < 5~\gevsq$ and $5 \leq  Q^2 \leq  60~\gevsq$.            
 }
\label{table:rlt-m}
 \end{center}
 \end{table}

 \begin{table}[H]
 \begin{center}
 \begin{tabular}{|c|lll|} 
 \hline
 $\langle\mpipi \rangle$ (\mbox{${\rm GeV}$}) & \multicolumn{3}{c|}{ $ b_{\gsrel} $ (\mbox{${\rm GeV}^{-2}$})} \\ 
 \hline
\hline
 \multicolumn{4}{|c|}{$\langle\qsq\rangle = 3.3$ \gevsq } \\
 \hline
 $0.67$ & $7.33$ & $\pm\ 0.51$ & $^{+0.38}_{-0.36}$ \\
 $0.87$ & $8.25$ & $\pm\ 0.40$ & $^{+0.26}_{-0.29}$ \\
 $1.07$ & $7.73$ & $\pm\ 0.51$ & $^{+0.46}_{-0.42}$ \\
 \hline
\hline
 \multicolumn{4}{|c|}{$\langle\qsq\rangle = 9.0$ \gevsq } \\
 \hline
 $0.67$ & $7.15$ & $\pm\ 0.48$ & $^{+0.27}_{-0.30}$ \\
 $0.87$ & $7.65$ & $\pm\ 0.43$ & $^{+0.30}_{-0.32}$ \\
 $1.07$ & $6.83$ & $\pm\ 0.45$ & $^{+0.27}_{-0.35}$ \\
 \hline
 \end{tabular}
 \caption{
  Dependence of the exponential $t$ slope for \rh\ elastic production,         
  as a function of the mass \mpp, for two domains in \qsq:                     
  $2.5 \leq \qsq < 5~\gevsq$ and $5 \leq \qsq \leq 60~\gevsq$                  
 }
 \label{table:rho_masst_bslp}
 \end{center}
 \end{table}

 \clearpage
 \begin{sidewaystable}[p]
 \begin{center}
 \begin{tabular}{|d|r|r|r|r||r|}
 \hline
 \multicolumn{1}{|c|}{$\langle\qsq\rangle$ (\gevsq)} &
 \multicolumn{1}{c|}{\ralpha                        }
  & 
 \multicolumn{1}{c|}{\rbeta                         }
  & 
 \multicolumn{1}{c|}{\rdelta                        }
  & 
 \multicolumn{1}{c|}{\reta                          }
  & 
 \multicolumn{1}{||c|}{$\cos (\phi_{11} - \phi_{00})$ } \\
 \hline
\hline
 \multicolumn{6}{|c|}{$\rho$ production} \\
 \hline
  3.1 & 
$ 0.786$ $\pm\ 0.019$ $^{+ 0.036}_{- 0.020}$ 
   & 
$ 0.177$ $\pm\ 0.019$ $^{+ 0.042}_{- 0.068}$ 
   & 
$-0.002$ $\pm\ 0.011$ $^{+ 0.016}_{- 0.005}$ 
   & 
$-0.019$ $\pm\ 0.014$ $^{+ 0.002}_{- 0.005}$ 
   & 
$ 0.949$ $\pm\ 0.024$ $^{+ 0.028}_{- 0.034}$ 
   \\ 
   4.8  & 
$ 0.619$ $\pm\ 0.021$ $^{+ 0.033}_{- 0.018}$ 
   & 
$ 0.169$ $\pm\ 0.022$ $^{+ 0.051}_{- 0.042}$ 
   & 
$-0.025$ $\pm\ 0.014$ $^{+ 0.008}_{- 0.009}$ 
   & 
$ 0.001$ $\pm\ 0.017$ $^{+ 0.002}_{- 0.005}$ 
   & 
$ 0.882$ $\pm\ 0.035$ $^{+ 0.028}_{- 0.038}$ 
   \\ 
   7.5  & 
$ 0.511$ $\pm\ 0.015$ $^{+ 0.017}_{- 0.011}$ 
   & 
$ 0.056$ $\pm\ 0.018$ $^{+ 0.038}_{- 0.038}$ 
   & 
$ 0.018$ $\pm\ 0.010$ $^{+ 0.006}_{- 0.005}$ 
   & 
$-0.034$ $\pm\ 0.014$ $^{+ 0.004}_{- 0.005}$ 
   & 
$ 0.997$ $\pm\ 0.038$ $^{+ 0.007}_{- 0.046}$ 
   \\ 
  12.2  & 
$ 0.444$ $\pm\ 0.018$ $^{+ 0.015}_{- 0.008}$ 
   & 
$ 0.106$ $\pm\ 0.022$ $^{+ 0.035}_{- 0.034}$ 
   & 
$-0.049$ $\pm\ 0.013$ $^{+ 0.006}_{- 0.007}$ 
   & 
$-0.038$ $\pm\ 0.017$ $^{+ 0.009}_{- 0.010}$ 
   & 
$ 0.900$ $\pm\ 0.049$ $^{+ 0.043}_{- 0.049}$ 
   \\ 
  23.1  & 
$ 0.287$ $\pm\ 0.020$ $^{+ 0.006}_{- 0.005}$ 
   & 
$ 0.090$ $\pm\ 0.025$ $^{+ 0.030}_{- 0.029}$ 
   & 
$ 0.047$ $\pm\ 0.016$ $^{+ 0.003}_{- 0.009}$ 
   & 
$ 0.029$ $\pm\ 0.020$ $^{+ 0.007}_{- 0.005}$ 
   & 
$ 0.879$ $\pm\ 0.094$ $^{+ 0.056}_{- 0.054}$ 
  \\ 
\hline
 \hline
 \multicolumn{6}{|c|}{$\phi$ production} \\
\hline
 \end{tabular}
 \caption{
  Ratios of the helicity amplitudes                                            
  (taken to be purely imaginary)                                               
  and phase difference between the $T_{11}$ and $T_{00}$ amplitudes            
  (the amplitude ratios \rdelta\ and \reta\ and the phase                      
  difference $\phi_{01} - \phi_{00}$ are taken to be 0)                        
  for \rh\ and \ph\ meson production, computed from global                     
  fits to the measurements of  the 15 spin density matrix elements,            
  as a function of \qsq\ (NPE is assumed).                                     
 }
\label{table:ampl-ratios-f-qsq}
 \end{center}
 \end{sidewaystable}

 \clearpage
 \begin{sidewaystable}[p]
 \begin{center}
 \begin{tabular}{|c|r|r|r|r||r|}
 \hline
 \multicolumn{1}{|c|}{$\langle\modt\rangle$ (\gevsq)} &
 \multicolumn{1}{c|}{\ralpha                        }
  & 
 \multicolumn{1}{c|}{\rbeta                         }
  & 
 \multicolumn{1}{c|}{\rdelta                        }
  & 
 \multicolumn{1}{c|}{\reta                          }
  & 
 \multicolumn{1}{||c|}{$\cos (\phi_{11} - \phi_{00})$ } \\
 \hline
 \hline
 \multicolumn{6}{|c|}{$\rho$ production} \\
 \hline
 \multicolumn{6}{|c|}{$\langle\qsq\rangle = 3.3$ \gevsq } \\
 \hline
$ 0.08$ & 
$ 0.773$ $\pm\ 0.020$ $^{+ 0.032}_{- 0.015}$ 
   & 
$ 0.128$ $\pm\ 0.020$ $^{+ 0.037}_{- 0.057}$ 
   & 
$-0.007$ $\pm\ 0.012$ $^{+ 0.015}_{- 0.004}$ 
   & 
$-0.004$ $\pm\ 0.015$ $^{+ 0.003}_{- 0.005}$ 
   & 
$ 0.939$ $\pm\ 0.026$ $^{+ 0.019}_{- 0.029}$ 
   \\ 
$  0.34$  & 
$ 0.769$ $\pm\ 0.031$ $^{+ 0.042}_{- 0.033}$ 
   & 
$ 0.281$ $\pm\ 0.034$ $^{+ 0.060}_{- 0.062}$ 
   & 
$-0.062$ $\pm\ 0.019$ $^{+ 0.012}_{- 0.010}$ 
   & 
$-0.015$ $\pm\ 0.026$ $^{+ 0.011}_{- 0.009}$ 
   & 
$ 0.851$ $\pm\ 0.040$ $^{+ 0.047}_{- 0.053}$ 
   \\ 
 $ 1.05$  & 
$ 0.673$ $\pm\ 0.067$ $^{+ 0.048}_{- 0.040}$ 
   & 
$ 0.248$ $\pm\ 0.070$ $^{+ 0.043}_{- 0.043}$ 
   & 
$ 0.138$ $\pm\ 0.046$ $^{+ 0.013}_{- 0.027}$ 
   & 
$-0.095$ $\pm\ 0.051$ $^{+ 0.036}_{- 0.015}$ 
   & 
$ 0.929$ $\pm\ 0.094$ $^{+ 0.088}_{- 0.293}$ 
  \\ 
 \hline
 \multicolumn{6}{|c|}{$\langle\qsq\rangle = 8.6$ \gevsq } \\
 \hline
$ 0.08$ & 
$ 0.559$ $\pm\ 0.013$ $^{+ 0.018}_{- 0.012}$ 
   & 
$ 0.059$ $\pm\ 0.014$ $^{+ 0.028}_{- 0.028}$ 
   & 
$ 0.012$ $\pm\ 0.009$ $^{+ 0.005}_{- 0.005}$ 
   & 
$-0.009$ $\pm\ 0.011$ $^{+ 0.002}_{- 0.001}$ 
   & 
$ 0.947$ $\pm\ 0.026$ $^{+ 0.023}_{- 0.031}$ 
   \\
\hline
 \end{tabular}
 \caption{
  Ratios of the helicity amplitudes                                            
  (taken to be purely imaginary)                                               
  and phase difference between the $T_{11}$ and $T_{00}$ amplitudes            
  (the amplitude ratios \rdelta\ and \reta\ and the phase                      
  difference $\phi_{01} - \phi_{00}$ are taken to be 0),                       
  computed from global                                                         
  fits to the measurements of  the 15 spin density matrix elements,            
  as a function of \modt,                                                      
  separately for $2.5 \leq  Q^2 < 5~\gevsq$ and $5 \leq  Q^2 \leq 60~\gevsq$   
  for \rh\ meson production and for $2.5 \leq  Q^2 \leq 60~\gevsq$             
  for \ph\ production                                                          
  (NPE is assumed).                                                            
 }
\label{table:ampl-ratios-f-t}
 \end{center}
 \end{sidewaystable}

 \clearpage
 \begin{sidewaystable}[p]
 \begin{center}
 \begin{tabular}{|c|r|r|r|r||r|}
 \hline
 $\langle\mpipi\rangle$ (\mbox{${\rm GeV}$}) &
 \multicolumn{1}{c|}{\ralpha                        } &
 \multicolumn{1}{c|}{\rbeta                         } &
 \multicolumn{1}{c|}{\rdelta                        } &
 \multicolumn{1}{c|}{\reta                          } &
 \multicolumn{1}{||c|}{$\cos (\phi_{11} - \phi_{00})$ } \\
 \hline
\hline
 \multicolumn{6}{|c|}{$\langle\qsq\rangle = 3.3$ \gevsq } \\
 \hline
$ 0.65$ & 
$ 0.618$ $\pm\ 0.029$ $^{+ 0.026}_{- 0.019}$ 
   & 
$ 0.138$ $\pm\ 0.034$ $^{+ 0.049}_{- 0.055}$ 
   & 
$ 0.020$ $\pm\ 0.021$ $^{+ 0.007}_{- 0.006}$ 
   & 
$-0.031$ $\pm\ 0.025$ $^{+ 0.015}_{- 0.006}$ 
   & 
$ 0.855$ $\pm\ 0.047$ $^{+ 0.078}_{- 0.070}$ 
   \\ 
$  0.75$  & 
$ 0.711$ $\pm\ 0.023$ $^{+ 0.022}_{- 0.018}$ 
   & 
$ 0.166$ $\pm\ 0.024$ $^{+ 0.056}_{- 0.037}$ 
   & 
$-0.006$ $\pm\ 0.014$ $^{+ 0.004}_{- 0.010}$ 
   & 
$-0.007$ $\pm\ 0.018$ $^{+ 0.006}_{- 0.006}$ 
   & 
$ 0.902$ $\pm\ 0.033$ $^{+ 0.022}_{- 0.034}$ 
   \\ 
$  0.85$  & 
$ 0.861$ $\pm\ 0.036$ $^{+ 0.052}_{- 0.020}$ 
   & 
$ 0.292$ $\pm\ 0.035$ $^{+ 0.032}_{- 0.212}$ 
   & 
$-0.058$ $\pm\ 0.020$ $^{+ 0.075}_{- 0.008}$ 
   & 
$-0.062$ $\pm\ 0.026$ $^{+ 0.010}_{- 0.006}$ 
   & 
$ 0.971$ $\pm\ 0.042$ $^{+ 0.019}_{- 0.057}$ 
   \\ 
$  0.95$  & 
$ 1.066$ $\pm\ 0.073$ $^{+ 0.040}_{- 0.038}$ 
   & 
$ 0.250$ $\pm\ 0.060$ $^{+ 0.053}_{- 0.049}$ 
   & 
$-0.018$ $\pm\ 0.035$ $^{+ 0.012}_{- 0.019}$ 
   & 
$ 0.009$ $\pm\ 0.048$ $^{+ 0.023}_{- 0.027}$ 
   & 
$ 0.898$ $\pm\ 0.065$ $^{+ 0.055}_{- 0.056}$ 
   \\ 
$  1.05$  & 
$ 0.981$ $\pm\ 0.106$ $^{+ 0.091}_{- 0.066}$ 
   & 
$ 0.275$ $\pm\ 0.090$ $^{+ 0.093}_{- 0.155}$ 
   & 
$-0.077$ $\pm\ 0.057$ $^{+ 0.073}_{- 0.043}$ 
   & 
$ 0.050$ $\pm\ 0.063$ $^{+ 0.043}_{- 0.051}$ 
   & 
$ 0.722$ $\pm\ 0.091$ $^{+ 0.084}_{- 0.106}$ 
  \\ 
 \hline
\hline
 \multicolumn{6}{|c|}{$\langle\qsq\rangle = 8.6$ \gevsq } \\
 \hline
$ 0.65$ & 
$ 0.429$ $\pm\ 0.015$ $^{+ 0.032}_{- 0.069}$ 
   & 
$ 0.050$ $\pm\ 0.022$ $^{+ 0.038}_{- 0.038}$ 
   & 
$-0.006$ $\pm\ 0.013$ $^{+ 0.010}_{- 0.009}$ 
   & 
$-0.095$ $\pm\ 0.017$ $^{+ 0.028}_{- 0.011}$ 
   & 
$ 0.997$ $\pm\ 0.052$ $^{+ 0.006}_{- 0.055}$ 
   \\ 
$  0.75$  & 
$ 0.472$ $\pm\ 0.013$ $^{+ 0.019}_{- 0.009}$ 
   & 
$ 0.069$ $\pm\ 0.015$ $^{+ 0.036}_{- 0.039}$ 
   & 
$ 0.001$ $\pm\ 0.009$ $^{+ 0.007}_{- 0.004}$ 
   & 
$ 0.006$ $\pm\ 0.012$ $^{+ 0.004}_{- 0.009}$ 
   & 
$ 0.923$ $\pm\ 0.034$ $^{+ 0.048}_{- 0.058}$ 
   \\ 
$  0.85$  & 
$ 0.558$ $\pm\ 0.019$ $^{+ 0.020}_{- 0.021}$ 
   & 
$ 0.145$ $\pm\ 0.023$ $^{+ 0.038}_{- 0.038}$ 
   & 
$-0.017$ $\pm\ 0.013$ $^{+ 0.006}_{- 0.009}$ 
   & 
$ 0.028$ $\pm\ 0.017$ $^{+ 0.007}_{- 0.014}$ 
   & 
$ 0.959$ $\pm\ 0.041$ $^{+ 0.061}_{- 0.069}$ 
   \\ 
$  0.95$  & 
$ 0.746$ $\pm\ 0.041$ $^{+ 0.071}_{- 0.022}$ 
   & 
$ 0.110$ $\pm\ 0.041$ $^{+ 0.037}_{- 0.042}$ 
   & 
$-0.066$ $\pm\ 0.027$ $^{+ 0.014}_{- 0.014}$ 
   & 
$-0.101$ $\pm\ 0.033$ $^{+ 0.011}_{- 0.007}$ 
   & 
$ 0.690$ $\pm\ 0.049$ $^{+ 0.077}_{- 0.064}$ 
   \\ 
$  1.05$  & 
$ 0.717$ $\pm\ 0.052$ $^{+ 0.062}_{- 0.043}$ 
   & 
$-0.062$ $\pm\ 0.045$ $^{+ 0.037}_{- 0.040}$ 
   & 
$ 0.261$ $\pm\ 0.038$ $^{+ 0.022}_{- 0.008}$ 
   & 
$-0.183$ $\pm\ 0.041$ $^{+ 0.019}_{- 0.029}$ 
   & 
$ 0.761$ $\pm\ 0.059$ $^{+ 0.072}_{- 0.065}$ 
  \\ 
 \hline
 \end{tabular}
 \caption{
  Ratios of the helicity amplitudes                                            
  (taken to be purely imaginary)                                               
  and phase difference between the $T_{11}$ and $T_{00}$ amplitudes            
  (the amplitude ratios \rdelta\ and \reta\ and the phase                      
  difference $\phi_{01} - \phi_{00}$ are taken to be 0),                       
  computed from global                                                         
  fits to the measurements of  the 15 spin density matrix elements,            
  as a function of \mpp\                                                       
  separately for $2.5 \leq  Q^2 < 5~\gevsq$ and $5 \leq  Q^2 \leq 60~\gevsq$   
  for \rh\ meson production and for $2.5 \leq  Q^2 \leq 60~\gevsq$             
  for \ph\ production                                                          
  (NPE is assumed).                                                            
 }
\label{table:ampl-ratios-f-m}
 \end{center}
 \end{sidewaystable}

\renewcommand{\arraystretch}{1.0}

\clearpage

\end{document}